\documentclass[12pt]{article}
\usepackage{jheppub}

\pdfoutput=1

\usepackage{amsmath,bbm,array,amsfonts,graphicx,wrapfig,lscape,float,mathtools,multirow,longtable}
\usepackage[dvipsnames]{xcolor}
\usepackage{array}
\usepackage{adjustbox}

\newcommand{\be}{\begin{equation}}
\newcommand{\ee}{\end{equation}}
\newcommand{\beq}{\begin{equation}}
\newcommand{\beql}[1]{\begin{equation}\label{#1}}
\newcommand{\eeq}{\end{equation}}
\newcommand{\ba}{\begin{array}}
\newcommand{\ea}{\end{array}}
\newcommand{\bea}{\begin{eqnarray}}
\newcommand{\beal}[1]{\begin{eqnarray}\label{#1}}
\newcommand{\eea}{\end{eqnarray}}
\newcommand{\ben}{\begin{enumerate}}
\newcommand{\een}{\end{enumerate}}
\newcommand{\bean}{\begin{eqnarray*}}
\newcommand{\eean}{\end{eqnarray*}}
\newcommand{\eref}[1]{(\ref{#1})}

\newcommand{\tref}[1]{Table~\ref{#1}}
\newcommand{\nn}{\nonumber}

\newcommand{\fref}[1]{Figure \ref{#1}}
\newcommand{\btab}[1]{\begin{tabular}{#1}}
\newcommand{\etab}{\end{tabular}}

\newcommand{\comment}[1]{}

\newcommand{\qed}{\nobreak \ifvmode \relax \else
      \ifdim\lastskip<1.5em \hskip-\lastskip
      \hskip1.5em plus0em minus0.5em \fi \nobreak
      \vrule height0.75em width0.5em depth0.25em\fi}

\definecolor{darkspringgreen}{rgb}{0.09, 0.45, 0.27}
\definecolor{forestgreen}{rgb}{0.13, 0.55, 0.13}

\usepackage{array}
\usepackage{physics}
\usepackage{subcaption}
\usepackage{tikz,tikz-3dplot}
\usetikzlibrary{decorations.markings, arrows.meta, shapes, calc, shapes.geometric}
\usepackage[colorlinks=true]{hyperref}

\newcolumntype{C}[1]{>{\centering\let\newline\\\arraybackslash\hspace{0pt}}m{#1}}

\title{Tilting Mutations and Quiver-Invariant Dualities in Brane Tilings} 
\author[a]{Seong-Jin Lee,}
\author[b,c]{Rak-Kyeong Seong,}
\author[b]{Benjamin Suzzoni}

\affiliation[a]{
Center for Geometry and Physics, Institute for Basic Science (IBS), Pohang 37673, South Korea
}

\affiliation[b]{
Department of Mathematical Sciences, and ${}^{c}$Department of Physics,
Ulsan National Institute of Science and Technology, 50 UNIST-gil, Ulsan 44919, South Korea
}

\emailAdd{seongjinlee@ibs.re.kr}
\emailAdd{seong@unist.ac.kr}
\emailAdd{b.suzzoni@benterre.com}

\preprint{
\begin{flushright}
UNIST-MTH-26-RS-05 \\
CGP26018
\end{flushright}
}

\abstract{
Quiver-invariant dualities relate distinct $4d$ $\mathcal{N}=1$ 
supersymmetric gauge theories 
that arise as worldvolume theories on a D3-brane probing the same 
toric Calabi-Yau 3-fold.
These dual theories share an identical quiver 
and differ only in their superpotentials.
We show that quiver-invariant dualities are realized by tilting mutations 
on the brane tilings that realize these theories.
Among the 42 toric phases of the $H_{1,1,2,1}$ model, 
we identify three general families of tilting mutations, 
all characterized by the reversal of the orientations of zig-zag paths 
within the mutation region of the brane tiling.
We present new examples of quiver-invariant dualities, 
given by five doublets and one triplet of toric phases with identical quivers, 
for which we verify that the brane tilings related by the tilting mutations 
correspond to the same toric Calabi-Yau 3-fold $H_{1,1,2,1}$.
}

\begin{document}

\maketitle

%%%%%%%%%%%%%%%%%%%%%%%%%%%%%%%%%%%%%%%%%%%%%%%%%%%%%%%%%%%%%
%%%%%%%%%%%%%%%%%%%%%%%%%%%%%%%%%%%%%%%%%%%%%%%%%%%%%%%%%%%%%

%=================================================================
\section{Introduction}
%=================================================================

String theory provides a powerful framework for studying the interplay between Calabi-Yau geometry 
and supersymmetric gauge theories. 
A prominent example is the large family of $4d$ $\mathcal{N}=1$ supersymmetric quiver gauge theories arising 
as worldvolume theories
on D3-branes probing toric Calabi-Yau 3-folds \cite{Douglas:1997de, Douglas:1996sw, Feng:2000mi, Feng:2001xr}.
These $4d$ $\mathcal{N}=1$ theories are fully captured by \textit{brane tilings} \cite{Franco:2005rj, Hanany:2005ve, Franco:2005sm}, also known as \textit{dimer models} \cite{2003math.....10326K,kasteleyn1967graph}, 
which encode Type IIB brane configurations that are T-dual to the original D3-brane setup. 
In this dual description, D5-branes are suspended within an NS5-brane wrapping a holomorphic curve $\Sigma$ 
defined by the Newton polynomial of the probed toric Calabi-Yau 3-fold.

Like many $4d$ $\mathcal{N}=1$ gauge theories, this family of supersymmetric gauge theories exhibits 
a non-perturbative equivalence known as \textit{Seiberg duality} \cite{Seiberg:1994pq}, 
also known as \textit{toric duality} \cite{Feng:2000mi, Feng:2001bn, Feng:2001xr, Feng:2002zw} in the context of brane tilings. 
Seiberg duality relates distinct $4d$ $\mathcal{N}=1$ gauge theories that flow to the same infrared fixed point.
A key feature of this duality is that it preserves the vacuum moduli space, namely the space of solutions to the F-term and D-term constraints modulo gauge equivalence. For brane tilings, this space is referred to as the mesonic moduli space \cite{Butti:2007jv, Forcella:2008bb, Forcella:2008eh}. 
For abelian theories with $U(1)$ gauge groups, the mesonic moduli space is exactly the probed toric Calabi-Yau 3-fold. 
Thus, a single toric Calabi-Yau 3-fold can serve as the mesonic moduli space for several distinct $4d$ $\mathcal{N}=1$ gauge theories. 
These theories are different \emph{toric phases} of the same underlying toric Calabi-Yau geometry \cite{fulton1993introduction, cox2011graduate}
related by Seiberg duality.

%---------------------------------------------------- 
\begin{figure}[htbp]
\centering
\includegraphics[width=0.6\textwidth]{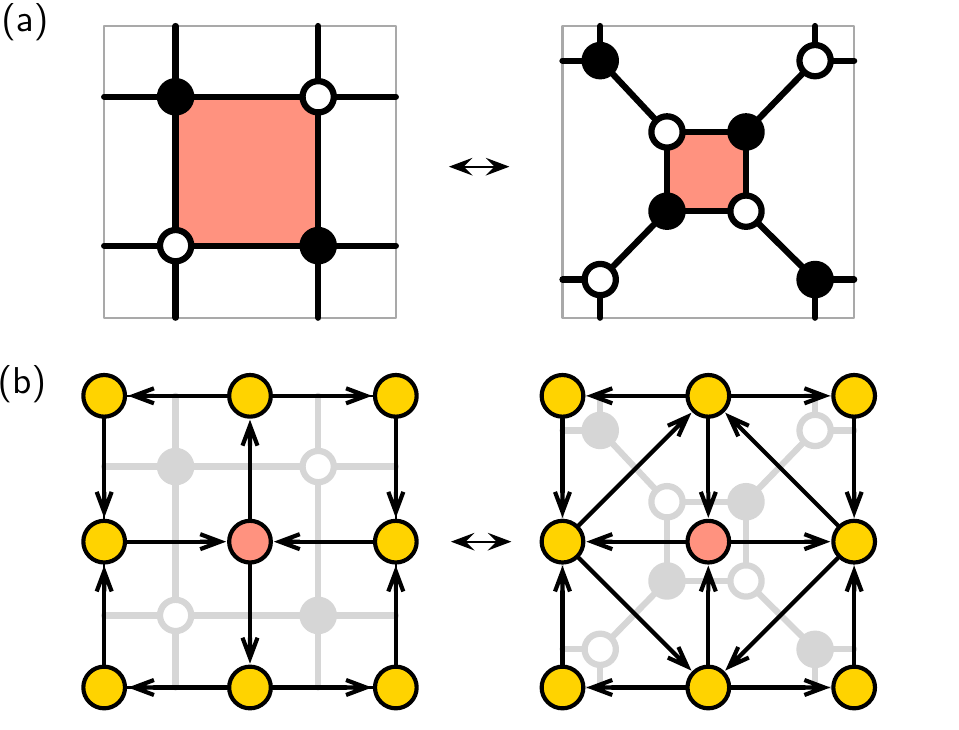}
\caption{Local mutation of (a) a brane tiling and (b) its corresponding periodic quiver, known as a spider move or urban renewal \cite{2011arXiv1107.5588G, CIUCU199834, 1999math......3025K}, that realizes Seiberg duality of the corresponding $4d$ $\mathcal{N}=1$ supersymmetric gauge theory.}
\label{fig_seiberg}
\end{figure}
%---------------------------------------------------- 

In terms of brane tilings, 
Seiberg duality is realized as a local mutation of the bipartite graph, 
originally referred to in the literature as \textit{urban renewal} or a \textit{spider move} \cite{2011arXiv1107.5588G, CIUCU199834, 1999math......3025K}. 
This mutation acts on quadrilateral faces of the brane tiling, 
which correspond to $U(N)$ gauge groups with $2N$ flavors in the quiver associated with the brane tiling, 
as illustrated in \fref{fig_seiberg}.
In general, the mutation changes the connectivity of the associated quiver. 
When all gauge groups are $U(N)$, the case we are considering in this work,
and Seiberg duality is applied to a quadrilateral face of the brane tiling, 
the rank of the dualized gauge group remains unchanged. 
Nevertheless, the mutation introduces new bifundamental chiral fields and conjugates the existing bifundamental fields attached to 
the dualized gauge node in the quiver, as illustrated in \fref{fig_seiberg}.

%---------------------------------------------------- 
\begin{figure}[htbp]
\centering
\includegraphics[width=0.67\textwidth]{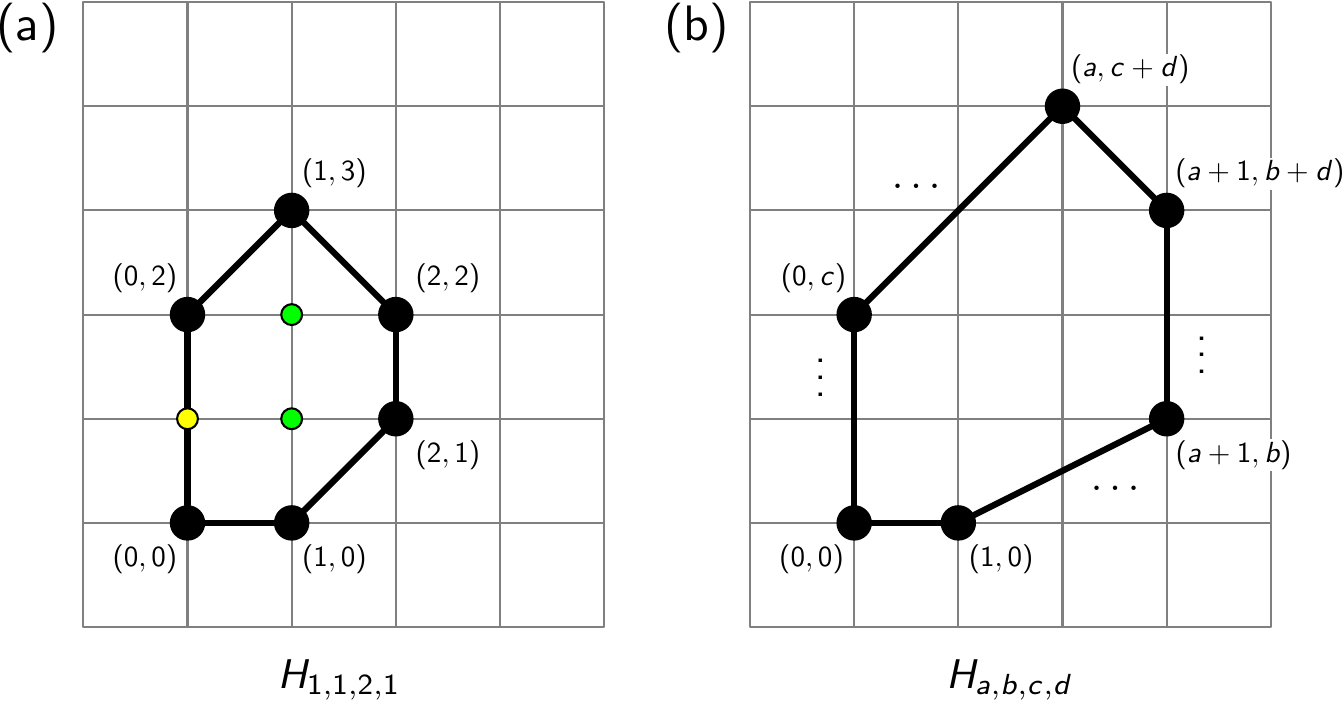}
\caption{(a) The toric diagram for the $H_{1,1,2,1}$ model, and (b) the general shape of the toric diagram for the family of $H_{a,b,c,d}$ models.}
\label{fig_model}
\end{figure}
%---------------------------------------------------- 

In many brane tilings, several gauge groups can correspond 
to quadrilateral faces with $2N$ flavors, allowing Seiberg duality to be applied iteratively. 
This process generates intricate duality trees, which have been extensively studied in the literature \cite{Franco:2025uap, Hanany:2012hi}. 
In this paper, we investigate the structure of the duality tree for the brane tiling 
associated with the toric Calabi-Yau 3-fold whose toric diagram is shown in \fref{fig_model}. 
This hexagonal toric diagram is part of a new family of toric Calabi-Yau 3-folds, which we name here the $H_{a,b,c,d}$ model with extremal vertices whose coordinates in $\mathbb{Z}^2$ are given by, 
\beal{es00a01}
(0,0)~,~
(1,0)~,~
(a+1,b)~,~
(a+1,b+d)~,~
(a,c+d)~,~
(0,c)~.~
\eea
Here, for the six vertices to be genuine extremal vertices of a strictly convex toric diagram, 
we have $a,b,c,d \in \mathbb{Z}_{>0}$ with $d+a(c-b) >0$ as illustrated in \fref{fig_model}.
In the following work, we study the $H_{1,1,2,1}$ model
whose toric diagram has extremal vertices given by,
\beal{es00a02}
(0,0)~,~
(1,0)~,~
(2,1)~,~
(2,2)~,~
(1,3)~,~
(0,2)~,~
\eea
and is shown in \fref{fig_model}.
We find that this model admits a total of 42 toric phases, 
all related by Seiberg duality. 
We provide a complete classification of all 42 brane tilings associated with $H_{1,1,2,1}$. 
We refer to this toric Calabi-Yau 3-fold, together with its 42 corresponding brane tilings, 
as the $H_{1,1,2,1}$ model. 

In \cite{Kho:2026geo}, it was shown that two brane tilings associated with $H_{1,1,2,1}$ 
have the same quiver, up to a relabelling of gauge nodes, 
but distinct superpotentials containing terms of different order.\footnote{Here, we note that similar examples have been studied
in relation to orientifold projections on brane tilings in \cite{Amariti:2022dui}.} 
This observation was surprising because the corresponding $4d$ $\mathcal{N}=1$ gauge theories 
are related by Seiberg duality, meaning that they share the same mesonic moduli space, given by $H_{1,1,2,1}$. 
As discussed above, Seiberg duality typically changes the structure of a quiver 
by reversing the directions of arrows connected to the dualized node 
and by introducing new arrows, as illustrated in \fref{fig_seiberg}. 
The example of \cite{Kho:2026geo} showed that, for sufficiently large toric Calabi-Yau 3-folds, 
the duality tree can contain sequences of Seiberg dualities whose endpoints have identical quivers up to relabelling of gauge nodes,
but different superpotentials.
We refer to such sequences as \textit{quiver-invariant duality chains}. 
Correspondingly, we say that two brane tilings associated with the same toric Calabi-Yau 3-fold 
are related by a \textit{quiver-invariant duality} if they have the same quiver, up to relabelling of gauge nodes, 
but distinct superpotentials.

In this work, we show that the 42 toric phases of the $H_{1,1,2,1}$ model exhibit several 
additional instances of quiver-invariant dual brane tilings. 
Among these toric phases, we identify five doublets and one triplet of brane tilings that share the same quiver but have distinct superpotentials. 
We demonstrate that these brane tilings are not only connected by quiver-invariant duality chains, 
but in fact can be related directly by a local mutation of the brane tiling
that keeps the quiver invariant without any relabelling of gauge nodes. 
In \cite{Kho:2026geo}, we introduced one such mutation under the name \textit{tilting mutation}. 
Here, we show that this mutation is part of a broader set of mutation families that realize quiver-invariant dualities.
\\

The paper is organized as follows.
Section~\ref{sec:background} provides a brief review of brane tilings,
Seiberg duality, and quiver-invariant dualities between $4d$ $\mathcal{N}=1$ supersymmetric gauge theories.
In Section~\ref{sec:classification}, we present the toric phases of the $H_{1,1,2,1}$ model that exhibit quiver-invariant dualities,
which form five doublets and one triplet of toric phases that share the same quiver.
In Section~\ref{sec:quiver-invariant-dualities}, we introduce three general families of tilting mutations
that realize the quiver-invariant dualities between these toric phases.
In Section~\ref{sec:examples}, we study each doublet and the triplet of toric phases in detail,
listing their quivers, brane tilings and superpotentials,
and identifying the tilting mutations that relate them,
together with the corresponding global symmetry charges and mesonic moduli spaces.
We conclude in Section~\ref{sec:conclusion} with a summary of our results and a discussion of future directions.
Appendix~\ref{appendix:models} contains the complete list of all 42 toric phases associated with
the $H_{1,1,2,1}$ model, including their quivers, superpotentials, and toric diagrams, with vertices labelled by the number of associated GLSM fields.
\\

%=================================================================
\section{Background \label{sec:background}}
%=================================================================

In this section, we briefly review 
brane tilings, Seiberg duality and quiver-invariant dualities as described in \cite{Kho:2026geo}.

%=================================================================
\subsection{Brane Tilings and Toric Calabi-Yau 3-Folds}
%=================================================================

\paragraph{Brane Tilings.}
Brane tilings \cite{Franco:2005rj, Hanany:2005ve, Franco:2005sm}, also known as dimer models \cite{2003math.....10326K,kasteleyn1967graph}, 
are bipartite periodic graphs on a 2-torus $T^2$
that encode the Lagrangian of a corresponding 
$4d$ $\mathcal{N}=1$ supersymmetric gauge theory.
Every edge in the brane tiling connects a white node with a black node of the bipartite graph, 
where edges are oriented clockwise around white nodes
and anti-clockwise around black nodes.

\begin{figure}[htbp]
\centering
\includegraphics[width=0.9\textwidth]{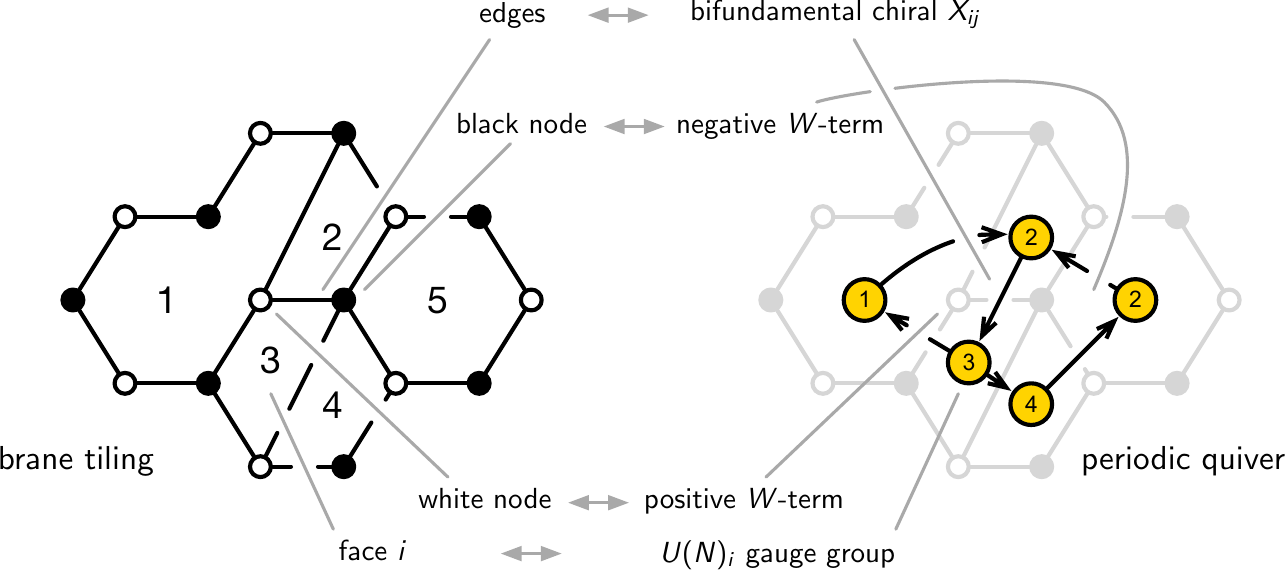}
\caption{The dictionary between a brane tiling on the 2-torus $T^2$ 
and the corresponding $4d$ $\mathcal{N}=1$ supersymmetric gauge theory, 
which is represented by the periodic quiver 
given by the graph dual of the brane tiling.
}
\label{fig_dict}
\end{figure}

The following dictionary allows us to translate between the brane tiling and the corresponding $4d$ $\mathcal{N}=1$
supersymmetric gauge theory:
\begin{itemize}
\item \textit{White and Black Nodes.}
The white and black nodes are oriented in a clockwise and anti-clockwise manner, respectively.
A white node corresponds to a positive term in the superpotential $W$ of the $4d$ $\mathcal{N}=1$ theory,
whereas a black node corresponds to a negative term.

\item \textit{Edges.}
Edges always connect a white node to a black node and are thus given an orientation that is inherited from that of the nodes.
They correspond to bifundamental chiral fields $X_{ij}$ in the corresponding 
$4d$ $\mathcal{N}=1$ supersymmetric gauge theory.
The orientation on the edge determines which gauge groups $U(N)_i$ the chiral field $X_{ij}$ transforms under, 
for both the fundamental and anti-fundamental representations.
The collection of edges connected to a certain node in the brane tiling
represents the gauge-invariant product of bifundamental chiral fields that forms the corresponding superpotential term.

\item \textit{Faces.}
Due to the bipartite nature of brane tilings, faces bounded by edges are always even-sided.
They represent $U(N)_i$ gauge groups in the corresponding $4d$ $\mathcal{N}=1$ supersymmetric gauge theory.
An edge associated with a bifundamental chiral multiplet $X_{ij}$
always neighbors two faces corresponding to two gauge groups $U(N)_i$ and $U(N)_j$
in which $X_{ij}$ transforms in the bifundamental representation. 
The orientation on the edge determines under which gauge groups the chiral field $X_{ij}$ transforms in the fundamental and anti-fundamental representations. 
\end{itemize}
\fref{fig_dict} illustrates the dictionary between a brane tiling on the 2-torus $T^2$ 
and the corresponding $4d$ $\mathcal{N}=1$ supersymmetric gauge theory.
We note that
due to the bipartite structure of the brane tiling, 
every edge is incident to precisely one white and one black node, 
such that every bifundamental chiral field appears exactly once 
in a positive and once in a negative term of the superpotential. 
This property is referred to as the \textit{toric condition} \cite{Feng:2000mi, Feng:2001xr, Feng:2002zw} of a brane tiling 
and implies that the $F$-term relations $\partial W / \partial X_{ij} = 0$ 
are all binomial relations among the chiral fields.

%---------------------------------------------------- 
%---------------------------------------------------- 

\paragraph{Brane Configuration.}
A brane tiling not only encodes the Lagrangian of the corresponding 
$4d$ $\mathcal{N}=1$ supersymmetric gauge theory, but also the Type IIB string configuration 
that realizes the gauge theory.
The brane configuration consists of 
D5-branes that are suspended from an NS5-brane, where the NS5-brane 
extends along the $(0123)$ directions and wraps
a holomorphic curve $\Sigma$ originating from the corresponding toric Calabi-Yau 3-fold.
$\Sigma$ is embedded into the $(4567)$ directions,
where $(45)$ and $(67)$ pairwise combine into complex variables $x,y \in \mathbb{C}^*$.
These parameterize the Newton polynomial $P(x,y)$
whose zero locus defines $\Sigma$
and is associated with the toric diagram $\Delta$ of the toric Calabi-Yau 3-fold.
The Newton polynomial for a given toric diagram $\Delta$ is defined as follows, 
\beal{es02a01}
P(x,y) = \sum_{(n_x,n_y) \in \Delta} c_{(n_x,n_y)} x^{n_x} y^{n_y} ~,~
\eea
where $c_{(n_x,n_y)} \in \mathbb{C}^*$. 
\tref{tab_configuration} summarizes the Type IIB brane configuration corresponding to a brane tiling.

%---------------------------------------------------- 
\begin{table}[htbp]
\centering
\begin{tabular}{|c|cccc|cccc|cc|}
\hline
\; & 0 & 1 & 2 & 3 & 4 & 5 & 6 & 7 & 8 & 9
\\
\hline
D5 & $\times$ & $\times$ & $\times$ & $\times$ & $\times$ & $\cdot$ & $\times$ & $\cdot$ & $\cdot$ & $\cdot$ 
\\
NS5 & $\times$ & $\times$ & $\times$ & $\times$ & \multicolumn{4}{c|}{------ $\Sigma$ ------ } & $\cdot$ & $\cdot$ 
\\
\hline
\end{tabular}
\caption{The Type IIB brane configuration of a brane tiling realizing a $4d$ $\mathcal{N}=1$ supersymmetric gauge theory.}
\label{tab_configuration}
\end{table}
%---------------------------------------------------- 

\paragraph{Mesonic Moduli Space.} 
The vacuum moduli space of a $4d$ $\mathcal{N}=1$ supersymmetric gauge theory described by a brane tiling is the space of solutions to the $F$- and $D$-term equations modulo gauge equivalence. 
Also referred to as the \textit{mesonic moduli space}, it can be described as the space of gauge-invariant operators subject to the $F$-term relations of the $4d$ $\mathcal{N}=1$ theory. 
When all gauge groups are abelian, namely $U(1)$, the mesonic moduli space is precisely the probed toric Calabi-Yau 3-fold 
associated with the brane tiling. It can be written as the quotient
\beal{es02a10}
\mathcal{M}^{mes} = \operatorname{Spec}
\left(\mathbb{C}[X_{ij}]/\mathcal{I}_{\mathrm{irr}}\right)//U(1)^{G-1}
~,~
\eea
where the coordinate ring $\mathbb{C}[X_{ij}]$ is generated by the chiral fields $X_{ij}$, 
and $\mathcal{I}_{\mathrm{irr}}$ denotes the irreducible component of the binomial ideal generated by the $F$-term relations. 
The quotient imposes invariance under the independent $U(1)^{G-1}$ gauge symmetries, 
where $G$ is the total number of $U(1)$ gauge groups and an overall $U(1)$ decouples. 
For a non-abelian $4d$ $\mathcal{N}=1$ theory with $U(N)$ gauge groups, 
the mesonic moduli space is given by the $N$-th symmetric product of the abelian mesonic moduli space,
\beal{es02a11}
\mathcal{M}^{mes}_N = \operatorname{Sym}^{N} \left(\mathcal{M}^{mes} \right)~.~
\eea

The mesonic moduli space $\mathcal{M}^{mes}$ of an abelian brane tiling, 
corresponding to a $4d$ $\mathcal{N}=1$ supersymmetric gauge theory with $U(1)$ gauge groups, 
can be described in terms of \textit{GLSM fields} \cite{Witten:1993yc}. 
These GLSM fields map to lattice points in the toric diagram of the associated toric Calabi-Yau 3-fold 
and are in one-to-one correspondence with \textit{perfect matchings} of the brane tiling \cite{2003math.....10326K, Kenyon:2003uj, Hanany:2006nm}. 
A perfect matching is a subset of edges in the bipartite graph such that each white and black node is incident to exactly one edge in the subset. 
Since edges of the brane tiling correspond to bifundamental chiral fields, 
each perfect matching specifies a collection of chiral fields associated with the corresponding GLSM field. 
We encode this relation in the $P$-matrix, which maps bifundamental chiral fields $X_k$ to GLSM fields $p_a$ as follows,
\beal{es02a12}
P_{ka} &= \begin{cases}
1 & \text{if } X_{k}\in p_a~,\\
0 & \text{if } X_{k}\notin p_a~,
\end{cases}
\eea
where $k=1,\dots,n_\chi$ runs over all chiral fields and $a=1,\dots,c$ runs over all GLSM fields of the brane tiling. 
Here, $n_\chi$ denotes the total number of chiral fields and $c$ denotes the number of GLSM fields. 
The $P$-matrix can be obtained directly from the $F$-term relations of the brane tiling using the \textit{forward algorithm} \cite{Feng:2000mi}.

After rewriting the chiral fields in terms of GLSM fields $p_a$, 
the $F$- and $D$-term constraints of the brane tiling are encoded by charge matrices acting on the GLSM fields. 
The $F$-term charges are summarized by the matrix $Q_F$, whose rows span the kernel of the $P$-matrix,
\beal{es02a15}
(Q_F)_{(c-G-2) \times c} = \ker P_{n_\chi \times c} ~.~
\eea
The $D$-term charges are encoded in the $Q_D$ matrix, 
which is obtained as follows, 
\beal{es02a16}
(\overline{d})_{(G-1) \times n_\chi} = (Q_D)_{(G-1) \times c} \cdot P^t_{c \times n_\chi} ~.~
\eea
Here, $\overline{d}$ denotes the reduced incidence matrix of the quiver, 
with the row associated with the decoupled overall $U(1)$ removed. 

In terms of the $F$- and $D$-term charge matrices, 
the mesonic moduli space in \eref{es02a10} can be written as the symplectic quotient,
\beal{es02a20}
\mathcal{M}^{mes} = \operatorname{Spec}
\left(\mathbb{C}[p_a]// Q_F \right)// Q_D ~,~
\eea
where $\mathbb{C}[p_a]$ is the coordinate ring generated by the GLSM fields $p_a$. 
Combining the $F$- and $D$-term charge matrices, we define the total charge matrix,
\beal{es02a21}
Q_t =
\begin{pmatrix}
Q_F \\
Q_D
\end{pmatrix}~.~
\eea
The toric diagram of the associated Calabi-Yau 3-fold is then obtained from the integer kernel of $Q_t$,
\beal{es02a22}
G_t = \ker(Q_t)~.
\eea
The columns of $G_t$ give the lattice coordinates associated with the GLSM fields, 
or equivalently the perfect matchings, of the brane tiling.
\\

\paragraph{Global Symmetries.} 
The mesonic moduli space $\mathcal{M}^{mes}$ carries the isometries of the associated toric Calabi-Yau 3-fold. 
For a toric Calabi-Yau 3-fold, 
the maximal torus of this isometry group has rank 3. 
In the corresponding $4d$ $\mathcal{N}=1$ gauge theory, this rank 3 symmetry is identified with the mesonic flavor symmetry $U(1)^2$, 
possibly enhanced to a non-abelian symmetry of the same rank, 
together with the $U(1)_R$ symmetry.

The superconformal $U(1)_R$ charges of the bifundamental chiral fields in the brane tiling can be determined 
by \textit{$a$-maximization} \cite{Intriligator:2003jj, Butti:2005vn, Butti:2005ps}. 
This procedure identifies the exact superconformal $U(1)_R$ symmetry by maximizing the trial central charge,
\beal{es02a30}
a = \frac{3}{32}\left(3\operatorname{Tr}(R^3)-\operatorname{Tr}(R)\right)~,~
\eea
over the space of trial $R$-symmetries of the form,
\beal{es02a31}
R_t = R_0 + \sum_m c_m \mathcal{F}_m~,~
\eea
where
$\mathcal{F}_m$ can be thought of as charges coming from the global symmetries, which are not the $U(1)_R$ symmetry.

We assign to each bifundamental chiral field $X_{ij}$ a trial $U(1)_R$ charge $R_{ij}$. 
These charges obey two sets of constraints. 
First, every term $W_k$ in the superpotential must have $U(1)_R$ charge $2$, giving
\beal{es02a35}
\sum_{X_{ij} \in W_k} R_{ij} = 2~.~
\eea
Second, the vanishing of the NSVZ beta function for each $U(N)$ gauge group imposes a constraint for every face $F_k$ of the brane tiling \cite{Butti:2005vn, Butti:2005ps, Franco:2025uap}. 
Since the boundary edges of $F_k$ correspond to bifundamental chiral fields transforming under the corresponding gauge group, 
their $U(1)_R$ charges satisfy,
\beal{es02a36}
\sum_{X_{ij} \in F_k} \left(1-R_{ij}\right) = 2~.~
\eea
For a brane tiling \cite{Butti:2005vn, Butti:2005ps, Franco:2025uap},
the trial $a$-function in \eref{es02a30} takes the following form, up to an overall factor of $N^2$,
\beal{es02a37}
a(R_{ij}) =
\frac{3}{32}
\left(
2G + 3\sum_{X_{ij}}
(R_{ij}-1)^3
\right)~,~
\eea
whose maximization subject to \eref{es02a35} and \eref{es02a36} determines the exact superconformal $U(1)_R$ charges.

Throughout, the brane tiling specifies the quiver and superpotential common
to both the abelian and non-abelian $4d$ $\mathcal{N}=1$ theories. 
We use the abelian theory, with gauge group $U(1)^G$, to compute the mesonic moduli space given by the toric Calabi-Yau 3-fold.
\\

%=================================================================
\subsection{Seiberg Duality}
%=================================================================

\textit{Seiberg duality} \cite{Seiberg:1994pq} relates distinct $4d$ $\mathcal{N}=1$ supersymmetric gauge theories 
that flow to the same IR fixed point under renormalization group flow. 
In the context of $4d$ $\mathcal{N}=1$ supersymmetric gauge theories realized by brane tilings and associated with toric Calabi-Yau 3-folds, 
this equivalence is often referred to as \textit{toric duality} \cite{Feng:2000mi, Feng:2001bn, Feng:2001xr, Feng:2002zw}. 
Dual brane tilings have the same mesonic moduli space and hence correspond to the same toric Calabi-Yau 3-fold. 
Consequently, a given toric Calabi-Yau 3-fold may admit multiple distinct brane tiling descriptions that are all related by Seiberg duality
and can form elaborate \textit{duality trees} as illustrated for the $H_{1,1,2,1}$ model in \fref{fig_duality_tree_all}.
We refer to brane tilings related by Seiberg duality and corresponding to the same toric Calabi-Yau 3-fold 
as \textit{toric phases}. 
\\

\paragraph{Seiberg Duality and Brane Tilings.}
Brane tilings describing $4d$ $\mathcal{N}=1$ supersymmetric gauge theories related by Seiberg duality are connected by a local mutation of the bipartite graph referred to as a \textit{spider move} or \textit{urban renewal} \cite{2011arXiv1107.5588G, CIUCU199834, 1999math......3025K}.
As illustrated in \fref{fig_seiberg}, this mutation acts on a $4$-sided face of the brane tiling corresponding to a $U(N_c)$ gauge group.
When all gauge groups have equal rank, the $4$ bifundamental fields incident on this face provide $N_f=2N_c$ flavors for the corresponding gauge node.
Consequently, the dual gauge group has rank,
\beal{es03a01}
N_f-N_c=N_c~,~
\eea
and therefore has the same rank as the original gauge group.

Let $k$ denote the gauge group on which Seiberg duality is performed.
The bifundamental chiral fields $X_{ak}$ and $X_{kb}$, transforming respectively in the fundamental and anti-fundamental representations of the gauge group labelled with $k$, are identified with the electric quarks of the original $4d$ $\mathcal{N}=1$ theory.
Under Seiberg duality, the corresponding arrows are reversed, giving rise to the dual quarks as follows,
\beal{es03a02}
X_{ak}\longrightarrow \widetilde{X}_{ka}~,~
X_{kb}\longrightarrow \widetilde{X}_{bk}~.~
\eea
In addition, each product of the form,
\beal{es03a03}
X_{ak}X_{kb}~,~
\eea
gives rise to a mesonic chiral field,
\beal{es03a04}
M_{ab}\sim X_{ak}X_{kb}
\eea
in the dual theory.

Under these transformations, the original superpotential $W$ is modified by replacing every occurrence of a product 
$X_{ak}X_{kb}$ with the corresponding meson $M_{ab}$.
In addition, the magnetic superpotential contains coupling terms of the form, 
\beal{es03a05}
W_{\text{mag}}
=
\sum_{a,b}
M_{ab}\widetilde{X}_{bk}\widetilde{X}_{ka}~.~
\eea
The superpotential of the dual theory can therefore be written as
\beal{es03a06}
W^\prime
=
W\Big|_{X_{ak}X_{kb}\,\rightarrow\,M_{ab}}
+
W_{\text{mag}}~.~
\eea
The resulting superpotential may contain massive quadratic terms of the form,
\beal{es03a07}
M_{ab}X_{ba}~,~
\eea
which can be integrated out by solving the corresponding F-term equations,
\beal{es03a08}
\frac{\partial W^\prime}{\partial M_{ab}}=0~,~
\qquad
\frac{\partial W^\prime}{\partial X_{ba}}=0~,~
\eea
and substituting their solutions back into $W^\prime$.

The complete transformation of the brane tiling 
and its corresponding $4d$ $\mathcal{N}=1$ supersymmetric gauge theory under Seiberg duality
is represented graphically by the spider move, or urban-renewal transformation, shown in \fref{fig_seiberg}. 
In the context of brane tilings, this realization of Seiberg duality is commonly referred to as \textit{toric duality} \cite{Feng:2000mi, Feng:2001bn, Feng:2001xr, Feng:2002zw}, 
and the brane tilings related by such duality transformations are referred to as distinct \textit{toric phases} corresponding to the same toric Calabi-Yau 3-fold.
\\

%=================================================================
\subsection{The Toric Phases of $H_{1,1,2,1}$ and Quiver-Invariant Dualities \label{sec:classification}}
%=================================================================

For certain toric Calabi-Yau 3-folds, 
the corresponding $4d$ $\mathcal{N}=1$ gauge theory admits a large number of distinct toric phases, 
each represented by a different brane tiling. 
These phases are connected by successive Seiberg-duality transformations and may form an intricate \textit{duality tree} \cite{Franco:2025uap, Hanany:2012hi}. 
A duality tree is a graph whose nodes correspond to distinct toric phases and whose edges represent elementary Seiberg-duality transformations between them. 
By convention, each edge is labelled by the gauge groups on which the corresponding Seiberg-duality transformation is performed.

\begin{figure}[htbp!!]
\centering
\includegraphics[width=0.25\textwidth]{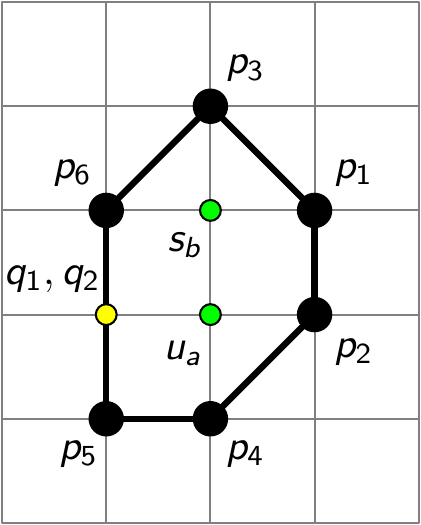}
\caption{The toric diagram for $H_{1,1,2,1}$ with the extremal vertices labelled by the corresponding GLSM fields $p_a$.}
\label{fig_toric_diagram}
\end{figure}

In this section, we examine the toric phases associated with the toric Calabi-Yau 3-fold whose toric diagram is shown in \fref{fig_toric_diagram}, 
which we refer to as the $H_{1,1,2,1}$ model. 
This model admits a total of 42 toric phases, each represented by a distinct brane tiling associated with the same underlying Calabi-Yau geometry. 
The complete list of all 42 toric phases, with their quivers and superpotentials, is provided in Appendix~\ref{appendix:models}.

%---------------------------------------------------- 
%---------------------------------------------------- 
An interesting observation for the $H_{1,1,2,1}$ model is that some distinct toric phases possess identical quivers, up to a relabelling of the gauge groups.
Consequently, distinct brane tilings, and hence distinct $4d$ $\mathcal{N}=1$ gauge theories, 
may be associated with the same toric Calabi-Yau 3-fold and have isomorphic quivers while differing in their superpotentials. 
The equivalence of their quivers implies that these theories have the same bifundamental chiral field content, 
although they form different superpotentials. 
Nevertheless, the resulting $4d$ $\mathcal{N}=1$ theories share the same mesonic moduli space, given by the common toric Calabi-Yau 3-fold. 
In previous work \cite{Kho:2026geo}, this relation between brane tilings was referred to as a \textit{quiver-invariant duality}.

\begin{figure}[htbp!!]
\centering
\includegraphics[width=0.8\textwidth]{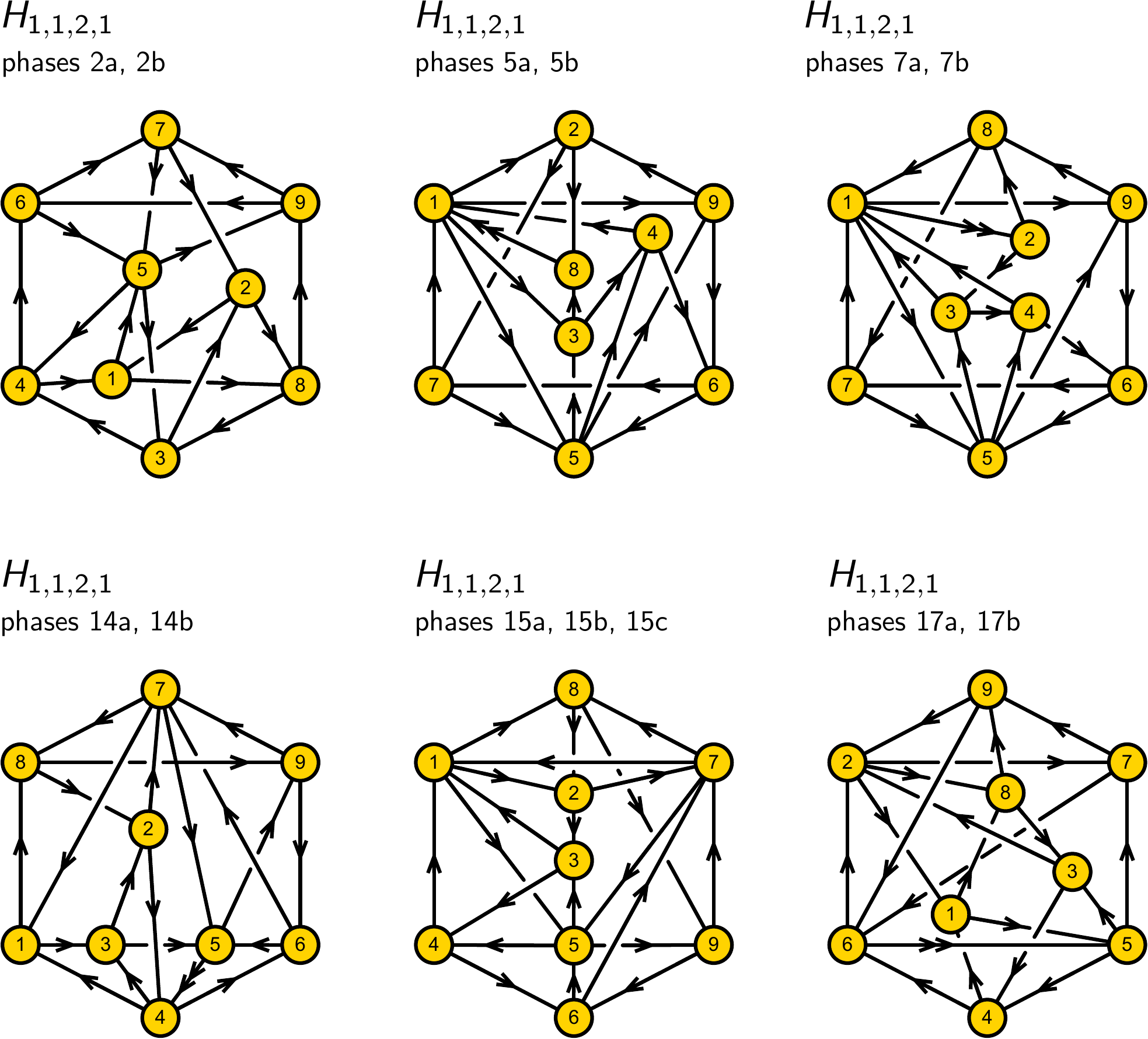}
\caption{Quiver diagrams shared between different phases of the $H_{1,1,2,1}$ model.}
\label{fig_quivers}
\end{figure}

Among the 42 toric phases of the $H_{1,1,2,1}$ model, 
we identify $5$ pairs of phases that share the same quiver up to a relabelling of the gauge groups.
We refer to each such pair as a \textit{doublet}. 
These include the pair originally presented in \cite{Kho:2026geo} as an example of quiver-invariant duality. 
In addition, we identify a \textit{triplet} consisting of $3$ distinct toric phases that share the same quiver.
The quiver diagrams shared by the toric phases of the doublets and of the triplet are illustrated in \fref{fig_quivers}. 
\tref{tab:quiver_degeneracy} summarizes the classification of the 42 toric phases of $H_{1,1,2,1}$
into singlets with unique quivers, and doublets and a triplet of toric phases that share the same quiver.

%---------------------------------------------------- 
\begin{table}[htbp!!]
\centering
\begin{tabular}{|r|cc|}
\hline
\; & quivers & toric phases \\
\hline\hline
singlets & 29 & 29 \\
doublets & 5 & 10 \\
triplet & 1 & 3 \\
\hline
total & 35 & 42 \\
\hline
\end{tabular}
\caption{The 42 toric phases for the $H_{1,1,2,1}$ model divided into singlets with unique quivers, and doublets and a triplet that share the same quivers.}
\label{tab:quiver_degeneracy}
\end{table}
%---------------------------------------------------- 

\begin{figure}[htbp!!]
\centering
\includegraphics[width=0.7\textwidth]{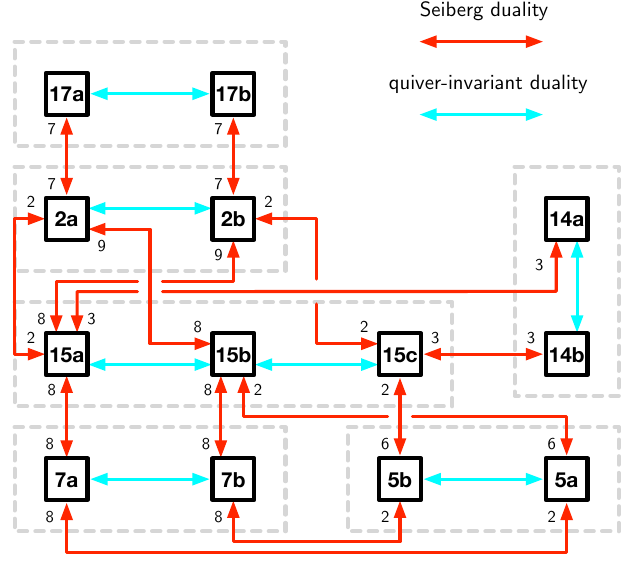}
\caption{
Part of the duality tree for $H_{1,1,2,1}$
containing the toric phases related by Seiberg duality (red) and quiver-invariant dualities (blue), 
including doublets and a triplet of toric phases that share the same quiver.
Each red arrow represents an elementary Seiberg duality, 
with the label indicating the gauge node of the toric phase at the tail of the arrow 
on which Seiberg duality is performed in order to obtain the toric phase at the head of the arrow.
}
\label{fig_duality_tree_all}
\end{figure}

Since the brane tilings and corresponding $4d$ $\mathcal{N}=1$ theories forming each doublet or triplet are toric phases associated with 
the same toric Calabi-Yau 3-fold, they are related by Seiberg duality. 
However, because their quivers are identical up to a relabelling of the gauge groups, 
the corresponding phases cannot be connected by a single Seiberg-duality transformation. 
Instead, they are related by a sequence of Seiberg dualities passing through intermediate toric phases, 
which we refer to as a \textit{quiver-invariant duality chain}. 
In \fref{fig_duality_tree_all}, we illustrate the subtree of the full duality tree of the $H_{1,1,2,1}$ model that contains all toric phases 
belonging to the identified doublets and triplet of quiver-invariant dual phases, 
together with the intermediate phases appearing along their quiver-invariant duality chains. 
Interestingly, every node in this subtree belongs to one of the identified doublets or to the triplet.
In particular, the intermediate toric phases appearing in a given quiver-invariant duality chain 
are themselves members of other doublets or of the triplet. 
Each red edge in \fref{fig_duality_tree_all} represents an elementary Seiberg duality transformation and its label indicates the gauge group on which the duality is performed.

%---------------------------------------------------- 
%---------------------------------------------------- 

As discussed in \cite{Kho:2026geo}, 
rather than viewing a quiver-invariant duality as a sequence of Seiberg-duality transformations 
forming a quiver-invariant duality chain, 
one may regard the corresponding brane tilings as being related directly by a single-step local mutation. 
For a doublet of brane tilings associated with the $H_{1,1,2,1}$ model 
and related by quiver-invariant duality, 
this local transformation was referred to in \cite{Kho:2026geo} as a \textit{tilting mutation}. 
In this work, the toric phases corresponding to this doublet presented in \cite{Kho:2026geo} are called phases 2a and 2b. 
In the following section, we summarize several families of local tilting mutations of brane tilings that realize quiver-invariant dualities
observed for the toric phases of the $H_{1,1,2,1}$ model.
\tref{tab:extremal_charges} summarizes the mesonic flavor charges and the $U(1)_R$ charges $R_a$
of the GLSM fields $p_a$ corresponding to the extremal points of the toric diagram of $H_{1,1,2,1}$,
which we use in the following sections to express the global symmetry charges of the chiral fields.
\\

%---------------------------------------------------- 
\begin{table}[htbp!!]
\centering
\begin{tabular}{|c|c|c|c|}
\hline
\; & $U(1)_{f_1}$ & $U(1)_{f_2}$ & $U(1)_R$ \\
\hline \hline
	$p_1$ &  1	&  0	& $R_1\simeq 0.273550 $ \\
	$p_2$ &  1 	&  0	& $R_2\simeq 0.320461$ \\
	$p_3$ & -1        &  1 	& $R_3\simeq 0.386171$ \\
	$p_4$ & -1        &  1 	& $R_4\simeq 0.253224$ \\
	$p_5$ &  0        &  -1	& $R_5\simeq 0.404904$ \\
	$p_6$ &  0        &  -1	& $R_6\simeq 0.361689$ \\
\hline
\end{tabular}
\caption{
The GLSM fields $p_a$ corresponding to extremal points of the toric diagram for $H_{1,1,2,1}$ 
with their mesonic flavor charges and $U(1)_R$ charges $R_a$, where $\sum_a R_a = 2$.
}
\label{tab:extremal_charges}
\end{table}
%---------------------------------------------------- 

%---------------------------------------------------- 
%---------------------------------------------------- 

%=================================================================
\section{Quiver-Invariant Dualities and Tilting Mutations}\label{sec:quiver-invariant-dualities}
%=================================================================

%---------------------------------------------------- 
\begin{figure}[htbp]
\centering
\includegraphics[width=0.85\textwidth]{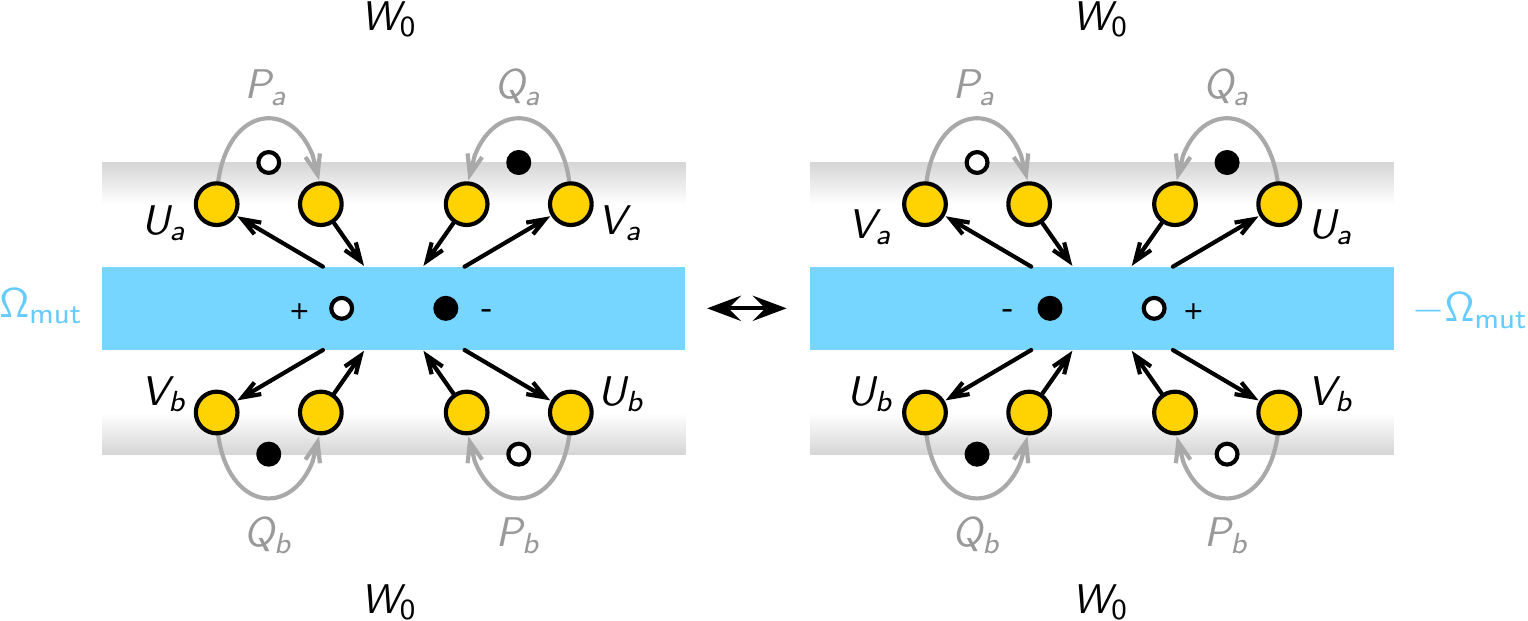}
\caption{
Schematic illustration of a general tilting mutation on a brane tiling and its periodic quiver,
exchanging the paths $U_a$ and $V_a$
and reversing the signs of all terms in $\Omega_{\text{mut}}$.
$W_0$ denotes the superpotential terms associated with the region of the brane tiling unaffected by the tilting mutation.
}
\label{fig_family_0}
\end{figure}
%---------------------------------------------------- 

\paragraph{Tilting Mutations and Brane Tilings.}
Quiver-invariant dualities between brane tilings can be realized as local
tilting mutations of their superpotentials. 
Let $U_a$ and $V_a$ denote two
oriented paths through the mutation region of the periodic quiver with the
same initial and terminal quiver nodes, 
and let $P_a$ and $Q_a$ denote the
complementary paths outside the mutation region that close them into
gauge-invariant cycles associated with white and black nodes of the brane
tiling, respectively, as illustrated in \fref{fig_family_0}. 
We write the superpotential in the following general form, 
\beal{es04a00a}
W
=
W_0
+\Omega_{\text{mut}}
+\sum_a
\left(
U_aP_a
-
V_aQ_a
\right)~,~
\eea
where $W_0$ collects all superpotential terms left invariant by the
tilting mutation, 
including all terms disjoint from the mutation region,
whereas $\Omega_{\text{mut}}$ denotes the signed sum of those
superpotential terms supported entirely within the mutation region whose
signs are reversed during the tilting mutation. 

The tilting mutation leaves the paths $P_a$ and $Q_a$
unchanged, while exchanging the paths $U_a$ and $V_a$ through the
mutation region and reversing the sign of every term in
$\Omega_{\mathrm{mut}}$.  This gives the mutated superpotential in the
following form,
\beal{es04a00b}
W^\prime
=
W_0
-\Omega_{\mathrm{mut}}
+\sum_a
\left(
V_aP_a
-
U_aQ_a
\right)~.~
\eea
Accordingly, each superpotential cycle passing through the mutation region
is rerouted along the path segments $U_a$ and $V_a$. The positive cycle
$U_aP_a$ and the negative cycle $-V_aQ_a$ appearing in $W$ are replaced
by $V_aP_a$ and $-U_aQ_a$, respectively, in $W^\prime$.  All terms
collected in $W_0$ remain unchanged.  Although the intermediate Seiberg
dualities may modify the quiver and generate massive fields, after these
fields are integrated out, the initial and final quivers are the same.
When $U_a=V_a$, the corresponding contribution $U_a(P_a-Q_a)$ is
unchanged by the tilting mutation.  

For the tilting mutations considered here, 
the mutated superpotential $W^\prime$ continues to satisfy the toric condition of a brane tiling. 
More explicitly, every monomial collected in $W_0$ is unchanged, whereas each affected contribution
of the form $U_a P_a - V_a Q_a$ is replaced by
$V_a P_a - U_a Q_a$
and every term in $\Omega_{\text{mut}}$ reverses its sign.
As a result, the paths $P_a$ and $Q_a$ remain fixed, 
while the path segments $U_a$ and $V_a$ are redistributed between the positive and negative superpotential terms. 
 A chiral field appearing in an unaffected term of $W_0$ retains that occurrence, 
 while its occurrence in an affected closed path, when present, is transferred to the corresponding rerouted closed path in $W^\prime$. 
 After integrating out all massive pairs generated during the tilting mutation, 
 every remaining chiral field therefore occurs in precisely two monomials of $W^\prime$, 
 once with a positive sign and once with a negative sign, satisfying the toric condition.

\begin{figure}[htbp]
\centering
\includegraphics[width=0.68\textwidth]{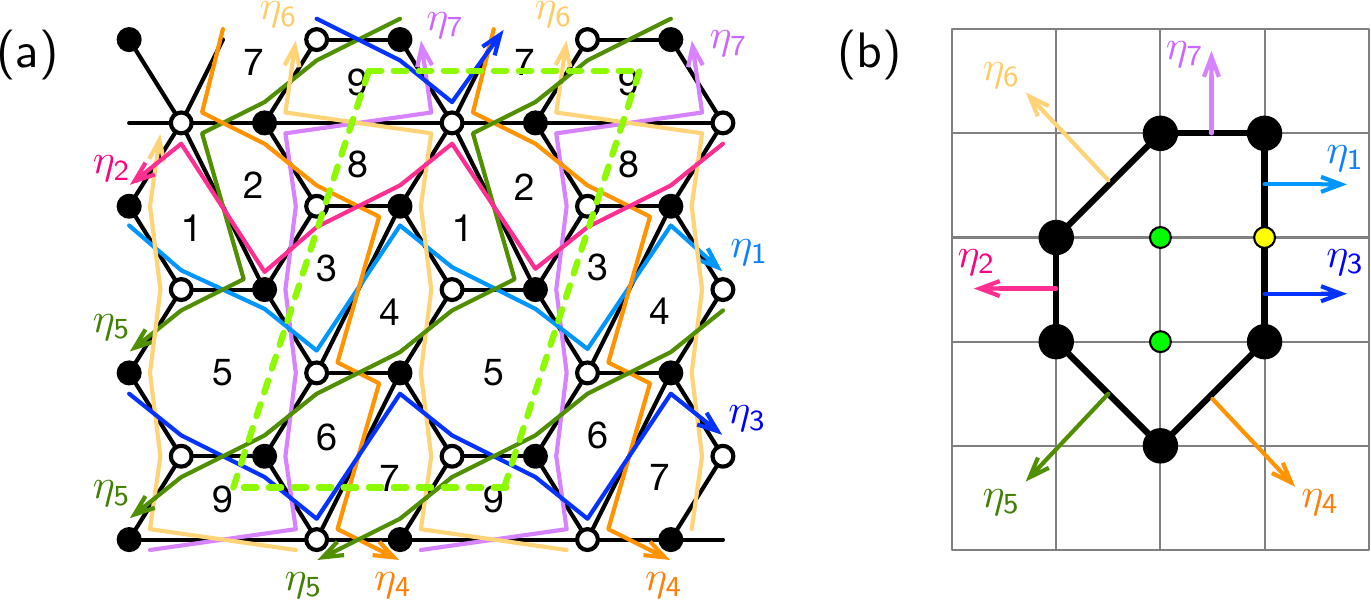}
\caption{
(a) The zig-zag paths in the brane tiling of toric phase 2a,
and (b) the corresponding outward normal vectors in the $(p,q)$-web diagram corresponding to the toric diagram for $H_{1,1,2,1}$.
}
\label{fig_tiling_pq_02_zz}

\end{figure}

\paragraph{Invariance of the Mesonic Moduli Space.}
In order for a tilting mutation to preserve the mesonic moduli space, and
hence the associated toric Calabi-Yau 3-fold, the set of
winding numbers of zig-zag paths \cite{Hanany:2005ss, 2003math.ph...5057K, Jejjala:2010vb, Hanany:2012vc} in the brane tiling has to be preserved under the tilting mutation
up to an overall change of basis under $GL(2,\mathbb{Z})$.  
Let the zig-zag paths of the brane tiling
associated with the superpotential $W$ be denoted by,
\beal{es04b01}
S_z(W)
=
\left\{
z_1,\ldots,z_{N_z}
\right\}
~,~
\eea
where $N_z$ is the total number of zig-zag paths.  
For a consistent brane
tiling, these zig-zag paths are in one-to-one correspondence
with the primitive boundary segments of the toric diagram
$\Delta$, or equivalently with the external legs of the corresponding
$(p,q)$-web diagram \cite{Aharony:1997bh, Aharony:1997ju, Leung:1997tw}.  
The oriented homology class of a zig-zag path $z_i$ on the
2-torus, which we refer to here as the winding number, is given by, 
\beal{es04b02}
w(z_i)
=
\left(
p_i,q_i
\right)
\in
\mathbb{Z}^2
~,~
\eea
which is a primitive outward normal vector to the corresponding boundary
edge of $\Delta$.  
\fref{fig_tiling_pq_02_zz} illustrates the zig-zag paths of phase 2a of the $H_{1,1,2,1}$ model
and the corresponding outward normal vectors in the $(p,q)$-web diagram corresponding to the toric diagram for $H_{1,1,2,1}$.
We note that the vectors $(p_i,q_i)$ are defined up to a common transformation
$G\in GL(2,\mathbb{Z})$ associated with a change of basis of the two
fundamental cycles of $T^2$. 

%---------------------------------------------------- 
%---------------------------------------------------- 

Let us here introduce the set of vectors $(p_i,q_i)$ corresponding to the winding numbers of zig-zag paths $z_i$ 
for a given brane tiling with superpotential $W$
as follows, 
\beal{es04b03}
S_w(W)
=
\left\{
w(z_i) ~|~ z_i \in S_z(W)
\right\}
~.~
\eea
For a tilting mutation relating $W$ and
$W^\prime$ to preserve the corresponding mesonic moduli space and hence the associated toric Calabi-Yau 3-fold, 
there must exist a single
transformation $G\in GL(2,\mathbb{Z})$ such that
\beal{es04b04}
S_w(W^\prime)
=
G\, S_w(W)
~.~
\eea
Thus, individual zig-zag paths may be relabelled or
permuted by the tilting mutation, 
but the complete collection of their winding
numbers and multiplicities must remain unchanged up to a common
transformation in $GL(2, \mathbb{Z})$.  
Provided that both the initial and mutated
brane tilings are consistent, this condition is equivalent to the
invariance of the toric diagram $\Delta$ up to $G\in GL(2,\mathbb{Z})$,
and consequently ensures that the tilting mutation preserves the toric Calabi-Yau 3-fold and hence the
mesonic moduli space of the mutated brane tiling.

We note that the condition in \eref{es04b04} imposes a nontrivial restriction 
on the choice of $W_0$ left invariant by the tilting mutation. 
While the tilting mutation acts locally on the mutation region, 
the zig-zag paths of the brane tiling are not confined to the mutation region. 
Zig-zag paths traversing the mutation region continue along edges 
associated with the exterior paths $P_a$ and $Q_a$ 
and with the chiral fields appearing in $W_0$, 
such that their winding numbers are determined 
by the choice of $W_0$ completing the brane tiling. 
Accordingly, not every choice of $W_0$ that closes the paths of the mutation region 
into a consistent brane tiling satisfies \eref{es04b04}. 
Only when the winding numbers of the zig-zag paths are preserved 
up to a common transformation $G \in GL(2,\mathbb{Z})$ 
does the tilting mutation realize a quiver-invariant duality 
that preserves the mesonic moduli space 
and hence the associated toric Calabi-Yau 3-fold. 
\\

In the remainder of this section, we present general families of tilting mutations
observed among the toric phases of the $H_{1,1,2,1}$ model.
\\

%=================================================================
\subsection{Tilting Mutation Family A \label{sec:mutfam_a}}
%=================================================================

%---------------------------------------------------- 
\begin{figure}[htbp]
\centering
\includegraphics[width=0.8\textwidth]{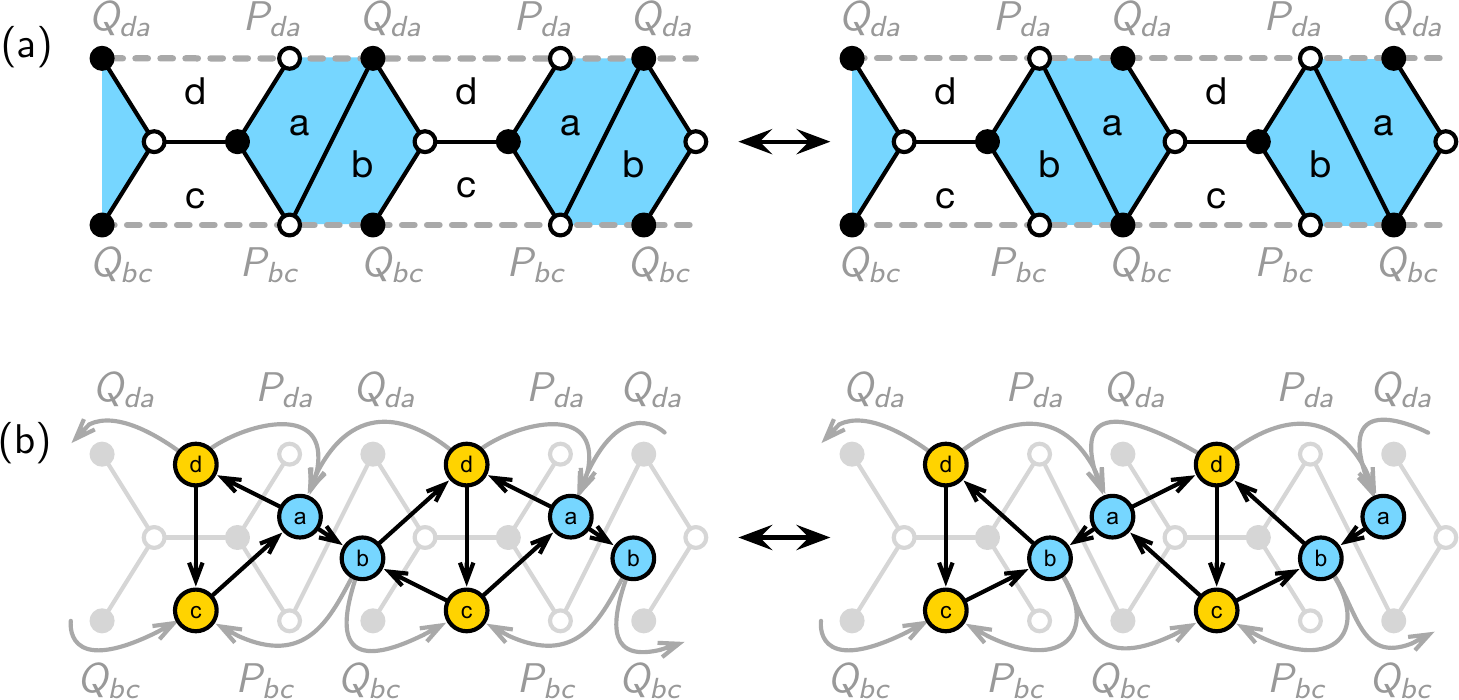}
\caption{
Tilting mutation family A on (a) a brane tiling and (b) its corresponding periodic quiver, resulting in a quiver-invariant duality.
}
\label{fig_family_1}

\end{figure}
%---------------------------------------------------- 

%---------------------------------------------------- 
\begin{figure}[htbp]
\centering
\includegraphics[width=0.8\textwidth]{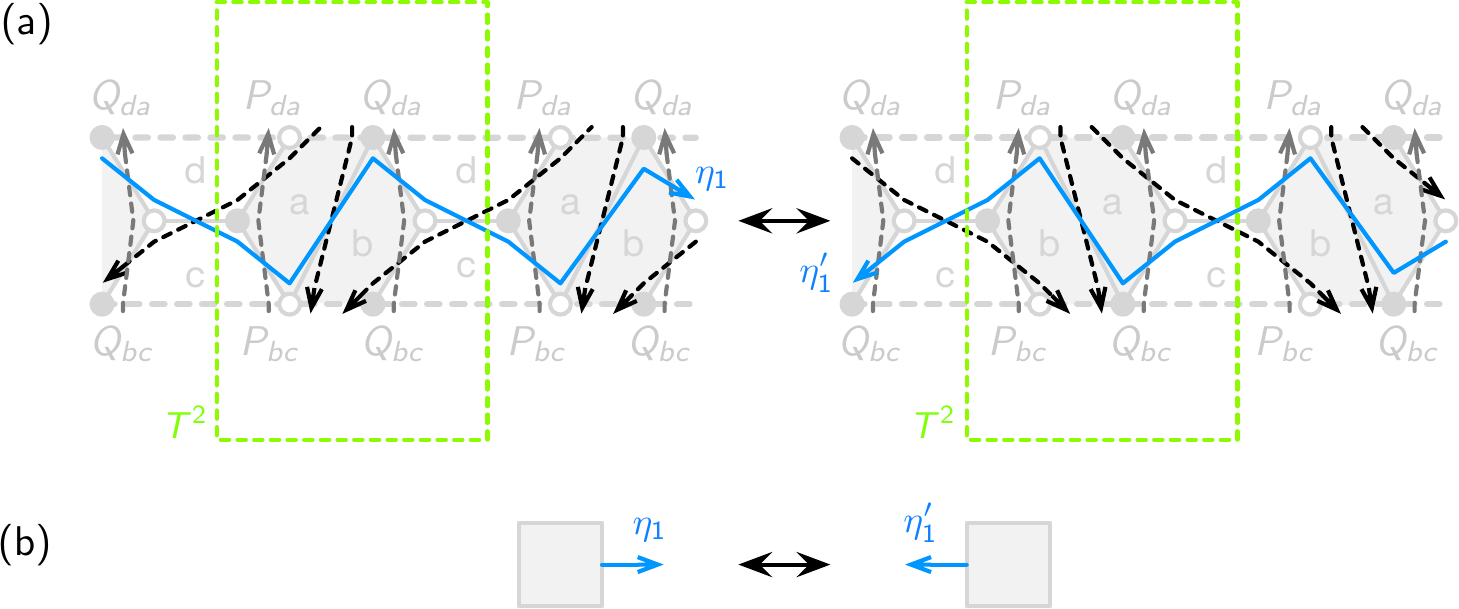}
\caption{
(a) The zig-zag path $\eta_1$ in the mutation region of tilting mutation family A.
(b) Under the tilting mutation, $\eta_1$ reverses its orientation, 
such that its winding number satisfies $w(\eta_1) = - w(\eta_1^\prime)$, 
with the corresponding external leg of the $(p,q)$-web diagram 
flipping its orientation accordingly.
}
\label{fig_family_1_zigzag}

\end{figure}
%---------------------------------------------------- 

The first family of tilting mutations that we identify among the toric phases 
of the $H_{1,1,2,1}$ model involves a mutation region containing 
four quiver nodes $a$, $b$, $c$ and $d$, 
as illustrated in \fref{fig_family_1}.
The superpotential takes the general form in \eref{es04a00a} with,
\beal{es04a01}
W = 
&&
X_{dc} X_{cb} X_{bd} 
- X_{dc} X_{ca} X_{ad}
\nn\\
&&
+ X_{ad} P_{da}
+ X_{ca} X_{ab} P_{bc}
- X_{ab} X_{bd} Q_{da}
- X_{cb} Q_{bc}
+ W_0
~,~
\eea
where $P_{da}$ and $Q_{da}$ denote the paths from node $d$ to node $a$, 
and $P_{bc}$ and $Q_{bc}$ those from node $b$ to node $c$ in the periodic quiver.
Under the tilting mutation, the superpotential in \eref{es04a01} 
is mapped to,
\beal{es04a02}
W^\prime = 
&& 
X_{dc} X_{ca} X_{ad}
- X_{dc} X_{cb} X_{bd}
\nn\\
&&
+ X_{ab} X_{bd} P_{da}
+ X_{cb} P_{bc}
- X_{ad} Q_{da}
- X_{ca} X_{ab} Q_{bc}
+ W_0
~,~
\eea
which is
in agreement with the general form in \eref{es04a00b}.
Here, the superpotential terms, which are 
entirely within the mutation region 
and whose signs are reversed under the tilting mutation, are given by,
\beal{es04a01b}
\Omega_{\text{mut}} = X_{dc} X_{cb} X_{bd} - X_{dc} X_{ca} X_{ad} ~.~
\eea
The paths in the periodic quiver that are exchanged by the tilting mutation take the following form,
\beal{es04a01c}
U_{da} = X_{ad} ~,~
U_{bc} =  X_{ca} X_{ab} ~,~
V_{da} = X_{ab} X_{bd} ~,~
V_{bc} = X_{cb} ~.~
\eea
We note that both $W$ and $W^\prime$ satisfy the toric condition, 
with every chiral field appearing in exactly two superpotential terms 
with opposite signs.

Based on \fref{fig_family_1_zigzag}, 
we also note that within the mutation region of the periodic quiver and brane tiling, 
we have one zig-zag path $\eta_1$ that flips its orientation under the tilting mutation such that,
\beal{es04a05}
w(\eta_1) = - w(\eta_1^\prime) ~.~
\eea
In order for the tilting mutation to preserve the mesonic moduli space and hence the associated toric Calabi-Yau 3-fold, 
the remaining zig-zag paths traversing the mutation region should have winding numbers such that 
there exists $G\in GL(2,\mathbb{Z})$ for which $S_w(W^\prime)=G\, S_w(W)$ as discussed in \eref{es04b04}.
This in turn restricts the choice of $W_0$ unaffected by the tilting mutation. 
\\

%=================================================================
\subsection{Tilting Mutation Family B \label{sec:mutfam_b}}
%=================================================================

%---------------------------------------------------- 
\begin{figure}[htbp]
\centering
\includegraphics[width=0.8\textwidth]{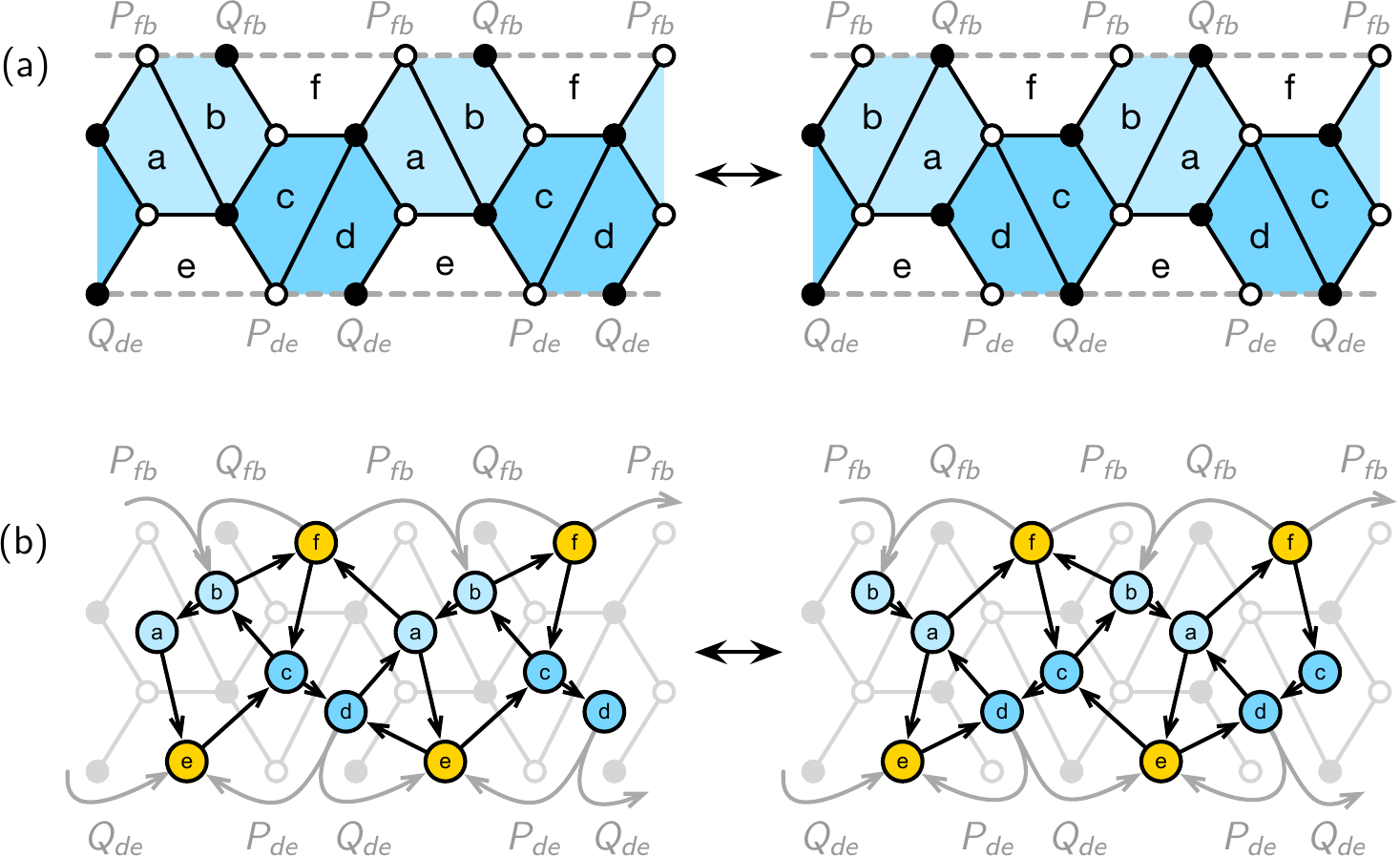}
\caption{
Tilting mutation family B on (a) a brane tiling and (b) its corresponding periodic quiver, resulting in a quiver-invariant duality.}
\label{fig_family_2}

\end{figure}
%---------------------------------------------------- 

%---------------------------------------------------- 
\begin{figure}[htbp]
\centering
\includegraphics[width=0.8\textwidth]{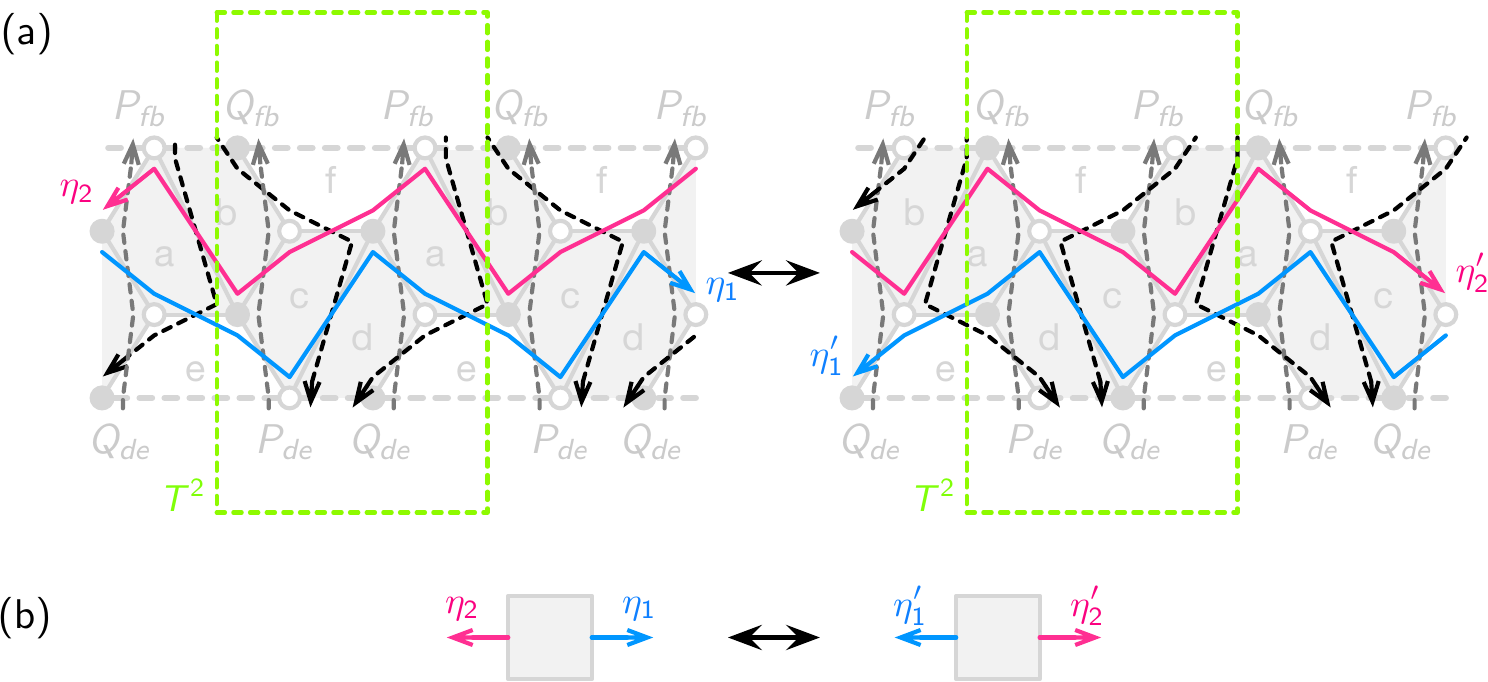}
\caption{
(a) The zig-zag paths $\eta_1$ and $\eta_2$ in the mutation region of tilting mutation family B.
(b) Under the tilting mutation, $\eta_1$ and  $\eta_2$ swap their orientation
with the corresponding external legs of the $(p,q)$-web diagram 
flipping their orientation accordingly.
}
\label{fig_family_2_zigzag}

\end{figure}
%---------------------------------------------------- 

The second family of tilting mutations that we identify among the toric phases
of the $H_{1,1,2,1}$ model involves a mutation region containing
six quiver nodes $a, \dots, f$
as illustrated in \fref{fig_family_2}.
The superpotential takes the general form in \eref{es04a00a} as follows,
\beal{es04a10}
W = 
&&
X_{ed} X_{da} X_{ae} 
+ X_{cb} X_{bf} X_{fc}
- X_{ec} X_{cb} X_{ba} X_{ae}
- X_{fc} X_{cd} X_{da} X_{af}
\nn\\
&&
+ X_{ba} X_{af} P_{fb}
+ X_{ec} X_{cd} P_{de}
- X_{bf} Q_{fb}
- X_{ed} Q_{de}
+ W_0
~,~
\eea
where $P_{fb}$ and $Q_{fb}$ denote the paths from node $f$ to node $b$,
and $P_{de}$ and $Q_{de}$ those from node $d$ to node $e$ in the periodic quiver.
Under the tilting mutation, the superpotential in \eref{es04a10}
is mapped to the following form,
\beal{es04a11}
W^\prime = 
&&
X_{ec} X_{cb} X_{ba} X_{ae} 
+ X_{fc} X_{cd} X_{da} X_{af} 
- X_{ed} X_{da} X_{ae}
- X_{cb} X_{bf} X_{fc} 
\nn\\
&&
+ X_{bf} P_{fb}
+ X_{ed} P_{de}
- X_{ba} X_{af} Q_{fb}
- X_{ec} X_{cd} Q_{de}
+ W_0
~,~
\eea
which is
in agreement with the general form in \eref{es04a00b}.
Here, the superpotential terms, which are within the mutation region and
whose signs are reversed under the tilting mutation, are given by,
\beal{es04a10b}
\Omega_{\text{mut}} = 
X_{ed} X_{da} X_{ae} 
+ X_{cb} X_{bf} X_{fc}
- X_{ec} X_{cb} X_{ba} X_{ae}
- X_{fc} X_{cd} X_{da} X_{af}
~.~
\eea
The paths in the periodic quiver exchanged by the tilting mutation take the following form,
\beal{es04a10c}
U_{fb} = X_{ba} X_{af}
~,~
U_{de} = X_{ec} X_{cd}
~,~
V_{fb} = X_{bf}
~,~
V_{de} = X_{ed}
~.~
\eea
We note that both $W$ and $W^\prime$ satisfy the toric condition,
with every chiral field appearing in exactly two superpotential terms
with opposite signs.

As illustrated in \fref{fig_family_2_zigzag}, 
we also note that within the mutation region of the periodic quiver and brane tiling, 
we have two zig-zag paths $\eta_1$ and $\eta_2$ that flip their orientation under the tilting mutation.
We note here that for the tilting mutation to preserve the mesonic moduli space and hence the associated toric Calabi-Yau 3-fold, 
the remaining zig-zag paths traversing the mutation region should have winding numbers such that 
there exists $G\in GL(2,\mathbb{Z})$ for which $S_w(W^\prime)=G\, S_w(W)$ as discussed in \eref{es04b04}.
This in turn restricts the choice of $W_0$ unaffected by the tilting mutation. 
\\

%=================================================================
\subsection{Tilting Mutation Family C \label{sec:mutfam_c}}
%=================================================================

%---------------------------------------------------- 
\begin{figure}[htbp]
\centering
\includegraphics[width=0.8\textwidth]{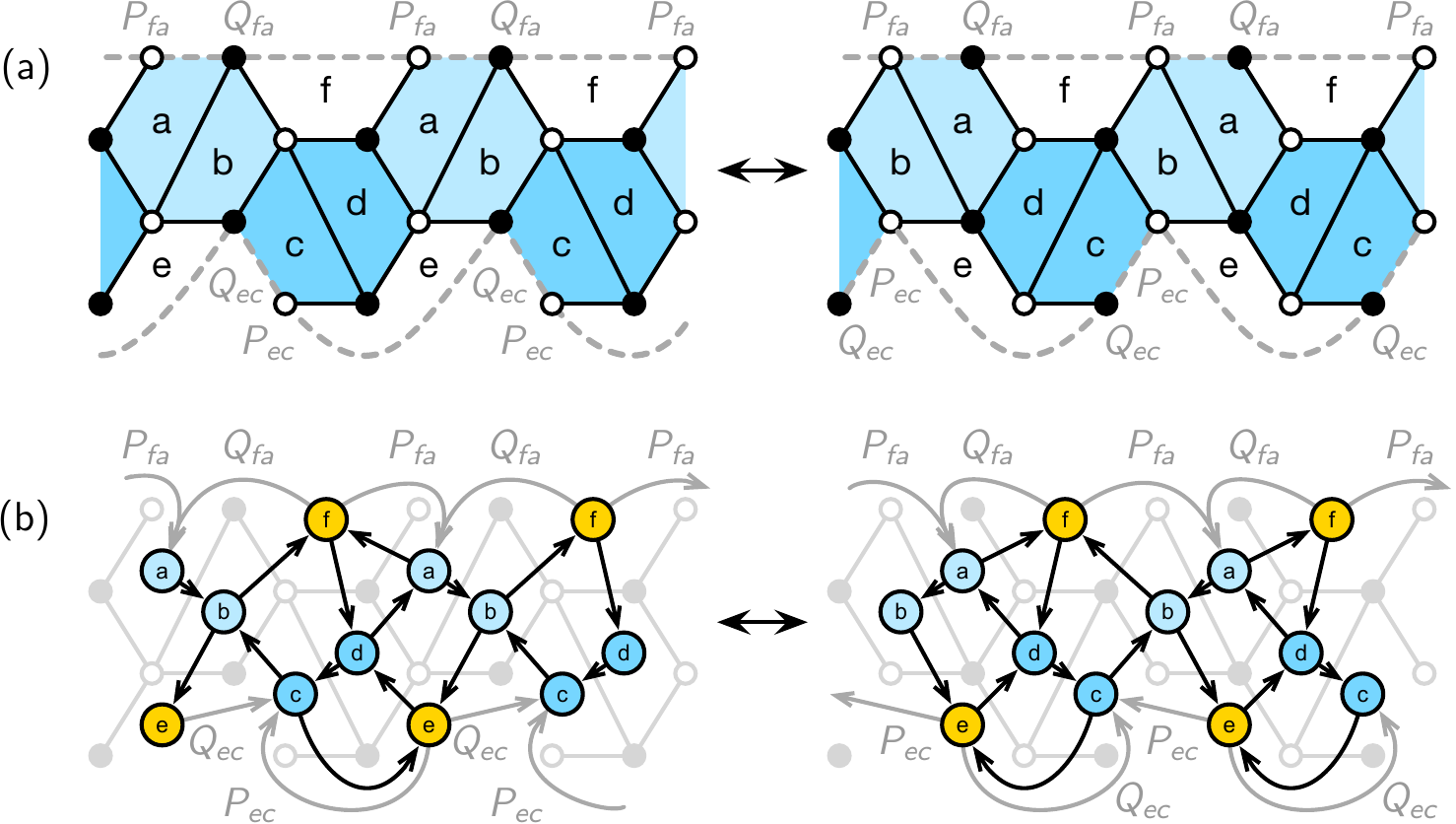}
\caption{
Tilting mutation family C on (a) a brane tiling and (b) its corresponding periodic quiver, resulting in a quiver-invariant duality.
}
\label{fig_family_4}
\end{figure}
%---------------------------------------------------- 
%---------------------------------------------------- 
\begin{figure}[htbp]
\centering
\includegraphics[width=0.8\textwidth]{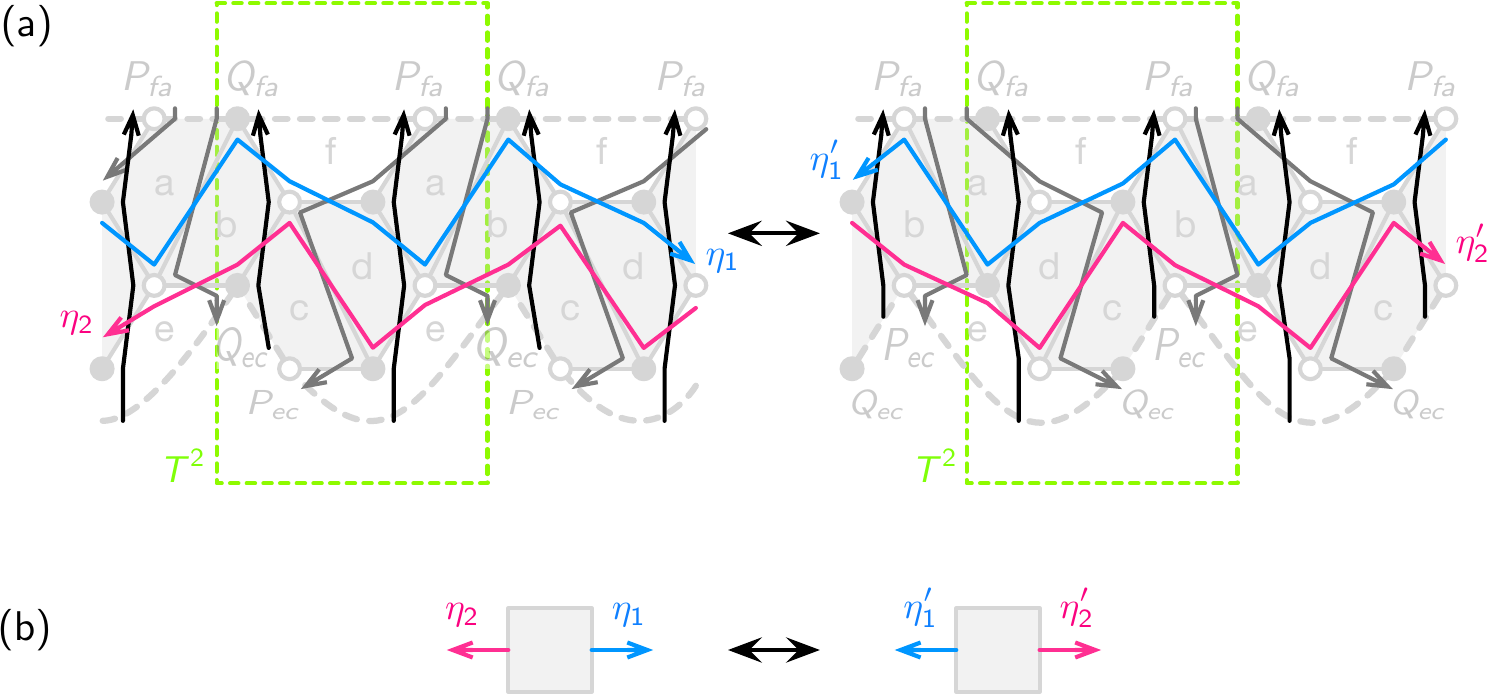}
\caption{
(a) The zig-zag paths $\eta_1$ and $\eta_2$ in the mutation region of tilting mutation family C.
(b) Under the tilting mutation, $\eta_1$ and  $\eta_2$ swap their orientation
with the corresponding external legs of the $(p,q)$-web diagram 
flipping their orientation accordingly.
}
\label{fig_family_4_zigzag}
\end{figure}
%---------------------------------------------------- 
The third family of tilting mutations 
that we identify among the toric phases
of the $H_{1,1,2,1}$ model 
involves a mutation region that contains six quiver nodes $a, \dots, f$, 
as illustrated in \fref{fig_family_4}.
The superpotential takes the general form in \eref{es04a00a} as follows, 
\beal{es04a30}
W = 
&&
X_{ab} X_{be} X_{ed} X_{da}
+ X_{fd} X_{dc} X_{cb} X_{bf}
- X_{dc} X_{ce} X_{ed}
- X_{fd} X_{da} X_{af}
\nn\\
&&
+ X_{af} P_{fa}
+ X_{ce} P_{ec}
- X_{ab} X_{bf} Q_{fa}
- X_{cb} X_{be} Q_{ec}
+ W_0
~,~
\eea
where $P_{fa}$ and $Q_{fa}$ denote the paths from node $f$ to node $a$, 
and $P_{ec}$ and $Q_{ec}$ those from node $e$ to node $c$ in the periodic quiver.
Under the tilting mutation, the superpotential in \eref{es04a30} 
is mapped to the following form,
\beal{es04a31}
W^\prime = 
&&
X_{dc} X_{ce} X_{ed}
+ X_{fd} X_{da} X_{af}
- X_{ab} X_{be} X_{ed} X_{da}
- X_{fd} X_{dc} X_{cb} X_{bf}
\nn\\
&&
+ X_{ab} X_{bf} P_{fa}
+ X_{cb} X_{be} P_{ec}
- X_{af} Q_{fa}
- X_{ce} Q_{ec}
+ W_0
~,~
\eea
which is in agreement with the general form in \eref{es04a00b}.
Here, the superpotential terms 
whose signs are reversed under the tilting mutation while their field content is left unchanged
are given by,
\beal{es04a30b}
\Omega_{\text{mut}} =
X_{ab} X_{be} X_{ed} X_{da}
+ X_{fd} X_{dc} X_{cb} X_{bf}
- X_{dc} X_{ce} X_{ed}
- X_{fd} X_{da} X_{af}
~,~
\eea
and the paths exchanged by the tilting mutation take the following form,
\beal{es04a30c}
U_{fa} = X_{af} 
~,~
U_{ec} = X_{ce} 
~,~
V_{fa} = X_{ab} X_{bf} 
~,~
V_{ec} = X_{cb} X_{be} 
~.~
\eea
We note that both $W$ and $W^\prime$ satisfy the toric condition,
with every chiral field appearing in exactly two superpotential terms
with opposite signs.

As illustrated in \fref{fig_family_4_zigzag}, we also note 
that the mutation region contains two zig-zag paths $\eta_1$ and $\eta_2$ 
that wind around the mutation region without crossing its boundaries.
Under the tilting mutation, $\eta_1$ and $\eta_2$ reverse their orientations, 
thereby exchanging their winding numbers $\pm(1,0)$ 
and the corresponding pair of external legs of the $(p,q)$-web diagram.
In order for the tilting mutation to preserve the mesonic moduli space and hence the associated toric Calabi-Yau 3-fold, 
the remaining zig-zag paths traversing the mutation region need to have winding numbers such that 
there exists $G\in GL(2,\mathbb{Z})$ for which $S_w(W^\prime)=G\, S_w(W)$ as discussed in \eref{es04b04}.
This in turn restricts the choice of $W_0$ unaffected by the tilting mutation. 
\\

%=================================================================
\section{Examples \label{sec:examples}}
%=================================================================

%=================================================================
\subsection{$H_{1,1,2,1}$ 2a and 2b} % family B
%=================================================================

%---------------------------------------------------- 
\begin{figure}[htbp]
\centering
\includegraphics[width=0.95\textwidth]{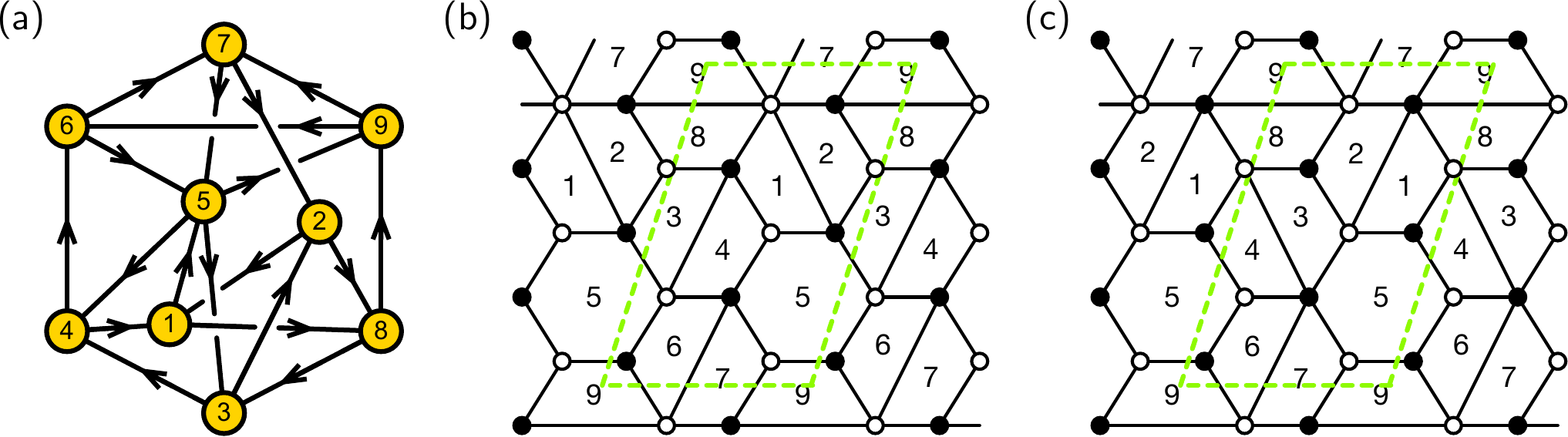}
\caption{
(a) The quiver shared between the two brane tilings in (b) and (c) corresponding to the two toric phases 2a and 2b of $H_{1,1,2,1}$,
respectively. 
}
\label{fig_tiling_quiver_02}
\end{figure}
%---------------------------------------------------- 

The two toric phases 2a and 2b of the $H_{1,1,2,1}$ model 
share the same quiver, which is shown in \fref{fig_tiling_quiver_02}(a).
The corresponding brane tilings, illustrated in 
\fref{fig_tiling_quiver_02}(b) and \fref{fig_tiling_quiver_02}(c), respectively,
are distinguished by their superpotentials, 
which take the following form,
\beal{es05a01}
W_{2a} = 
&&
X_{15} X_{54} X_{41} 
+ X_{18} X_{89} X_{96} X_{67} X_{72} X_{21} 
+ X_{28} X_{83} X_{32}
+ X_{34} X_{46} X_{65} X_{53}
\nn\\
&&
+ X_{59} X_{97} X_{75}
- X_{15} X_{53} X_{32} X_{21} 
- X_{18} X_{83} X_{34} X_{41}
- X_{28} X_{89} X_{97} X_{72}
\nn\\
&&
- X_{46} X_{67} X_{75} X_{54} 
- X_{59} X_{96} X_{65}
~,~
\eea
and
\beal{es05a02}
W_{2b} = 
&&
X_{15} X_{53} X_{32} X_{21} 
+ X_{18} X_{83} X_{34} X_{41} 
+ X_{28} X_{89} X_{96} X_{67} X_{72} 
+ X_{46} X_{65} X_{54}
\nn\\
&&
+ X_{59} X_{97} X_{75} 
- X_{15} X_{54} X_{41} 
- X_{18} X_{89} X_{97} X_{72} X_{21}
- X_{28} X_{83} X_{32}
\nn\\
&&
- X_{34} X_{46} X_{67} X_{75} X_{53} 
- X_{59} X_{96} X_{65}
~.~
\eea

\begin{figure}[htbp]
\centering
\includegraphics[width=0.7\textwidth]{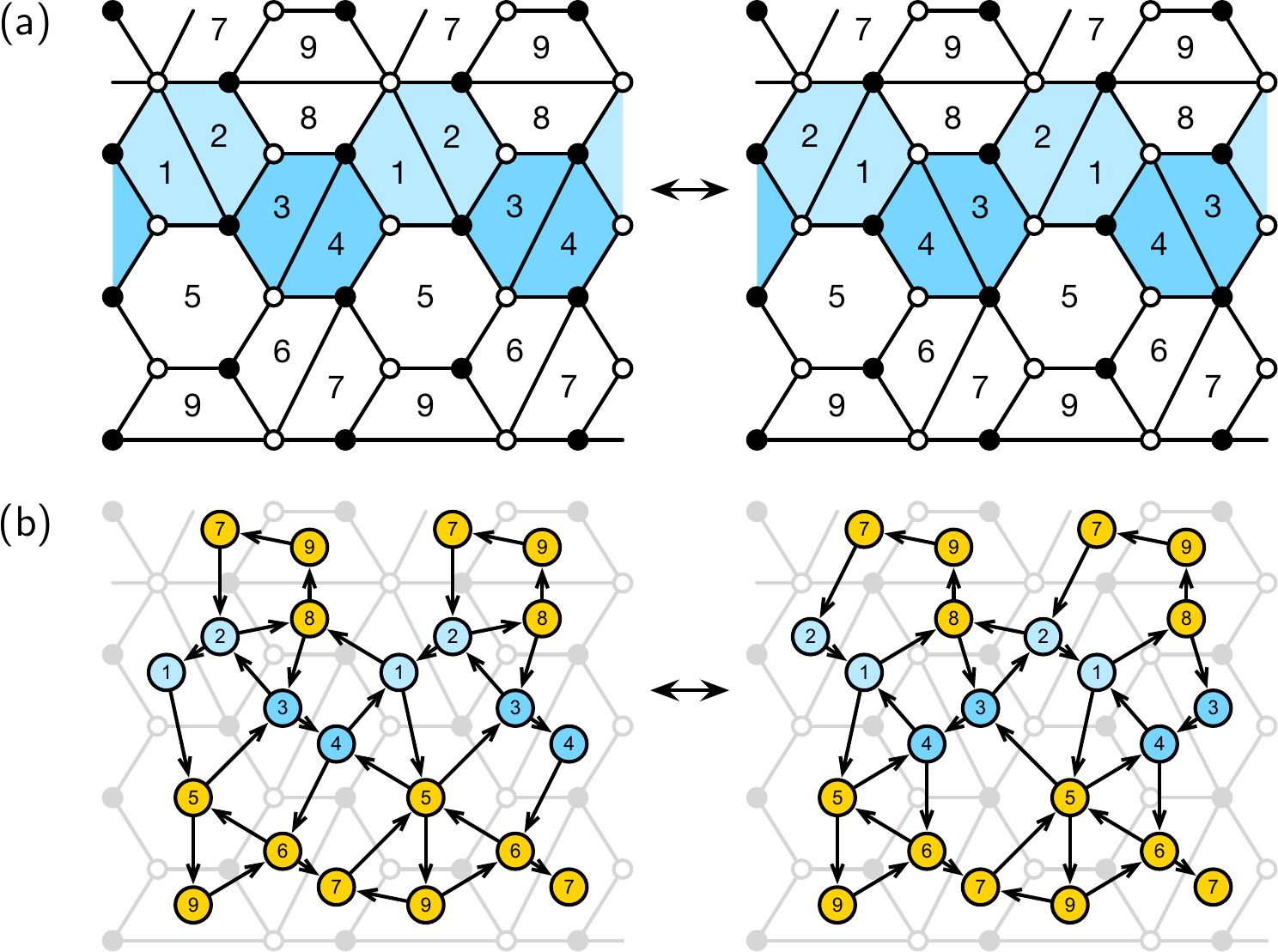}
\caption{
(a) The tilting mutation relating the brane tilings of the two toric phases 2a and 2b of $H_{1,1,2,1}$,
and (b) its effect on the corresponding periodic quivers. 
}
\label{fig_tiling_pq_02}
\end{figure}

\paragraph{Tilting Mutation.}
By comparing the superpotentials in \eref{es05a01} and \eref{es05a02},
we identify the tilting mutation relating toric phase 2a to 2b 
as belonging to tilting mutation family B introduced in Section~\ref{sec:mutfam_b},
where the quiver nodes $(a,b,c,d,e,f)$ of the mutation region 
are identified with the quiver nodes $(1,2,3,4,5,8)$ of the two toric phases.
Under this identification, the superpotential $W_{2a}$ takes the general form of $W$ 
for tilting mutation family B, while $W_{2b}$ takes the general form of $W^\prime$.
The part of the superpotential that is
unaffected by the tilting mutation
is given by,
\beal{es05a020}
&&
W_0 = X_{59} X_{97} X_{75} - X_{59} X_{96} X_{65}
~.~
\eea
The superpotential terms whose signs are reversed under the tilting mutation, 
are given by,
\beal{es05a03}
\Omega_{\text{mut}} =
X_{15} X_{54} X_{41} 
+ X_{28} X_{83} X_{32}
- X_{15} X_{53} X_{32} X_{21} 
- X_{18} X_{83} X_{34} X_{41}
~,~
\eea
and the exterior paths take the form,
\beal{es05a04}
&&
P_{82} = X_{89} X_{96} X_{67} X_{72} 
~,~
Q_{82} = X_{89} X_{97} X_{72} 
~,~
P_{45} = X_{46} X_{65} 
~,~
Q_{45} = X_{46} X_{67} X_{75} 
~.~
\nn\\
\eea
Accordingly, the paths exchanged by the tilting mutation are given by,
\beal{es05a05}
U_{82} = X_{21} X_{18}
~,~
U_{45} = X_{53} X_{34}
~,~
V_{82} = X_{28}
~,~
V_{45} = X_{54}
~,~
\eea
which is
in agreement with the general form in Section~\ref{sec:mutfam_b} for tilting mutation family B,
such that the tilting mutation maps the superpotential $W_{2a}$ in \eref{es05a01} 
to $W_{2b}$ in \eref{es05a02}.
The action of the tilting mutation on the brane tiling 
and on the corresponding periodic quiver 
is illustrated in \fref{fig_tiling_pq_02}.

%---------------------------------------------------- 
%---------------------------------------------------- 

\paragraph{Global Symmetry Charges.}
\tref{tab:models_2_Rcharges} summarizes the charges under the global symmetry
carried by the chiral fields of the two toric phases,
where the $U(1)_R$ charges are expressed in terms of
the $U(1)_R$ charges $R_a$ of the extremal GLSM fields $p_a$ in \tref{tab:extremal_charges}.
As shown in \tref{tab:models_2_Rcharges},
the chiral fields along the exterior paths
$P_{82}$, $Q_{82}$, $P_{45}$ and $Q_{45}$ in \eref{es05a04}
carry identical global symmetry charges in the two toric phases.
In contrast, the interior paths $U_{82}$ and $V_{82}$,
as well as $U_{45}$ and $V_{45}$ in \eref{es05a05},
exchange their global symmetry charges under the tilting mutation,
as can be seen from the charges carried by the chiral fields
in the mutation region in \tref{tab:models_2_Rcharges}.

%---------------------------------------------------- 
\begin{table}[h]
\setlength{\tabcolsep}{2pt}
\centering
\footnotesize{
\begin{tabular}[t]{c}
Phase 2a \\[4pt]
\begin{tabular}{| c | c | c | r |}
\hline
\; & $U(1)_{f_1}$ & $U(1)_{f_2}$ & $U(1)_R$\\\hline\hline
$X_{46}$ & $-1$ & $+1$ & $R_{3} \simeq 0.3862$\\
$X_{59}$ & $-1$ & $0$ & $R_{3} + R_{6} \simeq 0.7479$\\
$X_{65}$ & $+1$ & $+1$ & $R_{1} + R_{2} + R_{4} \simeq 0.8472$\\
$X_{67}$ & $0$ & $-1$ & $R_{6} \simeq 0.3617$\\
$X_{72}$ & $-1$ & $+1$ & $R_{3} \simeq 0.3862$\\
$X_{75}$ & $-1$ & $0$ & $R_{4} + R_{5} \simeq 0.6581$\\
$X_{89}$ & $-1$ & $+1$ & $R_{4} \simeq 0.2532$\\
$X_{96}$ & $0$ & $-1$ & $R_{5} \simeq 0.4049$\\
$X_{97}$ & $+2$ & $0$ & $R_{1} + R_{2} \simeq 0.5940$\\\hline
$X_{15}$ & $-1$ & $0$ & $R_{3} + R_{6} \simeq 0.7479$\\
$X_{18}$ & $+1$ & $0$ & $R_{2} \simeq 0.3205$\\
$X_{21}$ & $+1$ & $0$ & $R_{1} \simeq 0.2736$\\
$X_{28}$ & $0$ & $-2$ & $R_{5} + R_{6} \simeq 0.7666$\\
$X_{32}$ & $0$ & $+1$ & $R_{2} + R_{4} \simeq 0.5737$\\
$X_{34}$ & $0$ & $-1$ & $R_{6} \simeq 0.3617$\\
$X_{41}$ & $-1$ & $0$ & $R_{4} + R_{5} \simeq 0.6581$\\
$X_{53}$ & $0$ & $-1$ & $R_{5} \simeq 0.4049$\\
$X_{54}$ & $+2$ & $0$ & $R_{1} + R_{2} \simeq 0.5940$\\
$X_{83}$ & $0$ & $+1$ & $R_{1} + R_{3} \simeq 0.6597$\\\hline
\end{tabular}
\end{tabular}
}
\hspace{6pt}
\footnotesize{
\begin{tabular}[t]{c}
Phase 2b \\[4pt]
\begin{tabular}{| c | c | c | r |}
\hline
\; & $U(1)_{f_1}$ & $U(1)_{f_2}$ & $U(1)_R$\\\hline\hline
$X_{46}$ & $-1$ & $+1$ & $R_{3} \simeq 0.3862$\\
$X_{59}$ & $-1$ & $0$ & $R_{3} + R_{6} \simeq 0.7479$\\
$X_{65}$ & $+1$ & $+1$ & $R_{1} + R_{2} + R_{4} \simeq 0.8472$\\
$X_{67}$ & $0$ & $-1$ & $R_{6} \simeq 0.3617$\\
$X_{72}$ & $-1$ & $+1$ & $R_{3} \simeq 0.3862$\\
$X_{75}$ & $-1$ & $0$ & $R_{4} + R_{5} \simeq 0.6581$\\
$X_{89}$ & $-1$ & $+1$ & $R_{4} \simeq 0.2532$\\
$X_{96}$ & $0$ & $-1$ & $R_{5} \simeq 0.4049$\\
$X_{97}$ & $+2$ & $0$ & $R_{1} + R_{2} \simeq 0.5940$\\\hline
$X_{15}$ & $0$ & $+1$ & $R_{1} + R_{3} \simeq 0.6597$\\
$X_{18}$ & $0$ & $-1$ & $R_{5} \simeq 0.4049$\\
$X_{21}$ & $0$ & $-1$ & $R_{6} \simeq 0.3617$\\
$X_{28}$ & $+2$ & $0$ & $R_{1} + R_{2} \simeq 0.5940$\\
$X_{32}$ & $-1$ & $0$ & $R_{4} + R_{5} \simeq 0.6581$\\
$X_{34}$ & $+1$ & $0$ & $R_{1} \simeq 0.2736$\\
$X_{41}$ & $0$ & $+1$ & $R_{2} + R_{4} \simeq 0.5737$\\
$X_{53}$ & $+1$ & $0$ & $R_{2} \simeq 0.3205$\\
$X_{54}$ & $0$ & $-2$ & $R_{5} + R_{6} \simeq 0.7666$\\
$X_{83}$ & $-1$ & $0$ & $R_{3} + R_{6} \simeq 0.7479$\\\hline
\end{tabular}
\end{tabular}
}
\caption{
The charges under the global symmetry on the set of chiral fields 
shared between toric phases 2a and 2b of $H_{1,1,2,1}$, including the $U(1)_R$ charges, which are expressed in terms of $U(1)_R$ charges $R_a$
corresponding to the extremal GLSM fields $p_a$. 
}
\label{tab:models_2_Rcharges}
\end{table}
%----------------------------------------------------

\paragraph{Mesonic Moduli Spaces.}
Using the forward algorithm, 
we compute the $P$-matrices encoding the perfect matchings 
of the two brane tilings.
Both brane tilings admit $30$ perfect matchings, 
consisting of the extremal perfect matchings $p_1, \dots, p_6$ 
corresponding to the extremal points of the toric diagram of $H_{1,1,2,1}$,
as well as the non-extremal perfect matchings.
The $P$-matrices take the following form,
\beal{es05a06}
P^{(2a)} = \tiny{\setlength{\tabcolsep}{1pt}\left(\begin{tabular}{c | *{6}{c}| *{2}{c}| *{12}{c}| *{10}{c}}
& $p_{1}$ & $p_{2}$ & $p_{3}$ & $p_{4}$ & $p_{5}$ & $p_{6}$ & $q_{1}$ & $q_{2}$ & $u_{1}$ & $u_{2}$ & $u_{3}$ & $u_{4}$ & $u_{5}$ & $u_{6}$ & $u_{7}$ & $u_{8}$ & $u_{9}$ & $u_{10}$ & $u_{11}$ & $u_{12}$ & $s_{1}$ & $s_{2}$ & $s_{3}$ & $s_{4}$ & $s_{5}$ & $s_{6}$ & $s_{7}$ & $s_{8}$ & $s_{9}$ & $s_{10}$\\\hline
$X_{15}$ & 0 & 0 & 1 & 0 & 0 & 1 & 1 & 0 & 0 & 0 & 0 & 0 & 0 & 0 & 0 & 0 & 0 & 0 & 1 & 1 & 1 & 0 & 0 & 0 & 0 & 0 & 1 & 1 & 1 & 1 \\
$X_{18}$ & 0 & 1 & 0 & 0 & 0 & 0 & 0 & 0 & 0 & 0 & 0 & 1 & 0 & 0 & 0 & 0 & 0 & 0 & 1 & 0 & 0 & 0 & 0 & 0 & 0 & 0 & 1 & 0 & 0 & 0 \\
$X_{21}$ & 1 & 0 & 0 & 0 & 0 & 0 & 0 & 0 & 0 & 0 & 0 & 0 & 0 & 0 & 0 & 0 & 0 & 1 & 0 & 0 & 0 & 0 & 0 & 0 & 1 & 1 & 0 & 0 & 0 & 0 \\
$X_{28}$ & 0 & 0 & 0 & 0 & 1 & 1 & 1 & 1 & 0 & 0 & 0 & 1 & 0 & 0 & 0 & 0 & 0 & 1 & 1 & 0 & 0 & 0 & 0 & 0 & 1 & 1 & 1 & 0 & 0 & 0 \\
$X_{32}$ & 0 & 1 & 0 & 1 & 0 & 0 & 0 & 0 & 0 & 1 & 1 & 0 & 0 & 1 & 1 & 1 & 1 & 0 & 0 & 0 & 0 & 0 & 1 & 1 & 0 & 0 & 0 & 0 & 0 & 0 \\
$X_{34}$ & 0 & 0 & 0 & 0 & 0 & 1 & 1 & 0 & 0 & 1 & 0 & 0 & 0 & 1 & 0 & 0 & 0 & 0 & 0 & 0 & 0 & 0 & 1 & 0 & 1 & 0 & 0 & 0 & 0 & 0 \\
$X_{41}$ & 0 & 0 & 0 & 1 & 1 & 0 & 0 & 1 & 0 & 0 & 1 & 0 & 0 & 0 & 1 & 1 & 1 & 1 & 0 & 0 & 0 & 0 & 0 & 1 & 0 & 1 & 0 & 0 & 0 & 0 \\
$X_{46}$ & 0 & 0 & 1 & 0 & 0 & 0 & 0 & 0 & 0 & 0 & 1 & 0 & 0 & 0 & 1 & 0 & 0 & 0 & 0 & 0 & 1 & 0 & 0 & 1 & 0 & 1 & 1 & 1 & 0 & 0 \\
$X_{53}$ & 0 & 0 & 0 & 0 & 1 & 0 & 0 & 1 & 1 & 0 & 0 & 1 & 1 & 0 & 0 & 0 & 0 & 0 & 0 & 0 & 0 & 1 & 0 & 0 & 0 & 0 & 0 & 0 & 0 & 0 \\
$X_{54}$ & 1 & 1 & 0 & 0 & 0 & 0 & 0 & 0 & 1 & 1 & 0 & 1 & 1 & 1 & 0 & 0 & 0 & 0 & 0 & 0 & 0 & 1 & 1 & 0 & 1 & 0 & 0 & 0 & 0 & 0 \\
$X_{59}$ & 0 & 0 & 1 & 0 & 0 & 1 & 0 & 1 & 0 & 0 & 0 & 1 & 1 & 1 & 1 & 0 & 0 & 0 & 0 & 0 & 0 & 1 & 1 & 1 & 1 & 1 & 1 & 1 & 0 & 0 \\
$X_{65}$ & 1 & 1 & 0 & 1 & 0 & 0 & 0 & 0 & 0 & 0 & 0 & 0 & 0 & 0 & 0 & 1 & 1 & 1 & 1 & 1 & 0 & 0 & 0 & 0 & 0 & 0 & 0 & 0 & 1 & 1 \\
$X_{67}$ & 0 & 0 & 0 & 0 & 0 & 1 & 0 & 1 & 0 & 0 & 0 & 0 & 0 & 0 & 0 & 1 & 0 & 0 & 0 & 0 & 0 & 0 & 0 & 0 & 0 & 0 & 0 & 0 & 1 & 0 \\
$X_{72}$ & 0 & 0 & 1 & 0 & 0 & 0 & 0 & 0 & 0 & 0 & 0 & 0 & 0 & 0 & 0 & 0 & 1 & 0 & 0 & 0 & 0 & 1 & 1 & 1 & 0 & 0 & 0 & 0 & 0 & 1 \\
$X_{75}$ & 0 & 0 & 0 & 1 & 1 & 0 & 1 & 0 & 0 & 0 & 0 & 0 & 0 & 0 & 0 & 0 & 1 & 1 & 1 & 1 & 0 & 0 & 0 & 0 & 0 & 0 & 0 & 0 & 0 & 1 \\
$X_{83}$ & 1 & 0 & 1 & 0 & 0 & 0 & 0 & 0 & 1 & 0 & 0 & 0 & 1 & 0 & 0 & 0 & 0 & 0 & 0 & 1 & 1 & 1 & 0 & 0 & 0 & 0 & 0 & 1 & 1 & 1 \\
$X_{89}$ & 0 & 0 & 0 & 1 & 0 & 0 & 0 & 0 & 0 & 0 & 0 & 0 & 1 & 1 & 1 & 0 & 0 & 0 & 0 & 1 & 0 & 0 & 0 & 0 & 0 & 0 & 0 & 1 & 0 & 0 \\
$X_{96}$ & 0 & 0 & 0 & 0 & 1 & 0 & 1 & 0 & 1 & 1 & 1 & 0 & 0 & 0 & 0 & 0 & 0 & 0 & 0 & 0 & 1 & 0 & 0 & 0 & 0 & 0 & 0 & 0 & 0 & 0 \\
$X_{97}$ & 1 & 1 & 0 & 0 & 0 & 0 & 0 & 0 & 1 & 1 & 1 & 0 & 0 & 0 & 0 & 1 & 0 & 0 & 0 & 0 & 1 & 0 & 0 & 0 & 0 & 0 & 0 & 0 & 1 & 0 \\
\end{tabular}\right)}
\eea
and
\beal{es05a07}
P^{(2b)} = \tiny{\setlength{\tabcolsep}{1pt}\left(\begin{tabular}{c | *{6}{c}| *{2}{c}| *{12}{c}| *{10}{c}}
& $p_{1}$ & $p_{2}$ & $p_{3}$ & $p_{4}$ & $p_{5}$ & $p_{6}$ & $q_{1}$ & $q_{2}$ & $u_{1}$ & $u_{2}$ & $u_{3}$ & $u_{4}$ & $u_{5}$ & $u_{6}$ & $u_{7}$ & $u_{8}$ & $u_{9}$ & $u_{10}$ & $u_{11}$ & $u_{12}$ & $s_{1}$ & $s_{2}$ & $s_{3}$ & $s_{4}$ & $s_{5}$ & $s_{6}$ & $s_{7}$ & $s_{8}$ & $s_{9}$ & $s_{10}$\\\hline
$X_{15}$ & 1 & 0 & 1 & 0 & 0 & 0 & 0 & 0 & 0 & 0 & 0 & 0 & 0 & 0 & 0 & 0 & 0 & 0 & 1 & 1 & 0 & 0 & 0 & 1 & 0 & 0 & 1 & 1 & 1 & 1 \\
$X_{18}$ & 0 & 0 & 0 & 0 & 1 & 0 & 0 & 1 & 0 & 0 & 0 & 1 & 0 & 0 & 0 & 0 & 0 & 0 & 0 & 1 & 0 & 0 & 0 & 0 & 0 & 0 & 0 & 1 & 0 & 0 \\
$X_{21}$ & 0 & 0 & 0 & 0 & 0 & 1 & 1 & 0 & 0 & 0 & 0 & 0 & 0 & 0 & 0 & 0 & 1 & 0 & 0 & 0 & 0 & 1 & 0 & 0 & 1 & 0 & 0 & 0 & 0 & 0 \\
$X_{28}$ & 1 & 1 & 0 & 0 & 0 & 0 & 0 & 0 & 0 & 0 & 0 & 1 & 0 & 0 & 0 & 0 & 1 & 0 & 0 & 1 & 0 & 1 & 0 & 0 & 1 & 0 & 0 & 1 & 0 & 0 \\
$X_{32}$ & 0 & 0 & 0 & 1 & 1 & 0 & 0 & 1 & 0 & 1 & 0 & 0 & 1 & 1 & 1 & 1 & 0 & 1 & 0 & 0 & 0 & 0 & 1 & 0 & 0 & 1 & 0 & 0 & 0 & 0 \\
$X_{34}$ & 1 & 0 & 0 & 0 & 0 & 0 & 0 & 0 & 0 & 1 & 0 & 0 & 1 & 0 & 0 & 0 & 0 & 0 & 0 & 0 & 0 & 1 & 1 & 0 & 0 & 0 & 0 & 0 & 0 & 0 \\
$X_{41}$ & 0 & 1 & 0 & 1 & 0 & 0 & 0 & 0 & 0 & 0 & 0 & 0 & 0 & 1 & 1 & 1 & 1 & 1 & 0 & 0 & 0 & 0 & 0 & 0 & 1 & 1 & 0 & 0 & 0 & 0 \\
$X_{46}$ & 0 & 0 & 1 & 0 & 0 & 0 & 0 & 0 & 0 & 0 & 0 & 0 & 0 & 1 & 1 & 0 & 0 & 0 & 0 & 0 & 0 & 0 & 0 & 1 & 1 & 1 & 1 & 1 & 0 & 0 \\
$X_{53}$ & 0 & 1 & 0 & 0 & 0 & 0 & 0 & 0 & 1 & 0 & 1 & 1 & 0 & 0 & 0 & 0 & 0 & 0 & 0 & 0 & 1 & 0 & 0 & 0 & 0 & 0 & 0 & 0 & 0 & 0 \\
$X_{54}$ & 0 & 0 & 0 & 0 & 1 & 1 & 1 & 1 & 1 & 1 & 1 & 1 & 1 & 0 & 0 & 0 & 0 & 0 & 0 & 0 & 1 & 1 & 1 & 0 & 0 & 0 & 0 & 0 & 0 & 0 \\
$X_{59}$ & 0 & 0 & 1 & 0 & 0 & 1 & 0 & 1 & 0 & 0 & 1 & 1 & 1 & 0 & 1 & 0 & 0 & 0 & 0 & 0 & 1 & 1 & 1 & 0 & 1 & 1 & 1 & 1 & 0 & 0 \\
$X_{65}$ & 1 & 1 & 0 & 1 & 0 & 0 & 0 & 0 & 0 & 0 & 0 & 0 & 0 & 0 & 0 & 1 & 1 & 1 & 1 & 1 & 0 & 0 & 0 & 0 & 0 & 0 & 0 & 0 & 1 & 1 \\
$X_{67}$ & 0 & 0 & 0 & 0 & 0 & 1 & 0 & 1 & 0 & 0 & 0 & 0 & 0 & 0 & 0 & 1 & 0 & 0 & 0 & 0 & 0 & 0 & 0 & 0 & 0 & 0 & 0 & 0 & 1 & 0 \\
$X_{72}$ & 0 & 0 & 1 & 0 & 0 & 0 & 0 & 0 & 0 & 0 & 0 & 0 & 0 & 0 & 0 & 0 & 0 & 1 & 0 & 0 & 1 & 0 & 1 & 0 & 0 & 1 & 0 & 0 & 0 & 1 \\
$X_{75}$ & 0 & 0 & 0 & 1 & 1 & 0 & 1 & 0 & 0 & 0 & 0 & 0 & 0 & 0 & 0 & 0 & 1 & 1 & 1 & 1 & 0 & 0 & 0 & 0 & 0 & 0 & 0 & 0 & 0 & 1 \\
$X_{83}$ & 0 & 0 & 1 & 0 & 0 & 1 & 1 & 0 & 1 & 0 & 1 & 0 & 0 & 0 & 0 & 0 & 0 & 0 & 1 & 0 & 1 & 0 & 0 & 1 & 0 & 0 & 1 & 0 & 1 & 1 \\
$X_{89}$ & 0 & 0 & 0 & 1 & 0 & 0 & 0 & 0 & 0 & 0 & 1 & 0 & 1 & 0 & 1 & 0 & 0 & 0 & 1 & 0 & 0 & 0 & 0 & 0 & 0 & 0 & 1 & 0 & 0 & 0 \\
$X_{96}$ & 0 & 0 & 0 & 0 & 1 & 0 & 1 & 0 & 1 & 1 & 0 & 0 & 0 & 1 & 0 & 0 & 0 & 0 & 0 & 0 & 0 & 0 & 0 & 1 & 0 & 0 & 0 & 0 & 0 & 0 \\
$X_{97}$ & 1 & 1 & 0 & 0 & 0 & 0 & 0 & 0 & 1 & 1 & 0 & 0 & 0 & 1 & 0 & 1 & 0 & 0 & 0 & 0 & 0 & 0 & 0 & 1 & 0 & 0 & 0 & 0 & 1 & 0 \\
\end{tabular}\right)}
~.~
\eea
From the $P$-matrices in \eref{es05a06} and \eref{es05a07}, 
the forward algorithm yields identical $G_t$-matrices for the two toric phases, 
which take the form,
\beal{es05a08}
G_t^{(2a)} = G_t^{(2b)} = \small{\setlength{\tabcolsep}{1pt}\left(\begin{tabular}{ *{6}{c}| *{2}{c}| *{3}{c}| *{3}{c}}
$p_{1}$ & $p_{2}$ & $p_{3}$ & $p_{4}$ & $p_{5}$ & $p_{6}$ & $q_{1}$ & $q_{2}$ & $u_{1}$ & $\cdots$ & $u_{12}$ &$s_{1}$ & $\cdots$ & $s_{10}$\\\hline
2 & 2 & 1 & 1 & 0 & 0 & 0 & 0 & 1 & $\cdots$ & 1 & 1 & $\cdots$ & 1 \\
2 & 1 & 3 & 0 & 0 & 2 & 1 & 1 & 1 & $\cdots$ & 1 & 2 & $\cdots$ & 2 \\
1 & 1 & 1 & 1 & 1 & 1 & 1 & 1 & 1 & $\cdots$ & 1 & 1 & $\cdots$ & 1 \\
\end{tabular}\right)}
~.~
\eea
The toric diagrams obtained from the two brane tilings therefore coincide,
which verifies that the brane tilings of toric phases 2a and 2b 
correspond to the same toric Calabi-Yau 3-fold $H_{1,1,2,1}$, 
as required for the quiver-invariant duality realized by the tilting mutation.
\\

%=================================================================
\subsection{$H_{1,1,2,1}$ 5a and 5b} % family B
%=================================================================

%---------------------------------------------------- 
\begin{figure}[htbp]
\centering
\includegraphics[width=0.95\textwidth]{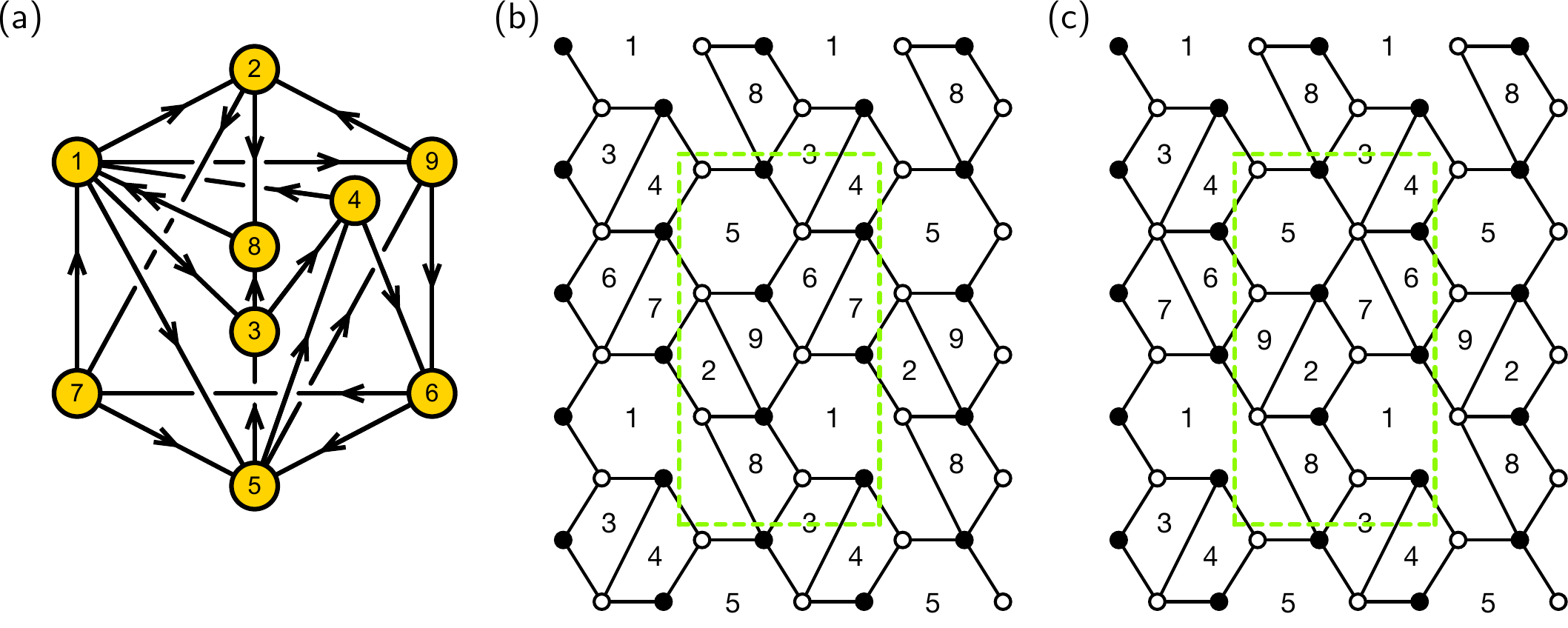}
\caption{
(a) The quiver shared between the two brane tilings in (b) and (c) corresponding to the two toric phases 5a and 5b of $H_{1,1,2,1}$,
respectively. 
}
\label{fig_doublet_5a_5b}
\end{figure}
%---------------------------------------------------- 

The two toric phases 5a and 5b of the $H_{1,1,2,1}$ model 
share the same quiver, which is shown in \fref{fig_doublet_5a_5b}(a).
The corresponding brane tilings are illustrated in 
\fref{fig_doublet_5a_5b}(b) and \fref{fig_doublet_5a_5b}(c), respectively,
and differ in their superpotentials, 
which are given by,
\beal{es05b01}
W_{5a} = 
&&
X_{12} X_{28} X_{81}^{2} 
+ X_{13} X_{38} X_{81}^{1}
+ X_{15} X_{54} X_{41}
+ X_{19} X_{96} X_{67} X_{71}
\nn\\
&&
+ X_{27} X_{75} X_{59} X_{92}
+ X_{34} X_{46} X_{65} X_{53}
- X_{12} X_{27} X_{71}
- X_{13} X_{34} X_{41}
\nn\\
&&
- X_{15} X_{53} X_{38} X_{81}^{2}
- X_{19} X_{92} X_{28} X_{81}^{1}
- X_{46} X_{67} X_{75} X_{54}
- X_{59} X_{96} X_{65}
~,~
\eea
and
\beal{es05b02}
W_{5b} = 
&&
X_{12} X_{27} X_{71}
+ X_{13} X_{38} X_{81}^{1}
+ X_{15} X_{54} X_{41}
+ X_{19} X_{92} X_{28} X_{81}^{2} 
\nn\\
&&
+ X_{34} X_{46} X_{67} X_{75} X_{53} 
+ X_{59} X_{96} X_{65} 
- X_{12} X_{28} X_{81}^{1}
- X_{13} X_{34} X_{41}
\nn\\
&&
- X_{15} X_{53} X_{38} X_{81}^{2}
- X_{19} X_{96} X_{67} X_{71}
- X_{27} X_{75} X_{59} X_{92}
- X_{46} X_{65} X_{54}
~.~
\eea

\begin{figure}[htbp]
\centering
\includegraphics[width=0.7\textwidth]{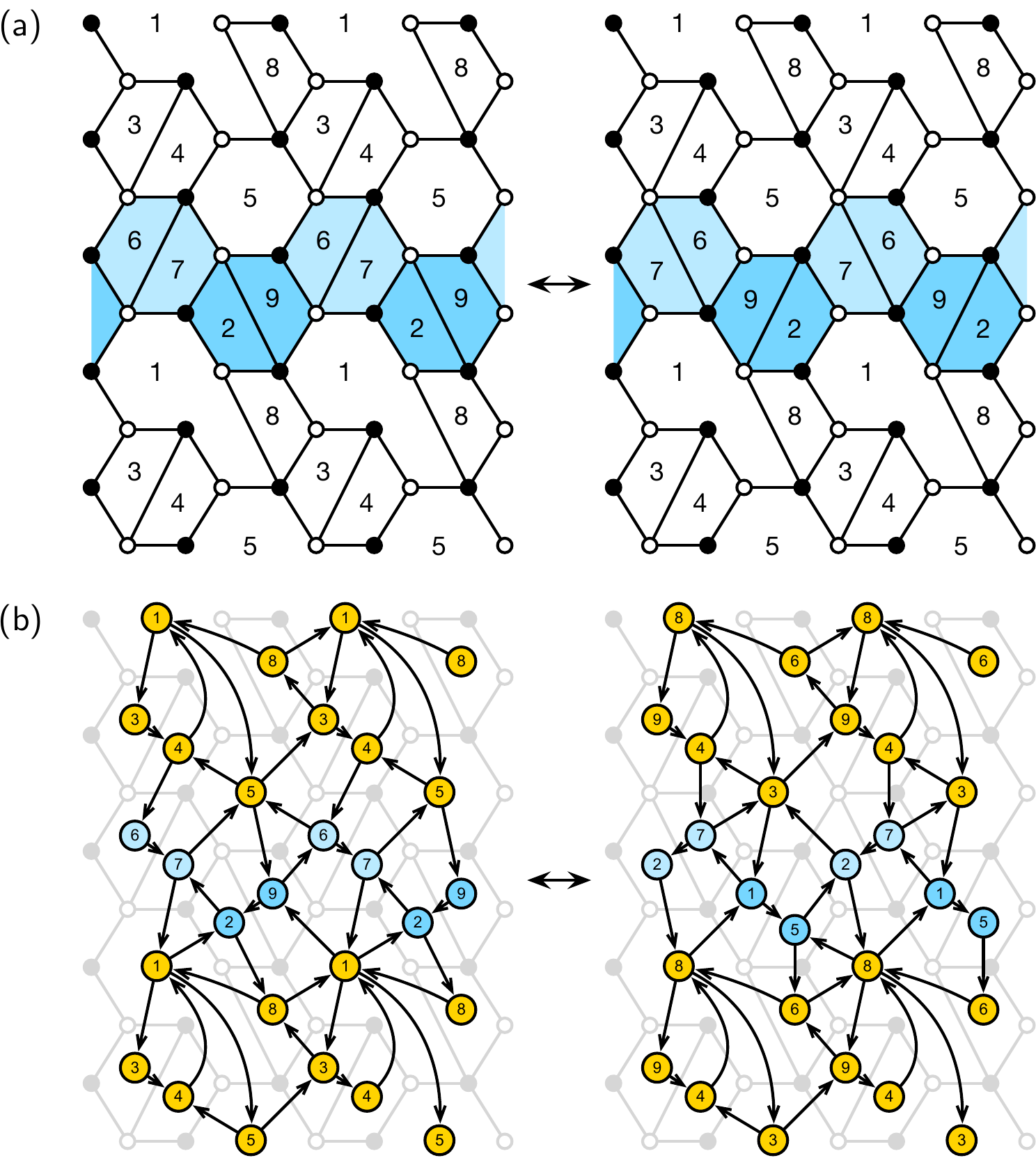}
\caption{
(a) The tilting mutation relating the brane tilings of the two toric phases 5a and 5b of $H_{1,1,2,1}$,
and (b) its effect on the corresponding periodic quivers. 
}
\label{fig_tilting_5a_5b}
\end{figure}

\paragraph{Tilting Mutation.}
By comparing the superpotentials in \eref{es05b01} and \eref{es05b02},
we identify the tilting mutation relating toric phase 5a to 5b 
as belonging to tilting mutation family B introduced in Section~\ref{sec:mutfam_b},
where the quiver nodes $(a,b,c,d,e,f)$ of the mutation region 
are identified with the quiver nodes $(7,6,9,2,1,5)$ of the two toric phases.
Under this identification, the superpotential $W_{5b}$ takes the general form of $W$ 
for tilting mutation family B, while $W_{5a}$ takes the general form of $W^\prime$.
The part of the superpotential that is
unaffected by the tilting mutation
is given by,
\beal{es05b03}
&&
W_0 = 
X_{13} X_{38} X_{81}^{1}
+ X_{15} X_{54} X_{41}
- X_{13} X_{34} X_{41}
- X_{15} X_{53} X_{38} X_{81}^{2}
~.~
\eea
The superpotential terms whose signs are reversed under the tilting mutation, 
with the signs taken as they appear in $W_{5b}$, 
are given by,
\beal{es05b04}
\Omega_{\text{mut}} =
X_{12} X_{27} X_{71}
+ X_{59} X_{96} X_{65}
- X_{19} X_{96} X_{67} X_{71}
- X_{27} X_{75} X_{59} X_{92}
~,~
\eea
and the exterior paths, which close the interior paths of the mutation region 
into gauge-invariant cycles, take the form,
\beal{es05b05}
&&
P_{56} = X_{53} X_{34} X_{46} 
~,~
Q_{56} = X_{54} X_{46} 
~,~
P_{21} = X_{28} X_{81}^{2} 
~,~
Q_{21} = X_{28} X_{81}^{1} 
~.~
\nn\\
\eea
Accordingly, the paths exchanged by the tilting mutation are given by,
\beal{es05b06}
U_{56} = X_{67} X_{75}
~,~
U_{21} = X_{19} X_{92}
~,~
V_{56} = X_{65}
~,~
V_{21} = X_{12}
~,~
\eea
which is
in agreement with the general form in Section~\ref{sec:mutfam_b} for tilting mutation family B,
such that the tilting mutation maps the superpotential $W_{5b}$ in \eref{es05b02} 
to $W_{5a}$ in \eref{es05b01}, and vice versa.
The action of the tilting mutation on the brane tiling 
and on the corresponding periodic quiver 
is illustrated in \fref{fig_tilting_5a_5b}.

%---------------------------------------------------- 
%---------------------------------------------------- 

\paragraph{Global Symmetry Charges.}
\tref{tab:models_5_Rcharges} summarizes the charges under the global symmetry
carried by the chiral fields of the two toric phases,
where the $U(1)_R$ charges are expressed in terms of
the $U(1)_R$ charges $R_a$ of the extremal GLSM fields $p_a$ in \tref{tab:extremal_charges}.
As shown in \tref{tab:models_5_Rcharges},
the chiral fields along the exterior paths
$P_{56}$, $Q_{56}$, $P_{21}$ and $Q_{21}$ in \eref{es05b05}
carry identical global symmetry charges in the two toric phases.
In contrast, the interior paths $U_{56}$ and $V_{56}$,
as well as $U_{21}$ and $V_{21}$ in \eref{es05b06},
exchange their global symmetry charges under the tilting mutation,
as can be seen from the charges carried by the chiral fields
in the mutation region in \tref{tab:models_5_Rcharges}.

%---------------------------------------------------- 
\begin{table}[h]
\setlength{\tabcolsep}{2pt}
\centering
\footnotesize{
\begin{tabular}[t]{c}
Phase 5a \\[4pt]
\begin{tabular}{| c | c | c | r |}
\hline
\; & $U(1)_{f_1}$ & $U(1)_{f_2}$ & $U(1)_R$\\\hline\hline
$X_{13}$ & $+1$ & $+1$ & $R_{1} + R_{2} + R_{3} \simeq 0.9802$\\
$X_{15}$ & $-1$ & $0$ & $R_{3} + R_{6} \simeq 0.7479$\\
$X_{28}$ & $-1$ & $+1$ & $R_{3} \simeq 0.3862$\\
$X_{34}$ & $0$ & $-1$ & $R_{6} \simeq 0.3617$\\
$X_{38}$ & $-1$ & $+1$ & $R_{4} \simeq 0.2532$\\
$X_{41}$ & $-1$ & $0$ & $R_{4} + R_{5} \simeq 0.6581$\\
$X_{46}$ & $-1$ & $+1$ & $R_{3} \simeq 0.3862$\\
$X_{53}$ & $0$ & $-1$ & $R_{5} \simeq 0.4049$\\
$X_{54}$ & $+2$ & $0$ & $R_{1} + R_{2} \simeq 0.5940$\\
$X^1_{81}$ & $0$ & $-2$ & $R_{5} + R_{6} \simeq 0.7666$\\
$X^2_{81}$ & $+2$ & $0$ & $R_{1} + R_{2} \simeq 0.5940$\\\hline
$X_{12}$ & $-1$ & $-1$ & $R_{4} + R_{5} + R_{6} \simeq 1.0198$\\
$X_{19}$ & $0$ & $+1$ & $R_{2} + R_{4} \simeq 0.5737$\\
$X_{27}$ & $+1$ & $0$ & $R_{2} \simeq 0.3205$\\
$X_{59}$ & $-1$ & $0$ & $R_{3} + R_{6} \simeq 0.7479$\\
$X_{65}$ & $+1$ & $+1$ & $R_{1} + R_{2} + R_{4} \simeq 0.8472$\\
$X_{67}$ & $0$ & $-1$ & $R_{6} \simeq 0.3617$\\
$X_{71}$ & $0$ & $+1$ & $R_{1} + R_{3} \simeq 0.6597$\\
$X_{75}$ & $-1$ & $0$ & $R_{4} + R_{5} \simeq 0.6581$\\
$X_{92}$ & $+1$ & $0$ & $R_{1} \simeq 0.2736$\\
$X_{96}$ & $0$ & $-1$ & $R_{5} \simeq 0.4049$\\\hline
\end{tabular}
\end{tabular}
}
\hspace{6pt}
\footnotesize{
\begin{tabular}[t]{c}
Phase 5b \\[4pt]
\begin{tabular}{| c | c | c | r |}
\hline
\; & $U(1)_{f_1}$ & $U(1)_{f_2}$ & $U(1)_R$\\\hline\hline
$X_{13}$ & $+1$ & $+1$ & $R_{1} + R_{2} + R_{3} \simeq 0.9802$\\
$X_{15}$ & $-1$ & $0$ & $R_{3} + R_{6} \simeq 0.7479$\\
$X_{28}$ & $-1$ & $+1$ & $R_{3} \simeq 0.3862$\\
$X_{34}$ & $0$ & $-1$ & $R_{6} \simeq 0.3617$\\
$X_{38}$ & $-1$ & $+1$ & $R_{4} \simeq 0.2532$\\
$X_{41}$ & $-1$ & $0$ & $R_{4} + R_{5} \simeq 0.6581$\\
$X_{46}$ & $-1$ & $+1$ & $R_{3} \simeq 0.3862$\\
$X_{53}$ & $0$ & $-1$ & $R_{5} \simeq 0.4049$\\
$X_{54}$ & $+2$ & $0$ & $R_{1} + R_{2} \simeq 0.5940$\\
$X^1_{81}$ & $0$ & $-2$ & $R_{5} + R_{6} \simeq 0.7666$\\
$X^2_{81}$ & $+2$ & $0$ & $R_{1} + R_{2} \simeq 0.5940$\\\hline
$X_{12}$ & $+1$ & $+1$ & $R_{1} + R_{2} + R_{4} \simeq 0.8472$\\
$X_{19}$ & $-1$ & $0$ & $R_{4} + R_{5} \simeq 0.6581$\\
$X_{27}$ & $0$ & $-1$ & $R_{5} \simeq 0.4049$\\
$X_{59}$ & $0$ & $+1$ & $R_{1} + R_{3} \simeq 0.6597$\\
$X_{65}$ & $-1$ & $-1$ & $R_{4} + R_{5} + R_{6} \simeq 1.0198$\\
$X_{67}$ & $+1$ & $0$ & $R_{1} \simeq 0.2736$\\
$X_{71}$ & $-1$ & $0$ & $R_{3} + R_{6} \simeq 0.7479$\\
$X_{75}$ & $0$ & $+1$ & $R_{2} + R_{4} \simeq 0.5737$\\
$X_{92}$ & $0$ & $-1$ & $R_{6} \simeq 0.3617$\\
$X_{96}$ & $+1$ & $0$ & $R_{2} \simeq 0.3205$\\\hline
\end{tabular}
\end{tabular}
}
\caption{
The charges under the global symmetry on the set of chiral fields 
shared between toric phases 5a and 5b of $H_{1,1,2,1}$, including the $U(1)_R$ charges, which are expressed in terms of $U(1)_R$ charges $R_a$
corresponding to the extremal GLSM fields $p_a$. 
}
\label{tab:models_5_Rcharges}
\end{table}
%---------------------------------------------------- 

\paragraph{Mesonic Moduli Spaces.}
Applying the forward algorithm to the two brane tilings, 
we compute the $P$-matrices encoding their perfect matchings.
Both brane tilings admit $36$ perfect matchings, 
consisting of the extremal perfect matchings $p_1, \dots, p_6$ 
corresponding to the extremal points of the toric diagram of $H_{1,1,2,1}$,
as well as the non-extremal perfect matchings.
The $P$-matrices take the following form,
\beal{es05b07}
P^{(5a)} = \resizebox{0.75\textwidth}{!}{$\left(\begin{tabular}{c | *{6}{c}| *{2}{c}| *{17}{c}| *{11}{c}}
& $p_{1}$ & $p_{2}$ & $p_{3}$ & $p_{4}$ & $p_{5}$ & $p_{6}$ & $q_{1}$ & $q_{2}$ & $u_{1}$ & $u_{2}$ & $u_{3}$ & $u_{4}$ & $u_{5}$ & $u_{6}$ & $u_{7}$ & $u_{8}$ & $u_{9}$ & $u_{10}$ & $u_{11}$ & $u_{12}$ & $u_{13}$ & $u_{14}$ & $u_{15}$ & $u_{16}$ & $u_{17}$ & $s_{1}$ & $s_{2}$ & $s_{3}$ & $s_{4}$ & $s_{5}$ & $s_{6}$ & $s_{7}$ & $s_{8}$ & $s_{9}$ & $s_{10}$ & $s_{11}$\\\hline
$X_{12}$ & 0 & 0 & 0 & 1 & 1 & 1 & 1 & 1 & 0 & 0 & 1 & 0 & 0 & 1 & 0 & 1 & 1 & 1 & 1 & 0 & 0 & 1 & 0 & 0 & 1 & 0 & 1 & 0 & 0 & 0 & 0 & 1 & 0 & 0 & 1 & 0\\
$X_{13}$ & 1 & 1 & 1 & 0 & 0 & 0 & 0 & 0 & 0 & 0 & 0 & 0 & 0 & 0 & 1 & 1 & 0 & 0 & 1 & 0 & 0 & 0 & 0 & 0 & 1 & 1 & 1 & 0 & 0 & 0 & 0 & 1 & 1 & 1 & 1 & 1\\
$X_{15}$ & 0 & 0 & 1 & 0 & 0 & 1 & 1 & 0 & 0 & 0 & 0 & 0 & 0 & 0 & 0 & 0 & 0 & 0 & 0 & 0 & 0 & 0 & 0 & 0 & 1 & 1 & 1 & 0 & 0 & 0 & 0 & 1 & 0 & 1 & 1 & 1\\
$X_{19}$ & 0 & 1 & 0 & 1 & 0 & 0 & 0 & 0 & 0 & 0 & 0 & 0 & 0 & 0 & 0 & 0 & 1 & 1 & 1 & 0 & 0 & 0 & 0 & 0 & 1 & 0 & 0 & 0 & 0 & 0 & 0 & 1 & 0 & 0 & 0 & 0\\
$X_{27}$ & 0 & 1 & 0 & 0 & 0 & 0 & 0 & 0 & 1 & 1 & 0 & 1 & 1 & 0 & 1 & 0 & 0 & 0 & 0 & 1 & 1 & 0 & 0 & 0 & 0 & 1 & 0 & 0 & 0 & 0 & 0 & 0 & 0 & 1 & 0 & 0\\
$X_{28}$ & 0 & 0 & 1 & 0 & 0 & 0 & 0 & 0 & 1 & 0 & 0 & 1 & 0 & 0 & 1 & 0 & 0 & 0 & 0 & 1 & 0 & 0 & 1 & 0 & 0 & 1 & 0 & 1 & 0 & 1 & 0 & 0 & 1 & 1 & 0 & 1\\
$X_{34}$ & 0 & 0 & 0 & 0 & 0 & 1 & 1 & 0 & 0 & 0 & 0 & 1 & 1 & 1 & 0 & 0 & 0 & 1 & 0 & 0 & 0 & 0 & 0 & 0 & 0 & 0 & 0 & 0 & 0 & 1 & 1 & 0 & 0 & 0 & 0 & 0\\
$X_{38}$ & 0 & 0 & 0 & 1 & 0 & 0 & 0 & 0 & 1 & 0 & 1 & 1 & 0 & 1 & 0 & 0 & 1 & 1 & 0 & 1 & 0 & 1 & 1 & 0 & 0 & 0 & 0 & 1 & 0 & 1 & 0 & 0 & 0 & 0 & 0 & 0\\
$X_{41}$ & 0 & 0 & 0 & 1 & 1 & 0 & 0 & 1 & 1 & 1 & 1 & 0 & 0 & 0 & 0 & 0 & 1 & 0 & 0 & 1 & 1 & 1 & 1 & 1 & 0 & 0 & 0 & 1 & 1 & 0 & 0 & 0 & 0 & 0 & 0 & 0\\
$X_{46}$ & 0 & 0 & 1 & 0 & 0 & 0 & 0 & 0 & 1 & 1 & 1 & 0 & 0 & 0 & 0 & 0 & 1 & 0 & 0 & 0 & 0 & 0 & 0 & 0 & 0 & 1 & 1 & 1 & 1 & 0 & 0 & 1 & 0 & 0 & 0 & 0\\
$X_{53}$ & 0 & 0 & 0 & 0 & 1 & 0 & 0 & 1 & 0 & 0 & 0 & 0 & 0 & 0 & 1 & 1 & 0 & 0 & 1 & 0 & 0 & 0 & 0 & 0 & 0 & 0 & 0 & 0 & 0 & 0 & 0 & 0 & 1 & 0 & 0 & 0\\
$X_{54}$ & 1 & 1 & 0 & 0 & 0 & 0 & 0 & 0 & 0 & 0 & 0 & 1 & 1 & 1 & 1 & 1 & 0 & 1 & 1 & 0 & 0 & 0 & 0 & 0 & 0 & 0 & 0 & 0 & 0 & 1 & 1 & 0 & 1 & 0 & 0 & 0\\
$X_{59}$ & 0 & 0 & 1 & 0 & 0 & 1 & 0 & 1 & 0 & 0 & 0 & 0 & 0 & 0 & 0 & 0 & 1 & 1 & 1 & 0 & 0 & 0 & 0 & 0 & 0 & 0 & 0 & 1 & 1 & 1 & 1 & 1 & 1 & 0 & 0 & 0\\
$X_{65}$ & 1 & 1 & 0 & 1 & 0 & 0 & 0 & 0 & 0 & 0 & 0 & 0 & 0 & 0 & 0 & 0 & 0 & 0 & 0 & 1 & 1 & 1 & 1 & 1 & 1 & 0 & 0 & 0 & 0 & 0 & 0 & 0 & 0 & 1 & 1 & 1\\
$X_{67}$ & 0 & 0 & 0 & 0 & 0 & 1 & 0 & 1 & 0 & 0 & 0 & 0 & 0 & 0 & 0 & 0 & 0 & 0 & 0 & 1 & 1 & 1 & 0 & 0 & 0 & 0 & 0 & 0 & 0 & 0 & 0 & 0 & 0 & 1 & 1 & 0\\
$X_{71}$ & 1 & 0 & 1 & 0 & 0 & 0 & 0 & 0 & 0 & 0 & 0 & 0 & 0 & 0 & 0 & 0 & 0 & 0 & 0 & 0 & 0 & 0 & 1 & 1 & 0 & 0 & 0 & 1 & 1 & 1 & 1 & 0 & 1 & 0 & 0 & 1\\
$X_{75}$ & 0 & 0 & 0 & 1 & 1 & 0 & 1 & 0 & 0 & 0 & 0 & 0 & 0 & 0 & 0 & 0 & 0 & 0 & 0 & 0 & 0 & 0 & 1 & 1 & 1 & 0 & 0 & 0 & 0 & 0 & 0 & 0 & 0 & 0 & 0 & 1\\
$X^1_{81}$ & 0 & 0 & 0 & 0 & 1 & 1 & 1 & 1 & 0 & 1 & 0 & 0 & 1 & 0 & 0 & 0 & 0 & 0 & 0 & 0 & 1 & 0 & 0 & 1 & 0 & 0 & 0 & 0 & 1 & 0 & 1 & 0 & 0 & 0 & 0 & 0\\
$X^2_{81}$ & 1 & 1 & 0 & 0 & 0 & 0 & 0 & 0 & 0 & 1 & 0 & 0 & 1 & 0 & 0 & 0 & 0 & 0 & 0 & 0 & 1 & 0 & 0 & 1 & 0 & 0 & 0 & 0 & 1 & 0 & 1 & 0 & 0 & 0 & 0 & 0\\
$X_{92}$ & 1 & 0 & 0 & 0 & 0 & 0 & 0 & 0 & 0 & 0 & 1 & 0 & 0 & 1 & 0 & 1 & 0 & 0 & 0 & 0 & 0 & 1 & 0 & 0 & 0 & 0 & 1 & 0 & 0 & 0 & 0 & 0 & 0 & 0 & 1 & 0\\
$X_{96}$ & 0 & 0 & 0 & 0 & 1 & 0 & 1 & 0 & 1 & 1 & 1 & 1 & 1 & 1 & 1 & 1 & 0 & 0 & 0 & 0 & 0 & 0 & 0 & 0 & 0 & 1 & 1 & 0 & 0 & 0 & 0 & 0 & 0 & 0 & 0 & 0\\
\end{tabular}\right)$}
\eea
and
\beal{es05b08}
P^{(5b)} = \resizebox{0.75\textwidth}{!}{$\left(\begin{tabular}{c | *{6}{c}| *{2}{c}| *{17}{c}| *{11}{c}}
& $p_{1}$ & $p_{2}$ & $p_{3}$ & $p_{4}$ & $p_{5}$ & $p_{6}$ & $q_{1}$ & $q_{2}$ & $u_{1}$ & $u_{2}$ & $u_{3}$ & $u_{4}$ & $u_{5}$ & $u_{6}$ & $u_{7}$ & $u_{8}$ & $u_{9}$ & $u_{10}$ & $u_{11}$ & $u_{12}$ & $u_{13}$ & $u_{14}$ & $u_{15}$ & $u_{16}$ & $u_{17}$ & $s_{1}$ & $s_{2}$ & $s_{3}$ & $s_{4}$ & $s_{5}$ & $s_{6}$ & $s_{7}$ & $s_{8}$ & $s_{9}$ & $s_{10}$ & $s_{11}$\\\hline
$X_{12}$ & 1 & 1 & 0 & 1 & 0 & 0 & 0 & 0 & 0 & 0 & 1 & 1 & 0 & 0 & 0 & 1 & 1 & 1 & 1 & 0 & 1 & 0 & 0 & 0 & 1 & 0 & 1 & 0 & 0 & 0 & 0 & 1 & 0 & 0 & 1 & 0\\
$X_{13}$ & 1 & 1 & 1 & 0 & 0 & 0 & 0 & 0 & 0 & 0 & 0 & 0 & 0 & 0 & 1 & 1 & 0 & 0 & 1 & 0 & 0 & 0 & 0 & 0 & 1 & 1 & 1 & 0 & 0 & 0 & 0 & 1 & 1 & 1 & 1 & 1\\
$X_{15}$ & 0 & 0 & 1 & 0 & 0 & 1 & 1 & 0 & 0 & 0 & 0 & 0 & 0 & 0 & 0 & 0 & 0 & 0 & 0 & 0 & 0 & 0 & 0 & 0 & 1 & 1 & 1 & 0 & 0 & 0 & 0 & 1 & 0 & 1 & 1 & 1\\
$X_{19}$ & 0 & 0 & 0 & 1 & 1 & 0 & 1 & 0 & 0 & 0 & 0 & 0 & 0 & 0 & 0 & 0 & 1 & 1 & 1 & 0 & 0 & 0 & 0 & 0 & 1 & 0 & 0 & 0 & 0 & 0 & 0 & 1 & 0 & 0 & 0 & 0\\
$X_{27}$ & 0 & 0 & 0 & 0 & 1 & 0 & 1 & 0 & 1 & 1 & 0 & 0 & 1 & 1 & 1 & 0 & 0 & 0 & 0 & 1 & 0 & 0 & 1 & 0 & 0 & 1 & 0 & 0 & 0 & 0 & 0 & 0 & 0 & 1 & 0 & 0\\
$X_{28}$ & 0 & 0 & 1 & 0 & 0 & 0 & 0 & 0 & 1 & 1 & 0 & 0 & 0 & 0 & 1 & 0 & 0 & 0 & 0 & 1 & 0 & 1 & 0 & 0 & 0 & 1 & 0 & 1 & 1 & 0 & 0 & 0 & 1 & 1 & 0 & 1\\
$X_{34}$ & 0 & 0 & 0 & 0 & 0 & 1 & 1 & 0 & 1 & 0 & 1 & 0 & 1 & 0 & 0 & 0 & 1 & 0 & 0 & 0 & 0 & 0 & 0 & 0 & 0 & 0 & 0 & 1 & 0 & 1 & 0 & 0 & 0 & 0 & 0 & 0\\
$X_{38}$ & 0 & 0 & 0 & 1 & 0 & 0 & 0 & 0 & 1 & 1 & 1 & 1 & 0 & 0 & 0 & 0 & 1 & 1 & 0 & 1 & 1 & 1 & 0 & 0 & 0 & 0 & 0 & 1 & 1 & 0 & 0 & 0 & 0 & 0 & 0 & 0\\
$X_{41}$ & 0 & 0 & 0 & 1 & 1 & 0 & 0 & 1 & 0 & 1 & 0 & 1 & 0 & 1 & 0 & 0 & 0 & 1 & 0 & 1 & 1 & 1 & 1 & 1 & 0 & 0 & 0 & 0 & 1 & 0 & 1 & 0 & 0 & 0 & 0 & 0\\
$X_{46}$ & 0 & 0 & 1 & 0 & 0 & 0 & 0 & 0 & 0 & 1 & 0 & 1 & 0 & 1 & 0 & 0 & 0 & 1 & 0 & 0 & 0 & 0 & 0 & 0 & 0 & 1 & 1 & 0 & 1 & 0 & 1 & 1 & 0 & 0 & 0 & 0\\
$X_{53}$ & 0 & 0 & 0 & 0 & 1 & 0 & 0 & 1 & 0 & 0 & 0 & 0 & 0 & 0 & 1 & 1 & 0 & 0 & 1 & 0 & 0 & 0 & 0 & 0 & 0 & 0 & 0 & 0 & 0 & 0 & 0 & 0 & 1 & 0 & 0 & 0\\
$X_{54}$ & 1 & 1 & 0 & 0 & 0 & 0 & 0 & 0 & 1 & 0 & 1 & 0 & 1 & 0 & 1 & 1 & 1 & 0 & 1 & 0 & 0 & 0 & 0 & 0 & 0 & 0 & 0 & 1 & 0 & 1 & 0 & 0 & 1 & 0 & 0 & 0\\
$X_{59}$ & 1 & 0 & 1 & 0 & 0 & 0 & 0 & 0 & 0 & 0 & 0 & 0 & 0 & 0 & 0 & 0 & 1 & 1 & 1 & 0 & 0 & 0 & 0 & 0 & 0 & 0 & 0 & 1 & 1 & 1 & 1 & 1 & 1 & 0 & 0 & 0\\
$X_{65}$ & 0 & 0 & 0 & 1 & 1 & 1 & 1 & 1 & 0 & 0 & 0 & 0 & 0 & 0 & 0 & 0 & 0 & 0 & 0 & 1 & 1 & 1 & 1 & 1 & 1 & 0 & 0 & 0 & 0 & 0 & 0 & 0 & 0 & 1 & 1 & 1\\
$X_{67}$ & 1 & 0 & 0 & 0 & 0 & 0 & 0 & 0 & 0 & 0 & 0 & 0 & 0 & 0 & 0 & 0 & 0 & 0 & 0 & 1 & 1 & 0 & 1 & 0 & 0 & 0 & 0 & 0 & 0 & 0 & 0 & 0 & 0 & 1 & 1 & 0\\
$X_{71}$ & 0 & 0 & 1 & 0 & 0 & 1 & 0 & 1 & 0 & 0 & 0 & 0 & 0 & 0 & 0 & 0 & 0 & 0 & 0 & 0 & 0 & 1 & 0 & 1 & 0 & 0 & 0 & 1 & 1 & 1 & 1 & 0 & 1 & 0 & 0 & 1\\
$X_{75}$ & 0 & 1 & 0 & 1 & 0 & 0 & 0 & 0 & 0 & 0 & 0 & 0 & 0 & 0 & 0 & 0 & 0 & 0 & 0 & 0 & 0 & 1 & 0 & 1 & 1 & 0 & 0 & 0 & 0 & 0 & 0 & 0 & 0 & 0 & 0 & 1\\
$X^1_{81}$ & 0 & 0 & 0 & 0 & 1 & 1 & 1 & 1 & 0 & 0 & 0 & 0 & 1 & 1 & 0 & 0 & 0 & 0 & 0 & 0 & 0 & 0 & 1 & 1 & 0 & 0 & 0 & 0 & 0 & 1 & 1 & 0 & 0 & 0 & 0 & 0\\
$X^2_{81}$ & 1 & 1 & 0 & 0 & 0 & 0 & 0 & 0 & 0 & 0 & 0 & 0 & 1 & 1 & 0 & 0 & 0 & 0 & 0 & 0 & 0 & 0 & 1 & 1 & 0 & 0 & 0 & 0 & 0 & 1 & 1 & 0 & 0 & 0 & 0 & 0\\
$X_{92}$ & 0 & 0 & 0 & 0 & 0 & 1 & 0 & 1 & 0 & 0 & 1 & 1 & 0 & 0 & 0 & 1 & 0 & 0 & 0 & 0 & 1 & 0 & 0 & 0 & 0 & 0 & 1 & 0 & 0 & 0 & 0 & 0 & 0 & 0 & 1 & 0\\
$X_{96}$ & 0 & 1 & 0 & 0 & 0 & 0 & 0 & 0 & 1 & 1 & 1 & 1 & 1 & 1 & 1 & 1 & 0 & 0 & 0 & 0 & 0 & 0 & 0 & 0 & 0 & 1 & 1 & 0 & 0 & 0 & 0 & 0 & 0 & 0 & 0 & 0\\
\end{tabular}\right)$}
~.~
\eea
From the $P$-matrices in \eref{es05b07} and \eref{es05b08}, 
the forward algorithm yields identical $G_t$-matrices for the two toric phases, 
which take the form,
\beal{es05b09}
G_t^{(5a)} = G_t^{(5b)} = \small{\setlength{\tabcolsep}{1pt}\left(\begin{tabular}{ *{6}{c}| *{2}{c}| *{3}{c}| *{3}{c}}
$p_{1}$ & $p_{2}$ & $p_{3}$ & $p_{4}$ & $p_{5}$ & $p_{6}$ & $q_{1}$ & $q_{2}$ & $u_{1}$ & $\cdots$ & $u_{17}$ &$s_{1}$ & $\cdots$ & $s_{11}$\\\hline
2 & 2 & 1 & 1 & 0 & 0 & 0 & 0 & 1 & $\cdots$ & 1 & 1 & $\cdots$ & 1 \\
2 & 1 & 3 & 0 & 0 & 2 & 1 & 1 & 1 & $\cdots$ & 1 & 2 & $\cdots$ & 2 \\
1 & 1 & 1 & 1 & 1 & 1 & 1 & 1 & 1 & $\cdots$ & 1 & 1 & $\cdots$ & 1 \\
\end{tabular}\right)}
~.~
\eea
The toric diagrams obtained from the two brane tilings therefore coincide,
which verifies that the brane tilings of toric phases 5a and 5b 
correspond to the same toric Calabi-Yau 3-fold $H_{1,1,2,1}$, 
as required for the quiver-invariant duality realized by the tilting mutation.
\\

%=================================================================
\subsection{$H_{1,1,2,1}$ 7a and 7b \label{sec:model7a7b}} % family C
%=================================================================

%---------------------------------------------------- 
\begin{figure}[htbp]
\centering
\includegraphics[width=\textwidth]{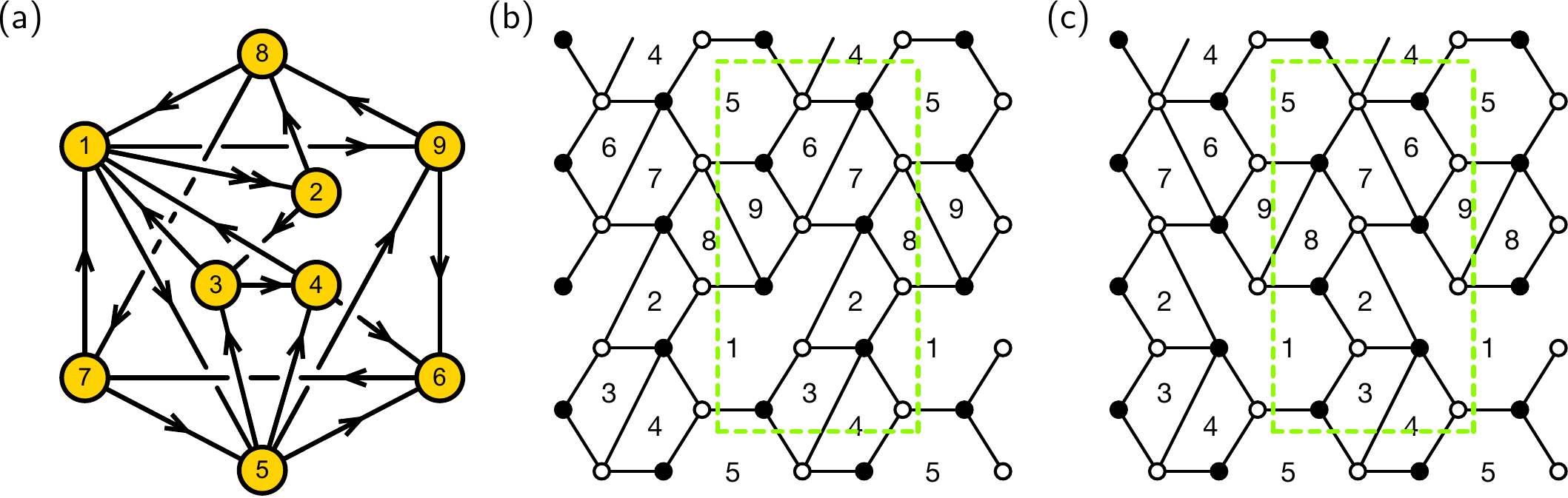}
\caption{
(a) The quiver shared between the two brane tilings in (b) and (c) corresponding to the two toric phases 7a and 7b of $H_{1,1,2,1}$,
respectively. 
}
\label{fig_doublet_7a_7b}
\end{figure}
%---------------------------------------------------- 

The two toric phases 7a and 7b of the $H_{1,1,2,1}$ model 
share the same quiver, which is shown in \fref{fig_doublet_7a_7b}(a).
The corresponding brane tilings are illustrated in 
\fref{fig_doublet_7a_7b}(b) and \fref{fig_doublet_7a_7b}(c), respectively,
and differ in their superpotentials, 
which are given by,
\beal{es05c01}
W_{7a} = 
&&
X_{12}^{2} X_{23} X_{31}
+ X_{12}^{1} X_{28} X_{81} 
+ X_{15} X_{54} X_{41}
+ X_{19} X_{96} X_{67} X_{71}
\nn\\
&&
+ X_{34} X_{46} X_{65} X_{53}
+ X_{59} X_{98} X_{87} X_{75}
- X_{12}^{2} X_{28} X_{87} X_{71}
- X_{12}^{1} X_{23} X_{34} X_{41}
\nn\\
&&
- X_{15} X_{53} X_{31}
- X_{19} X_{98} X_{81}
- X_{46} X_{67} X_{75} X_{54}
- X_{59} X_{96} X_{65}
~,~
\eea
and
\beal{es05c02}
W_{7b} = 
&&
X_{12}^{1} X_{28} X_{87} X_{71}
+ X_{12}^{2} X_{23} X_{31}
+ X_{15} X_{54} X_{41}
+ X_{19} X_{98} X_{81}
\nn\\
&&
+ X_{34} X_{46} X_{67} X_{75} X_{53}
+ X_{59} X_{96} X_{65}
- X_{12}^{1} X_{23} X_{34} X_{41}
- X_{12}^{2} X_{28} X_{81}
\nn\\
&&
- X_{15} X_{53} X_{31}
- X_{19} X_{96} X_{67} X_{71}
- X_{46} X_{65} X_{54}
- X_{59} X_{98} X_{87} X_{75}
~.~
\eea

\begin{figure}[htbp]
\centering
\includegraphics[width=0.7\textwidth]{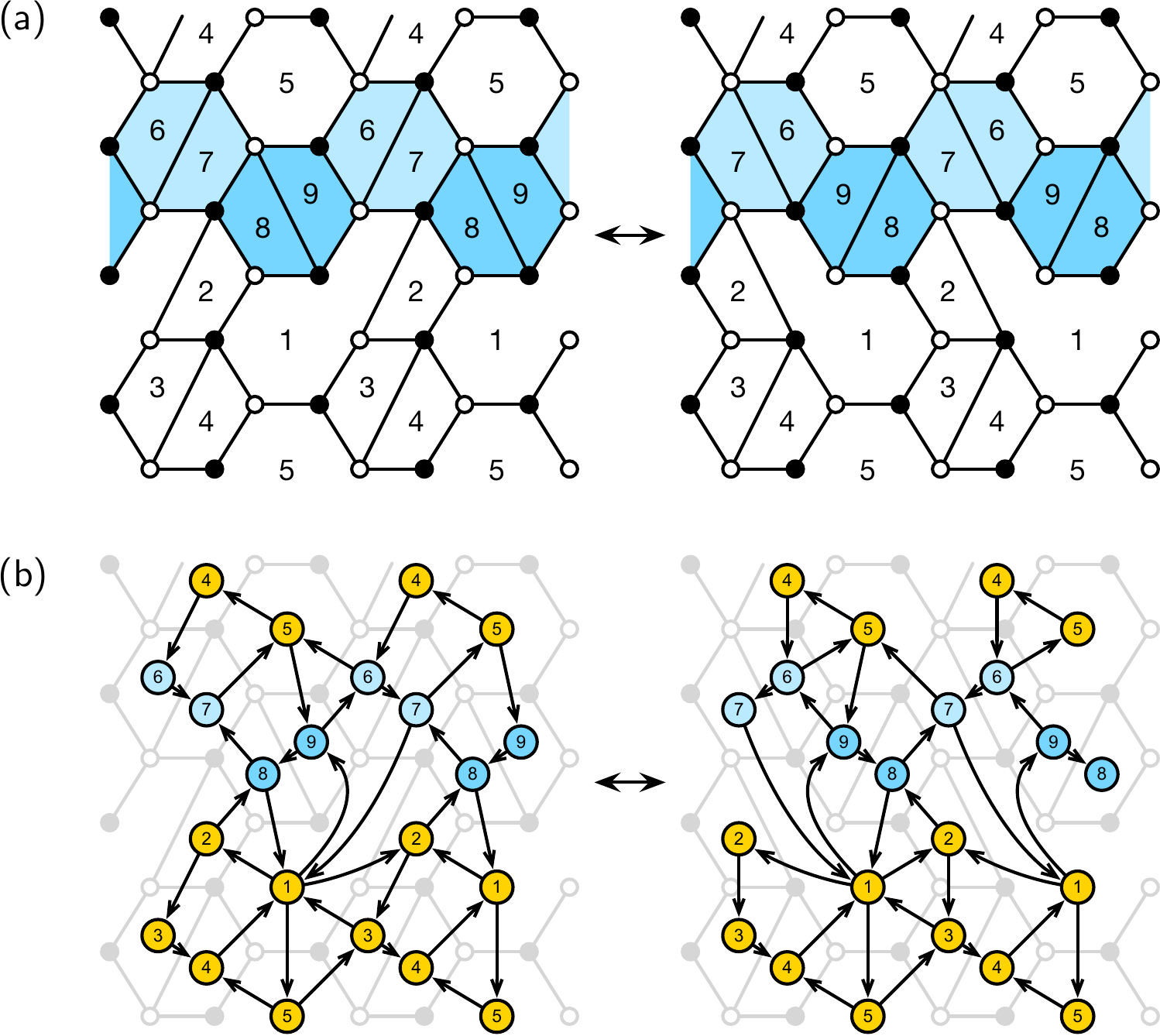}
\caption{
(a) The tilting mutation relating the brane tilings of the two toric phases 7a and 7b of $H_{1,1,2,1}$,
and (b) its effect on the corresponding periodic quivers. 
}
\label{fig_tilting_7a_7b}
\end{figure}

\paragraph{Tilting Mutation.}
By comparing the superpotentials in \eref{es05c01} and \eref{es05c02},
we identify the tilting mutation relating toric phase 7a to 7b 
as belonging to tilting mutation family C introduced in Section~\ref{sec:mutfam_c},
where the quiver nodes $(a,b,c,d,e,f)$ of the mutation region 
are identified with the quiver nodes $(6,7,8,9,1,5)$ of the two toric phases.
Under this identification, the superpotential $W_{7a}$ takes the general form of $W$ 
for tilting mutation family C, while $W_{7b}$ takes the general form of $W^\prime$.
The part of the superpotential that is
unaffected by the tilting mutation
is given by,
\beal{es05c03}
&&
W_0 = 
X_{12}^{2} X_{23} X_{31}
+X_{15} X_{54} X_{41}
- X_{15} X_{53} X_{31}
- X_{12}^{1} X_{23} X_{34} X_{41}
~.~
\eea
The superpotential terms whose signs are reversed under the tilting mutation, 
are given by,
\beal{es05c04}
\Omega_{\text{mut}} =
X_{19} X_{96} X_{67} X_{71}
+ X_{59} X_{98} X_{87} X_{75}
- X_{19} X_{98} X_{81}
- X_{59} X_{96} X_{65}
~,~
\eea
and the exterior paths take the form,
\beal{es05c05}
&&
P_{56} = X_{53} X_{34} X_{46} 
~,~
Q_{56} = X_{54} X_{46} 
~,~
P_{18} = X_{12}^{1} X_{28} 
~,~
Q_{18} = X_{12}^{2} X_{28} 
~.~
\eea
Accordingly, the paths exchanged by the tilting mutation are given by,
\beal{es05c06}
U_{56} = X_{65}
~,~
U_{18} = X_{81}
~,~
V_{56} = X_{67} X_{75}
~,~
V_{18} = X_{87} X_{71}
~,~
\eea
which is
in agreement with the general form in Section~\ref{sec:mutfam_c} for tilting mutation family C,
such that the tilting mutation maps the superpotential $W_{7a}$ in \eref{es05c01} 
to $W_{7b}$ in \eref{es05c02}, and vice versa.
The action of the tilting mutation on the brane tiling 
and on the corresponding periodic quiver 
is illustrated in \fref{fig_tilting_7a_7b}.

%---------------------------------------------------- 
%---------------------------------------------------- 

\paragraph{Global Symmetry Charges.}
\tref{tab:models_7_Rcharges} summarizes the charges under the global symmetry
carried by the chiral fields of the two toric phases,
where the $U(1)_R$ charges are expressed in terms of
the $U(1)_R$ charges $R_a$ of the extremal GLSM fields $p_a$ in \tref{tab:extremal_charges}.
As shown in \tref{tab:models_7_Rcharges},
the chiral fields along the exterior paths
$P_{56}$, $Q_{56}$, $P_{18}$ and $Q_{18}$ in \eref{es05c05}
carry identical global symmetry charges in the two toric phases.
In contrast, the interior paths $U_{56}$ and $V_{56}$,
as well as $U_{18}$ and $V_{18}$ in \eref{es05c06},
exchange their global symmetry charges under the tilting mutation,
as can be seen from the charges carried by the chiral fields
in the mutation region in \tref{tab:models_7_Rcharges}.

%---------------------------------------------------- 
\begin{table}[h]
\setlength{\tabcolsep}{2pt}
\centering
\footnotesize{
\begin{tabular}[t]{c}
Phase 7a \\[4pt]
\begin{tabular}{| c | c | c | r |}
\hline
\; & $U(1)_{f_1}$ & $U(1)_{f_2}$ & $U(1)_R$\\\hline\hline
$X_{15}$ & $-1$ & $0$ & $R_{3} + R_{6} \simeq 0.7479$\\
$X_{23}$ & $-1$ & $+1$ & $R_{3} \simeq 0.3862$\\
$X_{28}$ & $-1$ & $+1$ & $R_{4} \simeq 0.2532$\\
$X_{31}$ & $+1$ & $+1$ & $R_{1} + R_{2} + R_{4} \simeq 0.8472$\\
$X_{34}$ & $0$ & $-1$ & $R_{6} \simeq 0.3617$\\
$X_{41}$ & $-1$ & $0$ & $R_{4} + R_{5} \simeq 0.6581$\\
$X_{46}$ & $-1$ & $+1$ & $R_{3} \simeq 0.3862$\\
$X_{53}$ & $0$ & $-1$ & $R_{5} \simeq 0.4049$\\
$X_{54}$ & $+2$ & $0$ & $R_{1} + R_{2} \simeq 0.5940$\\
$X^1_{12}$ & $+2$ & $0$ & $R_{1} + R_{2} \simeq 0.5940$\\
$X^2_{12}$ & $0$ & $-2$ & $R_{5} + R_{6} \simeq 0.7666$\\
\hline
$X_{19}$ & $0$ & $+1$ & $R_{2} + R_{4} \simeq 0.5737$\\
$X_{59}$ & $-1$ & $0$ & $R_{3} + R_{6} \simeq 0.7479$\\
$X_{65}$ & $+1$ & $+1$ & $R_{1} + R_{2} + R_{4} \simeq 0.8472$\\
$X_{67}$ & $0$ & $-1$ & $R_{6} \simeq 0.3617$\\
$X_{71}$ & $0$ & $+1$ & $R_{1} + R_{3} \simeq 0.6597$\\
$X_{75}$ & $-1$ & $0$ & $R_{4} + R_{5} \simeq 0.6581$\\
$X_{81}$ & $-1$ & $-1$ & $R_{3} + R_{5} + R_{6} \simeq 1.1528$\\
$X_{87}$ & $+1$ & $0$ & $R_{2} \simeq 0.3205$\\
$X_{96}$ & $0$ & $-1$ & $R_{5} \simeq 0.4049$\\
$X_{98}$ & $+1$ & $0$ & $R_{1} \simeq 0.2736$\\\hline
\end{tabular}
\end{tabular}
}
\hspace{6pt}
\footnotesize{
\begin{tabular}[t]{c}
Phase 7b \\[4pt]
\begin{tabular}{| c | c | c | r |}
\hline
\; & $U(1)_{f_1}$ & $U(1)_{f_2}$ & $U(1)_R$\\\hline\hline
$X_{15}$ & $-1$ & $0$ & $R_{3} + R_{6} \simeq 0.7479$\\
$X_{23}$ & $-1$ & $+1$ & $R_{3} \simeq 0.3862$\\
$X_{28}$ & $-1$ & $+1$ & $R_{4} \simeq 0.2532$\\
$X_{31}$ & $+1$ & $+1$ & $R_{1} + R_{2} + R_{4} \simeq 0.8472$\\
$X_{34}$ & $0$ & $-1$ & $R_{6} \simeq 0.3617$\\
$X_{41}$ & $-1$ & $0$ & $R_{4} + R_{5} \simeq 0.6581$\\
$X_{46}$ & $-1$ & $+1$ & $R_{3} \simeq 0.3862$\\
$X_{53}$ & $0$ & $-1$ & $R_{5} \simeq 0.4049$\\
$X_{54}$ & $+2$ & $0$ & $R_{1} + R_{2} \simeq 0.5940$\\
$X^1_{12}$ & $+2$ & $0$ & $R_{1} + R_{2} \simeq 0.5940$\\
$X^2_{12}$ & $0$ & $-2$ & $R_{5} + R_{6} \simeq 0.7666$\\
\hline
$X_{19}$ & $-1$ & $0$ & $R_{4} + R_{5} \simeq 0.6581$\\
$X_{59}$ & $0$ & $+1$ & $R_{1} + R_{3} \simeq 0.6597$\\
$X_{65}$ & $-1$ & $-1$ & $R_{4} + R_{5} + R_{6} \simeq 1.0198$\\
$X_{67}$ & $+1$ & $0$ & $R_{1} \simeq 0.2736$\\
$X_{71}$ & $-1$ & $0$ & $R_{3} + R_{6} \simeq 0.7479$\\
$X_{75}$ & $0$ & $+1$ & $R_{2} + R_{4} \simeq 0.5737$\\
$X_{81}$ & $+1$ & $+1$ & $R_{1} + R_{2} + R_{3} \simeq 0.9802$\\
$X_{87}$ & $0$ & $-1$ & $R_{5} \simeq 0.4049$\\
$X_{96}$ & $+1$ & $0$ & $R_{2} \simeq 0.3205$\\
$X_{98}$ & $0$ & $-1$ & $R_{6} \simeq 0.3617$\\\hline
\end{tabular}
\end{tabular}
}
\caption{
The charges under the global symmetry on the set of chiral fields 
shared between toric phases 7a and 7b of $H_{1,1,2,1}$, including the $U(1)_R$ charges, which are expressed in terms of $U(1)_R$ charges $R_a$
corresponding to the extremal GLSM fields $p_a$. 
}
\label{tab:models_7_Rcharges}
\end{table}
%---------------------------------------------------- 

\paragraph{Mesonic Moduli Spaces.}
Applying the forward algorithm to the two brane tilings, 
we compute the $P$-matrices encoding their perfect matchings.
Both brane tilings admit $36$ perfect matchings, 
consisting of the extremal perfect matchings $p_1, \dots, p_6$ 
corresponding to the extremal points of the toric diagram of $H_{1,1,2,1}$,
as well as the non-extremal perfect matchings.
The $P$-matrices take the following form,
\beal{es05c07}
P^{(7a)} = \resizebox{0.75\textwidth}{!}{$\left(\begin{tabular}{c | *{6}{c}| *{2}{c}| *{16}{c}| *{12}{c}}
& $p_{1}$ & $p_{2}$ & $p_{3}$ & $p_{4}$ & $p_{5}$ & $p_{6}$ & $q_{1}$ & $q_{2}$ & $u_{1}$ & $u_{2}$ & $u_{3}$ & $u_{4}$ & $u_{5}$ & $u_{6}$ & $u_{7}$ & $u_{8}$ & $u_{9}$ & $u_{10}$ & $u_{11}$ & $u_{12}$ & $u_{13}$ & $u_{14}$ & $u_{15}$ & $u_{16}$ & $s_{1}$ & $s_{2}$ & $s_{3}$ & $s_{4}$ & $s_{5}$ & $s_{6}$ & $s_{7}$ & $s_{8}$ & $s_{9}$ & $s_{10}$ & $s_{11}$ & $s_{12}$\\\hline
$X^1_{12}$ & 1 & 1 & 0 & 0 & 0 & 0 & 0 & 0 & 1 & 0 & 0 & 0 & 0 & 0 & 0 & 0 & 0 & 0 & 0 & 1 & 0 & 0 & 1 & 0 & 1 & 0 & 0 & 0 & 0 & 1 & 0 & 1 & 0 & 0 & 0 & 0 \\
$X^2_{12}$ & 0 & 0 & 0 & 0 & 1 & 1 & 1 & 1 & 1 & 0 & 0 & 0 & 0 & 0 & 0 & 0 & 0 & 0 & 0 & 1 & 0 & 0 & 1 & 0 & 1 & 0 & 0 & 0 & 0 & 1 & 0 & 1 & 0 & 0 & 0 & 0 \\
$X_{15}$ & 0 & 0 & 1 & 0 & 0 & 1 & 1 & 0 & 1 & 1 & 0 & 0 & 0 & 0 & 0 & 0 & 0 & 0 & 0 & 0 & 0 & 0 & 0 & 0 & 1 & 1 & 1 & 1 & 1 & 1 & 1 & 1 & 1 & 0 & 0 & 0 \\
$X_{19}$ & 0 & 1 & 0 & 1 & 0 & 0 & 0 & 0 & 1 & 1 & 1 & 1 & 0 & 0 & 0 & 0 & 0 & 0 & 0 & 1 & 1 & 0 & 0 & 0 & 1 & 1 & 0 & 0 & 0 & 0 & 0 & 0 & 0 & 0 & 0 & 0 \\
$X_{23}$ & 0 & 0 & 1 & 0 & 0 & 0 & 0 & 0 & 0 & 1 & 0 & 0 & 0 & 0 & 0 & 0 & 0 & 0 & 0 & 0 & 1 & 1 & 0 & 1 & 0 & 1 & 1 & 1 & 1 & 0 & 1 & 0 & 1 & 0 & 0 & 1 \\
$X_{28}$ & 0 & 0 & 0 & 1 & 0 & 0 & 0 & 0 & 0 & 1 & 1 & 1 & 0 & 0 & 0 & 0 & 1 & 1 & 1 & 0 & 1 & 0 & 0 & 1 & 0 & 1 & 0 & 0 & 0 & 0 & 1 & 0 & 1 & 0 & 0 & 0 \\
$X_{31}$ & 1 & 1 & 0 & 1 & 0 & 0 & 0 & 0 & 0 & 0 & 1 & 1 & 1 & 1 & 1 & 1 & 1 & 1 & 1 & 0 & 0 & 0 & 0 & 0 & 0 & 0 & 0 & 0 & 0 & 0 & 0 & 0 & 0 & 1 & 1 & 0 \\
$X_{34}$ & 0 & 0 & 0 & 0 & 0 & 1 & 1 & 0 & 0 & 0 & 1 & 0 & 0 & 0 & 1 & 0 & 0 & 1 & 0 & 0 & 0 & 0 & 0 & 0 & 0 & 0 & 0 & 0 & 0 & 0 & 0 & 0 & 0 & 1 & 0 & 0 \\
$X_{41}$ & 0 & 0 & 0 & 1 & 1 & 0 & 0 & 1 & 0 & 0 & 0 & 1 & 1 & 1 & 0 & 1 & 1 & 0 & 1 & 0 & 0 & 0 & 0 & 0 & 0 & 0 & 0 & 0 & 0 & 0 & 0 & 0 & 0 & 0 & 1 & 0 \\
$X_{46}$ & 0 & 0 & 1 & 0 & 0 & 0 & 0 & 0 & 0 & 0 & 0 & 1 & 0 & 0 & 0 & 1 & 0 & 0 & 1 & 0 & 0 & 0 & 0 & 0 & 1 & 1 & 0 & 0 & 1 & 0 & 0 & 1 & 1 & 0 & 1 & 0 \\
$X_{53}$ & 0 & 0 & 0 & 0 & 1 & 0 & 0 & 1 & 0 & 0 & 0 & 0 & 0 & 0 & 0 & 0 & 0 & 0 & 0 & 1 & 1 & 1 & 1 & 1 & 0 & 0 & 0 & 0 & 0 & 0 & 0 & 0 & 0 & 0 & 0 & 1 \\
$X_{54}$ & 1 & 1 & 0 & 0 & 0 & 0 & 0 & 0 & 0 & 0 & 1 & 0 & 0 & 0 & 1 & 0 & 0 & 1 & 0 & 1 & 1 & 1 & 1 & 1 & 0 & 0 & 0 & 0 & 0 & 0 & 0 & 0 & 0 & 1 & 0 & 1 \\
$X_{59}$ & 0 & 0 & 1 & 0 & 0 & 1 & 0 & 1 & 0 & 0 & 1 & 1 & 0 & 0 & 0 & 0 & 0 & 0 & 0 & 1 & 1 & 0 & 0 & 0 & 1 & 1 & 0 & 0 & 0 & 0 & 0 & 0 & 0 & 1 & 1 & 1 \\
$X_{65}$ & 1 & 1 & 0 & 1 & 0 & 0 & 0 & 0 & 1 & 1 & 0 & 0 & 1 & 1 & 0 & 0 & 1 & 0 & 0 & 0 & 0 & 0 & 0 & 0 & 0 & 0 & 1 & 1 & 0 & 1 & 1 & 0 & 0 & 0 & 0 & 0 \\
$X_{67}$ & 0 & 0 & 0 & 0 & 0 & 1 & 0 & 1 & 0 & 0 & 0 & 0 & 0 & 1 & 0 & 0 & 1 & 0 & 0 & 0 & 0 & 0 & 0 & 0 & 0 & 0 & 0 & 1 & 0 & 1 & 1 & 0 & 0 & 0 & 0 & 0 \\
$X_{71}$ & 1 & 0 & 1 & 0 & 0 & 0 & 0 & 0 & 0 & 0 & 0 & 0 & 1 & 0 & 0 & 0 & 0 & 0 & 0 & 0 & 0 & 0 & 0 & 0 & 0 & 0 & 1 & 0 & 0 & 0 & 0 & 0 & 0 & 1 & 1 & 1 \\
$X_{75}$ & 0 & 0 & 0 & 1 & 1 & 0 & 1 & 0 & 1 & 1 & 0 & 0 & 1 & 0 & 0 & 0 & 0 & 0 & 0 & 0 & 0 & 0 & 0 & 0 & 0 & 0 & 1 & 0 & 0 & 0 & 0 & 0 & 0 & 0 & 0 & 0 \\
$X_{81}$ & 0 & 0 & 1 & 0 & 1 & 1 & 1 & 1 & 0 & 0 & 0 & 0 & 1 & 1 & 1 & 1 & 0 & 0 & 0 & 0 & 0 & 1 & 0 & 0 & 0 & 0 & 1 & 1 & 1 & 0 & 0 & 0 & 0 & 1 & 1 & 1 \\
$X_{87}$ & 0 & 1 & 0 & 0 & 0 & 0 & 0 & 0 & 0 & 0 & 0 & 0 & 0 & 1 & 1 & 1 & 0 & 0 & 0 & 0 & 0 & 1 & 0 & 0 & 0 & 0 & 0 & 1 & 1 & 0 & 0 & 0 & 0 & 0 & 0 & 0 \\
$X_{96}$ & 0 & 0 & 0 & 0 & 1 & 0 & 1 & 0 & 0 & 0 & 0 & 0 & 0 & 0 & 1 & 1 & 0 & 1 & 1 & 0 & 0 & 1 & 1 & 1 & 0 & 0 & 0 & 0 & 1 & 0 & 0 & 1 & 1 & 0 & 0 & 0 \\
$X_{98}$ & 1 & 0 & 0 & 0 & 0 & 0 & 0 & 0 & 0 & 0 & 0 & 0 & 0 & 0 & 0 & 0 & 1 & 1 & 1 & 0 & 0 & 0 & 1 & 1 & 0 & 0 & 0 & 0 & 0 & 1 & 1 & 1 & 1 & 0 & 0 & 0 \\
\end{tabular}\right)$}
\eea
and
\beal{es05c08}
P^{(7b)} = \resizebox{0.75\textwidth}{!}{$\left(\begin{tabular}{c | *{6}{c}| *{2}{c}| *{16}{c}| *{12}{c}}
& $p_{1}$ & $p_{2}$ & $p_{3}$ & $p_{4}$ & $p_{5}$ & $p_{6}$ & $q_{1}$ & $q_{2}$ & $u_{1}$ & $u_{2}$ & $u_{3}$ & $u_{4}$ & $u_{5}$ & $u_{6}$ & $u_{7}$ & $u_{8}$ & $u_{9}$ & $u_{10}$ & $u_{11}$ & $u_{12}$ & $u_{13}$ & $u_{14}$ & $u_{15}$ & $u_{16}$ & $s_{1}$ & $s_{2}$ & $s_{3}$ & $s_{4}$ & $s_{5}$ & $s_{6}$ & $s_{7}$ & $s_{8}$ & $s_{9}$ & $s_{10}$ & $s_{11}$ & $s_{12}$\\\hline
$X^1_{12}$ & 1 & 1 & 0 & 0 & 0 & 0 & 0 & 0 & 0 & 0 & 1 & 1 & 0 & 0 & 1 & 0 & 0 & 0 & 0 & 0 & 0 & 0 & 0 & 0 & 0 & 0 & 0 & 1 & 1 & 0 & 0 & 1 & 0 & 0 & 0 & 0\\
$X^2_{12}$ & 0 & 0 & 0 & 0 & 1 & 1 & 1 & 1 & 0 & 0 & 1 & 1 & 0 & 0 & 1 & 0 & 0 & 0 & 0 & 0 & 0 & 0 & 0 & 0 & 0 & 0 & 0 & 1 & 1 & 0 & 0 & 1 & 0 & 0 & 0 & 0\\
$X_{15}$ & 0 & 0 & 1 & 0 & 0 & 1 & 0 & 1 & 0 & 0 & 0 & 0 & 0 & 1 & 1 & 0 & 0 & 0 & 0 & 0 & 0 & 0 & 0 & 0 & 0 & 1 & 1 & 1 & 1 & 1 & 1 & 1 & 1 & 1 & 0 & 0\\
$X_{19}$ & 0 & 0 & 0 & 1 & 1 & 0 & 0 & 1 & 0 & 1 & 0 & 1 & 0 & 1 & 1 & 0 & 1 & 0 & 0 & 1 & 0 & 0 & 0 & 0 & 0 & 0 & 1 & 0 & 1 & 0 & 0 & 0 & 0 & 0 & 0 & 0\\
$X_{23}$ & 0 & 0 & 1 & 0 & 0 & 0 & 0 & 0 & 1 & 1 & 0 & 0 & 1 & 1 & 0 & 0 & 0 & 0 & 0 & 0 & 0 & 0 & 0 & 0 & 1 & 1 & 1 & 0 & 0 & 1 & 1 & 0 & 1 & 1 & 0 & 0\\
$X_{28}$ & 0 & 0 & 0 & 1 & 0 & 0 & 0 & 0 & 1 & 1 & 0 & 0 & 0 & 1 & 0 & 1 & 1 & 0 & 1 & 1 & 0 & 1 & 0 & 0 & 0 & 1 & 1 & 0 & 0 & 0 & 1 & 0 & 0 & 0 & 0 & 0\\
$X_{31}$ & 1 & 1 & 0 & 1 & 0 & 0 & 0 & 0 & 0 & 0 & 0 & 0 & 0 & 0 & 0 & 1 & 1 & 1 & 1 & 1 & 1 & 1 & 1 & 1 & 0 & 0 & 0 & 0 & 0 & 0 & 0 & 0 & 0 & 0 & 1 & 1\\
$X_{34}$ & 0 & 0 & 0 & 0 & 0 & 1 & 0 & 1 & 0 & 0 & 0 & 0 & 0 & 0 & 0 & 1 & 1 & 1 & 0 & 0 & 0 & 0 & 0 & 0 & 0 & 0 & 0 & 0 & 0 & 0 & 0 & 0 & 0 & 0 & 1 & 0\\
$X_{41}$ & 0 & 0 & 0 & 1 & 1 & 0 & 1 & 0 & 0 & 0 & 0 & 0 & 0 & 0 & 0 & 0 & 0 & 0 & 1 & 1 & 1 & 1 & 1 & 1 & 0 & 0 & 0 & 0 & 0 & 0 & 0 & 0 & 0 & 0 & 0 & 1\\
$X_{46}$ & 0 & 0 & 1 & 0 & 0 & 0 & 0 & 0 & 0 & 0 & 0 & 0 & 0 & 0 & 0 & 0 & 0 & 0 & 1 & 1 & 1 & 0 & 0 & 0 & 0 & 1 & 1 & 1 & 1 & 1 & 0 & 0 & 0 & 0 & 0 & 1\\
$X_{53}$ & 0 & 0 & 0 & 0 & 1 & 0 & 1 & 0 & 1 & 1 & 1 & 1 & 1 & 0 & 0 & 0 & 0 & 0 & 0 & 0 & 0 & 0 & 0 & 0 & 1 & 0 & 0 & 0 & 0 & 0 & 0 & 0 & 0 & 0 & 0 & 0\\
$X_{54}$ & 1 & 1 & 0 & 0 & 0 & 0 & 0 & 0 & 1 & 1 & 1 & 1 & 1 & 0 & 0 & 1 & 1 & 1 & 0 & 0 & 0 & 0 & 0 & 0 & 1 & 0 & 0 & 0 & 0 & 0 & 0 & 0 & 0 & 0 & 1 & 0\\
$X_{59}$ & 1 & 0 & 1 & 0 & 0 & 0 & 0 & 0 & 0 & 1 & 0 & 1 & 0 & 0 & 0 & 0 & 1 & 0 & 0 & 1 & 0 & 0 & 0 & 0 & 1 & 0 & 1 & 0 & 1 & 0 & 0 & 0 & 0 & 0 & 1 & 1\\
$X_{65}$ & 0 & 0 & 0 & 1 & 1 & 1 & 1 & 1 & 0 & 0 & 0 & 0 & 0 & 1 & 1 & 0 & 0 & 0 & 0 & 0 & 0 & 1 & 1 & 1 & 0 & 0 & 0 & 0 & 0 & 0 & 1 & 1 & 1 & 1 & 0 & 0\\
$X_{67}$ & 1 & 0 & 0 & 0 & 0 & 0 & 0 & 0 & 0 & 0 & 0 & 0 & 0 & 0 & 0 & 0 & 0 & 0 & 0 & 0 & 0 & 1 & 1 & 0 & 0 & 0 & 0 & 0 & 0 & 0 & 1 & 1 & 1 & 0 & 0 & 0\\
$X_{71}$ & 0 & 0 & 1 & 0 & 0 & 1 & 1 & 0 & 0 & 0 & 0 & 0 & 0 & 0 & 0 & 0 & 0 & 0 & 0 & 0 & 0 & 0 & 0 & 1 & 1 & 0 & 0 & 0 & 0 & 0 & 0 & 0 & 0 & 1 & 1 & 1\\
$X_{75}$ & 0 & 1 & 0 & 1 & 0 & 0 & 0 & 0 & 0 & 0 & 0 & 0 & 0 & 1 & 1 & 0 & 0 & 0 & 0 & 0 & 0 & 0 & 0 & 1 & 0 & 0 & 0 & 0 & 0 & 0 & 0 & 0 & 0 & 1 & 0 & 0\\
$X_{81}$ & 1 & 1 & 1 & 0 & 0 & 0 & 0 & 0 & 0 & 0 & 0 & 0 & 1 & 0 & 0 & 0 & 0 & 1 & 0 & 0 & 1 & 0 & 1 & 1 & 1 & 0 & 0 & 0 & 0 & 1 & 0 & 0 & 1 & 1 & 1 & 1\\
$X_{87}$ & 0 & 0 & 0 & 0 & 1 & 0 & 0 & 1 & 0 & 0 & 0 & 0 & 1 & 0 & 0 & 0 & 0 & 1 & 0 & 0 & 1 & 0 & 1 & 0 & 0 & 0 & 0 & 0 & 0 & 1 & 0 & 0 & 1 & 0 & 0 & 0\\
$X_{96}$ & 0 & 1 & 0 & 0 & 0 & 0 & 0 & 0 & 1 & 0 & 1 & 0 & 1 & 0 & 0 & 1 & 0 & 1 & 1 & 0 & 1 & 0 & 0 & 0 & 0 & 1 & 0 & 1 & 0 & 1 & 0 & 0 & 0 & 0 & 0 & 0\\
$X_{98}$ & 0 & 0 & 0 & 0 & 0 & 1 & 1 & 0 & 1 & 0 & 1 & 0 & 0 & 0 & 0 & 1 & 0 & 0 & 1 & 0 & 0 & 1 & 0 & 0 & 0 & 1 & 0 & 1 & 0 & 0 & 1 & 1 & 0 & 0 & 0 & 0\\
\end{tabular}\right)$}
~.~
\eea
From the $P$-matrices in \eref{es05c07} and \eref{es05c08}, 
the forward algorithm yields identical $G_t$-matrices for the two toric phases, 
which take the form,
\beal{es05c09}
G_t^{(7a)} = G_t^{(7b)} = \small{\setlength{\tabcolsep}{1pt}\left(\begin{tabular}{ *{6}{c}| *{2}{c}| *{3}{c}| *{3}{c}}
$p_{1}$ & $p_{2}$ & $p_{3}$ & $p_{4}$ & $p_{5}$ & $p_{6}$ & $q_{1}$ & $q_{2}$ & $u_{1}$ & $\cdots$ & $u_{16}$ &$s_{1}$ & $\cdots$ & $s_{12}$\\\hline
2 & 2 & 1 & 1 & 0 & 0 & 0 & 0 & 1 & $\cdots$ & 1 & 1 & $\cdots$ & 1 \\
2 & 1 & 3 & 0 & 0 & 2 & 1 & 1 & 1 & $\cdots$ & 1 & 2 & $\cdots$ & 2 \\
1 & 1 & 1 & 1 & 1 & 1 & 1 & 1 & 1 & $\cdots$ & 1 & 1 & $\cdots$ & 1 \\
\end{tabular}\right)}
~.~
\eea
The toric diagrams obtained from the two brane tilings therefore coincide,
which verifies that the brane tilings of toric phases 7a and 7b 
correspond to the same toric Calabi-Yau 3-fold $H_{1,1,2,1}$, 
as required for the quiver-invariant duality realized by the tilting mutation.
\\

%=================================================================
\subsection{$H_{1,1,2,1}$ 14a and 14b \label{sec:model14a14b}} % family C
%=================================================================

%---------------------------------------------------- 
\begin{figure}[htbp]
\centering
\includegraphics[width=\textwidth]{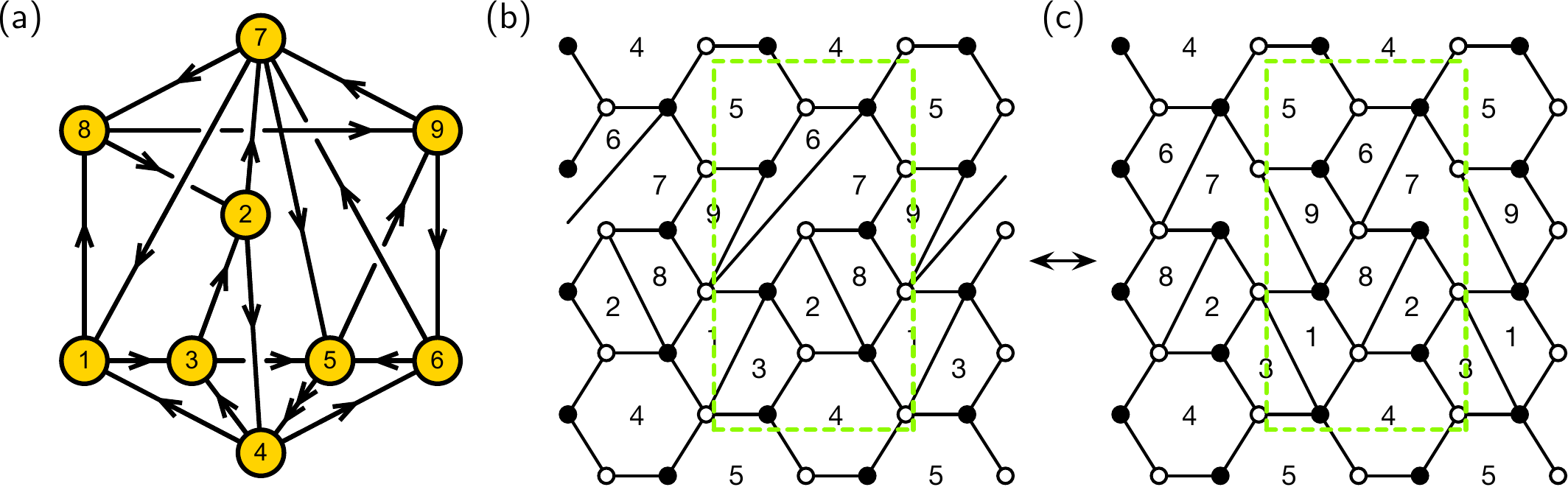}
\caption{
(a) The quiver shared between the two brane tilings in (b) and (c) corresponding to the two toric phases 14a and 14b of $H_{1,1,2,1}$,
respectively. 
}
\label{fig_doublet_14a_14b}
\end{figure}
%---------------------------------------------------- 

The two toric phases 14a and 14b of the $H_{1,1,2,1}$ model 
share the same quiver, which is shown in \fref{fig_doublet_14a_14b}(a).
The corresponding brane tilings are illustrated in 
\fref{fig_doublet_14a_14b}(b) and \fref{fig_doublet_14a_14b}(c), respectively,
and differ in their superpotentials, 
which are given by,
\beal{es05d01}
W_{14a} = 
&&
X_{13} X_{35} X_{54}^{1} X_{41}
+ X_{18} X_{89} X_{96} X_{67} X_{71}
+ X_{24} X_{43} X_{32}
+ X_{27} X_{78} X_{82}
\nn\\
&&
+ X_{46} X_{65} X_{54}^{2}
+ X_{59} X_{97} X_{75}
- X_{13} X_{32} X_{27} X_{71}
- X_{18} X_{82} X_{24} X_{41}
\nn\\
&&
- X_{35} X_{54}^{2} X_{43} 
- X_{46} X_{67} X_{75} X_{54}^{1}
- X_{59} X_{96} X_{65} 
- X_{78} X_{89} X_{97}
~,~
\eea
and
\beal{es05d02}
W_{14b} = 
&&
X_{13} X_{32} X_{27} X_{71}
+ X_{18} X_{82} X_{24} X_{41}
+ X_{35} X_{54}^{1} X_{43}
+ X_{46} X_{65} X_{54}^{2}
\nn\\
&&
+ X_{59} X_{97} X_{75}
+ X_{67} X_{78} X_{89} X_{96}
- X_{13} X_{35} X_{54}^{2} X_{41}
- X_{18} X_{89} X_{97} X_{71}
\nn\\
&&
- X_{24} X_{43} X_{32}
- X_{27} X_{78} X_{82}
- X_{46} X_{67} X_{75} X_{54}^{1}
- X_{59} X_{96} X_{65}
~.~
\eea

\begin{figure}[htbp]
\centering
\includegraphics[width=0.7\textwidth]{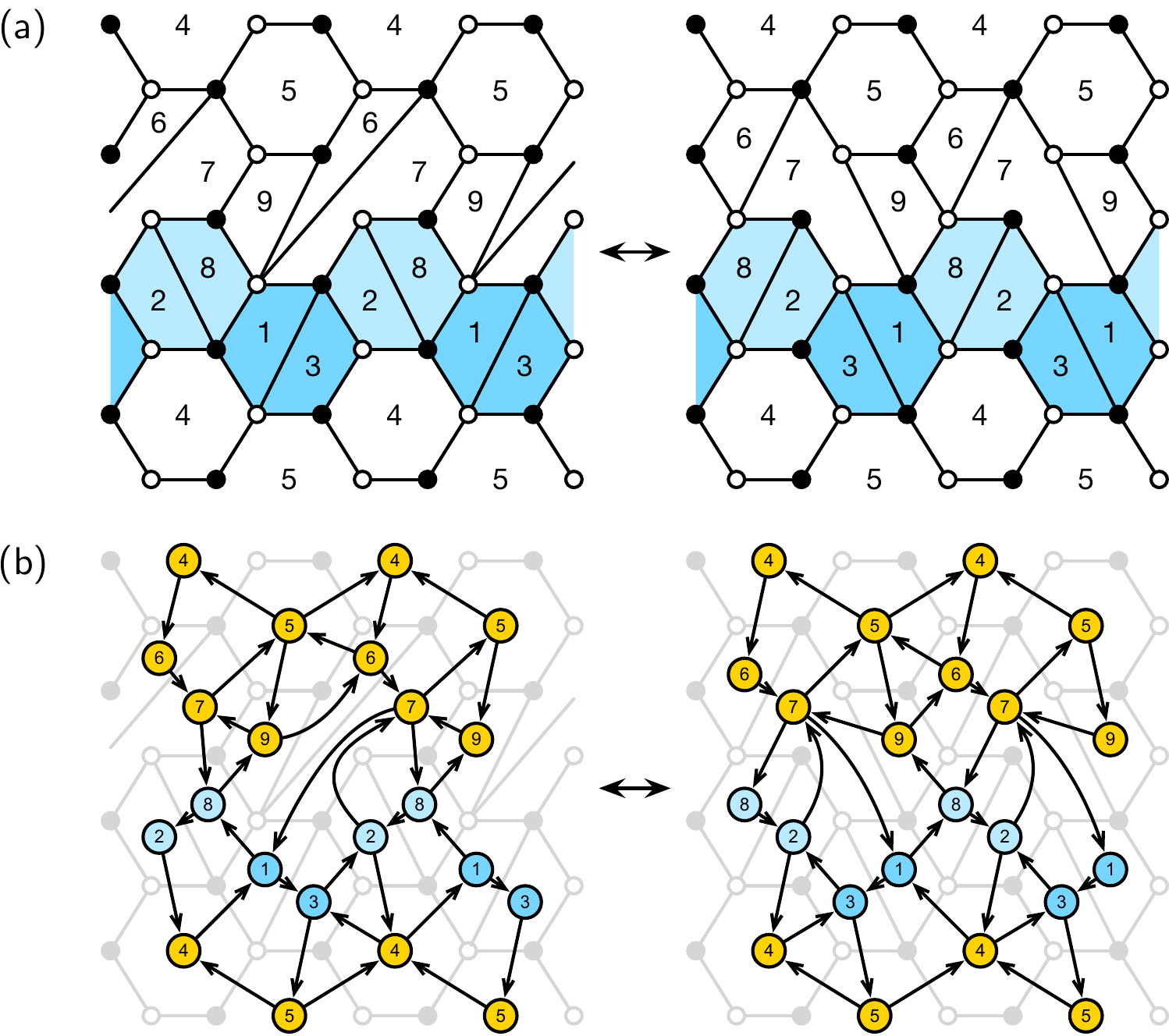}
\caption{
(a) The tilting mutation relating the brane tilings of the two toric phases 14a and 14b of $H_{1,1,2,1}$,
and (b) its effect on the corresponding periodic quivers. 
}
\label{fig_tilting_14a_14b}
\end{figure}

\paragraph{Tilting Mutation.}
By comparing the superpotentials in \eref{es05d01} and \eref{es05d02},
we identify the tilting mutation relating toric phase 14a to 14b 
as belonging to tilting mutation family C introduced in Section~\ref{sec:mutfam_c},
where the quiver nodes $(a,b,c,d,e,f)$ of the mutation region 
are identified with the quiver nodes $(4,1,7,2,8,3)$ of the two toric phases.
Under this identification, the superpotential $W_{14a}$ takes the general form of $W$ 
for tilting mutation family C, while $W_{14b}$ takes the general form of $W^\prime$,
up to the overall sign of the superpotential.
The part of the superpotential that is
unaffected by the tilting mutation
is given by,
\beal{es05d03}
&&
W_0 = 
X_{46} X_{65} X_{54}^{2}
+ X_{59} X_{97} X_{75}
- X_{46} X_{67} X_{75} X_{54}^{1}
- X_{59} X_{96} X_{65}
~.~
\eea
The superpotential terms whose signs are reversed under the tilting mutation, 
with the signs taken as they appear in $W_{14a}$, 
are given by,
\beal{es05d04}
\Omega_{\text{mut}} =
X_{24} X_{43} X_{32}
+ X_{27} X_{78} X_{82}
- X_{13} X_{32} X_{27} X_{71}
- X_{18} X_{82} X_{24} X_{41}
~,~
\eea
and the exterior paths, which close the interior paths of the mutation region 
into gauge-invariant cycles, take the form,
\beal{es05d05}
&&
P_{34} = X_{35} X_{54}^{2} 
~,~
Q_{34} = X_{35} X_{54}^{1} 
~,~
P_{87} = X_{89} X_{97} 
~,~
Q_{87} = X_{89} X_{96} X_{67} 
~.~
\nn\\
\eea
Accordingly, the paths exchanged by the tilting mutation are given by,
\beal{es05d06}
U_{34} = X_{43}
~,~
U_{87} = X_{78}
~,~
V_{34} = X_{41} X_{13}
~,~
V_{87} = X_{71} X_{18}
~,~
\eea
which is
in agreement with the general form in Section~\ref{sec:mutfam_c} for tilting mutation family C,
such that the tilting mutation maps the superpotential $W_{14a}$ in \eref{es05d01} 
to $W_{14b}$ in \eref{es05d02}, and vice versa.
The action of the tilting mutation on the brane tiling 
and on the corresponding periodic quiver 
is illustrated in \fref{fig_tilting_14a_14b}.

%---------------------------------------------------- 
%---------------------------------------------------- 

\paragraph{Global Symmetry Charges.}
\tref{tab:models_14_Rcharges} summarizes the charges under the global symmetry
carried by the chiral fields of the two toric phases,
where the $U(1)_R$ charges are expressed in terms of
the $U(1)_R$ charges $R_a$ of the extremal GLSM fields $p_a$ in \tref{tab:extremal_charges}.
As shown in \tref{tab:models_14_Rcharges},
the chiral fields along the exterior paths
$P_{34}$, $Q_{34}$, $P_{87}$ and $Q_{87}$ in \eref{es05d05}
carry identical global symmetry charges in the two toric phases.
In contrast, the interior paths $U_{34}$ and $V_{34}$,
as well as $U_{87}$ and $V_{87}$ in \eref{es05d06},
exchange their global symmetry charges under the tilting mutation,
as can be seen from the charges carried by the chiral fields
in the mutation region in \tref{tab:models_14_Rcharges}.

%---------------------------------------------------- 
\begin{table}[h]
\setlength{\tabcolsep}{2pt}
\centering
\footnotesize{
\begin{tabular}[t]{c}
Phase 14a \\[4pt]
\begin{tabular}{| c | c | c | r |}
\hline
\; & $U(1)_{f_1}$ & $U(1)_{f_2}$ & $U(1)_R$\\\hline\hline
$X_{35}$ & $-1$ & $+1$ & $R_{3} \simeq 0.3862$\\
$X_{46}$ & $-1$ & $+1$ & $R_{3} \simeq 0.3862$\\
$X^1_{54}$ & $+2$ & $0$ & $R_{1} + R_{2} \simeq 0.5940$\\
$X^2_{54}$ & $0$ & $-2$ & $R_{5} + R_{6} \simeq 0.7666$\\
$X_{59}$ & $-1$ & $0$ & $R_{3} + R_{6} \simeq 0.7479$\\
$X_{65}$ & $+1$ & $+1$ & $R_{1} + R_{2} + R_{4} \simeq 0.8472$\\
$X_{67}$ & $0$ & $-1$ & $R_{6} \simeq 0.3617$\\
$X_{75}$ & $-1$ & $0$ & $R_{4} + R_{5} \simeq 0.6581$\\
$X_{89}$ & $-1$ & $+1$ & $R_{4} \simeq 0.2532$\\
$X_{96}$ & $0$ & $-1$ & $R_{5} \simeq 0.4049$\\
$X_{97}$ & $+2$ & $0$ & $R_{1} + R_{2} \simeq 0.5940$\\\hline
$X_{13}$ & $0$ & $-1$ & $R_{6} \simeq 0.3617$\\
$X_{18}$ & $+1$ & $0$ & $R_{2} \simeq 0.3205$\\
$X_{24}$ & $-1$ & $0$ & $R_{3} + R_{6} \simeq 0.7479$\\
$X_{27}$ & $0$ & $+1$ & $R_{2} + R_{4} \simeq 0.5737$\\
$X_{32}$ & $0$ & $-1$ & $R_{5} \simeq 0.4049$\\
$X_{41}$ & $-1$ & $0$ & $R_{4} + R_{5} \simeq 0.6581$\\
$X_{43}$ & $+1$ & $+1$ & $R_{1} + R_{2} + R_{4} \simeq 0.8472$\\
$X_{71}$ & $0$ & $+1$ & $R_{1} + R_{3} \simeq 0.6597$\\
$X_{78}$ & $-1$ & $-1$ & $R_{3} + R_{5} + R_{6} \simeq 1.1528$\\
$X_{82}$ & $+1$ & $0$ & $R_{1} \simeq 0.2736$\\\hline
\end{tabular}
\end{tabular}
}
\hspace{6pt}
\footnotesize{
\begin{tabular}[t]{c}
Phase 14b \\[4pt]
\begin{tabular}{| c | c | c | r |}
\hline
\; & $U(1)_{f_1}$ & $U(1)_{f_2}$ & $U(1)_R$\\\hline\hline
$X_{35}$ & $-1$ & $+1$ & $R_{3} \simeq 0.3862$\\
$X_{46}$ & $-1$ & $+1$ & $R_{3} \simeq 0.3862$\\
$X^1_{54}$ & $+2$ & $0$ & $R_{1} + R_{2} \simeq 0.5940$\\
$X^2_{54}$ & $0$ & $-2$ & $R_{5} + R_{6} \simeq 0.7666$\\
$X_{59}$ & $-1$ & $0$ & $R_{3} + R_{6} \simeq 0.7479$\\
$X_{65}$ & $+1$ & $+1$ & $R_{1} + R_{2} + R_{4} \simeq 0.8472$\\
$X_{67}$ & $0$ & $-1$ & $R_{6} \simeq 0.3617$\\
$X_{75}$ & $-1$ & $0$ & $R_{4} + R_{5} \simeq 0.6581$\\
$X_{89}$ & $-1$ & $+1$ & $R_{4} \simeq 0.2532$\\
$X_{96}$ & $0$ & $-1$ & $R_{5} \simeq 0.4049$\\
$X_{97}$ & $+2$ & $0$ & $R_{1} + R_{2} \simeq 0.5940$\\\hline
$X_{13}$ & $+1$ & $0$ & $R_{1} \simeq 0.2736$\\
$X_{18}$ & $0$ & $-1$ & $R_{5} \simeq 0.4049$\\
$X_{24}$ & $0$ & $+1$ & $R_{1} + R_{3} \simeq 0.6597$\\
$X_{27}$ & $-1$ & $0$ & $R_{4} + R_{5} \simeq 0.6581$\\
$X_{32}$ & $+1$ & $0$ & $R_{2} \simeq 0.3205$\\
$X_{41}$ & $0$ & $+1$ & $R_{2} + R_{4} \simeq 0.5737$\\
$X_{43}$ & $-1$ & $-1$ & $R_{4} + R_{5} + R_{6} \simeq 1.0198$\\
$X_{71}$ & $-1$ & $0$ & $R_{3} + R_{6} \simeq 0.7479$\\
$X_{78}$ & $+1$ & $+1$ & $R_{1} + R_{2} + R_{3} \simeq 0.9802$\\
$X_{82}$ & $0$ & $-1$ & $R_{6} \simeq 0.3617$\\\hline
\end{tabular}
\end{tabular}
}
\caption{
The charges under the global symmetry on the set of chiral fields 
shared between toric phases 14a and 14b of $H_{1,1,2,1}$, including the $U(1)_R$ charges, which are expressed in terms of $U(1)_R$ charges $R_a$
corresponding to the extremal GLSM fields $p_a$. 
}
\label{tab:models_14_Rcharges}
\end{table}
%---------------------------------------------------- 

\paragraph{Mesonic Moduli Spaces.}
Applying the forward algorithm to the two brane tilings, 
we compute the $P$-matrices encoding their perfect matchings.
Both brane tilings admit $36$ perfect matchings, 
consisting of the extremal perfect matchings $p_1, \dots, p_6$ 
corresponding to the extremal points of the toric diagram of $H_{1,1,2,1}$,
as well as the non-extremal perfect matchings.
The $P$-matrices take the following form,
\beal{es05d07}
P^{(14a)} = \resizebox{0.75\textwidth}{!}{$\left(\begin{tabular}{c | *{6}{c}| *{2}{c}| *{14}{c}| *{14}{c}}
& $p_{1}$ & $p_{2}$ & $p_{3}$ & $p_{4}$ & $p_{5}$ & $p_{6}$ & $q_{1}$ & $q_{2}$ & $u_{1}$ & $u_{2}$ & $u_{3}$ & $u_{4}$ & $u_{5}$ & $u_{6}$ & $u_{7}$ & $u_{8}$ & $u_{9}$ & $u_{10}$ & $u_{11}$ & $u_{12}$ & $u_{13}$ & $u_{14}$ & $s_{1}$ & $s_{2}$ & $s_{3}$ & $s_{4}$ & $s_{5}$ & $s_{6}$ & $s_{7}$ & $s_{8}$ & $s_{9}$ & $s_{10}$ & $s_{11}$ & $s_{12}$ & $s_{13}$ & $s_{14}$\\\hline
$X_{13}$ & 0 & 0 & 0 & 0 & 0 & 1 & 1 & 0 & 0 & 0 & 0 & 0 & 0 & 0 & 0 & 0 & 0 & 0 & 1 & 0 & 0 & 1 & 0 & 0 & 1 & 0 & 0 & 1 & 0 & 0 & 1 & 0 & 0 & 0 & 1 & 0\\
$X_{18}$ & 0 & 1 & 0 & 0 & 0 & 0 & 0 & 0 & 0 & 0 & 0 & 0 & 0 & 0 & 1 & 0 & 0 & 0 & 0 & 1 & 0 & 1 & 0 & 0 & 0 & 0 & 0 & 0 & 1 & 0 & 1 & 0 & 0 & 0 & 0 & 0\\
$X_{24}$ & 0 & 0 & 1 & 0 & 0 & 1 & 1 & 0 & 0 & 0 & 1 & 0 & 0 & 1 & 0 & 0 & 0 & 1 & 0 & 0 & 0 & 0 & 0 & 1 & 0 & 0 & 1 & 0 & 0 & 0 & 0 & 1 & 0 & 1 & 0 & 1\\
$X_{27}$ & 0 & 1 & 0 & 1 & 0 & 0 & 0 & 0 & 1 & 0 & 1 & 1 & 0 & 1 & 0 & 1 & 0 & 1 & 0 & 0 & 0 & 0 & 0 & 1 & 0 & 0 & 1 & 0 & 0 & 0 & 0 & 0 & 0 & 1 & 0 & 0\\
$X_{32}$ & 0 & 0 & 0 & 0 & 1 & 0 & 0 & 1 & 0 & 1 & 0 & 0 & 1 & 0 & 1 & 0 & 1 & 0 & 0 & 1 & 0 & 0 & 1 & 0 & 0 & 1 & 0 & 0 & 1 & 0 & 0 & 0 & 1 & 0 & 0 & 0\\
$X_{35}$ & 0 & 0 & 1 & 0 & 0 & 0 & 0 & 0 & 0 & 0 & 0 & 0 & 0 & 0 & 0 & 0 & 1 & 1 & 0 & 1 & 0 & 0 & 1 & 1 & 0 & 1 & 1 & 0 & 1 & 0 & 0 & 0 & 1 & 1 & 0 & 1\\
$X_{41}$ & 0 & 0 & 0 & 1 & 1 & 0 & 0 & 1 & 1 & 0 & 0 & 1 & 0 & 0 & 0 & 1 & 0 & 0 & 0 & 0 & 1 & 0 & 0 & 0 & 0 & 0 & 0 & 0 & 0 & 1 & 0 & 0 & 0 & 0 & 0 & 0\\
$X_{43}$ & 1 & 1 & 0 & 1 & 0 & 0 & 0 & 0 & 1 & 0 & 0 & 1 & 0 & 0 & 0 & 1 & 0 & 0 & 1 & 0 & 1 & 1 & 0 & 0 & 1 & 0 & 0 & 1 & 0 & 1 & 1 & 0 & 0 & 0 & 1 & 0\\
$X_{46}$ & 0 & 0 & 1 & 0 & 0 & 0 & 0 & 0 & 1 & 0 & 0 & 1 & 0 & 0 & 0 & 0 & 0 & 0 & 0 & 0 & 0 & 0 & 1 & 1 & 1 & 1 & 1 & 1 & 1 & 1 & 1 & 0 & 0 & 0 & 0 & 0\\
$X^1_{54}$ & 1 & 1 & 0 & 0 & 0 & 0 & 0 & 0 & 0 & 1 & 1 & 0 & 1 & 1 & 1 & 0 & 0 & 0 & 0 & 0 & 0 & 0 & 0 & 0 & 0 & 0 & 0 & 0 & 0 & 0 & 0 & 1 & 0 & 0 & 0 & 0\\
$X^2_{54}$ & 0 & 0 & 0 & 0 & 1 & 1 & 1 & 1 & 0 & 1 & 1 & 0 & 1 & 1 & 1 & 0 & 0 & 0 & 0 & 0 & 0 & 0 & 0 & 0 & 0 & 0 & 0 & 0 & 0 & 0 & 0 & 1 & 0 & 0 & 0 & 0\\
$X_{59}$ & 0 & 0 & 1 & 0 & 0 & 1 & 0 & 1 & 0 & 0 & 0 & 1 & 1 & 1 & 1 & 0 & 0 & 0 & 0 & 0 & 0 & 0 & 0 & 0 & 0 & 1 & 1 & 1 & 1 & 1 & 1 & 1 & 0 & 0 & 0 & 0\\
$X_{65}$ & 1 & 1 & 0 & 1 & 0 & 0 & 0 & 0 & 0 & 0 & 0 & 0 & 0 & 0 & 0 & 1 & 1 & 1 & 1 & 1 & 1 & 1 & 0 & 0 & 0 & 0 & 0 & 0 & 0 & 0 & 0 & 0 & 1 & 1 & 1 & 1\\
$X_{67}$ & 0 & 0 & 0 & 0 & 0 & 1 & 0 & 1 & 0 & 0 & 0 & 0 & 0 & 0 & 0 & 1 & 0 & 0 & 0 & 0 & 0 & 0 & 0 & 0 & 0 & 0 & 0 & 0 & 0 & 0 & 0 & 0 & 1 & 1 & 1 & 0\\
$X_{71}$ & 1 & 0 & 1 & 0 & 0 & 0 & 0 & 0 & 0 & 0 & 0 & 0 & 0 & 0 & 0 & 0 & 0 & 0 & 0 & 0 & 1 & 0 & 0 & 0 & 0 & 0 & 0 & 0 & 0 & 1 & 0 & 1 & 0 & 0 & 0 & 1\\
$X_{75}$ & 0 & 0 & 0 & 1 & 1 & 0 & 1 & 0 & 0 & 0 & 0 & 0 & 0 & 0 & 0 & 0 & 1 & 1 & 1 & 1 & 1 & 1 & 0 & 0 & 0 & 0 & 0 & 0 & 0 & 0 & 0 & 0 & 0 & 0 & 0 & 1\\
$X_{78}$ & 0 & 0 & 1 & 0 & 1 & 1 & 1 & 1 & 0 & 0 & 0 & 0 & 0 & 0 & 1 & 0 & 0 & 0 & 0 & 1 & 1 & 1 & 0 & 0 & 0 & 0 & 0 & 0 & 1 & 1 & 1 & 1 & 0 & 0 & 0 & 1\\
$X_{82}$ & 1 & 0 & 0 & 0 & 0 & 0 & 0 & 0 & 0 & 1 & 0 & 0 & 1 & 0 & 0 & 0 & 1 & 0 & 1 & 0 & 0 & 0 & 1 & 0 & 1 & 1 & 0 & 1 & 0 & 0 & 0 & 0 & 1 & 0 & 1 & 0\\
$X_{89}$ & 0 & 0 & 0 & 1 & 0 & 0 & 0 & 0 & 0 & 0 & 0 & 1 & 1 & 1 & 0 & 0 & 1 & 1 & 1 & 0 & 0 & 0 & 0 & 0 & 0 & 1 & 1 & 1 & 0 & 0 & 0 & 0 & 0 & 0 & 0 & 0\\
$X_{96}$ & 0 & 0 & 0 & 0 & 1 & 0 & 1 & 0 & 1 & 1 & 1 & 0 & 0 & 0 & 0 & 0 & 0 & 0 & 0 & 0 & 0 & 0 & 1 & 1 & 1 & 0 & 0 & 0 & 0 & 0 & 0 & 0 & 0 & 0 & 0 & 0\\
$X_{97}$ & 1 & 1 & 0 & 0 & 0 & 0 & 0 & 0 & 1 & 1 & 1 & 0 & 0 & 0 & 0 & 1 & 0 & 0 & 0 & 0 & 0 & 0 & 1 & 1 & 1 & 0 & 0 & 0 & 0 & 0 & 0 & 0 & 1 & 1 & 1 & 0\\
\end{tabular}\right)$}
\eea
and
\beal{es05d08}
P^{(14b)} = \resizebox{0.75\textwidth}{!}{$\left(\begin{tabular}{c | *{6}{c}| *{2}{c}| *{14}{c}| *{14}{c}}
& $p_{1}$ & $p_{2}$ & $p_{3}$ & $p_{4}$ & $p_{5}$ & $p_{6}$ & $q_{1}$ & $q_{2}$ & $u_{1}$ & $u_{2}$ & $u_{3}$ & $u_{4}$ & $u_{5}$ & $u_{6}$ & $u_{7}$ & $u_{8}$ & $u_{9}$ & $u_{10}$ & $u_{11}$ & $u_{12}$ & $u_{13}$ & $u_{14}$ & $s_{1}$ & $s_{2}$ & $s_{3}$ & $s_{4}$ & $s_{5}$ & $s_{6}$ & $s_{7}$ & $s_{8}$ & $s_{9}$ & $s_{10}$ & $s_{11}$ & $s_{12}$ & $s_{13}$ & $s_{14}$\\\hline
$X_{13}$ & 1 & 0 & 0 & 0 & 0 & 0 & 0 & 0 & 0 & 0 & 0 & 0 & 0 & 0 & 1 & 0 & 1 & 0 & 0 & 0 & 0 & 0 & 0 & 0 & 1 & 0 & 0 & 0 & 1 & 0 & 1 & 0 & 0 & 1 & 0 & 0\\
$X_{18}$ & 0 & 0 & 0 & 0 & 1 & 0 & 0 & 1 & 0 & 0 & 0 & 0 & 0 & 1 & 0 & 0 & 1 & 0 & 0 & 0 & 0 & 1 & 0 & 0 & 0 & 0 & 0 & 1 & 0 & 0 & 1 & 0 & 0 & 0 & 0 & 0\\
$X_{24}$ & 1 & 0 & 1 & 0 & 0 & 0 & 0 & 0 & 0 & 0 & 0 & 0 & 1 & 0 & 0 & 0 & 0 & 0 & 1 & 0 & 1 & 0 & 0 & 1 & 0 & 0 & 1 & 0 & 0 & 0 & 0 & 0 & 1 & 0 & 1 & 1\\
$X_{27}$ & 0 & 0 & 0 & 1 & 1 & 0 & 0 & 1 & 1 & 1 & 1 & 0 & 1 & 0 & 0 & 0 & 0 & 0 & 1 & 0 & 1 & 0 & 0 & 1 & 0 & 0 & 1 & 0 & 0 & 0 & 0 & 0 & 1 & 0 & 0 & 0\\
$X_{32}$ & 0 & 1 & 0 & 0 & 0 & 0 & 0 & 0 & 0 & 0 & 0 & 1 & 0 & 1 & 0 & 0 & 0 & 1 & 0 & 1 & 0 & 1 & 1 & 0 & 0 & 1 & 0 & 1 & 0 & 0 & 0 & 1 & 0 & 0 & 0 & 0\\
$X_{35}$ & 0 & 0 & 1 & 0 & 0 & 0 & 0 & 0 & 0 & 0 & 0 & 1 & 1 & 1 & 0 & 0 & 0 & 0 & 0 & 0 & 0 & 0 & 1 & 1 & 0 & 1 & 1 & 1 & 0 & 0 & 0 & 1 & 1 & 0 & 1 & 0\\
$X_{41}$ & 0 & 1 & 0 & 1 & 0 & 0 & 0 & 0 & 1 & 1 & 1 & 0 & 0 & 0 & 0 & 1 & 0 & 0 & 0 & 0 & 0 & 0 & 0 & 0 & 0 & 0 & 0 & 0 & 0 & 1 & 0 & 0 & 0 & 0 & 0 & 0\\
$X_{43}$ & 0 & 0 & 0 & 1 & 1 & 1 & 1 & 1 & 1 & 1 & 1 & 0 & 0 & 0 & 1 & 1 & 1 & 0 & 0 & 0 & 0 & 0 & 0 & 0 & 1 & 0 & 0 & 0 & 1 & 1 & 1 & 0 & 0 & 1 & 0 & 0\\
$X_{46}$ & 0 & 0 & 1 & 0 & 0 & 0 & 0 & 0 & 1 & 1 & 0 & 0 & 0 & 0 & 0 & 0 & 0 & 0 & 0 & 0 & 0 & 0 & 1 & 1 & 1 & 1 & 1 & 1 & 1 & 1 & 1 & 0 & 0 & 0 & 0 & 0\\
$X^1_{54}$ & 1 & 1 & 0 & 0 & 0 & 0 & 0 & 0 & 0 & 0 & 0 & 0 & 0 & 0 & 0 & 0 & 0 & 1 & 1 & 1 & 1 & 1 & 0 & 0 & 0 & 0 & 0 & 0 & 0 & 0 & 0 & 0 & 0 & 0 & 0 & 1\\
$X^2_{54}$ & 0 & 0 & 0 & 0 & 1 & 1 & 1 & 1 & 0 & 0 & 0 & 0 & 0 & 0 & 0 & 0 & 0 & 1 & 1 & 1 & 1 & 1 & 0 & 0 & 0 & 0 & 0 & 0 & 0 & 0 & 0 & 0 & 0 & 0 & 0 & 1\\
$X_{59}$ & 0 & 0 & 1 & 0 & 0 & 1 & 0 & 1 & 0 & 1 & 0 & 0 & 0 & 0 & 0 & 0 & 0 & 0 & 0 & 1 & 1 & 1 & 0 & 0 & 0 & 1 & 1 & 1 & 1 & 1 & 1 & 0 & 0 & 0 & 0 & 1\\
$X_{65}$ & 1 & 1 & 0 & 1 & 0 & 0 & 0 & 0 & 0 & 0 & 1 & 1 & 1 & 1 & 1 & 1 & 1 & 0 & 0 & 0 & 0 & 0 & 0 & 0 & 0 & 0 & 0 & 0 & 0 & 0 & 0 & 1 & 1 & 1 & 1 & 0\\
$X_{67}$ & 0 & 0 & 0 & 0 & 0 & 1 & 0 & 1 & 0 & 0 & 1 & 0 & 0 & 0 & 0 & 0 & 0 & 0 & 0 & 0 & 0 & 0 & 0 & 0 & 0 & 0 & 0 & 0 & 0 & 0 & 0 & 1 & 1 & 1 & 0 & 0\\
$X_{71}$ & 0 & 0 & 1 & 0 & 0 & 1 & 1 & 0 & 0 & 0 & 0 & 0 & 0 & 0 & 0 & 1 & 0 & 0 & 0 & 0 & 0 & 0 & 0 & 0 & 0 & 0 & 0 & 0 & 0 & 1 & 0 & 0 & 0 & 0 & 1 & 1\\
$X_{75}$ & 0 & 0 & 0 & 1 & 1 & 0 & 1 & 0 & 0 & 0 & 0 & 1 & 1 & 1 & 1 & 1 & 1 & 0 & 0 & 0 & 0 & 0 & 0 & 0 & 0 & 0 & 0 & 0 & 0 & 0 & 0 & 0 & 0 & 0 & 1 & 0\\
$X_{78}$ & 1 & 1 & 1 & 0 & 0 & 0 & 0 & 0 & 0 & 0 & 0 & 0 & 0 & 1 & 0 & 1 & 1 & 0 & 0 & 0 & 0 & 1 & 0 & 0 & 0 & 0 & 0 & 1 & 0 & 1 & 1 & 0 & 0 & 0 & 1 & 1\\
$X_{82}$ & 0 & 0 & 0 & 0 & 0 & 1 & 1 & 0 & 0 & 0 & 0 & 1 & 0 & 0 & 1 & 0 & 0 & 1 & 0 & 1 & 0 & 0 & 1 & 0 & 1 & 1 & 0 & 0 & 1 & 0 & 0 & 1 & 0 & 1 & 0 & 0\\
$X_{89}$ & 0 & 0 & 0 & 1 & 0 & 0 & 0 & 0 & 0 & 1 & 0 & 1 & 1 & 0 & 1 & 0 & 0 & 0 & 0 & 1 & 1 & 0 & 0 & 0 & 0 & 1 & 1 & 0 & 1 & 0 & 0 & 0 & 0 & 0 & 0 & 0\\
$X_{96}$ & 0 & 0 & 0 & 0 & 1 & 0 & 1 & 0 & 1 & 0 & 0 & 0 & 0 & 0 & 0 & 0 & 0 & 1 & 1 & 0 & 0 & 0 & 1 & 1 & 1 & 0 & 0 & 0 & 0 & 0 & 0 & 0 & 0 & 0 & 0 & 0\\
$X_{97}$ & 1 & 1 & 0 & 0 & 0 & 0 & 0 & 0 & 1 & 0 & 1 & 0 & 0 & 0 & 0 & 0 & 0 & 1 & 1 & 0 & 0 & 0 & 1 & 1 & 1 & 0 & 0 & 0 & 0 & 0 & 0 & 1 & 1 & 1 & 0 & 0\\
\end{tabular}\right)$}
~.~
\eea
From the $P$-matrices in \eref{es05d07} and \eref{es05d08}, 
the forward algorithm yields identical $G_t$-matrices for the two toric phases, 
which take the form,
\beal{es05d09}
G_t^{(14a)} = G_t^{(14b)} = \small{\setlength{\tabcolsep}{1pt}\left(\begin{tabular}{ *{6}{c}| *{2}{c}| *{3}{c}| *{3}{c}}
$p_{1}$ & $p_{2}$ & $p_{3}$ & $p_{4}$ & $p_{5}$ & $p_{6}$ & $q_{1}$ & $q_{2}$ & $u_{1}$ & $\cdots$ & $u_{14}$ &$s_{1}$ & $\cdots$ & $s_{14}$\\\hline
2 & 2 & 1 & 1 & 0 & 0 & 0 & 0 & 1 & $\cdots$ & 1 & 1 & $\cdots$ & 1 \\
2 & 1 & 3 & 0 & 0 & 2 & 1 & 1 & 1 & $\cdots$ & 1 & 2 & $\cdots$ & 2 \\
1 & 1 & 1 & 1 & 1 & 1 & 1 & 1 & 1 & $\cdots$ & 1 & 1 & $\cdots$ & 1 \\
\end{tabular}\right)}
~.~
\eea
The toric diagrams obtained from the two brane tilings therefore coincide,
which verifies that the brane tilings of toric phases 14a and 14b 
correspond to the same toric Calabi-Yau 3-fold $H_{1,1,2,1}$, 
as required for the quiver-invariant duality realized by the tilting mutation.
\\

%=================================================================
\subsection{$H_{1,1,2,1}$ 15a, 15b and 15c \label{sec:model15abc}} % family A
%=================================================================

%---------------------------------------------------- 
\begin{figure}[htbp]
\centering
\includegraphics[width=\textwidth]{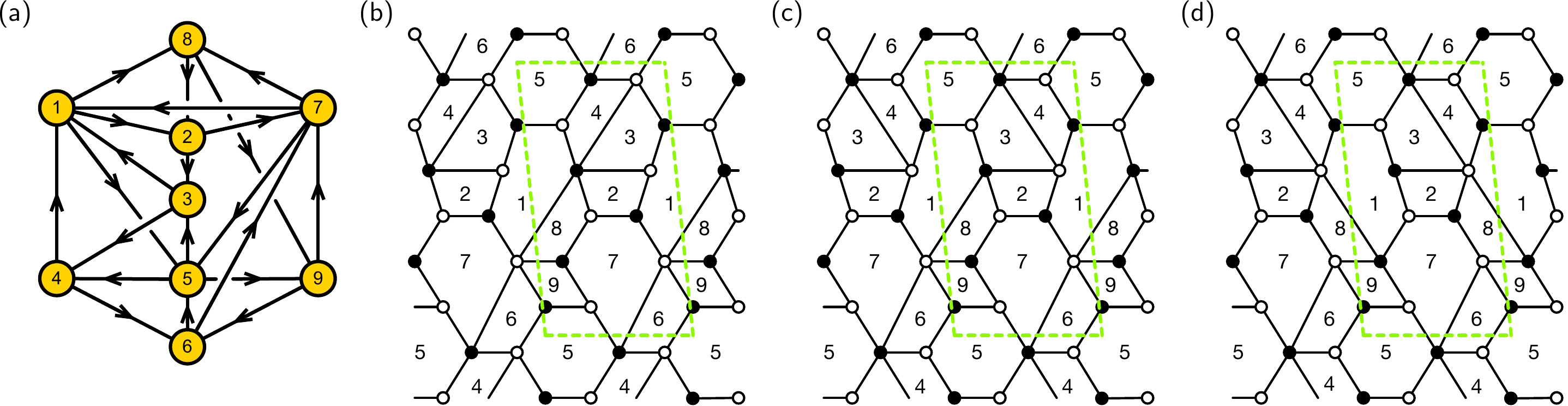}
\caption{
(a) The quiver shared between the three brane tilings in (b), (c) and (d) corresponding to the three toric phases 15a, 15b and 15c of $H_{1,1,2,1}$,
respectively. 
}
\label{fig_quiver_15abc}
\end{figure}
%---------------------------------------------------- 

The three toric phases 15a, 15b and 15c of the $H_{1,1,2,1}$ model 
share the same quiver, which is shown in \fref{fig_quiver_15abc}.
In contrast to the doublets of toric phases discussed above, 
these three toric phases exhibit a sequence of quiver-invariant dualities.
Their superpotentials are given by,
\beal{es05f01}
W_{15a} = 
&&
X_{12} X_{23} X_{31}
+ X_{15} X_{54} X_{41}
+ X_{18} X_{89} X_{96} X_{67} X_{71}
+ X_{27} X_{78} X_{82}
\nn\\
&&
+ X_{34} X_{46} X_{65} X_{53}
+ X_{59} X_{97} X_{75} 
- X_{12} X_{27} X_{71}
- X_{15} X_{53} X_{31} 
\nn\\
&&
- X_{18} X_{82} X_{23} X_{34} X_{41}
- X_{46} X_{67} X_{75} X_{54}
- X_{59} X_{96} X_{65}
- X_{78} X_{89} X_{97}
~,~
\nn\\
\eea
\beal{es05f02}
W_{15b} = 
&&
X_{12} X_{23} X_{34} X_{41}
+ X_{15} X_{53} X_{31}
+ X_{18} X_{89} X_{96} X_{67} X_{71}
+ X_{27} X_{78} X_{82}
\nn\\
&&
+ X_{46} X_{65} X_{54}
+ X_{59} X_{97} X_{75}
- X_{12} X_{27} X_{71}
- X_{15} X_{54} X_{41}
- X_{18} X_{82} X_{23} X_{31}
\nn\\
&&
- X_{34} X_{46} X_{67} X_{75} X_{53}
- X_{59} X_{96} X_{65}
- X_{78} X_{89} X_{97}
~,~
\eea
and
\beal{es05f03}
W_{15c} = 
&&
X_{12} X_{27} X_{71} 
+ X_{15} X_{53} X_{31} 
+ X_{18} X_{82} X_{23} X_{34} X_{41} 
+ X_{46} X_{65} X_{54} 
\nn\\
&&
+ X_{59} X_{97} X_{75} 
+ X_{67} X_{78} X_{89} X_{96}
- X_{12} X_{23} X_{31} 
- X_{15} X_{54} X_{41} 
- X_{18} X_{89} X_{97} X_{71} 
\nn\\
&&
- X_{27} X_{78} X_{82}
- X_{34} X_{46} X_{67} X_{75} X_{53}
- X_{59} X_{96} X_{65}
~.~
\eea

\begin{figure}[htbp]
\centering
\includegraphics[width=0.7\textwidth]{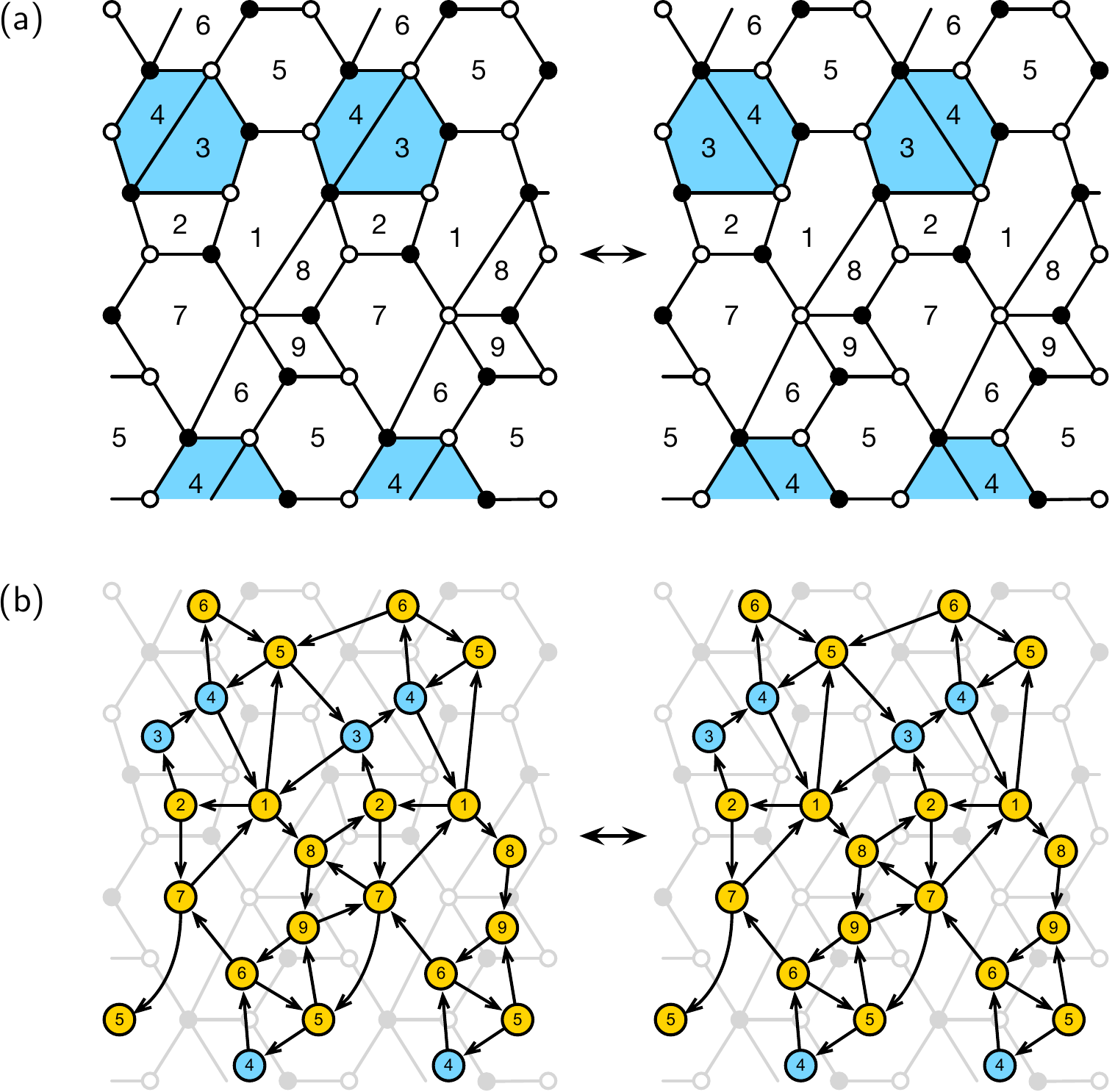}
\caption{
(a) The tilting mutation relating the brane tilings of the two toric phases 15a and 15b of $H_{1,1,2,1}$,
and (b) its effect on the corresponding periodic quivers. 
}
\label{fig_tilting_15a_15b}
\end{figure}

\paragraph{Tilting Mutations.}
By comparing the superpotentials in \eref{es05f01} and \eref{es05f02},
we identify the tilting mutation relating toric phase 15a to 15b 
as belonging to tilting mutation family A introduced in Section~\ref{sec:mutfam_a},
where the quiver nodes $(a,b,c,d)$ of the mutation region 
are identified with the quiver nodes $(3,4,5,1)$ of the two toric phases.
Under this identification, the superpotential $W_{15a}$ takes the general form of $W$ 
for tilting mutation family A, while $W_{15b}$ takes the general form of $W^\prime$.
The part of the superpotential that is
unaffected by the tilting mutation
is given by,
\beal{es05f04}
&&
W_0 = 
X_{18} X_{89} X_{96} X_{67} X_{71}
+ X_{27} X_{78} X_{82}
+ X_{59} X_{97} X_{75}
\nn\\
&&
\hspace{1cm}
- X_{12} X_{27} X_{71}
- X_{59} X_{96} X_{65}
- X_{78} X_{89} X_{97}
~.~
\eea
The superpotential terms whose signs are reversed under the tilting mutation, 
are given by,
\beal{es05f05}
\Omega_{\text{mut}} =
X_{15} X_{54} X_{41}
- X_{15} X_{53} X_{31}
~,~
\eea
and the exterior paths, which close the interior paths of the mutation region 
into gauge-invariant cycles, take the form,
\beal{es05f06}
&&
P_{13} = X_{12} X_{23} 
~,~
Q_{13} = X_{18} X_{82} X_{23} 
~,~
P_{45} = X_{46} X_{65} 
~,~
Q_{45} = X_{46} X_{67} X_{75} 
~.~
\eea
Accordingly, the paths exchanged by the tilting mutation are given by,
\beal{es05f07}
U_{13} = X_{31}
~,~
U_{45} = X_{53} X_{34}
~,~
V_{13} = X_{34} X_{41}
~,~
V_{45} = X_{54}
~,~
\eea
which is
in agreement with the general form in Section~\ref{sec:mutfam_a} for tilting mutation family A,
such that this tilting mutation maps the superpotential $W_{15a}$ in \eref{es05f01} 
to $W_{15b}$ in \eref{es05f02}, and vice versa.
The action of this tilting mutation on the brane tiling 
is illustrated in \fref{fig_tilting_15a_15b}.

%---------------------------------------------------- 
%---------------------------------------------------- 

\begin{figure}[htbp]
\centering
\includegraphics[width=0.7\textwidth]{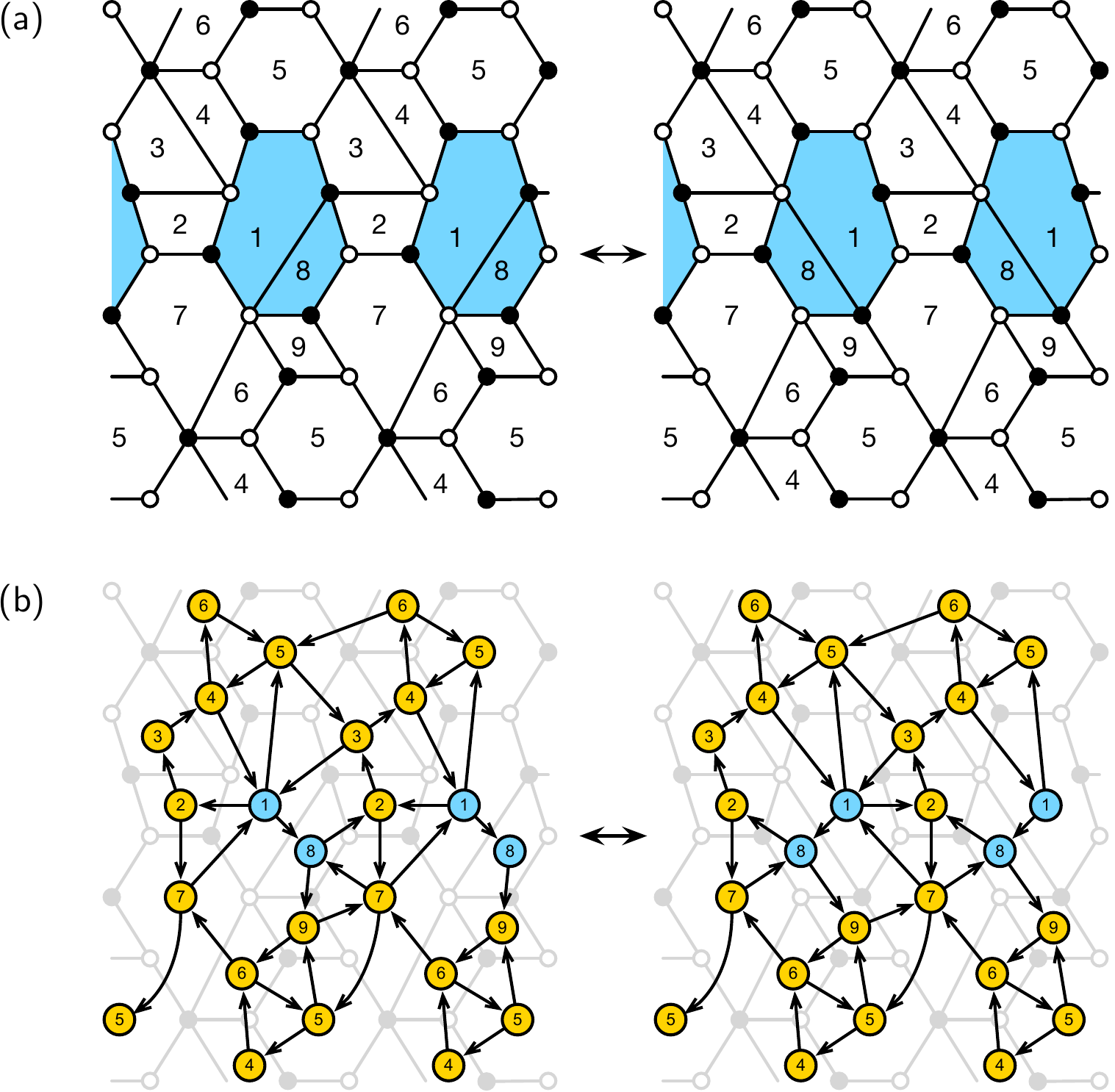}
\caption{
(a) The tilting mutation relating the brane tilings of the two toric phases 15b and 15c of $H_{1,1,2,1}$,
and (b) its effect on the corresponding periodic quivers. 
}
\label{fig_tilting_15b_15c}
\end{figure}

Similarly, by comparing the superpotentials in \eref{es05f02} and \eref{es05f03},
we identify the tilting mutation relating toric phase 15b to 15c 
as a second tilting mutation of family A,
where the quiver nodes $(a,b,c,d)$ of the mutation region 
are identified with the quiver nodes $(1,8,7,2)$ of the two toric phases.
Under this identification, the superpotential $W_{15b}$ takes the general form of $W$ 
for tilting mutation family A, while $W_{15c}$ takes the general form of $W^\prime$.
The part of the superpotential that is
unaffected by the tilting mutation
is given by,
\beal{es05f08}
&&
W_0 = 
X_{15} X_{53} X_{31}
+ X_{46} X_{65} X_{54}
+ X_{59} X_{97} X_{75}
\nn\\
&&
\hspace{1cm}
- X_{15} X_{54} X_{41}
- X_{34} X_{46} X_{67} X_{75} X_{53}
- X_{59} X_{96} X_{65}
~.~
\eea
The superpotential terms whose signs are reversed under the tilting mutation, 
are given by,
\beal{es05f09}
\Omega_{\text{mut}} =
X_{27} X_{78} X_{82}
- X_{27} X_{71} X_{12}
~,~
\eea
and the exterior paths take the form,
\beal{es05f10}
&&
P_{21} = X_{23} X_{34} X_{41} 
~,~
Q_{21} = X_{23} X_{31} 
~,~
P_{87} = X_{89} X_{96} X_{67} 
~,~
Q_{87} = X_{89} X_{97} 
~,~
\eea
with the paths exchanged by the tilting mutation given by,
\beal{es05f11}
U_{21} = X_{12}
~,~
U_{87} = X_{71} X_{18}
~,~
V_{21} = X_{18} X_{82}
~,~
V_{87} = X_{78}
~,~
\eea
in agreement with the general form in Section~\ref{sec:mutfam_a} for tilting mutation family A,
such that this tilting mutation maps the superpotential $W_{15b}$ in \eref{es05f02} 
to $W_{15c}$ in \eref{es05f03}, and vice versa.
The action of this tilting mutation on the brane tiling 
is illustrated in \fref{fig_tilting_15b_15c}.

%---------------------------------------------------- 
%---------------------------------------------------- 

\paragraph{Global Symmetry Charges.}
\tref{tab:models_15ab_Rcharges} and \tref{tab:models_15bc_Rcharges} summarize the charges under the global symmetry 
carried by the chiral fields of the three toric phases,
where the $U(1)_R$ charges are expressed in terms of the $U(1)_R$ charges $R_a$
of the extremal GLSM fields $p_a$ in \tref{tab:extremal_charges}.
As shown in \tref{tab:models_15ab_Rcharges}, 
for the tilting mutation from toric phase 15a to 15b
the chiral fields along the interior paths
$U_{13}$, $V_{13}$, $U_{45}$ and $V_{45}$ in \eref{es05f07}
carry identical global symmetry charges in the two toric phases.
Similarly, 
as shown in \tref{tab:models_15bc_Rcharges}
for the tilting mutation from toric phase 15b to 15c
the chiral fields along the interior paths
$U_{21}$, $V_{21}$, $U_{87}$ and $V_{87}$ in \eref{es05f11}
carry identical global symmetry charges in the two toric phases.
In contrast, for both tilting mutations, the exterior paths 
exchange their global symmetry charges.
This sets tilting mutation family A apart from the other tilting mutation families B and C. 
%---------------------------------------------------- 
\begin{table}[h]
\setlength{\tabcolsep}{2pt}
\centering
\footnotesize{
\begin{tabular}[t]{c}
Phase 15a \\[4pt]
\begin{tabular}{| c | c | c | r |}
\hline
\; & $U(1)_{f_1}$ & $U(1)_{f_2}$ & $U(1)_R$\\\hline\hline
$X_{15}$ & $-1$ & $0$ & $R_{3} + R_{6} \simeq 0.7479$\\
$X_{23}$ & $-1$ & $+1$ & $R_{3} \simeq 0.3862$\\
$X_{31}$ & $+1$ & $+1$ & $R_{1} + R_{2} + R_{4} \simeq 0.8472$\\
$X_{34}$ & $0$ & $-1$ & $R_{6} \simeq 0.3617$\\
$X_{41}$ & $-1$ & $0$ & $R_{4} + R_{5} \simeq 0.6581$\\
$X_{46}$ & $-1$ & $+1$ & $R_{3} \simeq 0.3862$\\
$X_{53}$ & $0$ & $-1$ & $R_{5} \simeq 0.4049$\\
$X_{54}$ & $+2$ & $0$ & $R_{1} + R_{2} \simeq 0.5940$\\
$X_{89}$ & $-1$ & $+1$ & $R_{4} \simeq 0.2532$\\\hline
$X_{12}$ & $0$ & $-2$ & $R_{5} + R_{6} \simeq 0.7666$\\
$X_{18}$ & $+1$ & $0$ & $R_{2} \simeq 0.3205$\\
$X_{27}$ & $0$ & $+1$ & $R_{2} + R_{4} \simeq 0.5737$\\
$X_{59}$ & $-1$ & $0$ & $R_{3} + R_{6} \simeq 0.7479$\\
$X_{65}$ & $+1$ & $+1$ & $R_{1} + R_{2} + R_{4} \simeq 0.8472$\\
$X_{67}$ & $0$ & $-1$ & $R_{6} \simeq 0.3617$\\
$X_{71}$ & $0$ & $+1$ & $R_{1} + R_{3} \simeq 0.6597$\\
$X_{75}$ & $-1$ & $0$ & $R_{4} + R_{5} \simeq 0.6581$\\
$X_{78}$ & $-1$ & $-1$ & $R_{3} + R_{5} + R_{6} \simeq 1.1528$\\
$X_{82}$ & $+1$ & $0$ & $R_{1} \simeq 0.2736$\\
$X_{96}$ & $0$ & $-1$ & $R_{5} \simeq 0.4049$\\
$X_{97}$ & $+2$ & $0$ & $R_{1} + R_{2} \simeq 0.5940$\\\hline
\end{tabular}
\end{tabular}
}
\hspace{6pt}
\footnotesize{
\begin{tabular}[t]{c}
Phase 15b \\[4pt]
\begin{tabular}{| c | c | c | r |}
\hline
\; & $U(1)_{f_1}$ & $U(1)_{f_2}$ & $U(1)_R$\\\hline\hline
$X_{15}$ & $-1$ & $0$ & $R_{3} + R_{6} \simeq 0.7479$\\
$X_{23}$ & $-1$ & $+1$ & $R_{3} \simeq 0.3862$\\
$X_{31}$ & $+1$ & $+1$ & $R_{1} + R_{2} + R_{4} \simeq 0.8472$\\
$X_{34}$ & $0$ & $-1$ & $R_{6} \simeq 0.3617$\\
$X_{41}$ & $-1$ & $0$ & $R_{4} + R_{5} \simeq 0.6581$\\
$X_{46}$ & $-1$ & $+1$ & $R_{3} \simeq 0.3862$\\
$X_{53}$ & $0$ & $-1$ & $R_{5} \simeq 0.4049$\\
$X_{54}$ & $+2$ & $0$ & $R_{1} + R_{2} \simeq 0.5940$\\
$X_{89}$ & $-1$ & $+1$ & $R_{4} \simeq 0.2532$\\\hline
$X_{12}$ & $+2$ & $0$ & $R_{1} + R_{2} \simeq 0.5940$\\
$X_{18}$ & $0$ & $-1$ & $R_{5} \simeq 0.4049$\\
$X_{27}$ & $-1$ & $0$ & $R_{4} + R_{5} \simeq 0.6581$\\
$X_{59}$ & $0$ & $+1$ & $R_{1} + R_{3} \simeq 0.6597$\\
$X_{65}$ & $-1$ & $-1$ & $R_{4} + R_{5} + R_{6} \simeq 1.0198$\\
$X_{67}$ & $+1$ & $0$ & $R_{1} \simeq 0.2736$\\
$X_{71}$ & $-1$ & $0$ & $R_{3} + R_{6} \simeq 0.7479$\\
$X_{75}$ & $0$ & $+1$ & $R_{2} + R_{4} \simeq 0.5737$\\
$X_{78}$ & $+1$ & $+1$ & $R_{1} + R_{2} + R_{3} \simeq 0.9802$\\
$X_{82}$ & $0$ & $-1$ & $R_{6} \simeq 0.3617$\\
$X_{96}$ & $+1$ & $0$ & $R_{2} \simeq 0.3205$\\
$X_{97}$ & $0$ & $-2$ & $R_{5} + R_{6} \simeq 0.7666$\\\hline
\end{tabular}
\end{tabular}
}
\caption{
The charges under the global symmetry on the set of chiral fields 
shared between toric phases 15a and 15b of $H_{1,1,2,1}$, including the $U(1)_R$ charges, which are expressed in terms of $U(1)_R$ charges $R_a$
corresponding to the extremal GLSM fields $p_a$. 
}
\label{tab:models_15ab_Rcharges}
\end{table}
%---------------------------------------------------- 

%---------------------------------------------------- 
\begin{table}[h]
\setlength{\tabcolsep}{2pt}
\centering
\footnotesize{
\begin{tabular}[t]{c}
Phase 15b \\[4pt]
\begin{tabular}{| c | c | c | r |}
\hline
\; & $U(1)_{f_1}$ & $U(1)_{f_2}$ & $U(1)_R$\\\hline\hline
$X_{12}$ & $+2$ & $0$ & $R_{1} + R_{2} \simeq 0.5940$\\
$X_{18}$ & $0$ & $-1$ & $R_{5} \simeq 0.4049$\\
$X_{23}$ & $-1$ & $+1$ & $R_{3} \simeq 0.3862$\\
$X_{27}$ & $-1$ & $0$ & $R_{4} + R_{5} \simeq 0.6581$\\
$X_{46}$ & $-1$ & $+1$ & $R_{3} \simeq 0.3862$\\
$X_{71}$ & $-1$ & $0$ & $R_{3} + R_{6} \simeq 0.7479$\\
$X_{78}$ & $+1$ & $+1$ & $R_{1} + R_{2} + R_{3} \simeq 0.9802$\\
$X_{82}$ & $0$ & $-1$ & $R_{6} \simeq 0.3617$\\
$X_{89}$ & $-1$ & $+1$ & $R_{4} \simeq 0.2532$\\\hline
$X_{15}$ & $-1$ & $0$ & $R_{3} + R_{6} \simeq 0.7479$\\
$X_{31}$ & $+1$ & $+1$ & $R_{1} + R_{2} + R_{4} \simeq 0.8472$\\
$X_{34}$ & $0$ & $-1$ & $R_{6} \simeq 0.3617$\\
$X_{41}$ & $-1$ & $0$ & $R_{4} + R_{5} \simeq 0.6581$\\
$X_{53}$ & $0$ & $-1$ & $R_{5} \simeq 0.4049$\\
$X_{54}$ & $+2$ & $0$ & $R_{1} + R_{2} \simeq 0.5940$\\
$X_{59}$ & $0$ & $+1$ & $R_{1} + R_{3} \simeq 0.6597$\\
$X_{65}$ & $-1$ & $-1$ & $R_{4} + R_{5} + R_{6} \simeq 1.0198$\\
$X_{67}$ & $+1$ & $0$ & $R_{1} \simeq 0.2736$\\
$X_{75}$ & $0$ & $+1$ & $R_{2} + R_{4} \simeq 0.5737$\\
$X_{96}$ & $+1$ & $0$ & $R_{2} \simeq 0.3205$\\
$X_{97}$ & $0$ & $-2$ & $R_{5} + R_{6} \simeq 0.7666$\\\hline
\end{tabular}
\end{tabular}
}
\hspace{6pt}
\footnotesize{
\begin{tabular}[t]{c}
Phase 15c \\[4pt]
\begin{tabular}{| c | c | c | r |}
\hline
\; & $U(1)_{f_1}$ & $U(1)_{f_2}$ & $U(1)_R$\\\hline\hline
$X_{12}$ & $+2$ & $0$ & $R_{1} + R_{2} \simeq 0.5940$\\
$X_{18}$ & $0$ & $-1$ & $R_{5} \simeq 0.4049$\\
$X_{23}$ & $-1$ & $+1$ & $R_{3} \simeq 0.3862$\\
$X_{27}$ & $-1$ & $0$ & $R_{4} + R_{5} \simeq 0.6581$\\
$X_{46}$ & $-1$ & $+1$ & $R_{3} \simeq 0.3862$\\
$X_{71}$ & $-1$ & $0$ & $R_{3} + R_{6} \simeq 0.7479$\\
$X_{78}$ & $+1$ & $+1$ & $R_{1} + R_{2} + R_{3} \simeq 0.9802$\\
$X_{82}$ & $0$ & $-1$ & $R_{6} \simeq 0.3617$\\
$X_{89}$ & $-1$ & $+1$ & $R_{4} \simeq 0.2532$\\\hline
$X_{15}$ & $0$ & $+1$ & $R_{1} + R_{3} \simeq 0.6597$\\
$X_{31}$ & $-1$ & $-1$ & $R_{4} + R_{5} + R_{6} \simeq 1.0198$\\
$X_{34}$ & $+1$ & $0$ & $R_{1} \simeq 0.2736$\\
$X_{41}$ & $0$ & $+1$ & $R_{2} + R_{4} \simeq 0.5737$\\
$X_{53}$ & $+1$ & $0$ & $R_{2} \simeq 0.3205$\\
$X_{54}$ & $0$ & $-2$ & $R_{5} + R_{6} \simeq 0.7666$\\
$X_{59}$ & $-1$ & $0$ & $R_{3} + R_{6} \simeq 0.7479$\\
$X_{65}$ & $+1$ & $+1$ & $R_{1} + R_{2} + R_{4} \simeq 0.8472$\\
$X_{67}$ & $0$ & $-1$ & $R_{6} \simeq 0.3617$\\
$X_{75}$ & $-1$ & $0$ & $R_{4} + R_{5} \simeq 0.6581$\\
$X_{96}$ & $0$ & $-1$ & $R_{5} \simeq 0.4049$\\
$X_{97}$ & $+2$ & $0$ & $R_{1} + R_{2} \simeq 0.5940$\\\hline
\end{tabular}
\end{tabular}
}
\caption{
The charges under the global symmetry on the set of chiral fields 
shared between toric phases 15b and 15c of $H_{1,1,2,1}$, including the $U(1)_R$ charges, which are expressed in terms of $U(1)_R$ charges $R_a$
corresponding to the extremal GLSM fields $p_a$. 
}
\label{tab:models_15bc_Rcharges}
\end{table}
%---------------------------------------------------- 

\paragraph{Mesonic Moduli Spaces.}
Applying the forward algorithm to the three brane tilings, 
we compute the $P$-matrices encoding their perfect matchings.
All three brane tilings admit $33$ perfect matchings, 
consisting of the extremal perfect matchings $p_1, \dots, p_6$ 
corresponding to the extremal points of the toric diagram of $H_{1,1,2,1}$,
as well as the non-extremal perfect matchings.
The $P$-matrices take the following form,
\beal{es05f13}
P^{(15a)} = \resizebox{0.75\textwidth}{!}{$\left(\begin{tabular}{c | *{6}{c}| *{2}{c}| *{14}{c}| *{11}{c}}
 & $p_{1}$ & $p_{2}$ & $p_{3}$ & $p_{4}$ & $p_{5}$ & $p_{6}$ & $q_{1}$ & $q_{2}$ & $u_{1}$ & $u_{2}$ & $u_{3}$ & $u_{4}$ & $u_{5}$ & $u_{6}$ & $u_{7}$ & $u_{8}$ & $u_{9}$ & $u_{10}$ & $u_{11}$ & $u_{12}$ & $u_{13}$ & $u_{14}$ & $s_{1}$ & $s_{2}$ & $s_{3}$ & $s_{4}$ & $s_{5}$ & $s_{6}$ & $s_{7}$ & $s_{8}$ & $s_{9}$ & $s_{10}$ & $s_{11}$\\\hline
$X_{12}$ & 0 & 0 & 0 & 0 & 1 & 1 & 1 & 1 & 1 & 1 & 1 & 1 & 1 & 0 & 0 & 0 & 0 & 0 & 0 & 0 & 0 & 0 & 1 & 1 & 1 & 1 & 0 & 0 & 0 & 0 & 0 & 0 & 0\\
$X_{15}$ & 0 & 0 & 1 & 0 & 0 & 1 & 1 & 0 & 1 & 1 & 0 & 0 & 0 & 1 & 0 & 0 & 0 & 0 & 0 & 0 & 0 & 0 & 1 & 1 & 1 & 1 & 1 & 1 & 1 & 1 & 0 & 0 & 0\\
$X_{18}$ & 0 & 1 & 0 & 0 & 0 & 0 & 0 & 0 & 1 & 0 & 1 & 0 & 0 & 0 & 0 & 0 & 0 & 0 & 0 & 0 & 0 & 0 & 1 & 0 & 0 & 0 & 0 & 0 & 0 & 0 & 0 & 0 & 0\\
$X_{23}$ & 0 & 0 & 1 & 0 & 0 & 0 & 0 & 0 & 0 & 0 & 0 & 0 & 0 & 1 & 1 & 1 & 0 & 0 & 0 & 0 & 0 & 0 & 0 & 0 & 0 & 0 & 1 & 1 & 1 & 1 & 1 & 0 & 0\\
$X_{27}$ & 0 & 1 & 0 & 1 & 0 & 0 & 0 & 0 & 0 & 0 & 0 & 0 & 0 & 1 & 1 & 1 & 1 & 1 & 1 & 1 & 1 & 0 & 0 & 0 & 0 & 0 & 1 & 1 & 1 & 0 & 0 & 0 & 0\\
$X_{31}$ & 1 & 1 & 0 & 1 & 0 & 0 & 0 & 0 & 0 & 0 & 0 & 0 & 0 & 0 & 0 & 0 & 1 & 1 & 1 & 1 & 1 & 1 & 0 & 0 & 0 & 0 & 0 & 0 & 0 & 0 & 0 & 1 & 1\\
$X_{34}$ & 0 & 0 & 0 & 0 & 0 & 1 & 1 & 0 & 0 & 0 & 0 & 0 & 0 & 0 & 0 & 0 & 1 & 1 & 0 & 0 & 0 & 0 & 0 & 0 & 0 & 0 & 0 & 0 & 0 & 0 & 0 & 1 & 0\\
$X_{41}$ & 0 & 0 & 0 & 1 & 1 & 0 & 0 & 1 & 0 & 0 & 0 & 0 & 0 & 0 & 0 & 0 & 0 & 0 & 1 & 1 & 1 & 1 & 0 & 0 & 0 & 0 & 0 & 0 & 0 & 0 & 0 & 0 & 1\\
$X_{46}$ & 0 & 0 & 1 & 0 & 0 & 0 & 0 & 0 & 0 & 0 & 0 & 0 & 0 & 0 & 0 & 0 & 0 & 0 & 1 & 1 & 0 & 0 & 1 & 1 & 1 & 0 & 1 & 1 & 0 & 0 & 0 & 0 & 1\\
$X_{53}$ & 0 & 0 & 0 & 0 & 1 & 0 & 0 & 1 & 0 & 0 & 1 & 1 & 1 & 0 & 1 & 1 & 0 & 0 & 0 & 0 & 0 & 0 & 0 & 0 & 0 & 0 & 0 & 0 & 0 & 0 & 1 & 0 & 0\\
$X_{54}$ & 1 & 1 & 0 & 0 & 0 & 0 & 0 & 0 & 0 & 0 & 1 & 1 & 1 & 0 & 1 & 1 & 1 & 1 & 0 & 0 & 0 & 0 & 0 & 0 & 0 & 0 & 0 & 0 & 0 & 0 & 1 & 1 & 0\\
$X_{59}$ & 0 & 0 & 1 & 0 & 0 & 1 & 0 & 1 & 0 & 0 & 1 & 1 & 0 & 0 & 1 & 0 & 1 & 0 & 1 & 0 & 0 & 0 & 1 & 1 & 0 & 0 & 1 & 0 & 0 & 0 & 1 & 1 & 1\\
$X_{65}$ & 1 & 1 & 0 & 1 & 0 & 0 & 0 & 0 & 1 & 1 & 0 & 0 & 0 & 1 & 0 & 0 & 0 & 0 & 0 & 0 & 1 & 1 & 0 & 0 & 0 & 1 & 0 & 0 & 1 & 1 & 0 & 0 & 0\\
$X_{67}$ & 0 & 0 & 0 & 0 & 0 & 1 & 0 & 1 & 0 & 0 & 0 & 0 & 0 & 0 & 0 & 0 & 0 & 0 & 0 & 0 & 1 & 0 & 0 & 0 & 0 & 1 & 0 & 0 & 1 & 0 & 0 & 0 & 0\\
$X_{71}$ & 1 & 0 & 1 & 0 & 0 & 0 & 0 & 0 & 0 & 0 & 0 & 0 & 0 & 0 & 0 & 0 & 0 & 0 & 0 & 0 & 0 & 1 & 0 & 0 & 0 & 0 & 0 & 0 & 0 & 1 & 1 & 1 & 1\\
$X_{75}$ & 0 & 0 & 0 & 1 & 1 & 0 & 1 & 0 & 1 & 1 & 0 & 0 & 0 & 1 & 0 & 0 & 0 & 0 & 0 & 0 & 0 & 1 & 0 & 0 & 0 & 0 & 0 & 0 & 0 & 1 & 0 & 0 & 0\\
$X_{78}$ & 0 & 0 & 1 & 0 & 1 & 1 & 1 & 1 & 1 & 0 & 1 & 0 & 0 & 0 & 0 & 0 & 0 & 0 & 0 & 0 & 0 & 1 & 1 & 0 & 0 & 0 & 0 & 0 & 0 & 1 & 1 & 1 & 1\\
$X_{82}$ & 1 & 0 & 0 & 0 & 0 & 0 & 0 & 0 & 0 & 1 & 0 & 1 & 1 & 0 & 0 & 0 & 0 & 0 & 0 & 0 & 0 & 0 & 0 & 1 & 1 & 1 & 0 & 0 & 0 & 0 & 0 & 0 & 0\\
$X_{89}$ & 0 & 0 & 0 & 1 & 0 & 0 & 0 & 0 & 0 & 1 & 0 & 1 & 0 & 1 & 1 & 0 & 1 & 0 & 1 & 0 & 0 & 0 & 0 & 1 & 0 & 0 & 1 & 0 & 0 & 0 & 0 & 0 & 0\\
$X_{96}$ & 0 & 0 & 0 & 0 & 1 & 0 & 1 & 0 & 0 & 0 & 0 & 0 & 1 & 0 & 0 & 1 & 0 & 1 & 0 & 1 & 0 & 0 & 0 & 0 & 1 & 0 & 0 & 1 & 0 & 0 & 0 & 0 & 0\\
$X_{97}$ & 1 & 1 & 0 & 0 & 0 & 0 & 0 & 0 & 0 & 0 & 0 & 0 & 1 & 0 & 0 & 1 & 0 & 1 & 0 & 1 & 1 & 0 & 0 & 0 & 1 & 1 & 0 & 1 & 1 & 0 & 0 & 0 & 0\\
\end{tabular}\right)$}
\eea
and
\beal{es05f14}
P^{(15b)} = \resizebox{0.75\textwidth}{!}{$\left(\begin{tabular}{c | *{6}{c}| *{2}{c}| *{14}{c}| *{11}{c}}
 & $p_{1}$ & $p_{2}$ & $p_{3}$ & $p_{4}$ & $p_{5}$ & $p_{6}$ & $q_{1}$ & $q_{2}$ & $u_{1}$ & $u_{2}$ & $u_{3}$ & $u_{4}$ & $u_{5}$ & $u_{6}$ & $u_{7}$ & $u_{8}$ & $u_{9}$ & $u_{10}$ & $u_{11}$ & $u_{12}$ & $u_{13}$ & $u_{14}$ & $s_{1}$ & $s_{2}$ & $s_{3}$ & $s_{4}$ & $s_{5}$ & $s_{6}$ & $s_{7}$ & $s_{8}$ & $s_{9}$ & $s_{10}$ & $s_{11}$\\\hline
$X_{12}$ & 1 & 1 & 0 & 0 & 0 & 0 & 0 & 0 & 1 & 1 & 1 & 1 & 1 & 0 & 0 & 0 & 0 & 0 & 0 & 0 & 0 & 0 & 1 & 1 & 1 & 1 & 0 & 0 & 0 & 0 & 0 & 0 & 0\\
$X_{15}$ & 0 & 0 & 1 & 0 & 0 & 1 & 1 & 0 & 1 & 1 & 0 & 0 & 0 & 1 & 0 & 0 & 0 & 0 & 0 & 0 & 0 & 0 & 1 & 1 & 1 & 1 & 1 & 1 & 1 & 1 & 0 & 0 & 0\\
$X_{18}$ & 0 & 0 & 0 & 0 & 1 & 0 & 1 & 0 & 1 & 0 & 1 & 0 & 0 & 0 & 0 & 0 & 0 & 0 & 0 & 0 & 0 & 0 & 1 & 0 & 0 & 0 & 0 & 0 & 0 & 0 & 0 & 0 & 0\\
$X_{23}$ & 0 & 0 & 1 & 0 & 0 & 0 & 0 & 0 & 0 & 0 & 0 & 0 & 0 & 1 & 1 & 1 & 0 & 0 & 0 & 0 & 0 & 0 & 0 & 0 & 0 & 0 & 1 & 1 & 1 & 1 & 1 & 0 & 0\\
$X_{27}$ & 0 & 0 & 0 & 1 & 1 & 0 & 1 & 0 & 0 & 0 & 0 & 0 & 0 & 1 & 1 & 1 & 1 & 1 & 1 & 1 & 1 & 0 & 0 & 0 & 0 & 0 & 1 & 1 & 1 & 0 & 0 & 0 & 0\\
$X_{31}$ & 1 & 1 & 0 & 1 & 0 & 0 & 0 & 0 & 0 & 0 & 0 & 0 & 0 & 0 & 0 & 0 & 1 & 1 & 1 & 1 & 1 & 1 & 0 & 0 & 0 & 0 & 0 & 0 & 0 & 0 & 0 & 1 & 1\\
$X_{34}$ & 0 & 0 & 0 & 0 & 0 & 1 & 1 & 0 & 0 & 0 & 0 & 0 & 0 & 0 & 0 & 0 & 1 & 1 & 0 & 0 & 0 & 0 & 0 & 0 & 0 & 0 & 0 & 0 & 0 & 0 & 0 & 1 & 0\\
$X_{41}$ & 0 & 0 & 0 & 1 & 1 & 0 & 0 & 1 & 0 & 0 & 0 & 0 & 0 & 0 & 0 & 0 & 0 & 0 & 1 & 1 & 1 & 1 & 0 & 0 & 0 & 0 & 0 & 0 & 0 & 0 & 0 & 0 & 1\\
$X_{46}$ & 0 & 0 & 1 & 0 & 0 & 0 & 0 & 0 & 0 & 0 & 0 & 0 & 0 & 0 & 0 & 0 & 0 & 0 & 1 & 1 & 0 & 0 & 1 & 1 & 1 & 0 & 1 & 1 & 0 & 0 & 0 & 0 & 1\\
$X_{53}$ & 0 & 0 & 0 & 0 & 1 & 0 & 0 & 1 & 0 & 0 & 1 & 1 & 1 & 0 & 1 & 1 & 0 & 0 & 0 & 0 & 0 & 0 & 0 & 0 & 0 & 0 & 0 & 0 & 0 & 0 & 1 & 0 & 0\\
$X_{54}$ & 1 & 1 & 0 & 0 & 0 & 0 & 0 & 0 & 0 & 0 & 1 & 1 & 1 & 0 & 1 & 1 & 1 & 1 & 0 & 0 & 0 & 0 & 0 & 0 & 0 & 0 & 0 & 0 & 0 & 0 & 1 & 1 & 0\\
$X_{59}$ & 1 & 0 & 1 & 0 & 0 & 0 & 0 & 0 & 0 & 0 & 1 & 1 & 0 & 0 & 1 & 0 & 1 & 0 & 1 & 0 & 0 & 0 & 1 & 1 & 0 & 0 & 1 & 0 & 0 & 0 & 1 & 1 & 1\\
$X_{65}$ & 0 & 0 & 0 & 1 & 1 & 1 & 1 & 1 & 1 & 1 & 0 & 0 & 0 & 1 & 0 & 0 & 0 & 0 & 0 & 0 & 1 & 1 & 0 & 0 & 0 & 1 & 0 & 0 & 1 & 1 & 0 & 0 & 0\\
$X_{67}$ & 1 & 0 & 0 & 0 & 0 & 0 & 0 & 0 & 0 & 0 & 0 & 0 & 0 & 0 & 0 & 0 & 0 & 0 & 0 & 0 & 1 & 0 & 0 & 0 & 0 & 1 & 0 & 0 & 1 & 0 & 0 & 0 & 0\\
$X_{71}$ & 0 & 0 & 1 & 0 & 0 & 1 & 0 & 1 & 0 & 0 & 0 & 0 & 0 & 0 & 0 & 0 & 0 & 0 & 0 & 0 & 0 & 1 & 0 & 0 & 0 & 0 & 0 & 0 & 0 & 1 & 1 & 1 & 1\\
$X_{75}$ & 0 & 1 & 0 & 1 & 0 & 0 & 0 & 0 & 1 & 1 & 0 & 0 & 0 & 1 & 0 & 0 & 0 & 0 & 0 & 0 & 0 & 1 & 0 & 0 & 0 & 0 & 0 & 0 & 0 & 1 & 0 & 0 & 0\\
$X_{78}$ & 1 & 1 & 1 & 0 & 0 & 0 & 0 & 0 & 1 & 0 & 1 & 0 & 0 & 0 & 0 & 0 & 0 & 0 & 0 & 0 & 0 & 1 & 1 & 0 & 0 & 0 & 0 & 0 & 0 & 1 & 1 & 1 & 1\\
$X_{82}$ & 0 & 0 & 0 & 0 & 0 & 1 & 0 & 1 & 0 & 1 & 0 & 1 & 1 & 0 & 0 & 0 & 0 & 0 & 0 & 0 & 0 & 0 & 0 & 1 & 1 & 1 & 0 & 0 & 0 & 0 & 0 & 0 & 0\\
$X_{89}$ & 0 & 0 & 0 & 1 & 0 & 0 & 0 & 0 & 0 & 1 & 0 & 1 & 0 & 1 & 1 & 0 & 1 & 0 & 1 & 0 & 0 & 0 & 0 & 1 & 0 & 0 & 1 & 0 & 0 & 0 & 0 & 0 & 0\\
$X_{96}$ & 0 & 1 & 0 & 0 & 0 & 0 & 0 & 0 & 0 & 0 & 0 & 0 & 1 & 0 & 0 & 1 & 0 & 1 & 0 & 1 & 0 & 0 & 0 & 0 & 1 & 0 & 0 & 1 & 0 & 0 & 0 & 0 & 0\\
$X_{97}$ & 0 & 0 & 0 & 0 & 1 & 1 & 1 & 1 & 0 & 0 & 0 & 0 & 1 & 0 & 0 & 1 & 0 & 1 & 0 & 1 & 1 & 0 & 0 & 0 & 1 & 1 & 0 & 1 & 1 & 0 & 0 & 0 & 0\\
\end{tabular}\right)$}
\eea
and
\beal{es05f15}
P^{(15c)} = \resizebox{0.75\textwidth}{!}{$\left(\begin{tabular}{c | *{6}{c}| *{2}{c}| *{14}{c}| *{11}{c}}
 & $p_{1}$ & $p_{2}$ & $p_{3}$ & $p_{4}$ & $p_{5}$ & $p_{6}$ & $q_{1}$ & $q_{2}$ & $u_{1}$ & $u_{2}$ & $u_{3}$ & $u_{4}$ & $u_{5}$ & $u_{6}$ & $u_{7}$ & $u_{8}$ & $u_{9}$ & $u_{10}$ & $u_{11}$ & $u_{12}$ & $u_{13}$ & $u_{14}$ & $s_{1}$ & $s_{2}$ & $s_{3}$ & $s_{4}$ & $s_{5}$ & $s_{6}$ & $s_{7}$ & $s_{8}$ & $s_{9}$ & $s_{10}$ & $s_{11}$\\\hline
$X_{12}$ & 1 & 1 & 0 & 0 & 0 & 0 & 0 & 0 & 1 & 1 & 1 & 1 & 1 & 0 & 0 & 0 & 0 & 0 & 0 & 0 & 0 & 0 & 1 & 1 & 1 & 1 & 0 & 0 & 0 & 0 & 0 & 0 & 0\\
$X_{15}$ & 1 & 0 & 1 & 0 & 0 & 0 & 0 & 0 & 1 & 1 & 0 & 0 & 0 & 1 & 0 & 0 & 0 & 0 & 0 & 0 & 0 & 0 & 1 & 1 & 1 & 1 & 1 & 1 & 1 & 1 & 0 & 0 & 0\\
$X_{18}$ & 0 & 0 & 0 & 0 & 1 & 0 & 1 & 0 & 1 & 0 & 1 & 0 & 0 & 0 & 0 & 0 & 0 & 0 & 0 & 0 & 0 & 0 & 1 & 0 & 0 & 0 & 0 & 0 & 0 & 0 & 0 & 0 & 0\\
$X_{23}$ & 0 & 0 & 1 & 0 & 0 & 0 & 0 & 0 & 0 & 0 & 0 & 0 & 0 & 1 & 1 & 1 & 0 & 0 & 0 & 0 & 0 & 0 & 0 & 0 & 0 & 0 & 1 & 1 & 1 & 1 & 1 & 0 & 0\\
$X_{27}$ & 0 & 0 & 0 & 1 & 1 & 0 & 1 & 0 & 0 & 0 & 0 & 0 & 0 & 1 & 1 & 1 & 1 & 1 & 1 & 1 & 1 & 0 & 0 & 0 & 0 & 0 & 1 & 1 & 1 & 0 & 0 & 0 & 0\\
$X_{31}$ & 0 & 0 & 0 & 1 & 1 & 1 & 1 & 1 & 0 & 0 & 0 & 0 & 0 & 0 & 0 & 0 & 1 & 1 & 1 & 1 & 1 & 1 & 0 & 0 & 0 & 0 & 0 & 0 & 0 & 0 & 0 & 1 & 1\\
$X_{34}$ & 1 & 0 & 0 & 0 & 0 & 0 & 0 & 0 & 0 & 0 & 0 & 0 & 0 & 0 & 0 & 0 & 1 & 1 & 0 & 0 & 0 & 0 & 0 & 0 & 0 & 0 & 0 & 0 & 0 & 0 & 0 & 1 & 0\\
$X_{41}$ & 0 & 1 & 0 & 1 & 0 & 0 & 0 & 0 & 0 & 0 & 0 & 0 & 0 & 0 & 0 & 0 & 0 & 0 & 1 & 1 & 1 & 1 & 0 & 0 & 0 & 0 & 0 & 0 & 0 & 0 & 0 & 0 & 1\\
$X_{46}$ & 0 & 0 & 1 & 0 & 0 & 0 & 0 & 0 & 0 & 0 & 0 & 0 & 0 & 0 & 0 & 0 & 0 & 0 & 1 & 1 & 0 & 0 & 1 & 1 & 1 & 0 & 1 & 1 & 0 & 0 & 0 & 0 & 1\\
$X_{53}$ & 0 & 1 & 0 & 0 & 0 & 0 & 0 & 0 & 0 & 0 & 1 & 1 & 1 & 0 & 1 & 1 & 0 & 0 & 0 & 0 & 0 & 0 & 0 & 0 & 0 & 0 & 0 & 0 & 0 & 0 & 1 & 0 & 0\\
$X_{54}$ & 0 & 0 & 0 & 0 & 1 & 1 & 1 & 1 & 0 & 0 & 1 & 1 & 1 & 0 & 1 & 1 & 1 & 1 & 0 & 0 & 0 & 0 & 0 & 0 & 0 & 0 & 0 & 0 & 0 & 0 & 1 & 1 & 0\\
$X_{59}$ & 0 & 0 & 1 & 0 & 0 & 1 & 1 & 0 & 0 & 0 & 1 & 1 & 0 & 0 & 1 & 0 & 1 & 0 & 1 & 0 & 0 & 0 & 1 & 1 & 0 & 0 & 1 & 0 & 0 & 0 & 1 & 1 & 1\\
$X_{65}$ & 1 & 1 & 0 & 1 & 0 & 0 & 0 & 0 & 1 & 1 & 0 & 0 & 0 & 1 & 0 & 0 & 0 & 0 & 0 & 0 & 1 & 1 & 0 & 0 & 0 & 1 & 0 & 0 & 1 & 1 & 0 & 0 & 0\\
$X_{67}$ & 0 & 0 & 0 & 0 & 0 & 1 & 1 & 0 & 0 & 0 & 0 & 0 & 0 & 0 & 0 & 0 & 0 & 0 & 0 & 0 & 1 & 0 & 0 & 0 & 0 & 1 & 0 & 0 & 1 & 0 & 0 & 0 & 0\\
$X_{71}$ & 0 & 0 & 1 & 0 & 0 & 1 & 0 & 1 & 0 & 0 & 0 & 0 & 0 & 0 & 0 & 0 & 0 & 0 & 0 & 0 & 0 & 1 & 0 & 0 & 0 & 0 & 0 & 0 & 0 & 1 & 1 & 1 & 1\\
$X_{75}$ & 0 & 0 & 0 & 1 & 1 & 0 & 0 & 1 & 1 & 1 & 0 & 0 & 0 & 1 & 0 & 0 & 0 & 0 & 0 & 0 & 0 & 1 & 0 & 0 & 0 & 0 & 0 & 0 & 0 & 1 & 0 & 0 & 0\\
$X_{78}$ & 1 & 1 & 1 & 0 & 0 & 0 & 0 & 0 & 1 & 0 & 1 & 0 & 0 & 0 & 0 & 0 & 0 & 0 & 0 & 0 & 0 & 1 & 1 & 0 & 0 & 0 & 0 & 0 & 0 & 1 & 1 & 1 & 1\\
$X_{82}$ & 0 & 0 & 0 & 0 & 0 & 1 & 0 & 1 & 0 & 1 & 0 & 1 & 1 & 0 & 0 & 0 & 0 & 0 & 0 & 0 & 0 & 0 & 0 & 1 & 1 & 1 & 0 & 0 & 0 & 0 & 0 & 0 & 0\\
$X_{89}$ & 0 & 0 & 0 & 1 & 0 & 0 & 0 & 0 & 0 & 1 & 0 & 1 & 0 & 1 & 1 & 0 & 1 & 0 & 1 & 0 & 0 & 0 & 0 & 1 & 0 & 0 & 1 & 0 & 0 & 0 & 0 & 0 & 0\\
$X_{96}$ & 0 & 0 & 0 & 0 & 1 & 0 & 0 & 1 & 0 & 0 & 0 & 0 & 1 & 0 & 0 & 1 & 0 & 1 & 0 & 1 & 0 & 0 & 0 & 0 & 1 & 0 & 0 & 1 & 0 & 0 & 0 & 0 & 0\\
$X_{97}$ & 1 & 1 & 0 & 0 & 0 & 0 & 0 & 0 & 0 & 0 & 0 & 0 & 1 & 0 & 0 & 1 & 0 & 1 & 0 & 1 & 1 & 0 & 0 & 0 & 1 & 1 & 0 & 1 & 1 & 0 & 0 & 0 & 0\\
\end{tabular}\right)$}
~.~
\eea

From the $P$-matrices in \eref{es05f13}, \eref{es05f14} and \eref{es05f15}, 
the forward algorithm yields identical $G_t$-matrices for the three toric phases, 
which take the form,
\beal{es05f16}
G_t^{(15a)} = G_t^{(15b)} = G_t^{(15c)} = \small{\setlength{\tabcolsep}{1pt}\left(\begin{tabular}{ *{6}{c}| *{2}{c}| *{3}{c}| *{3}{c}}
$p_{1}$ & $p_{2}$ & $p_{3}$ & $p_{4}$ & $p_{5}$ & $p_{6}$ & $q_{1}$ & $q_{2}$ & $u_{1}$ & $\cdots$ & $u_{14}$ &$s_{1}$ & $\cdots$ & $s_{11}$\\\hline
2 & 2 & 1 & 1 & 0 & 0 & 0 & 0 & 1 & $\cdots$ & 1 & 1 & $\cdots$ & 1 \\
2 & 1 & 3 & 0 & 0 & 2 & 1 & 1 & 1 & $\cdots$ & 1 & 2 & $\cdots$ & 2 \\
1 & 1 & 1 & 1 & 1 & 1 & 1 & 1 & 1 & $\cdots$ & 1 & 1 & $\cdots$ & 1 \\
\end{tabular}\right)}
~.~
\eea
The toric diagrams obtained from the three brane tilings therefore coincide,
which verifies that the brane tilings of toric phases 15a, 15b and 15c 
correspond to the same toric Calabi-Yau 3-fold $H_{1,1,2,1}$, 
as required for the quiver-invariant dualities realized by the tilting mutations.
\\

%=================================================================
\subsection{$H_{1,1,2,1}$ 17a and 17b} % family A
%=================================================================

%---------------------------------------------------- 
\begin{figure}[htbp]
\centering
\includegraphics[width=\textwidth]{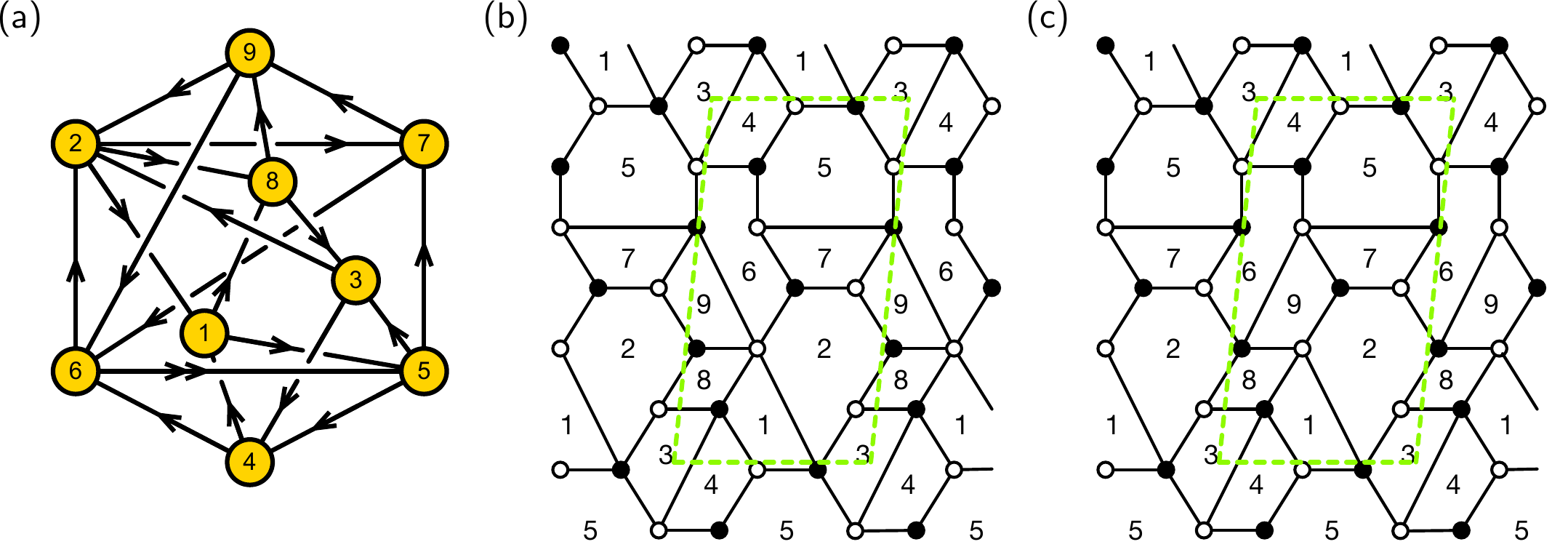}
\caption{
(a) The quiver shared between the two brane tilings in (b) and (c) corresponding to the two toric phases 17a and 17b of $H_{1,1,2,1}$,
respectively. 
}
\label{fig_doublet_17a_17b}
\end{figure}
%---------------------------------------------------- 

The two toric phases 17a and 17b of the $H_{1,1,2,1}$ model 
share the same quiver, which is shown in \fref{fig_doublet_17a_17b}(a).
The corresponding brane tilings are illustrated in 
\fref{fig_doublet_17a_17b}(b) and \fref{fig_doublet_17a_17b}(c), respectively,
and differ in their superpotentials, 
which are given by,
\beal{es05e01}
W_{17a} = 
&&
X_{15} X_{54} X_{41}
+ X_{18} X_{89} X_{96} X_{62} X_{21}
+ X_{27} X_{79} X_{92}
+ X_{28} X_{83} X_{32}
\nn\\
&&
+ X_{34} X_{46} X_{65}^{1} X_{53}
+ X_{57} X_{76} X_{65}^{2}
- X_{15} X_{53} X_{32} X_{21}
- X_{18} X_{83} X_{34} X_{41}
\nn\\
&&
- X_{27} X_{76} X_{62}
- X_{28} X_{89} X_{92}
- X_{46} X_{65}^{2} X_{54}
- X_{57} X_{79} X_{96} X_{65}^{1}
~,~
\eea
and
\beal{es05e02}
W_{17b} = 
&&
X_{15} X_{54} X_{41}
+ X_{18} X_{89} X_{92} X_{21}
+ X_{27} X_{76} X_{62}
+ X_{28} X_{83} X_{32}
\nn\\
&&
+ X_{34} X_{46} X_{65}^{1} X_{53}
+ X_{57} X_{79} X_{96} X_{65}^{2}
- X_{15} X_{53} X_{32} X_{21}
- X_{18} X_{83} X_{34} X_{41}
\nn\\
&&
- X_{27} X_{79} X_{92}
- X_{28} X_{89} X_{96} X_{62}
- X_{46} X_{65}^{2} X_{54}
- X_{57} X_{76} X_{65}^{1}
~.~
\eea

\begin{figure}[htbp]
\centering
\includegraphics[width=0.7\textwidth]{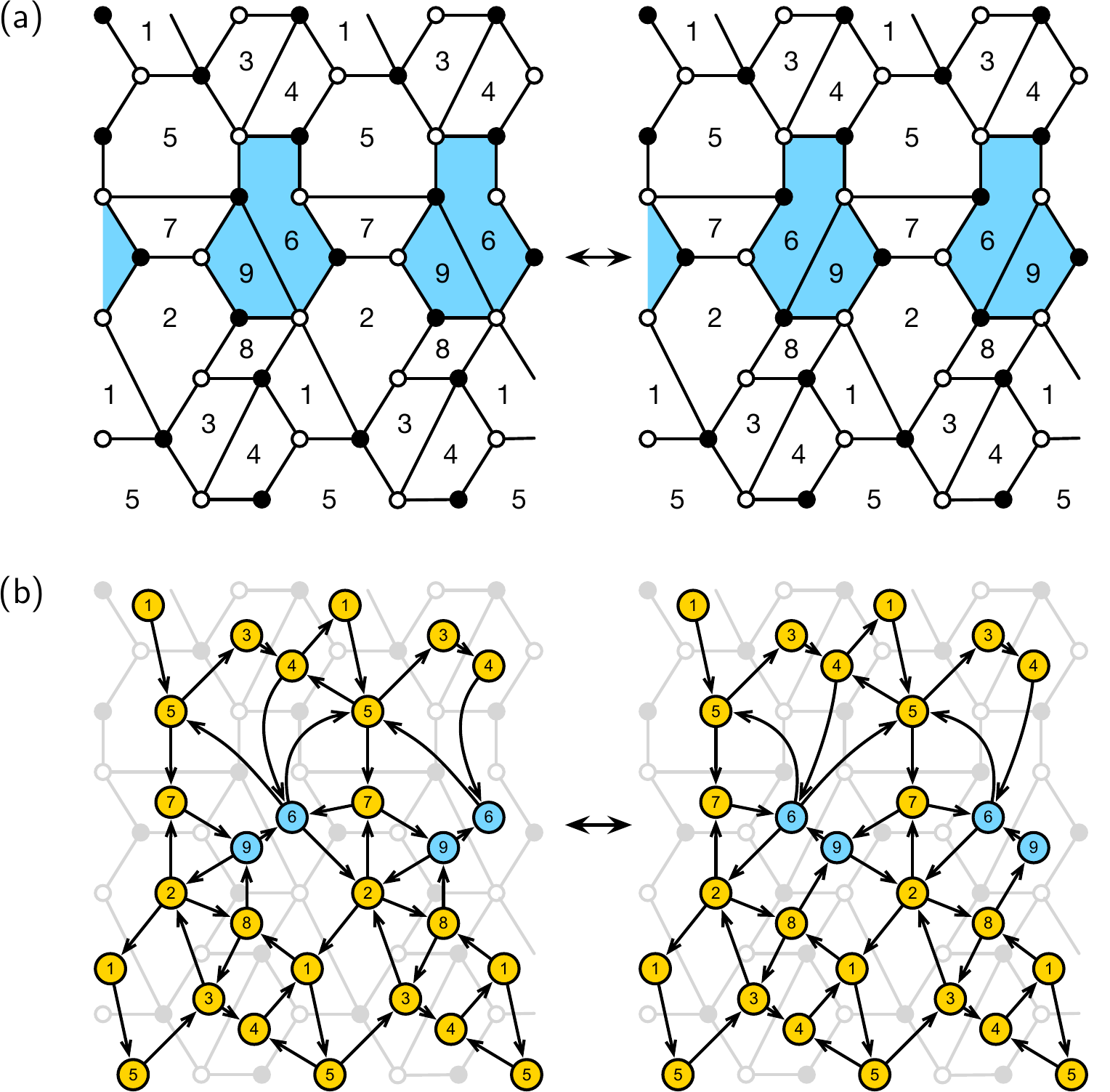}
\caption{
(a) The tilting mutation relating the brane tilings of the two toric phases 17a and 17b of $H_{1,1,2,1}$,
and (b) its effect on the corresponding periodic quivers. 
}
\label{fig_tilting_17a_17b}
\end{figure}

\paragraph{Tilting Mutation.}
By comparing the superpotentials in \eref{es05e01} and \eref{es05e02},
we identify the tilting mutation relating toric phase 17a to 17b 
as belonging to tilting mutation family A introduced in Section~\ref{sec:mutfam_a},
where the quiver nodes $(a,b,c,d)$ of the mutation region 
are identified with the quiver nodes $(9,6,7,2)$ of the two toric phases.
Under this identification, the superpotential $W_{17b}$ takes the general form of $W$ 
for tilting mutation family A, while $W_{17a}$ takes the general form of $W^\prime$.
The part of the superpotential that is
unaffected by the tilting mutation
is given by,
\beal{es05e03}
&&
W_0 = 
X_{15} X_{54} X_{41}
+ X_{28} X_{83} X_{32}
+ X_{34} X_{46} X_{65}^{1} X_{53}
\nn\\
&&
\hspace{1cm}
- X_{15} X_{53} X_{32} X_{21}
- X_{18} X_{83} X_{34} X_{41}
- X_{46} X_{65}^{2} X_{54}
~.~
\eea
The superpotential terms whose signs are reversed under the tilting mutation, 
with the signs taken as they appear in $W_{17b}$, 
are given by,
\beal{es05e04}
\Omega_{\text{mut}} =
X_{27} X_{76} X_{62}
- X_{27} X_{79} X_{92}
~,~
\eea
and the exterior paths take the form,
\beal{es05e05}
&&
P_{29} = X_{21} X_{18} X_{89} 
~,~
Q_{29} = X_{28} X_{89} 
~,~
P_{67} = X_{65}^{2} X_{57} 
~,~
Q_{67} = X_{65}^{1} X_{57} 
~.~
\eea
Accordingly, the paths exchanged by the tilting mutation are given by,
\beal{es05e06}
U_{29} = X_{92}
~,~
U_{67} = X_{79} X_{96}
~,~
V_{29} = X_{96} X_{62}
~,~
V_{67} = X_{76}
~,~
\eea
which is
in agreement with the general form in Section~\ref{sec:mutfam_a} for tilting mutation family A,
such that the tilting mutation maps the superpotential $W_{17b}$ in \eref{es05e02} 
to $W_{17a}$ in \eref{es05e01}, and vice versa.
We note that the two chiral fields $X_{65}^{1}$ and $X_{65}^{2}$ 
associated with the double arrow of the quiver 
lie outside the mutation region,
with the exterior paths $P_{67}$ and $Q_{67}$ in \eref{es05e05} 
differing precisely by which of the two chiral fields they contain,
such that the tilting mutation exchanges the roles of 
$X_{65}^{1}$ and $X_{65}^{2}$ between $W_{17a}$ and $W_{17b}$.
The action of the tilting mutation on the brane tiling 
and on the corresponding periodic quiver 
is illustrated in \fref{fig_tilting_17a_17b}.

%---------------------------------------------------- 
%---------------------------------------------------- 

\paragraph{Global Symmetry Charges.}
\tref{tab:models_17_Rcharges} summarizes the charges under the global symmetry
carried by the chiral fields of the two toric phases,
where the $U(1)_R$ charges are expressed in terms of
the $U(1)_R$ charges $R_a$ of the extremal GLSM fields $p_a$ in \tref{tab:extremal_charges}.
As shown in \tref{tab:models_17_Rcharges},
the chiral fields along the interior paths
$U_{29}$, $V_{29}$, $U_{67}$ and $V_{67}$ in \eref{es05e06}
carry identical global symmetry charges in the two toric phases.
In contrast, the exterior paths $P_{29}$ and $Q_{29}$,
as well as $P_{67}$ and $Q_{67}$ in \eref{es05e05},
exchange their global symmetry charges under the tilting mutation.
In particular, the exchange of charges between the two chiral fields
$X_{65}^{1}$ and $X_{65}^{2}$ associated with the double arrow of the quiver
reflects the exchange of the exterior paths $P_{67}$ and $Q_{67}$.
This sets tilting mutation family A apart from the other tilting mutation families B and C. 

%---------------------------------------------------- 
\begin{table}[h]
\setlength{\tabcolsep}{2pt}
\centering
\footnotesize{
\begin{tabular}[t]{c}
Phase 17a \\[4pt]
\begin{tabular}{| c | c | c | r |}
\hline
\; & $U(1)_{f_1}$ & $U(1)_{f_2}$ & $U(1)_R$\\\hline\hline
$X_{27}$ & $-1$ & $0$ & $R_{4} + R_{5} \simeq 0.6581$\\
$X_{46}$ & $-1$ & $+1$ & $R_{3} \simeq 0.3862$\\
$X_{57}$ & $-1$ & $+1$ & $R_{3} \simeq 0.3862$\\
$X_{62}$ & $-1$ & $0$ & $R_{3} + R_{6} \simeq 0.7479$\\
$X_{76}$ & $+2$ & $0$ & $R_{1} + R_{2} \simeq 0.5940$\\
$X_{79}$ & $0$ & $-1$ & $R_{6} \simeq 0.3617$\\
$X_{89}$ & $-1$ & $+1$ & $R_{4} \simeq 0.2532$\\
$X_{92}$ & $+1$ & $+1$ & $R_{1} + R_{2} + R_{3} \simeq 0.9802$\\
$X_{96}$ & $0$ & $-1$ & $R_{5} \simeq 0.4049$\\\hline
$X_{15}$ & $-1$ & $0$ & $R_{3} + R_{6} \simeq 0.7479$\\
$X_{18}$ & $+1$ & $0$ & $R_{2} \simeq 0.3205$\\
$X_{21}$ & $+1$ & $0$ & $R_{1} \simeq 0.2736$\\
$X_{28}$ & $0$ & $-2$ & $R_{5} + R_{6} \simeq 0.7666$\\
$X_{32}$ & $0$ & $+1$ & $R_{2} + R_{4} \simeq 0.5737$\\
$X_{34}$ & $0$ & $-1$ & $R_{6} \simeq 0.3617$\\
$X_{41}$ & $-1$ & $0$ & $R_{4} + R_{5} \simeq 0.6581$\\
$X_{53}$ & $0$ & $-1$ & $R_{5} \simeq 0.4049$\\
$X_{54}$ & $+2$ & $0$ & $R_{1} + R_{2} \simeq 0.5940$\\
$X^1_{65}$ & $+1$ & $+1$ & $R_{1} + R_{2} + R_{4} \simeq 0.8472$\\
$X^2_{65}$ & $-1$ & $-1$ & $R_{4} + R_{5} + R_{6} \simeq 1.0198$\\
$X_{83}$ & $0$ & $+1$ & $R_{1} + R_{3} \simeq 0.6597$\\\hline
\end{tabular}
\end{tabular}
}
\hspace{6pt}
\footnotesize{
\begin{tabular}[t]{c}
Phase 17b \\[4pt]
\begin{tabular}{| c | c | c | r |}
\hline
\; & $U(1)_{f_1}$ & $U(1)_{f_2}$ & $U(1)_R$\\\hline\hline
$X_{27}$ & $-1$ & $0$ & $R_{4} + R_{5} \simeq 0.6581$\\
$X_{46}$ & $-1$ & $+1$ & $R_{3} \simeq 0.3862$\\
$X_{57}$ & $-1$ & $+1$ & $R_{3} \simeq 0.3862$\\
$X_{62}$ & $-1$ & $0$ & $R_{3} + R_{6} \simeq 0.7479$\\
$X_{76}$ & $+2$ & $0$ & $R_{1} + R_{2} \simeq 0.5940$\\
$X_{79}$ & $0$ & $-1$ & $R_{6} \simeq 0.3617$\\
$X_{89}$ & $-1$ & $+1$ & $R_{4} \simeq 0.2532$\\
$X_{92}$ & $+1$ & $+1$ & $R_{1} + R_{2} + R_{3} \simeq 0.9802$\\
$X_{96}$ & $0$ & $-1$ & $R_{5} \simeq 0.4049$\\\hline
$X_{15}$ & $0$ & $+1$ & $R_{1} + R_{3} \simeq 0.6597$\\
$X_{18}$ & $0$ & $-1$ & $R_{5} \simeq 0.4049$\\
$X_{21}$ & $0$ & $-1$ & $R_{6} \simeq 0.3617$\\
$X_{28}$ & $+2$ & $0$ & $R_{1} + R_{2} \simeq 0.5940$\\
$X_{32}$ & $-1$ & $0$ & $R_{4} + R_{5} \simeq 0.6581$\\
$X_{34}$ & $+1$ & $0$ & $R_{1} \simeq 0.2736$\\
$X_{41}$ & $0$ & $+1$ & $R_{2} + R_{4} \simeq 0.5737$\\
$X_{53}$ & $+1$ & $0$ & $R_{2} \simeq 0.3205$\\
$X_{54}$ & $0$ & $-2$ & $R_{5} + R_{6} \simeq 0.7666$\\
$X^1_{65}$ & $-1$ & $-1$ & $R_{4} + R_{5} + R_{6} \simeq 1.0198$\\
$X^2_{65}$ & $+1$ & $+1$ & $R_{1} + R_{2} + R_{4} \simeq 0.8472$\\
$X_{83}$ & $-1$ & $0$ & $R_{3} + R_{6} \simeq 0.7479$\\\hline
\end{tabular}
\end{tabular}
}
\caption{
The charges under the global symmetry on the set of chiral fields 
shared between toric phases 17a and 17b of $H_{1,1,2,1}$, including the $U(1)_R$ charges, which are expressed in terms of $U(1)_R$ charges $R_a$
corresponding to the extremal GLSM fields $p_a$. 
}
\label{tab:models_17_Rcharges}
\end{table}
%---------------------------------------------------- 

\paragraph{Mesonic Moduli Spaces.}
Applying the forward algorithm to the two brane tilings, 
we compute the $P$-matrices encoding their perfect matchings.
Both brane tilings admit $36$ perfect matchings, 
consisting of the extremal perfect matchings $p_1, \dots, p_6$ 
corresponding to the extremal points of the toric diagram of $H_{1,1,2,1}$,
as well as the non-extremal perfect matchings.
The $P$-matrices take the following form,
\beal{es05e07}
P^{(17a)} = \resizebox{0.75\textwidth}{!}{$\left(\begin{tabular}{c | *{6}{c}| *{2}{c}| *{15}{c}| *{13}{c}}
 & $p_{1}$ & $p_{2}$ & $p_{3}$ & $p_{4}$ & $p_{5}$ & $p_{6}$ & $q_{1}$ & $q_{2}$ & $u_{1}$ & $u_{2}$ & $u_{3}$ & $u_{4}$ & $u_{5}$ & $u_{6}$ & $u_{7}$ & $u_{8}$ & $u_{9}$ & $u_{10}$ & $u_{11}$ & $u_{12}$ & $u_{13}$ & $u_{14}$ & $u_{15}$ & $s_{1}$ & $s_{2}$ & $s_{3}$ & $s_{4}$ & $s_{5}$ & $s_{6}$ & $s_{7}$ & $s_{8}$ & $s_{9}$ & $s_{10}$ & $s_{11}$ & $s_{12}$ & $s_{13}$\\\hline
$X_{15}$ & 0 & 0 & 1 & 0 & 0 & 1 & 1 & 0 & 1 & 1 & 0 & 0 & 0 & 0 & 0 & 0 & 0 & 0 & 0 & 0 & 0 & 0 & 0 & 1 & 1 & 1 & 1 & 1 & 1 & 0 & 0 & 0 & 0 & 0 & 0 & 0\\
$X_{18}$ & 0 & 1 & 0 & 0 & 0 & 0 & 0 & 0 & 1 & 0 & 1 & 1 & 0 & 0 & 0 & 0 & 0 & 0 & 0 & 0 & 0 & 0 & 0 & 1 & 1 & 0 & 0 & 0 & 0 & 0 & 0 & 0 & 0 & 0 & 0 & 0\\
$X_{21}$ & 1 & 0 & 0 & 0 & 0 & 0 & 0 & 0 & 0 & 0 & 0 & 0 & 1 & 0 & 0 & 0 & 0 & 0 & 0 & 0 & 0 & 0 & 0 & 0 & 0 & 0 & 0 & 0 & 0 & 1 & 1 & 1 & 1 & 0 & 0 & 0\\
$X_{27}$ & 0 & 0 & 0 & 1 & 1 & 0 & 1 & 0 & 1 & 1 & 1 & 0 & 1 & 1 & 1 & 1 & 0 & 0 & 0 & 0 & 0 & 0 & 0 & 1 & 0 & 1 & 0 & 0 & 0 & 1 & 1 & 0 & 0 & 0 & 0 & 0\\
$X_{28}$ & 0 & 0 & 0 & 0 & 1 & 1 & 1 & 1 & 1 & 0 & 1 & 1 & 1 & 0 & 0 & 0 & 0 & 0 & 0 & 0 & 0 & 0 & 0 & 1 & 1 & 0 & 0 & 0 & 0 & 1 & 1 & 1 & 1 & 0 & 0 & 0\\
$X_{32}$ & 0 & 1 & 0 & 1 & 0 & 0 & 0 & 0 & 0 & 0 & 0 & 0 & 0 & 1 & 1 & 0 & 1 & 1 & 1 & 1 & 1 & 0 & 0 & 0 & 0 & 0 & 0 & 0 & 0 & 0 & 0 & 0 & 0 & 1 & 1 & 0\\
$X_{34}$ & 0 & 0 & 0 & 0 & 0 & 1 & 1 & 0 & 0 & 0 & 0 & 0 & 0 & 1 & 0 & 0 & 1 & 1 & 0 & 0 & 0 & 0 & 0 & 0 & 0 & 0 & 0 & 0 & 0 & 1 & 0 & 1 & 0 & 1 & 0 & 0\\
$X_{41}$ & 0 & 0 & 0 & 1 & 1 & 0 & 0 & 1 & 0 & 0 & 0 & 0 & 1 & 0 & 1 & 0 & 0 & 0 & 1 & 1 & 1 & 0 & 0 & 0 & 0 & 0 & 0 & 0 & 0 & 0 & 1 & 0 & 1 & 0 & 1 & 0\\
$X_{46}$ & 0 & 0 & 1 & 0 & 0 & 0 & 0 & 0 & 0 & 0 & 0 & 0 & 0 & 0 & 1 & 0 & 0 & 0 & 1 & 1 & 0 & 0 & 0 & 1 & 1 & 1 & 1 & 1 & 0 & 0 & 1 & 0 & 1 & 0 & 1 & 0\\
$X_{53}$ & 0 & 0 & 0 & 0 & 1 & 0 & 0 & 1 & 0 & 0 & 1 & 1 & 0 & 0 & 0 & 1 & 0 & 0 & 0 & 0 & 0 & 1 & 1 & 0 & 0 & 0 & 0 & 0 & 0 & 0 & 0 & 0 & 0 & 0 & 0 & 1\\
$X_{54}$ & 1 & 1 & 0 & 0 & 0 & 0 & 0 & 0 & 0 & 0 & 1 & 1 & 0 & 1 & 0 & 1 & 1 & 1 & 0 & 0 & 0 & 1 & 1 & 0 & 0 & 0 & 0 & 0 & 0 & 1 & 0 & 1 & 0 & 1 & 0 & 1\\
$X_{57}$ & 0 & 0 & 1 & 0 & 0 & 0 & 0 & 0 & 0 & 0 & 1 & 0 & 0 & 1 & 1 & 1 & 0 & 0 & 0 & 0 & 0 & 0 & 0 & 1 & 0 & 1 & 0 & 0 & 0 & 1 & 1 & 0 & 0 & 1 & 1 & 1\\
$X_{62}$ & 0 & 0 & 1 & 0 & 0 & 1 & 0 & 1 & 0 & 0 & 0 & 0 & 0 & 0 & 0 & 0 & 0 & 0 & 0 & 0 & 1 & 0 & 0 & 0 & 0 & 0 & 0 & 0 & 1 & 0 & 0 & 0 & 0 & 1 & 1 & 1\\
$X^1_{65}$ & 1 & 1 & 0 & 1 & 0 & 0 & 0 & 0 & 1 & 1 & 0 & 0 & 1 & 0 & 0 & 0 & 0 & 0 & 0 & 0 & 1 & 0 & 0 & 0 & 0 & 0 & 0 & 0 & 1 & 0 & 0 & 0 & 0 & 0 & 0 & 0\\
$X^2_{65}$ & 0 & 0 & 0 & 1 & 1 & 1 & 1 & 1 & 1 & 1 & 0 & 0 & 1 & 0 & 0 & 0 & 0 & 0 & 0 & 0 & 1 & 0 & 0 & 0 & 0 & 0 & 0 & 0 & 1 & 0 & 0 & 0 & 0 & 0 & 0 & 0\\
$X_{76}$ & 1 & 1 & 0 & 0 & 0 & 0 & 0 & 0 & 0 & 0 & 0 & 1 & 0 & 0 & 0 & 0 & 1 & 1 & 1 & 1 & 0 & 1 & 1 & 0 & 1 & 0 & 1 & 1 & 0 & 0 & 0 & 1 & 1 & 0 & 0 & 0\\
$X_{79}$ & 0 & 0 & 0 & 0 & 0 & 1 & 0 & 1 & 0 & 0 & 0 & 1 & 0 & 0 & 0 & 0 & 1 & 0 & 1 & 0 & 0 & 1 & 0 & 0 & 1 & 0 & 1 & 0 & 0 & 0 & 0 & 1 & 1 & 0 & 0 & 0\\
$X_{83}$ & 1 & 0 & 1 & 0 & 0 & 0 & 0 & 0 & 0 & 1 & 0 & 0 & 0 & 0 & 0 & 1 & 0 & 0 & 0 & 0 & 0 & 1 & 1 & 0 & 0 & 1 & 1 & 1 & 1 & 0 & 0 & 0 & 0 & 0 & 0 & 1\\
$X_{89}$ & 0 & 0 & 0 & 1 & 0 & 0 & 0 & 0 & 0 & 1 & 0 & 0 & 0 & 1 & 1 & 1 & 1 & 0 & 1 & 0 & 0 & 1 & 0 & 0 & 0 & 1 & 1 & 0 & 0 & 0 & 0 & 0 & 0 & 0 & 0 & 0\\
$X_{92}$ & 1 & 1 & 1 & 0 & 0 & 0 & 0 & 0 & 0 & 0 & 0 & 0 & 0 & 0 & 0 & 0 & 0 & 1 & 0 & 1 & 1 & 0 & 1 & 0 & 0 & 0 & 0 & 1 & 1 & 0 & 0 & 0 & 0 & 1 & 1 & 1\\
$X_{96}$ & 0 & 0 & 0 & 0 & 1 & 0 & 1 & 0 & 0 & 0 & 0 & 0 & 0 & 0 & 0 & 0 & 0 & 1 & 0 & 1 & 0 & 0 & 1 & 0 & 0 & 0 & 0 & 1 & 0 & 0 & 0 & 0 & 0 & 0 & 0 & 0\\
\end{tabular}\right)$}
\eea
and
\beal{es05e08}
P^{(17b)} = \resizebox{0.75\textwidth}{!}{$\left(\begin{tabular}{c | *{6}{c}| *{2}{c}| *{15}{c}| *{13}{c}}
 & $p_{1}$ & $p_{2}$ & $p_{3}$ & $p_{4}$ & $p_{5}$ & $p_{6}$ & $q_{1}$ & $q_{2}$ & $u_{1}$ & $u_{2}$ & $u_{3}$ & $u_{4}$ & $u_{5}$ & $u_{6}$ & $u_{7}$ & $u_{8}$ & $u_{9}$ & $u_{10}$ & $u_{11}$ & $u_{12}$ & $u_{13}$ & $u_{14}$ & $u_{15}$ & $s_{1}$ & $s_{2}$ & $s_{3}$ & $s_{4}$ & $s_{5}$ & $s_{6}$ & $s_{7}$ & $s_{8}$ & $s_{9}$ & $s_{10}$ & $s_{11}$ & $s_{12}$ & $s_{13}$\\\hline
$X_{15}$ & 1 & 0 & 1 & 0 & 0 & 0 & 0 & 0 & 1 & 1 & 0 & 0 & 0 & 0 & 0 & 0 & 0 & 0 & 0 & 0 & 0 & 0 & 0 & 1 & 1 & 1 & 1 & 1 & 1 & 0 & 0 & 0 & 0 & 0 & 0 & 0\\
$X_{18}$ & 0 & 0 & 0 & 0 & 1 & 0 & 1 & 0 & 1 & 0 & 1 & 1 & 0 & 0 & 0 & 0 & 0 & 0 & 0 & 0 & 0 & 0 & 0 & 1 & 1 & 0 & 0 & 0 & 0 & 0 & 0 & 0 & 0 & 0 & 0 & 0\\
$X_{21}$ & 0 & 0 & 0 & 0 & 0 & 1 & 0 & 1 & 0 & 0 & 0 & 0 & 1 & 0 & 0 & 0 & 0 & 0 & 0 & 0 & 0 & 0 & 0 & 0 & 0 & 0 & 0 & 0 & 0 & 1 & 1 & 1 & 1 & 0 & 0 & 0\\
$X_{27}$ & 0 & 0 & 0 & 1 & 1 & 0 & 0 & 1 & 1 & 1 & 1 & 0 & 1 & 1 & 1 & 1 & 0 & 0 & 0 & 0 & 0 & 0 & 0 & 1 & 0 & 1 & 0 & 0 & 0 & 1 & 1 & 0 & 0 & 0 & 0 & 0\\
$X_{28}$ & 1 & 1 & 0 & 0 & 0 & 0 & 0 & 0 & 1 & 0 & 1 & 1 & 1 & 0 & 0 & 0 & 0 & 0 & 0 & 0 & 0 & 0 & 0 & 1 & 1 & 0 & 0 & 0 & 0 & 1 & 1 & 1 & 1 & 0 & 0 & 0\\
$X_{32}$ & 0 & 0 & 0 & 1 & 1 & 0 & 1 & 0 & 0 & 0 & 0 & 0 & 0 & 1 & 1 & 0 & 1 & 1 & 1 & 1 & 1 & 0 & 0 & 0 & 0 & 0 & 0 & 0 & 0 & 0 & 0 & 0 & 0 & 1 & 1 & 0\\
$X_{34}$ & 1 & 0 & 0 & 0 & 0 & 0 & 0 & 0 & 0 & 0 & 0 & 0 & 0 & 1 & 0 & 0 & 1 & 1 & 0 & 0 & 0 & 0 & 0 & 0 & 0 & 0 & 0 & 0 & 0 & 1 & 0 & 1 & 0 & 1 & 0 & 0\\
$X_{41}$ & 0 & 1 & 0 & 1 & 0 & 0 & 0 & 0 & 0 & 0 & 0 & 0 & 1 & 0 & 1 & 0 & 0 & 0 & 1 & 1 & 1 & 0 & 0 & 0 & 0 & 0 & 0 & 0 & 0 & 0 & 1 & 0 & 1 & 0 & 1 & 0\\
$X_{46}$ & 0 & 0 & 1 & 0 & 0 & 0 & 0 & 0 & 0 & 0 & 0 & 0 & 0 & 0 & 1 & 0 & 0 & 0 & 1 & 1 & 0 & 0 & 0 & 1 & 1 & 1 & 1 & 1 & 0 & 0 & 1 & 0 & 1 & 0 & 1 & 0\\
$X_{53}$ & 0 & 1 & 0 & 0 & 0 & 0 & 0 & 0 & 0 & 0 & 1 & 1 & 0 & 0 & 0 & 1 & 0 & 0 & 0 & 0 & 0 & 1 & 1 & 0 & 0 & 0 & 0 & 0 & 0 & 0 & 0 & 0 & 0 & 0 & 0 & 1\\
$X_{54}$ & 0 & 0 & 0 & 0 & 1 & 1 & 1 & 1 & 0 & 0 & 1 & 1 & 0 & 1 & 0 & 1 & 1 & 1 & 0 & 0 & 0 & 1 & 1 & 0 & 0 & 0 & 0 & 0 & 0 & 1 & 0 & 1 & 0 & 1 & 0 & 1\\
$X_{57}$ & 0 & 0 & 1 & 0 & 0 & 0 & 0 & 0 & 0 & 0 & 1 & 0 & 0 & 1 & 1 & 1 & 0 & 0 & 0 & 0 & 0 & 0 & 0 & 1 & 0 & 1 & 0 & 0 & 0 & 1 & 1 & 0 & 0 & 1 & 1 & 1\\
$X_{62}$ & 0 & 0 & 1 & 0 & 0 & 1 & 1 & 0 & 0 & 0 & 0 & 0 & 0 & 0 & 0 & 0 & 0 & 0 & 0 & 0 & 1 & 0 & 0 & 0 & 0 & 0 & 0 & 0 & 1 & 0 & 0 & 0 & 0 & 1 & 1 & 1\\
$X^1_{65}$ & 0 & 0 & 0 & 1 & 1 & 1 & 1 & 1 & 1 & 1 & 0 & 0 & 1 & 0 & 0 & 0 & 0 & 0 & 0 & 0 & 1 & 0 & 0 & 0 & 0 & 0 & 0 & 0 & 1 & 0 & 0 & 0 & 0 & 0 & 0 & 0\\
$X^2_{65}$ & 1 & 1 & 0 & 1 & 0 & 0 & 0 & 0 & 1 & 1 & 0 & 0 & 1 & 0 & 0 & 0 & 0 & 0 & 0 & 0 & 1 & 0 & 0 & 0 & 0 & 0 & 0 & 0 & 1 & 0 & 0 & 0 & 0 & 0 & 0 & 0\\
$X_{76}$ & 1 & 1 & 0 & 0 & 0 & 0 & 0 & 0 & 0 & 0 & 0 & 1 & 0 & 0 & 0 & 0 & 1 & 1 & 1 & 1 & 0 & 1 & 1 & 0 & 1 & 0 & 1 & 1 & 0 & 0 & 0 & 1 & 1 & 0 & 0 & 0\\
$X_{79}$ & 0 & 0 & 0 & 0 & 0 & 1 & 1 & 0 & 0 & 0 & 0 & 1 & 0 & 0 & 0 & 0 & 1 & 0 & 1 & 0 & 0 & 1 & 0 & 0 & 1 & 0 & 1 & 0 & 0 & 0 & 0 & 1 & 1 & 0 & 0 & 0\\
$X_{83}$ & 0 & 0 & 1 & 0 & 0 & 1 & 0 & 1 & 0 & 1 & 0 & 0 & 0 & 0 & 0 & 1 & 0 & 0 & 0 & 0 & 0 & 1 & 1 & 0 & 0 & 1 & 1 & 1 & 1 & 0 & 0 & 0 & 0 & 0 & 0 & 1\\
$X_{89}$ & 0 & 0 & 0 & 1 & 0 & 0 & 0 & 0 & 0 & 1 & 0 & 0 & 0 & 1 & 1 & 1 & 1 & 0 & 1 & 0 & 0 & 1 & 0 & 0 & 0 & 1 & 1 & 0 & 0 & 0 & 0 & 0 & 0 & 0 & 0 & 0\\
$X_{92}$ & 1 & 1 & 1 & 0 & 0 & 0 & 0 & 0 & 0 & 0 & 0 & 0 & 0 & 0 & 0 & 0 & 0 & 1 & 0 & 1 & 1 & 0 & 1 & 0 & 0 & 0 & 0 & 1 & 1 & 0 & 0 & 0 & 0 & 1 & 1 & 1\\
$X_{96}$ & 0 & 0 & 0 & 0 & 1 & 0 & 0 & 1 & 0 & 0 & 0 & 0 & 0 & 0 & 0 & 0 & 0 & 1 & 0 & 1 & 0 & 0 & 1 & 0 & 0 & 0 & 0 & 1 & 0 & 0 & 0 & 0 & 0 & 0 & 0 & 0\\
\end{tabular}\right)$}
~.~
\eea
From the $P$-matrices in \eref{es05e07} and \eref{es05e08}, 
the forward algorithm yields identical $G_t$-matrices for the two toric phases, 
which take the form,
\beal{es05e09}
G_t^{(17a)} = G_t^{(17b)} = \small{\setlength{\tabcolsep}{1pt}\left(\begin{tabular}{ *{6}{c}| *{2}{c}| *{3}{c}| *{3}{c}}
$p_{1}$ & $p_{2}$ & $p_{3}$ & $p_{4}$ & $p_{5}$ & $p_{6}$ & $q_{1}$ & $q_{2}$ & $u_{1}$ & $\cdots$ & $u_{15}$ &$s_{1}$ & $\cdots$ & $s_{13}$\\\hline
2 & 2 & 1 & 1 & 0 & 0 & 0 & 0 & 1 & $\cdots$ & 1 & 1 & $\cdots$ & 1 \\
2 & 1 & 3 & 0 & 0 & 2 & 1 & 1 & 1 & $\cdots$ & 1 & 2 & $\cdots$ & 2 \\
1 & 1 & 1 & 1 & 1 & 1 & 1 & 1 & 1 & $\cdots$ & 1 & 1 & $\cdots$ & 1 \\
\end{tabular}\right)}
~.~
\eea
The toric diagrams obtained from the two brane tilings therefore coincide,
which verifies that the brane tilings of toric phases 17a and 17b 
correspond to the same toric Calabi-Yau 3-fold $H_{1,1,2,1}$, 
as required for the quiver-invariant duality realized by the tilting mutation.
\\

%=================================================================
\section{Conclusion \label{sec:conclusion}}
%=================================================================

In this work, we have studied quiver-invariant dualities 
between $4d$ $\mathcal{N}=1$ supersymmetric gauge theories realized by brane tilings, 
which relate distinct toric phases of the same toric Calabi-Yau 3-fold 
that share an identical quiver 
and differ only in their superpotentials.
In contrast to Seiberg duality, 
which generically changes the quiver of the gauge theory, 
these dualities leave the gauge group structure and the matter content invariant, 
such that they are entirely invisible at the level of the quiver.
Following the discovery of quiver-invariant dualities in \cite{Kho:2026geo}, 
we have substantially extended the set of known examples 
among the toric phases of the $H_{1,1,2,1}$ model 
and identified the general structures that generate them:
tilting mutations, 
which act locally on a mutation region of the brane tiling 
by reversing the signs of the superpotential terms $\Omega_{\text{mut}}$ 
supported within the mutation region 
and by exchanging pairs of interior paths $U$ and $V$ of different lengths 
between the exterior paths $P$ and $Q$ 
that close them into gauge-invariant superpotential terms, 
while leaving the remaining part $W_0$ of the superpotential invariant.

Among the toric phases of the $H_{1,1,2,1}$ model, 
we have identified three families A, B and C of tilting mutations.
These families are distinguished by the structure of their reversed superpotential terms 
and of their exchanged paths in the periodic quiver.
On the brane tiling, each tilting mutation reverses the orientation 
of the zig-zag paths confined to the mutation region, 
thereby flipping or exchanging the corresponding external legs 
of the $(p,q)$-web diagram.
The requirement that the tilting mutation preserves the mesonic moduli space, 
and hence the associated toric Calabi-Yau 3-fold, 
translates into the condition $S_w(W^\prime) = G \, S_w(W)$ with $G \in GL(2,\mathbb{Z})$ 
on the winding numbers of the zig-zag paths, 
which restricts the choice of the invariant part $W_0$ of the superpotential.

We have presented explicit realizations of all three families 
among the toric phases of the $H_{1,1,2,1}$ model, 
given by five doublets and one triplet of toric phases with identical quivers.
For each realization, we have verified with the forward algorithm 
that the brane tilings related by the tilting mutation 
yield identical $G_t$-matrices and hence the same toric Calabi-Yau 3-fold.
It would be interesting to extend the classification of tilting mutations 
and their quiver-invariant dualities 
to the toric phases of other toric Calabi-Yau 3-folds \cite{}, 
and to clarify their relation to sequences of Seiberg dualities, 
to specular duality \cite{Hanany:2012vc, Cremonesi:2013aba} and to birational transformations \cite{2012arXiv1212.1785A, 2022SIGMA..18..030H, Ghim:2024asj, Ghim:2025zhs, Kho:2025fmp}.
It would moreover be interesting to investigate the action of tilting mutations 
on dimer integrable systems associated with brane tilings \cite{Eager:2011dp, 2011arXiv1107.5588G, Kho:2025fmp, Kho:2025jxk, Kho:2026zwc}, 
brane tilings associated with generalized toric polygons (GTPs) \cite{Benini:2009gi, vanBeest:2020kou, Bourget:2023wlb, CarrenoBolla:2024fxy, Franco:2023flw, Arias-Tamargo:2024fjt},
bipartite field theories (BFTs) \cite{Franco:2012mm, Franco:2012wv, Kho:2024wcw},
and brane brick models \cite{Franco:2015tna, Franco:2015tya, Franco:2016nwv, Franco:2016qxh, Franco:2016tcm, Franco:2017cjj, Franco:2022gvl, Franco:2022isw, Kho:2023dcm,  Franco:2023tyf, Carcamo:2025shw, Kwon:2026stu}, 
where analogous quiver-invariant dualities may relate 
toric phases of $2d$ $(0,2)$ gauge theories 
associated with toric Calabi-Yau 4-folds.
These investigations can be conducted with the help of machine learning techniques \cite{Krefl:2017yox, Choi:2023rqg, Seong:2023njx, Seong:2024wkt, Bao:2020nbi, Capuozzo:2024vdw}.
We plan to report on these directions in future work.
\\

%=================================================================
\section*{Acknowledgements}
We would like to thank
Jiakang Bao, Eugene Choi, Sebastian Franco, Dongwook Ghim, Amihay Hanany, Minsung Kho, Juno Kwon, Kimyeong Lee, Norton Lee, Sangmin Lee,
and Masahito Yamazaki
for discussions on related topics.
R.-K. S. would like to thank the
Simons Center for Geometry and Physics at Stony Brook University,
the Korea Institute for Advanced Study (KIAS),
the Shanghai Institute for Mathematics and Interdisciplinary Sciences (SIMIS),
the Banff International Research Station for Mathematical Innovation and Discovery (BIRS),
the Gwangju Institute of Science and Technology (GIST),
the Kyungpook National University,
as well as the Institute for Basic Science Center for Geometry and Physics (IBS-CGP)
for their kind and generous
hospitality during various stages of this work. 
S.-J. L. is supported by the Institute for Basic Science (IBS) under Project No. IBS-R003-D1. 
R.-K. S. is supported by an Outstanding Young Scientist Grant (RS-2025-00516583) of the National Research Foundation
of Korea (NRF). He is also partly supported by the BK21 Program (“Next Generation Education Program for Mathematical Sciences”, 4299990414089) funded by the
Ministry of Education in Korea and the National Research Foundation of Korea (NRF).
B.S. is also supported by the National Research Foundation of Korea (NRF) grant RS-2025-00516583. 
\\

%=================================================================
\appendix
%=================================================================

%=================================================================
\section{Toric phases of $H_{1,1,2,1}$}
%=================================================================

\label{appendix:models}
In this appendix, we present the complete list of the 42 distinct toric phases of $H_{1,1,2,1}$, 
detailing their quiver, superpotential and toric diagram with GLSM multiplicities.
\\

%---------------------------------------------------- 
\begin{longtable}{| C{0.05\textwidth} | C{0.14\textwidth} | C{0.57\textwidth} | C{0.12\textwidth} |}
\hline
 & Quiver & Superpotential & Toric Diagram\\
\hline
\endfirsthead

\hline
 & Quiver & Superpotential & Toric Diagram\\
\hline
\endhead

\hline
\endfoot

\hline
\endlastfoot

1 & \raisebox{-0.5\height}{\includegraphics[width=0.9\linewidth]{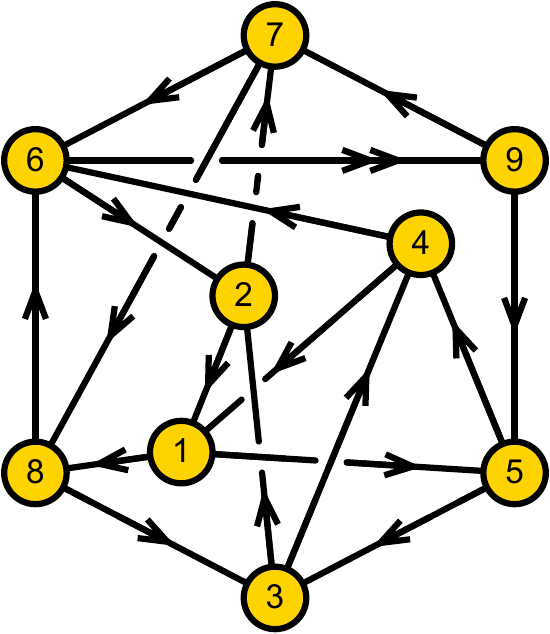}} & 
{\scriptsize $\begin{aligned}
W_1 &= X_{15} X_{54} X_{41}
+ X_{18} X_{86} X_{62} X_{21}
+ X_{27} X_{78} X_{83} X_{32}\\
&+ X_{34} X_{46} X_{69}^{1} X_{95} X_{53}
+ X_{69}^{2} X_{97} X_{76}
- X_{15} X_{53} X_{32} X_{21}\\ 
&- X_{18} X_{83} X_{34} X_{41}
- X_{27} X_{76} X_{62} 
- X_{46} X_{69}^{2} X_{95} X_{54}\\ 
&- X_{69}^{1} X_{97} X_{78} X_{86}
\end{aligned}$} & \resizebox{\linewidth}{!}{
\adjustbox{valign=m}{
\includegraphics[width=\linewidth]{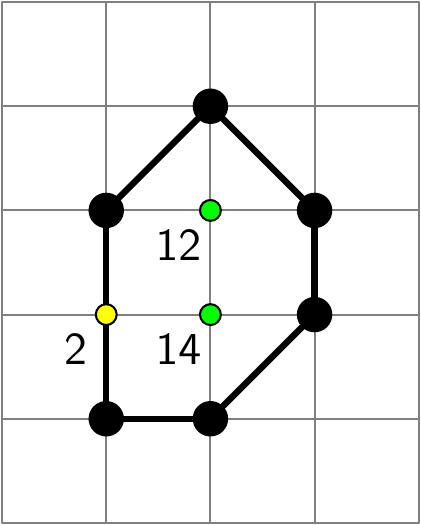}
}} \\ \hline
2a &
\multirow{2}{*}[-4ex]{
\adjustbox{valign=m}{\includegraphics[width=0.9\linewidth]{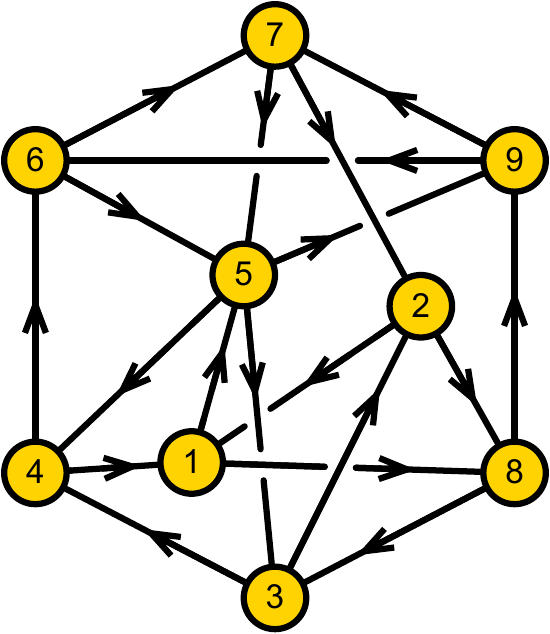}}
}
& {\scriptsize $\begin{aligned}
&\\
W_{2a} =&X_{15} X_{54} X_{41} 
+ X_{18} X_{89} X_{96} X_{67} X_{72} X_{21}
+ X_{28} X_{83} X_{32}\\
&+ X_{34} X_{46} X_{65} X_{53} 
+ X_{59} X_{97} X_{75} 
- X_{15} X_{53} X_{32} X_{21} \\
&- X_{18} X_{83} X_{34} X_{41} 
- X_{28} X_{89} X_{97} X_{72}  \\
&- X_{46} X_{67} X_{75} X_{54} 
- X_{59} X_{96} X_{65}\\
&
\end{aligned}$} & \multirow{2}{=}[-4ex]{
\resizebox{\linewidth}{!}{\adjustbox{valign=m}{
\includegraphics[width=\linewidth]{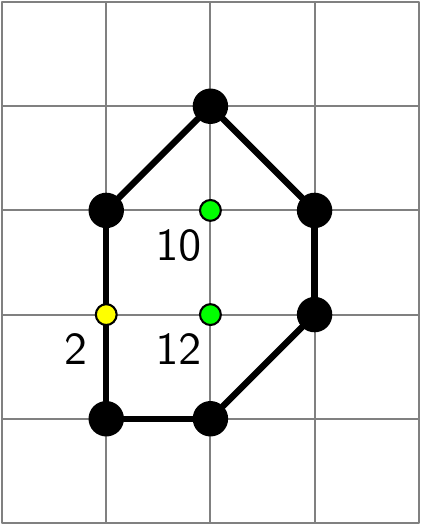}
}}
}\\

\cline{1-1} \cline{3-3}

2b & 
&
{\scriptsize $\begin{aligned}
&\\
W_{2b} = &X_{15} X_{53} X_{32} X_{21} 
+ X_{18} X_{83} X_{34} X_{41} \\
&+ X_{28} X_{89} X_{96} X_{67} X_{72}
+ X_{46} X_{65} X_{54}
+ X_{59} X_{97} X_{75} \\
&- X_{15} X_{54} X_{41}
- X_{18} X_{89} X_{97} X_{72} X_{21} 
- X_{28} X_{83} X_{32}\\
&- X_{34} X_{46} X_{67} X_{75} X_{53}
- X_{59} X_{96} X_{65} \\
&
\end{aligned}$} &\\ \hline
3 & \adjustbox{valign=m}{\includegraphics[width=0.9\linewidth]{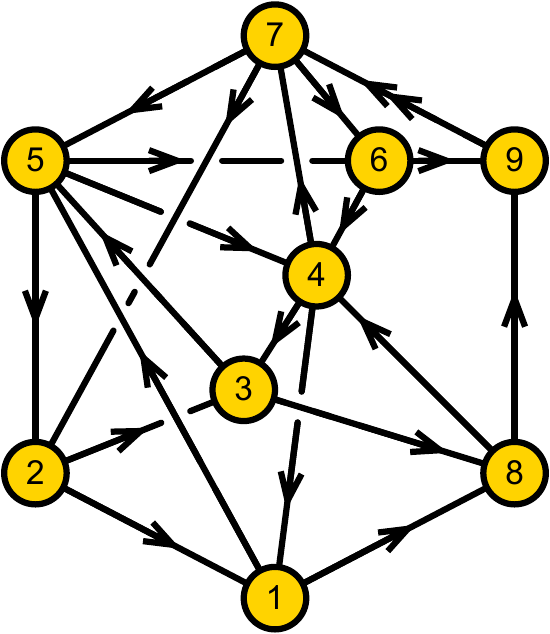} }&
{\scriptsize $\begin{aligned}
W_3 &= X_{15} X_{54} X_{41}
+ X_{18} X_{89} X_{97}^{2} X_{72} X_{21}
+ X_{23} X_{35} X_{52}\\
&+ X_{38} X_{84} X_{43}
+ X_{47} X_{76} X_{64}
+ X_{56} X_{69} X_{97}^{1} X_{75}\\
&- X_{15} X_{52} X_{21}
- X_{18} X_{84} X_{41} 
- X_{23} X_{38} X_{89} X_{97}^{1} X_{72} \\
&- X_{35} X_{56} X_{64} X_{43}
- X_{47} X_{75} X_{54} 
- X_{69} X_{97}^{2} X_{76}
\end{aligned}$} & \resizebox{\linewidth}{!}{\adjustbox{valign=m}{
\includegraphics[width=\linewidth]{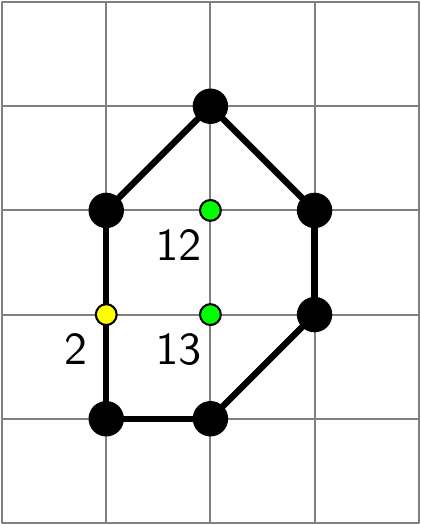}
}}\\ \hline
4 & \adjustbox{valign=m}{\includegraphics[width=0.9\linewidth]{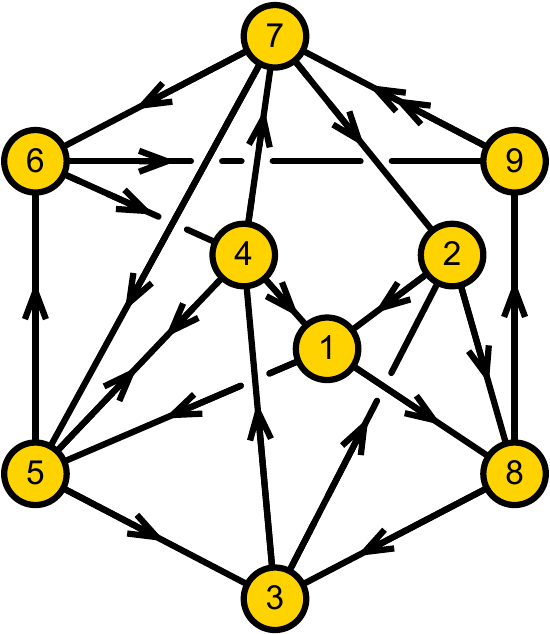} }& 
{\scriptsize $\begin{aligned}
W_4 &= X_{15} X_{54} X_{41}
+ X_{18} X_{89} X_{97}^{2} X_{72} X_{21}
+ X_{28} X_{83} X_{32}\\
&+ X_{34} X_{45} X_{53}
+ X_{47} X_{76} X_{64} 
+ X_{56} X_{69} X_{97}^{1} X_{75}\\
&- X_{15} X_{53} X_{32} X_{21} 
- X_{18} X_{83} X_{34} X_{41}
- X_{28} X_{89} X_{97}^{1} X_{72} \\
&- X_{45} X_{56} X_{64}
- X_{47} X_{75} X_{54} 
- X_{69} X_{97}^{2} X_{76}
\end{aligned}$} & \resizebox{\linewidth}{!}{\adjustbox{valign=m}{
\includegraphics[width=\linewidth]{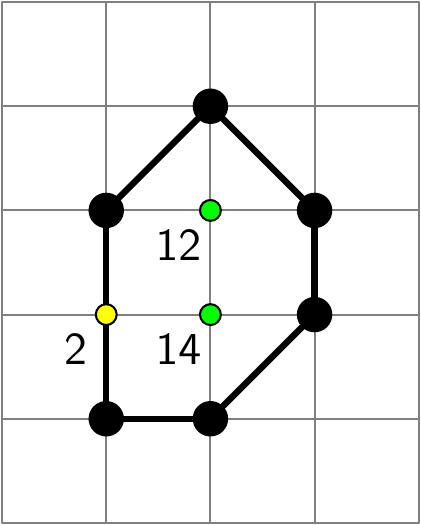}
}}\\ \hline

\pagebreak

5a &
\multirow{2}{*}[-4ex]{
\adjustbox{valign=m}{\includegraphics[width=0.9\linewidth]{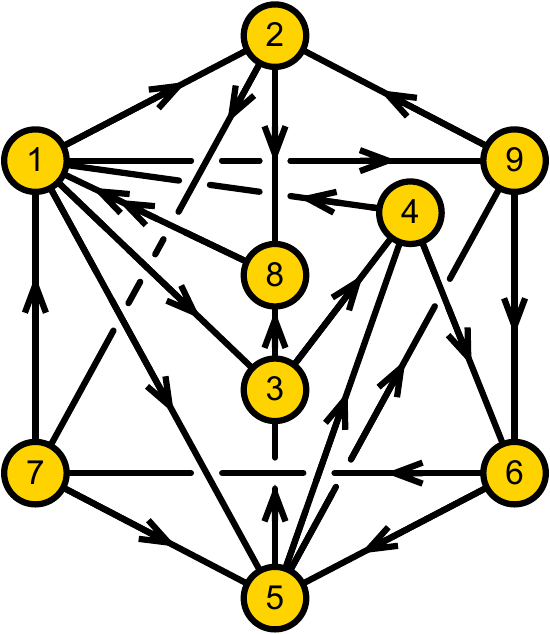}}
} 
&{\scriptsize $\begin{aligned}
&\\
W_{5a} &= X_{12} X_{28} X_{81}^{2}
+ X_{13} X_{38} X_{81}^{1}
+ X_{15} X_{54} X_{41}\\
&+ X_{19} X_{96} X_{67} X_{71} 
+ X_{27} X_{75} X_{59} X_{92}
+ X_{34} X_{46} X_{65} X_{53}\\
&- X_{12} X_{27} X_{71}
- X_{13} X_{34} X_{41} 
- X_{15} X_{53} X_{38} X_{81}^{2}\\ 
&- X_{19} X_{92} X_{28} X_{81}^{1}
- X_{46} X_{67} X_{75} X_{54}  
- X_{59} X_{96} X_{65} \\
&
\end{aligned}$} & \multirow{2}{=}[-4ex]{
\resizebox{\linewidth}{!}{
\adjustbox{valign=m}{
\includegraphics[width=\linewidth]{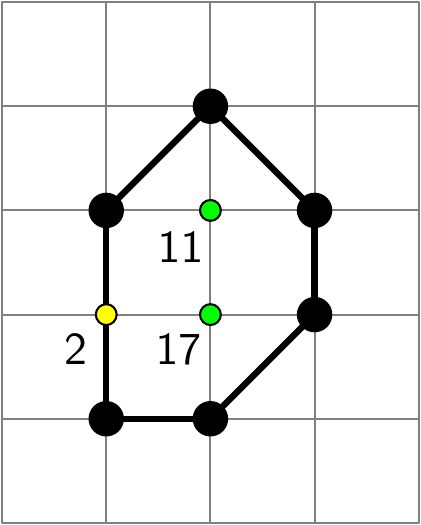}
}
}}\\ 
\cline{1-1} \cline{3-3}
5b &  &
{\scriptsize$\begin{aligned}
&\\
W_{5b}=&X_{12} X_{27} X_{71}
+ X_{13} X_{38} X_{81}^{1}
+ X_{15} X_{54} X_{41}\\
&+ X_{19} X_{92} X_{28} X_{81}^{2}
+ X_{34} X_{46} X_{67} X_{75} X_{53}
+ X_{59} X_{96} X_{65}\\
&- X_{12} X_{28} X_{81}^{1} 
- X_{13} X_{34} X_{41}
- X_{15} X_{53} X_{38} X_{81}^{2}\\
&- X_{19} X_{96} X_{67} X_{71}
- X_{27} X_{75} X_{59} X_{92}
- X_{46} X_{65} X_{54}\\
&
\end{aligned}$} &\\ \hline
6 &\adjustbox{valign=m}{ \includegraphics[width=0.9\linewidth]{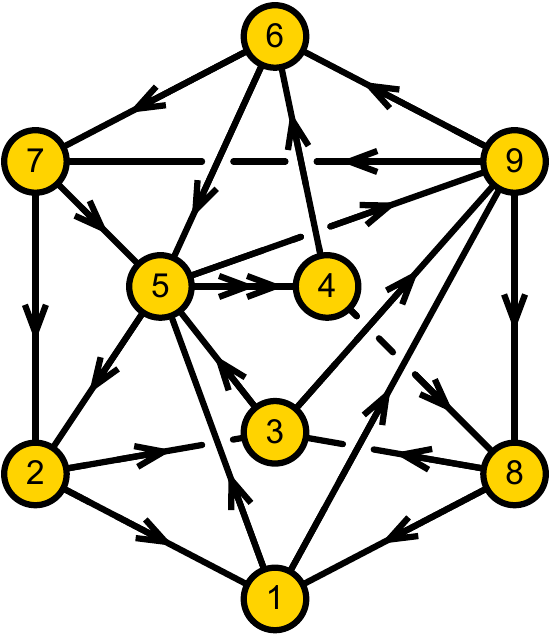} }&
{\scriptsize $\begin{aligned}
W_6 &= X_{15} X_{54}^{1} X_{48} X_{81}
+ X_{19} X_{96} X_{67} X_{72} X_{21}
+ X_{23} X_{35} X_{52}\\
&+ X_{39} X_{98} X_{83}
+ X_{46} X_{65} X_{54}^{2}
+ X_{59} X_{97} X_{75}\\
&- X_{15} X_{52} X_{21} 
- X_{19} X_{98} X_{81} 
- X_{23} X_{39} X_{97} X_{72}\\
&- X_{35} X_{54}^{2} X_{48} X_{83} 
- X_{46} X_{67} X_{75} X_{54}^{1}
- X_{59} X_{96} X_{65}
\end{aligned}$} & \resizebox{\linewidth}{!}{\adjustbox{valign=m}{
\includegraphics[width=\linewidth]{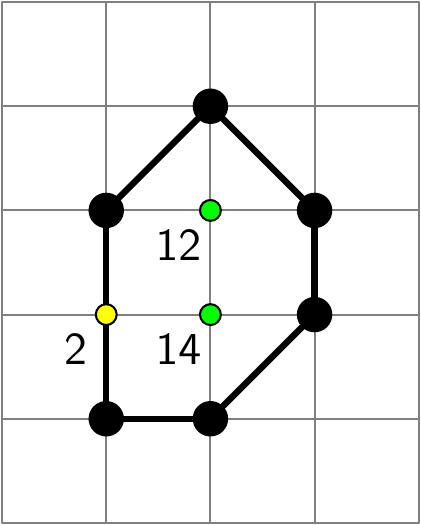}
}}\\ \hline

7a & \multirow{2}{*}[-4ex]{
\adjustbox{valign=m}{\includegraphics[width=0.9\linewidth]{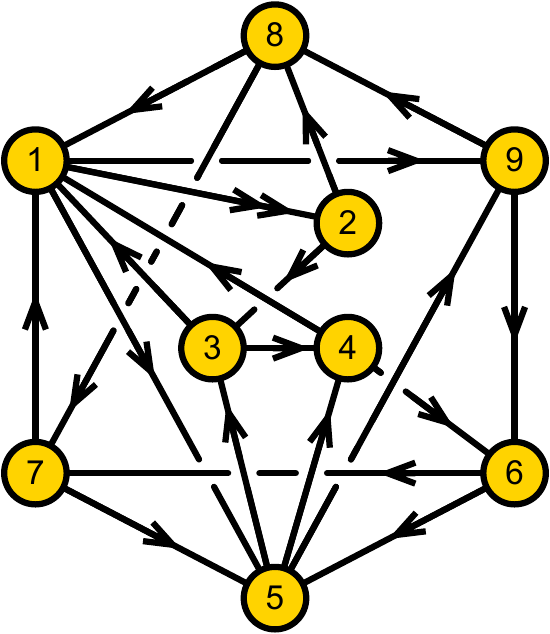}}} &
{\scriptsize $\begin{aligned}
&\\
W_{7a} &= X_{12}^{1} X_{28} X_{81}
+ X_{12}^{2} X_{23} X_{31}
+ X_{15} X_{54} X_{41}\\
&+ X_{19} X_{96} X_{67} X_{71} 
+ X_{34} X_{46} X_{65} X_{53}
+ X_{59} X_{98} X_{87} X_{75} \\
&- X_{12}^{2} X_{28} X_{87} X_{71} 
- X_{12}^{1} X_{23} X_{34} X_{41} 
- X_{15} X_{53} X_{31}\\
&- X_{19} X_{98} X_{81}
- X_{46} X_{67} X_{75} X_{54}
- X_{59} X_{96} X_{65}\\
&
\end{aligned}$} & \multirow{2}{=}[-4ex]{
\resizebox{\linewidth}{!}{
\adjustbox{valign=m}{
\includegraphics[width=\linewidth]{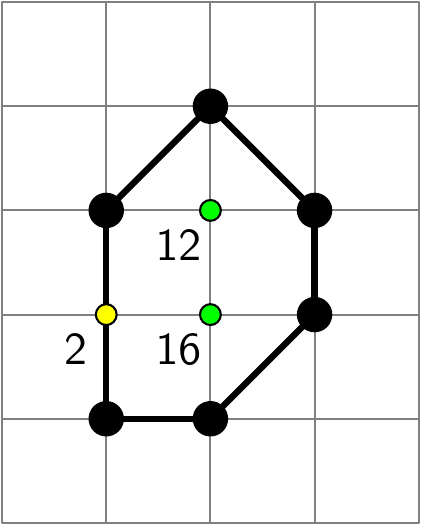}
}
}}\\ 
\cline{1-1} \cline{3-3}
7b &  & 
{\scriptsize $\begin{aligned}
&\\
W_{7b}=&X_{12}^{2} X_{23} X_{31}
+ X_{12}^{1} X_{28} X_{87} X_{71}
+ X_{15} X_{54} X_{41}\\
&+ X_{19} X_{98} X_{81} 
+ X_{34} X_{46} X_{67} X_{75} X_{53}
+ X_{59} X_{96} X_{65}\\
&- X_{12}^{1} X_{23} X_{34} X_{41}
- X_{12}^{2} X_{28} X_{81}
- X_{15} X_{53} X_{31}\\
&- X_{19} X_{96} X_{67} X_{71} 
- X_{46} X_{65} X_{54}
- X_{59} X_{98} X_{87} X_{75} \\
&
\end{aligned}$} &\\ \hline
8 &\adjustbox{valign=m}{\includegraphics[width=0.9\linewidth]{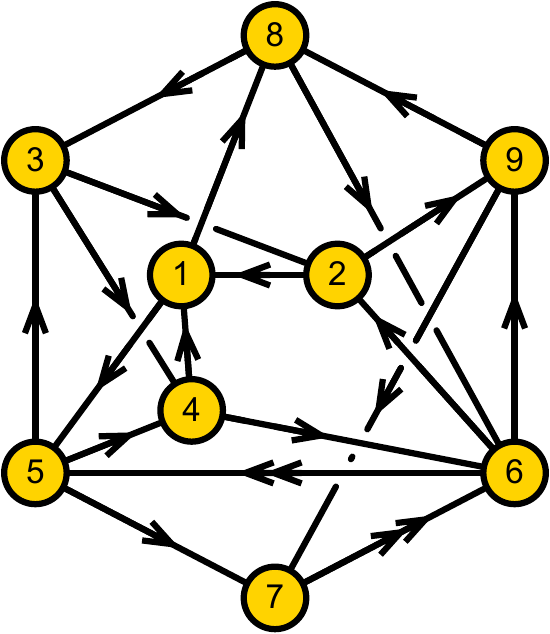} }& 
{\scriptsize $\begin{aligned}
W_8 &= X_{15} X_{54} X_{41}
+ X_{18} X_{86} X_{62} X_{21}
+ X_{29} X_{98} X_{83} X_{32} \\
&+ X_{34} X_{46} X_{65}^{1} X_{53}
+ X_{57} X_{76}^{1} X_{65}^{2} 
+ X_{69} X_{97} X_{76}^{2}\\
&- X_{15} X_{53} X_{32} X_{21} 
- X_{18} X_{83} X_{34} X_{41}
- X_{29} X_{97} X_{76}^{1} X_{62} \\
&- X_{46} X_{65}^{2} X_{54}
- X_{57} X_{76}^{2} X_{65}^{1}
- X_{69} X_{98} X_{86}
\end{aligned}$} & \resizebox{\linewidth}{!}{\adjustbox{valign=m}{
\includegraphics[width=\linewidth]{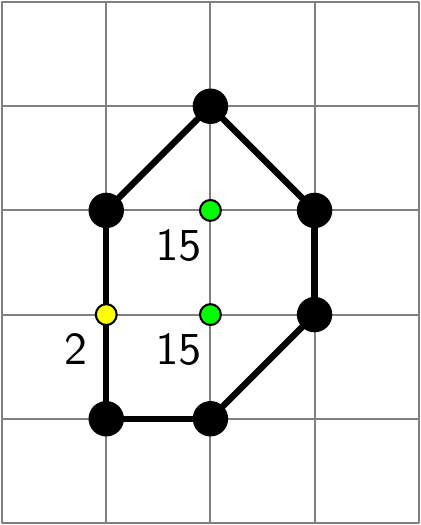}
}}\\ \hline
9 & \adjustbox{valign=m}{\includegraphics[width=0.9\linewidth]{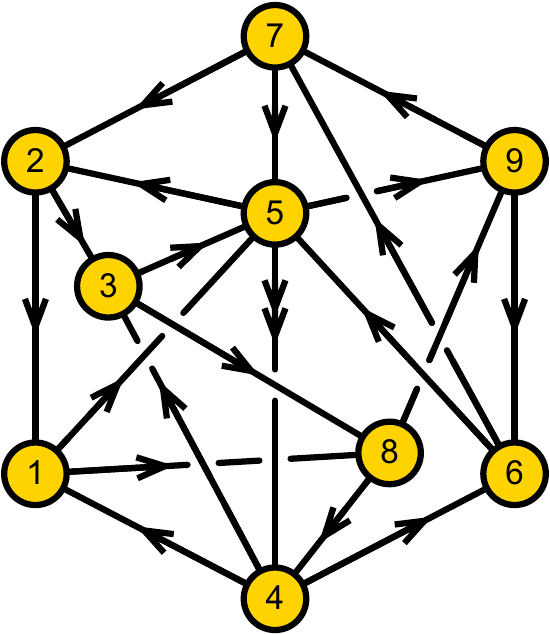}} & 
{\scriptsize $\begin{aligned}
W_9 &= X_{15} X_{54}^{1} X_{41}
+ X_{18} X_{89} X_{96} X_{67} X_{72} X_{21}
+ X_{23} X_{35} X_{52}\\
&+ X_{38} X_{84} X_{43}
+ X_{46} X_{65} X_{54}^{2}
+ X_{59} X_{97} X_{75}\\
&- X_{15} X_{52} X_{21}
- X_{18} X_{84} X_{41} 
- X_{23} X_{38} X_{89} X_{97} X_{72}\\
&- X_{35} X_{54}^{2} X_{43} 
- X_{46} X_{67} X_{75} X_{54}^{1} 
- X_{59} X_{96} X_{65}
\end{aligned}$} & \resizebox{\linewidth}{!}{\adjustbox{valign=m}{
\includegraphics[width=\linewidth]{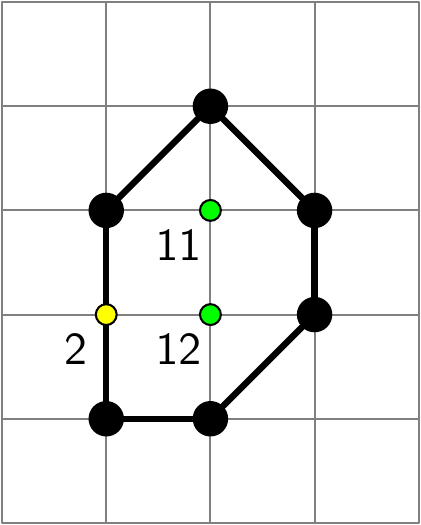}
}}\\ \hline
10 &  \adjustbox{valign=m}{\includegraphics[width=0.9\linewidth]{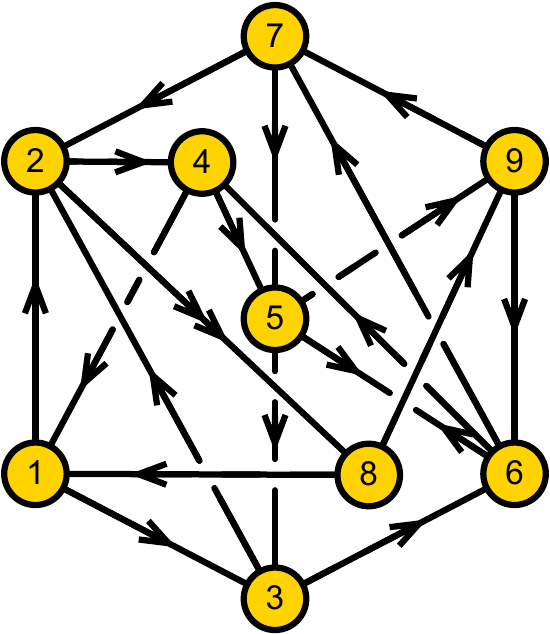} }&
{\scriptsize $\begin{aligned}
W_{10} &= X_{12} X_{24} X_{41} + X_{13} X_{32} X_{28}^{1} X_{81} + X_{28}^{2} X_{89} X_{96} X_{67} X_{72} \\
&\quad + X_{36} X_{65} X_{53} + X_{45} X_{56} X_{64} + X_{59} X_{97} X_{75} \\
&\quad - X_{12} X_{28}^{2} X_{81} - X_{13} X_{36} X_{64} X_{41} - X_{24} X_{45} X_{53} X_{32} \\
&\quad - X_{28}^{1} X_{89} X_{97} X_{72} - X_{56} X_{67} X_{75} - X_{59} X_{96} X_{65} 
\end{aligned}$} & \resizebox{\linewidth}{!}{\adjustbox{valign=m}{
\includegraphics[width=\linewidth]{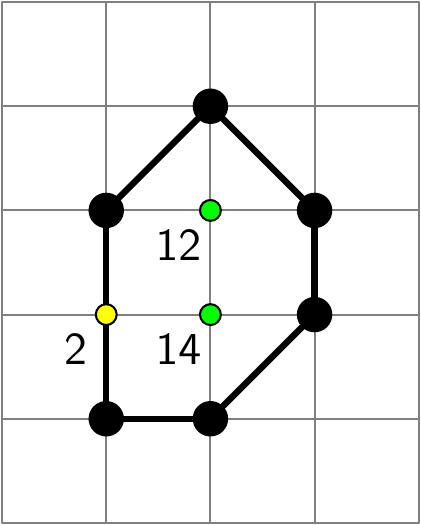}
}}\\ \hline

11 & \adjustbox{valign=m}{\includegraphics[width=0.9\linewidth]{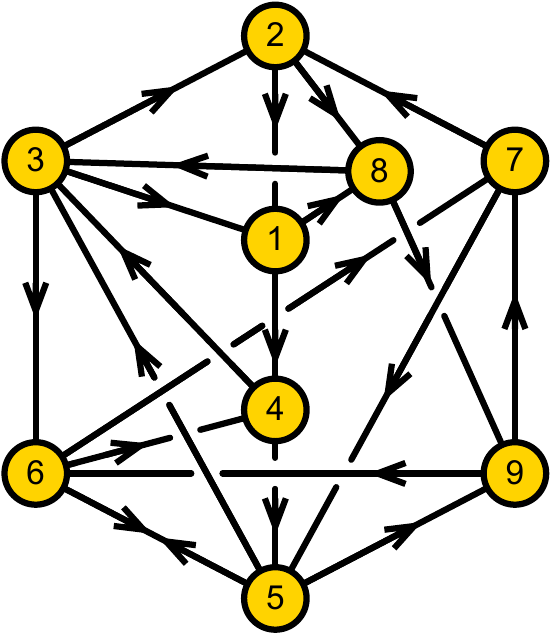}} &
{\scriptsize $\begin{aligned}
W_{11} &= X_{14} X_{43} X_{31} + X_{18} X_{89} X_{96} X_{67} X_{72} X_{21} + X_{28} X_{83} X_{32} \\
&\quad + X_{36} X_{65} X_{53} + X_{45} X_{56} X_{64} + X_{59} X_{97} X_{75} \\
&\quad - X_{14} X_{45} X_{53} X_{32} X_{21} - X_{18} X_{83} X_{31} - X_{28} X_{89} X_{97} X_{72} \\
&\quad - X_{36} X_{64} X_{43} - X_{56} X_{67} X_{75} - X_{59} X_{96} X_{65}
\end{aligned}$} & \resizebox{\linewidth}{!}{\adjustbox{valign=m}{
\includegraphics[width=\linewidth]{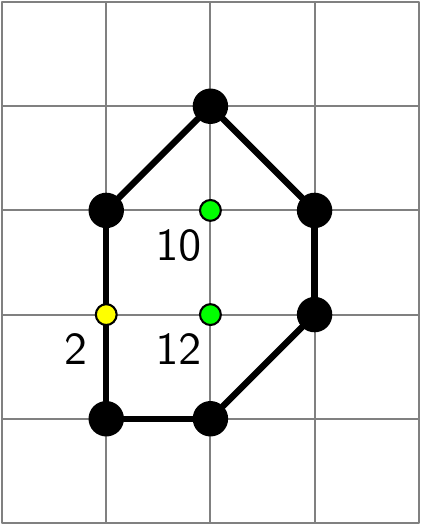}
}}\\ \hline

12 &  \adjustbox{valign=m}{\includegraphics[width=0.9\linewidth]{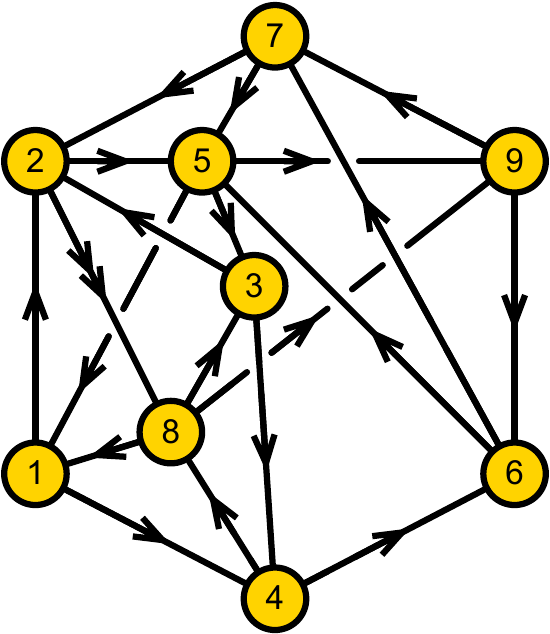} }&
{\scriptsize $\begin{aligned}
W_{12} &= X_{12} X_{25} X_{51} + X_{14} X_{48} X_{81} + X_{28}^{2} X_{89} X_{96} X_{67} X_{72} \\
&\quad + X_{28}^{1} X_{83} X_{32} + X_{34} X_{46} X_{65} X_{53} + X_{59} X_{97} X_{75} \\
&\quad - X_{12} X_{28}^{2} X_{81} - X_{14} X_{46} X_{67} X_{75} X_{51} - X_{25} X_{53} X_{32} \\
&\quad - X_{28}^{1} X_{89} X_{97} X_{72} - X_{34} X_{48} X_{83} - X_{59} X_{96} X_{65} 
\end{aligned}$} & \resizebox{\linewidth}{!}{\adjustbox{valign=m}{
\includegraphics[width=\linewidth]{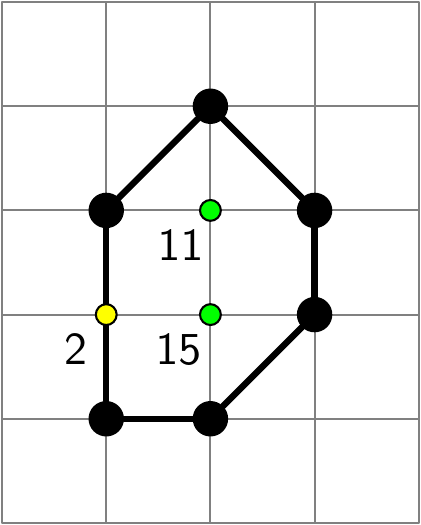}
}}\\ \hline
13 & \adjustbox{valign=m}{\includegraphics[width=0.9\linewidth]{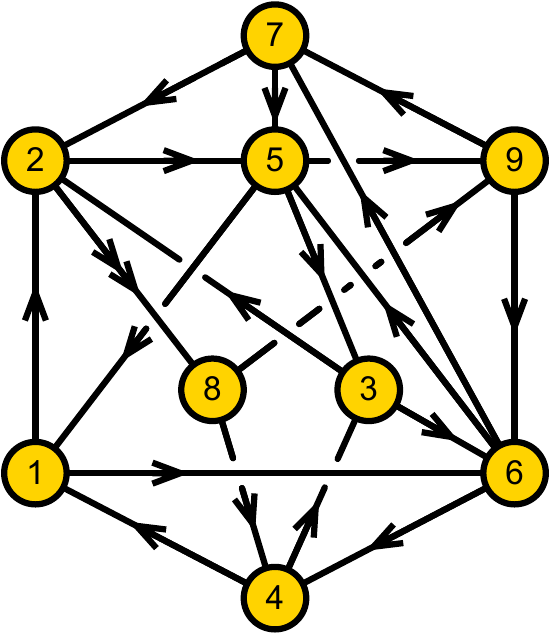}} &
{\scriptsize $\begin{aligned}
W_{13} &= X_{12} X_{25} X_{51} + X_{16} X_{64} X_{41} + X_{28}^{2} X_{89} X_{96} X_{67} X_{72} \\
&\quad + X_{28}^{1} X_{84} X_{43} X_{32} + X_{36} X_{65} X_{53} + X_{59} X_{97} X_{75} \\
&\quad - X_{12} X_{28}^{2} X_{84} X_{41} - X_{16} X_{67} X_{75} X_{51} - X_{25} X_{53} X_{32} \\
&\quad - X_{28}^{1} X_{89} X_{97} X_{72} - X_{36} X_{64} X_{43} - X_{59} X_{96} X_{65}  
\end{aligned}$} & \resizebox{\linewidth}{!}{\adjustbox{valign=m}{
\includegraphics[width=\linewidth]{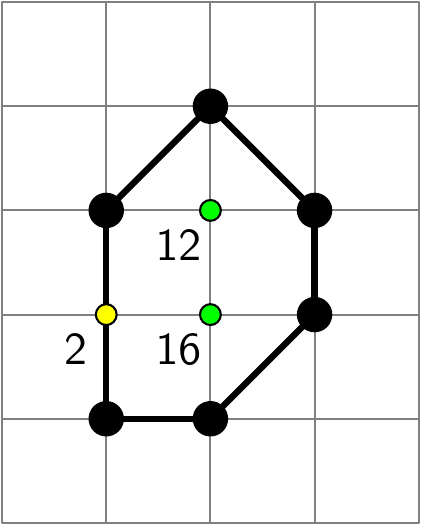}
}}\\ \hline

14a &\multirow{2}{*}[-6ex]{
\adjustbox{valign=m}{\includegraphics[width=0.9\linewidth]{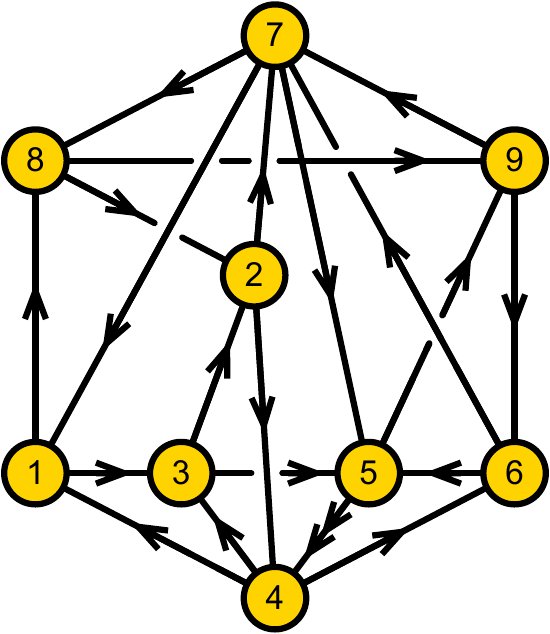}}}
&{\scriptsize $\begin{aligned}
&\\
W_{14a} &= X_{13} X_{35} X_{54}^{1} X_{41} + X_{18} X_{89} X_{96} X_{67} X_{71} + X_{24} X_{43} X_{32} \\
&\quad + X_{27} X_{78} X_{82} + X_{46} X_{65} X_{54}^{2} + X_{59} X_{97} X_{75} \\
&\quad - X_{13} X_{32} X_{27} X_{71} - X_{18} X_{82} X_{24} X_{41} - X_{35} X_{54}^{2} X_{43} \\
&\quad - X_{46} X_{67} X_{75} X_{54}^{1} - X_{59} X_{96} X_{65} - X_{78} X_{89} X_{97} \\
&
\end{aligned}$} & \multirow{2}{=}[-4ex]{
\resizebox{\linewidth}{!}{\adjustbox{valign=m}{
\includegraphics[width=\linewidth]{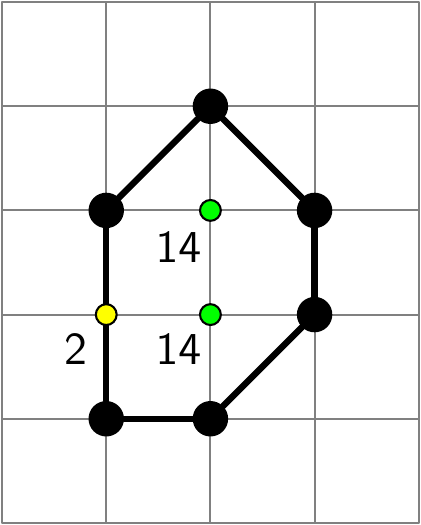}
}
}}\\ 
\cline{1-1} \cline{3-3}
14b & & {\scriptsize $\begin{aligned}
&\\
W_{14b} &= X_{13} X_{32} X_{27} X_{71} + X_{18} X_{82} X_{24} X_{41} + X_{35} X_{54}^{1} X_{43} \\
&\quad + X_{46} X_{65} X_{54}^{2} + X_{59} X_{97} X_{75} + X_{67} X_{78} X_{89} X_{96} \\
&\quad - X_{13} X_{35} X_{54}^{2} X_{41} - X_{18} X_{89} X_{97} X_{71} - X_{24} X_{43} X_{32} \\
&\quad - X_{27} X_{78} X_{82} - X_{46} X_{67} X_{75} X_{54}^{1} - X_{59} X_{96} X_{65}\\
&
\end{aligned}$} &\\ \hline

\pagebreak

15a &
\multirow{3}{=}[-13ex]{
\adjustbox{valign=m}{\includegraphics[width=0.9\linewidth]{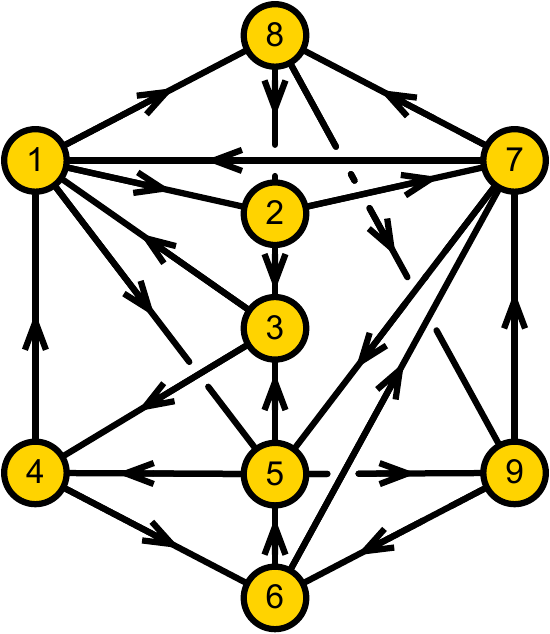}}
}
&
{\scriptsize $\begin{aligned}
&\\
W_{15a} &= X_{12} X_{23} X_{31} + X_{15} X_{54} X_{41} + X_{18} X_{89} X_{96} X_{67} X_{71} \\
&+ X_{27} X_{78} X_{82} + X_{34} X_{46} X_{65} X_{53} + X_{59} X_{97} X_{75} \\
& - X_{12} X_{27} X_{71} - X_{15} X_{53} X_{31} - X_{18} X_{82} X_{23} X_{34} X_{41} \\
& - X_{46} X_{67} X_{75} X_{54} - X_{59} X_{96} X_{65} - X_{78} X_{89} X_{97}\\
&
\end{aligned}$}  & \multirow{3}{=}[-12ex]{
\resizebox{\linewidth}{!}{
\adjustbox{valign=m}{
\includegraphics[width=\linewidth]{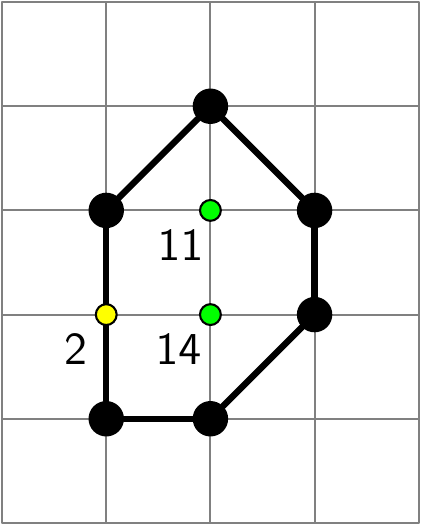}
}
}}\\

\cline{1-1} \cline{3-3}

15b & 
&
{\scriptsize $\begin{aligned}
&\\
W_{15b}=&X_{12} X_{23} X_{34} X_{41}
+ X_{15} X_{53} X_{31}
+ X_{18} X_{89} X_{96} X_{67} X_{71}\\
&+ X_{27} X_{78} X_{82}
+ X_{46} X_{65} X_{54}
+ X_{59} X_{97} X_{75}\\
&- X_{12} X_{27} X_{71}
- X_{15} X_{54} X_{41}
- X_{18} X_{82} X_{23} X_{31}\\
&- X_{34} X_{46} X_{67} X_{75} X_{53}
- X_{59} X_{96} X_{65}
- X_{78} X_{89} X_{97}\\
&
\end{aligned}$} &\\

\cline{1-1} \cline{3-3}

15c & 
&
{\scriptsize $\begin{aligned}
&\\
W_{15c} =& X_{12} X_{27} X_{71} 
+ X_{15} X_{53} X_{31} 
+ X_{18} X_{82} X_{23} X_{34} X_{41} \\
&+ X_{46} X_{65} X_{54} 
+ X_{59} X_{97} X_{75}
+ X_{67} X_{78} X_{89} X_{96} \\
&- X_{12} X_{23} X_{31} 
- X_{15} X_{54} X_{41}
- X_{18} X_{89} X_{97} X_{71} \\
&- X_{27} X_{78} X_{82}
- X_{34} X_{46} X_{67} X_{75} X_{53}
- X_{59} X_{96} X_{65}\\
&
\end{aligned}$} &\\ \hline
16 & \adjustbox{valign=m}{\includegraphics[width=0.9\linewidth]{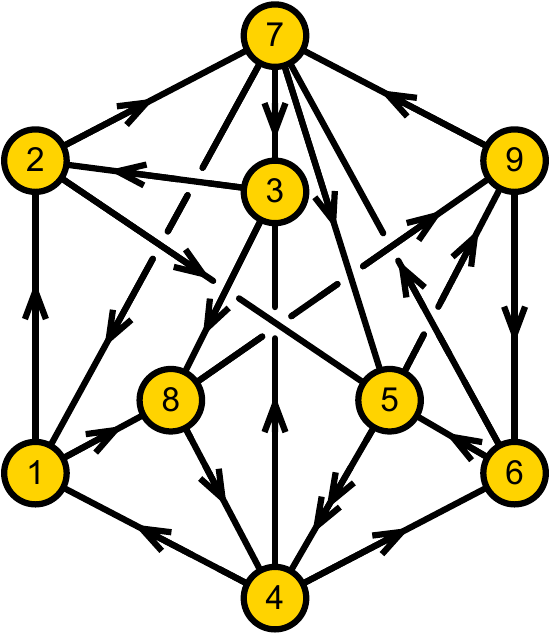} }&
{\scriptsize $\begin{aligned}
W_{16} &= X_{12} X_{25} X_{54}^{1} X_{41} + X_{18} X_{89} X_{96} X_{67} X_{71} + X_{27} X_{73} X_{32} \\
&\quad + X_{38} X_{84} X_{43} + X_{46} X_{65} X_{54}^{2} + X_{59} X_{97} X_{75} \\
&\quad - X_{12} X_{27} X_{71} - X_{18} X_{84} X_{41} - X_{25} X_{54}^{2} X_{43} X_{32} \\
&\quad - X_{38} X_{89} X_{97} X_{73} - X_{46} X_{67} X_{75} X_{54}^{1} - X_{59} X_{96} X_{65}
\end{aligned}$} & \resizebox{\linewidth}{!}{\adjustbox{valign=m}{
\includegraphics[width=\linewidth]{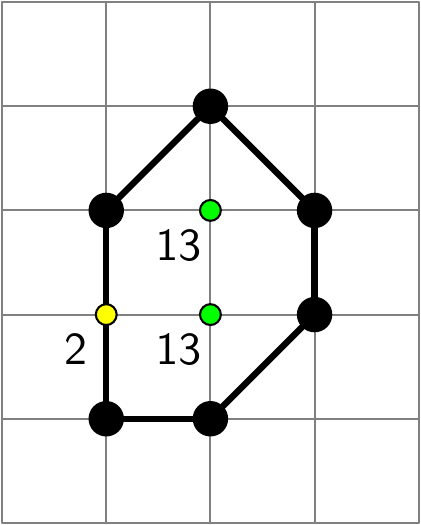}
}}\\ \hline

17a &\multirow{2}{=}{
\adjustbox{valign=m}{\includegraphics[width=0.9\linewidth]{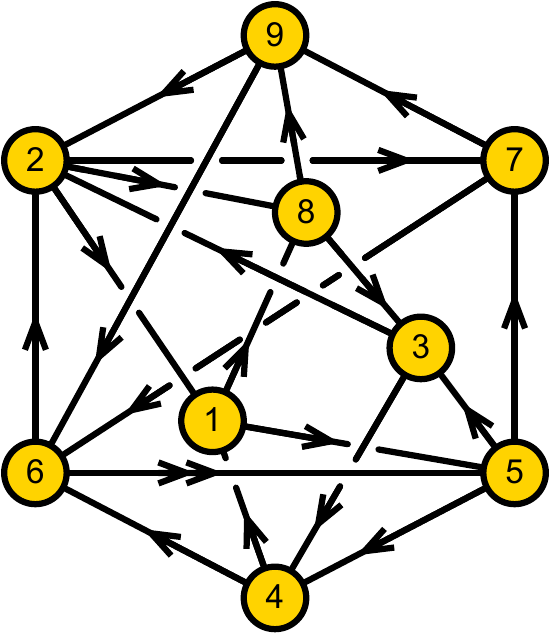}}} &
{\scriptsize $\begin{aligned}
W_{17a} &= X_{15} X_{54} X_{41} + X_{18} X_{89} X_{96} X_{62} X_{21} + X_{27} X_{79} X_{92} \\
&\quad + X_{28} X_{83} X_{32} + X_{34} X_{46} X_{65}^{1} X_{53} + X_{57} X_{76} X_{65}^{2} \\
&\quad - X_{15} X_{53} X_{32} X_{21} - X_{18} X_{83} X_{34} X_{41} - X_{27} X_{76} X_{62} \\
&\quad - X_{28} X_{89} X_{92} - X_{46} X_{65}^{2} X_{54} - X_{57} X_{79} X_{96} X_{65}^{1}
\end{aligned}$}  & \multirow{2}{=}{
\resizebox{\linewidth}{!}{\adjustbox{valign=m}{
\includegraphics[width=\linewidth]{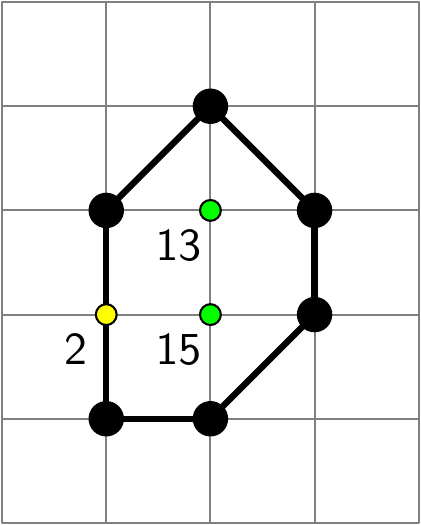}
}
}}\\ 
\cline{1-1} \cline{3-3}
17b && 
{\scriptsize $\begin{aligned}
W_{17b} &= X_{15} X_{54} X_{41} 
+ X_{18} X_{89} X_{92} X_{21}
+ X_{27} X_{76} X_{62} \\
&\quad + X_{28} X_{83} X_{32}
+ X_{34} X_{46} X_{65}^{1} X_{53}
+ X_{57} X_{79} X_{96} X_{65}^{2}\\
&\quad-X_{15} X_{53} X_{32} X_{21} 
-X_{18} X_{83} X_{34} X_{41} 
- X_{27} X_{79} X_{92} \\
&\quad - X_{28} X_{89} X_{96} X_{62} 
-X_{46} X_{65}^{2} X_{54}
 - X_{57} X_{76} X_{65}^{1} 
\end{aligned}$} &\\ \hline
18 &\adjustbox{valign=m}{\includegraphics[width=0.9\linewidth]{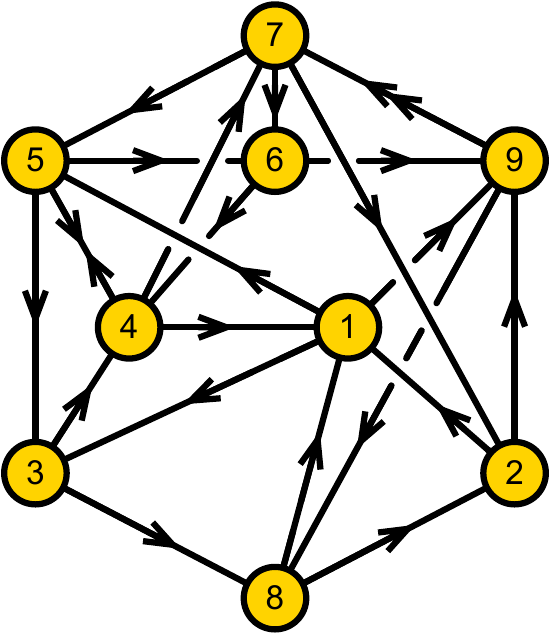} }& 
{\scriptsize $\begin{aligned}
W_{18} &= X_{13} X_{38} X_{81} + X_{15} X_{54} X_{41} + X_{19} X_{97}^{2} X_{72} X_{21} \\
&\quad + X_{29} X_{98} X_{82} + X_{34} X_{45} X_{53} + X_{47} X_{76} X_{64} \\
&\quad + X_{56} X_{69} X_{97}^{1} X_{75} - X_{13} X_{34} X_{41} - X_{15} X_{53} X_{38} X_{82} X_{21} \\
&\quad - X_{19} X_{98} X_{81} - X_{29} X_{97}^{1} X_{72} - X_{45} X_{56} X_{64} \\
&\quad - X_{47} X_{75} X_{54} - X_{69} X_{97}^{2} X_{76}
\end{aligned}$} & \resizebox{\linewidth}{!}{\adjustbox{valign=m}{
\includegraphics[width=\linewidth]{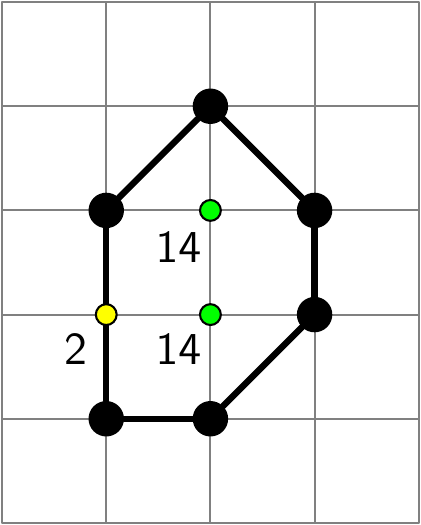}
}}\\ \hline

19 &\adjustbox{valign=m}{\includegraphics[width=0.9\linewidth]{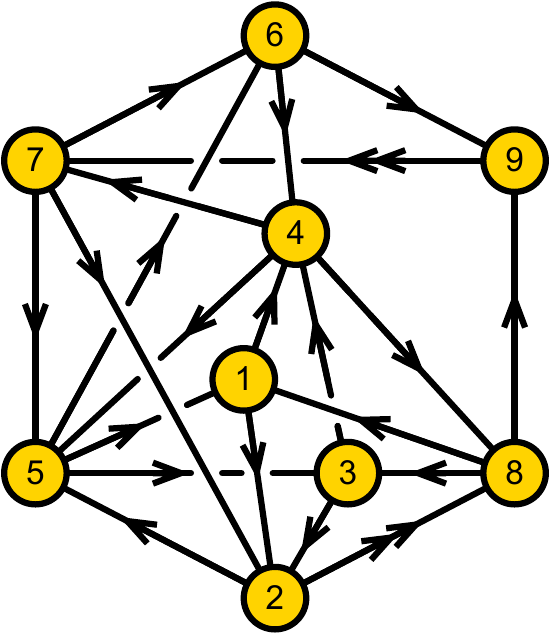}} & 
{\scriptsize $\begin{aligned}
W_{19} &= X_{12} X_{25} X_{51} + X_{14} X_{48} X_{81} + X_{28}^{2} X_{89} X_{97}^{2} X_{72} \\
&\quad + X_{28}^{1} X_{83} X_{32} + X_{34} X_{45} X_{53} + X_{47} X_{76} X_{64} \\
&\quad + X_{56} X_{69} X_{97}^{1} X_{75} - X_{12} X_{28}^{2} X_{81} - X_{14} X_{47} X_{75} X_{51} \\
&\quad - X_{25} X_{53} X_{32} - X_{28}^{1} X_{89} X_{97}^{1} X_{72} - X_{34} X_{48} X_{83} \\
&\quad - X_{45} X_{56} X_{64} - X_{69} X_{97}^{2} X_{76} 
\end{aligned}$} & \resizebox{\linewidth}{!}{\adjustbox{valign=m}{
\includegraphics[width=\linewidth]{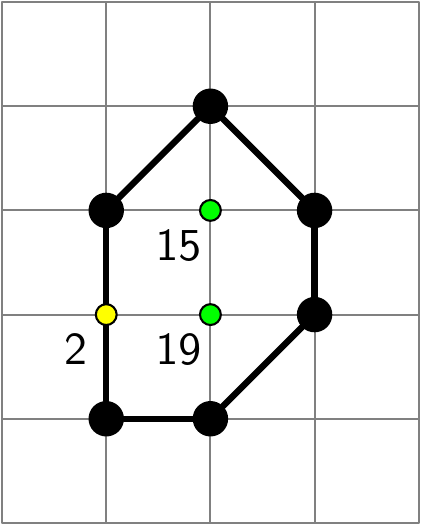}
}}\\ \hline

20 &\adjustbox{valign=m}{ \includegraphics[width=0.9\linewidth]{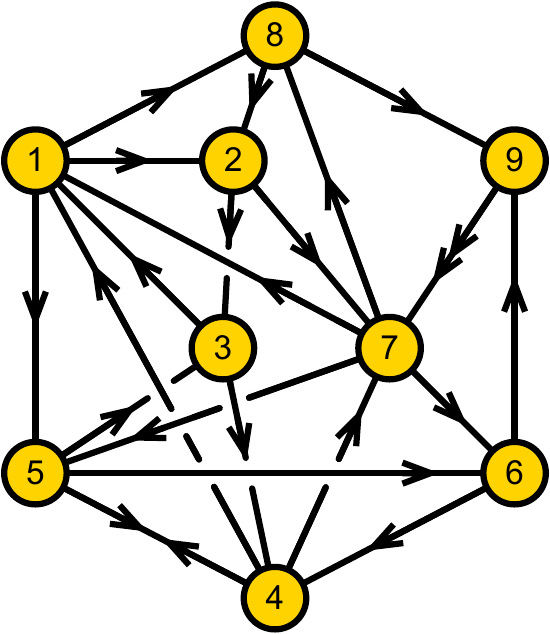}} &
{\scriptsize $\begin{aligned}
W_{20} &= X_{12} X_{23} X_{31} + X_{15} X_{54} X_{41} + X_{18} X_{89} X_{97}^{2} X_{71} \\
&\quad + X_{27} X_{78} X_{82} + X_{34} X_{45} X_{53} + X_{47} X_{76} X_{64} \\
&\quad + X_{56} X_{69} X_{97}^{1} X_{75} - X_{12} X_{27} X_{71} - X_{15} X_{53} X_{31} \\
&\quad - X_{18} X_{82} X_{23} X_{34} X_{41} - X_{45} X_{56} X_{64} - X_{47} X_{75} X_{54} \\
&\quad - X_{69} X_{97}^{2} X_{76} - X_{78} X_{89} X_{97}^{1}
\end{aligned}$} & \resizebox{\linewidth}{!}{\adjustbox{valign=m}{
\includegraphics[width=\linewidth]{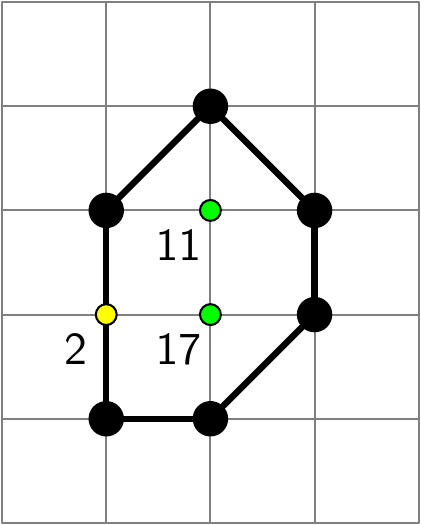}
}}\\ \hline
21 & \adjustbox{valign=m}{\includegraphics[width=0.9\linewidth]{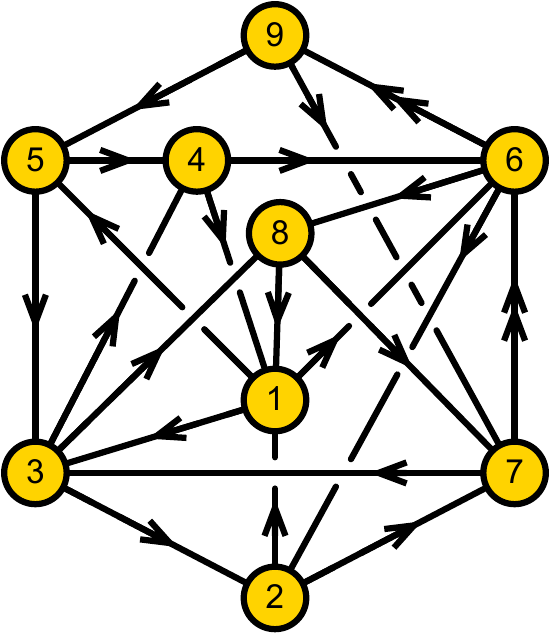}} & 
{\scriptsize $\begin{aligned}
W_{21} &= X_{13} X_{38} X_{81} + X_{15} X_{54} X_{41} + X_{16} X_{62} X_{21} \\
&\quad + X_{27} X_{73} X_{32} + X_{34} X_{46} X_{69}^{1} X_{95} X_{53} + X_{68} X_{87} X_{76}^{2} \\
&\quad + X_{69}^{2} X_{97} X_{76}^{1} - X_{13} X_{34} X_{41} - X_{15} X_{53} X_{32} X_{21} \\
&\quad - X_{16} X_{68} X_{81} - X_{27} X_{76}^{1} X_{62} - X_{38} X_{87} X_{73} \\
&\quad - X_{46} X_{69}^{2} X_{95} X_{54} - X_{69}^{1} X_{97} X_{76}^{2}
\end{aligned}$} & \resizebox{\linewidth}{!}{\adjustbox{valign=m}{
\includegraphics[width=\linewidth]{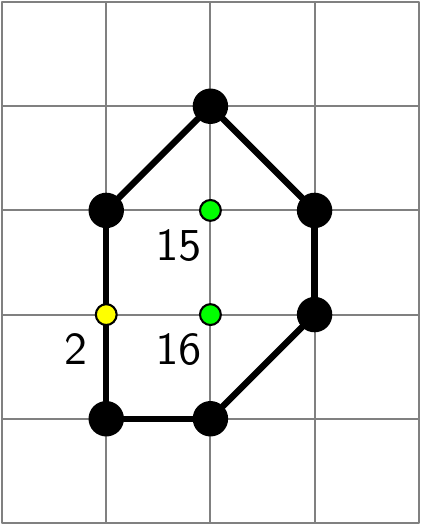}
}}\\ \hline
22 & \adjustbox{valign=m}{\includegraphics[width=0.9\linewidth]{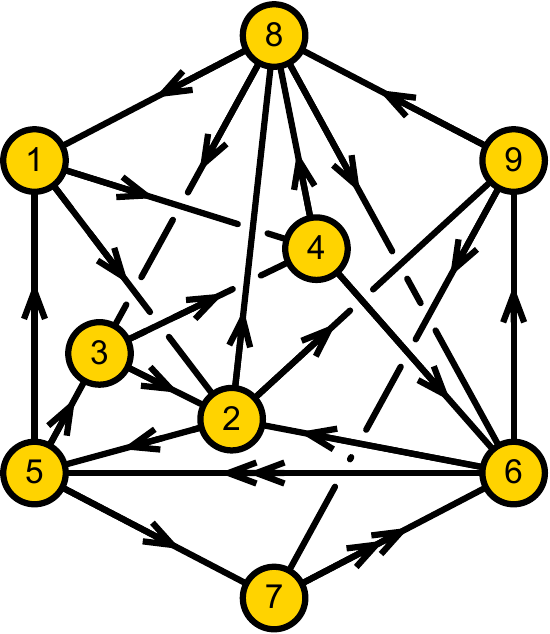}} &
{\scriptsize $\begin{aligned}
W_{22} &= X_{12} X_{25} X_{51} + X_{14} X_{48} X_{81} + X_{28} X_{86} X_{62} \\
&\quad + X_{29} X_{98} X_{83} X_{32} + X_{34} X_{46} X_{65}^{1} X_{53} + X_{57} X_{76}^{1} X_{65}^{2} \\
&\quad + X_{69} X_{97} X_{76}^{2} - X_{12} X_{28} X_{81} - X_{14} X_{46} X_{65}^{2} X_{51} \\
&\quad - X_{25} X_{53} X_{32} - X_{29} X_{97} X_{76}^{1} X_{62} - X_{34} X_{48} X_{83} \\
&\quad - X_{57} X_{76}^{2} X_{65}^{1} - X_{69} X_{98} X_{86} 
\end{aligned}$} & \resizebox{\linewidth}{!}{\adjustbox{valign=m}{
\includegraphics[width=\linewidth]{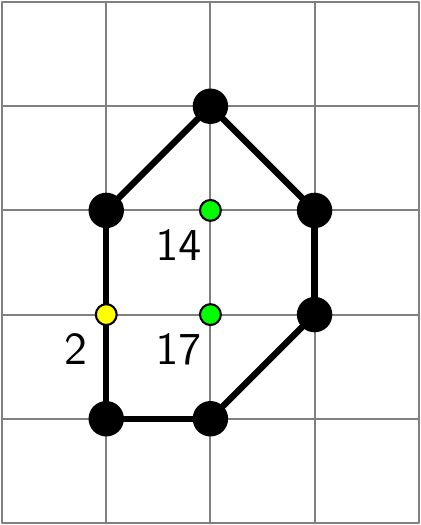}
}}\\ \hline
23 &\adjustbox{valign=m}{ \includegraphics[width=0.9\linewidth]{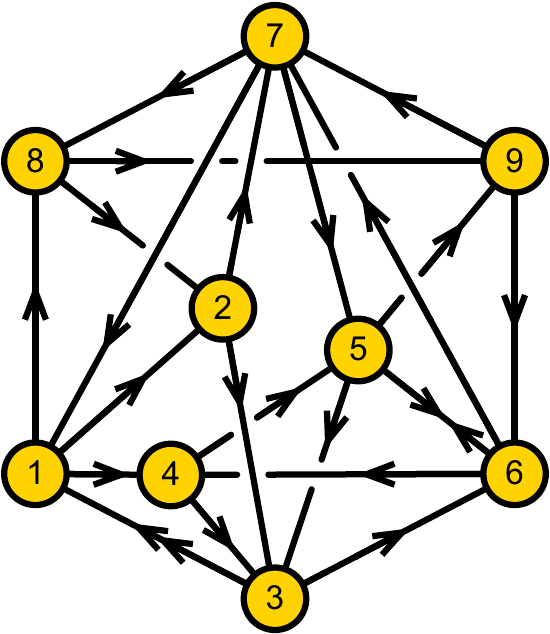}} & 
{\scriptsize $\begin{aligned}
W_{23} &= X_{12} X_{23} X_{31}^{2} + X_{14} X_{43} X_{31}^{1} + X_{18} X_{89} X_{96} X_{67} X_{71} \\
&\quad + X_{27} X_{78} X_{82} + X_{36} X_{65} X_{53} + X_{45} X_{56} X_{64} \\
&\quad + X_{59} X_{97} X_{75} - X_{12} X_{27} X_{71} - X_{14} X_{45} X_{53} X_{31}^{2} \\
&\quad - X_{18} X_{82} X_{23} X_{31}^{1} - X_{36} X_{64} X_{43} - X_{56} X_{67} X_{75} \\
&\quad - X_{59} X_{96} X_{65} - X_{78} X_{89} X_{97}
\end{aligned}$} & \resizebox{\linewidth}{!}{\adjustbox{valign=m}{
\includegraphics[width=\linewidth]{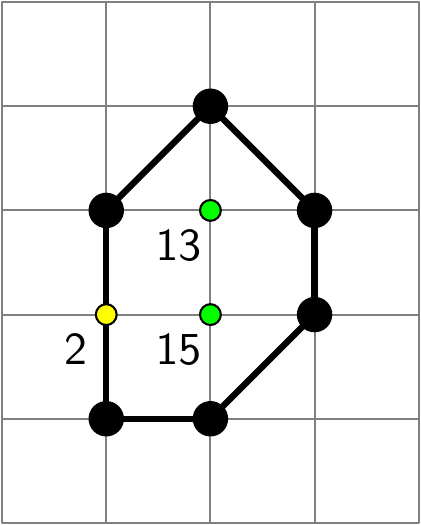}
}}\\ \hline
24 &\adjustbox{valign=m}{\includegraphics[width=0.9\linewidth]{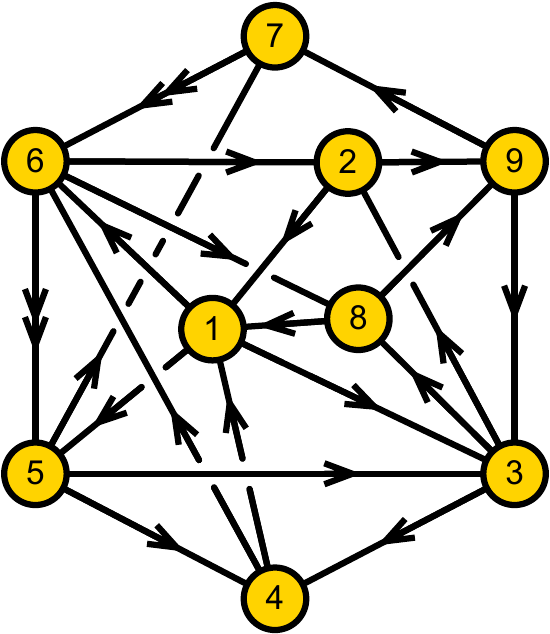}}&
{\scriptsize $\begin{aligned}
W_{24} &= X_{13} X_{38} X_{81} + X_{15} X_{54} X_{41} + X_{16} X_{62} X_{21} \\
&\quad + X_{29} X_{93} X_{32} + X_{34} X_{46} X_{65}^{1} X_{53} + X_{57} X_{76}^{1} X_{65}^{2} \\
&\quad + X_{68} X_{89} X_{97} X_{76}^{2} - X_{13} X_{34} X_{41} - X_{15} X_{53} X_{32} X_{21} \\
&\quad - X_{16} X_{68} X_{81} - X_{29} X_{97} X_{76}^{1} X_{62} - X_{38} X_{89} X_{93} \\
&\quad - X_{46} X_{65}^{2} X_{54} - X_{57} X_{76}^{2} X_{65}^{1}
\end{aligned}$} & \resizebox{\linewidth}{!}{\adjustbox{valign=m}{
\includegraphics[width=\linewidth]{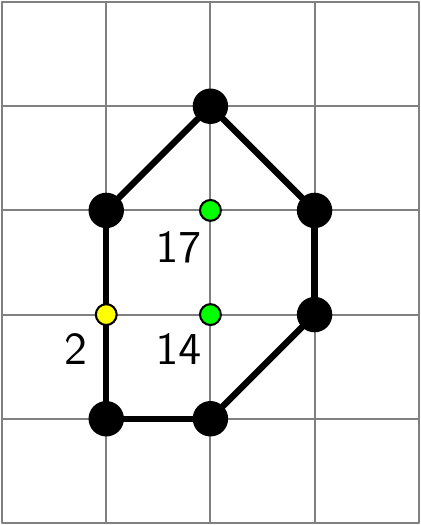}
}}\\ \hline
25 &\adjustbox{valign=m}{\includegraphics[width=0.9\linewidth]{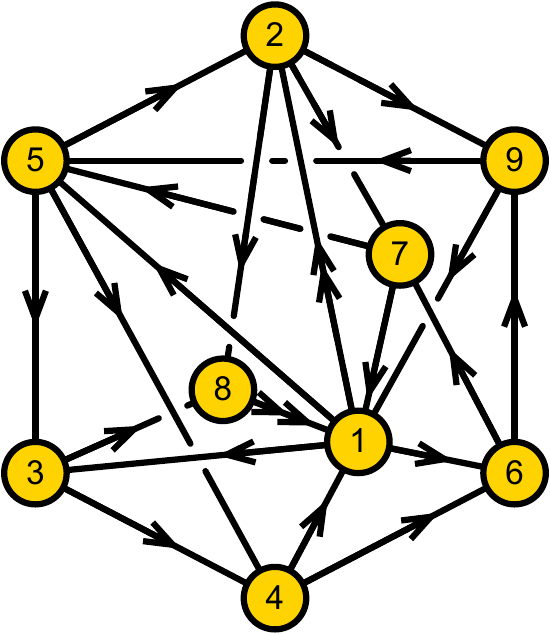}} & 
{\scriptsize $\begin{aligned}
W_{25} &= X_{12}^{1} X_{28} X_{81}^{2} + X_{12}^{2} X_{29} X_{91} + X_{13} X_{38} X_{81}^{1} \\
&\quad + X_{15} X_{54} X_{41} + X_{16} X_{67} X_{71} + X_{27} X_{75} X_{52} \\
&\quad + X_{34} X_{46} X_{69} X_{95} X_{53} - X_{12}^{1} X_{27} X_{71} - X_{12}^{2} X_{28} X_{81}^{1} \\
&\quad - X_{13} X_{34} X_{41} - X_{15} X_{53} X_{38} X_{81}^{2} - X_{16} X_{69} X_{91} \\
&\quad - X_{29} X_{95} X_{52} - X_{46} X_{67} X_{75} X_{54}
\end{aligned}$} & \resizebox{\linewidth}{!}{\adjustbox{valign=m}{
\includegraphics[width=\linewidth]{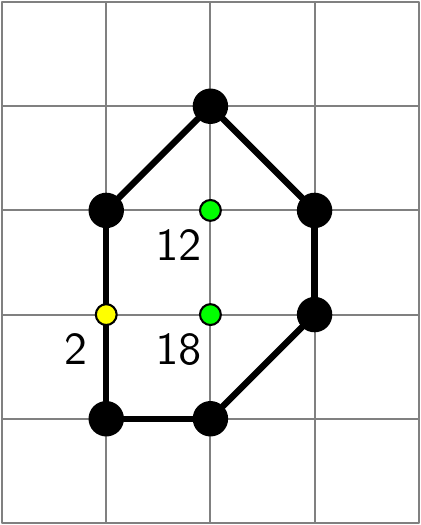}
}}\\ \hline

26 & \adjustbox{valign=m}{\includegraphics[width=0.9\linewidth]{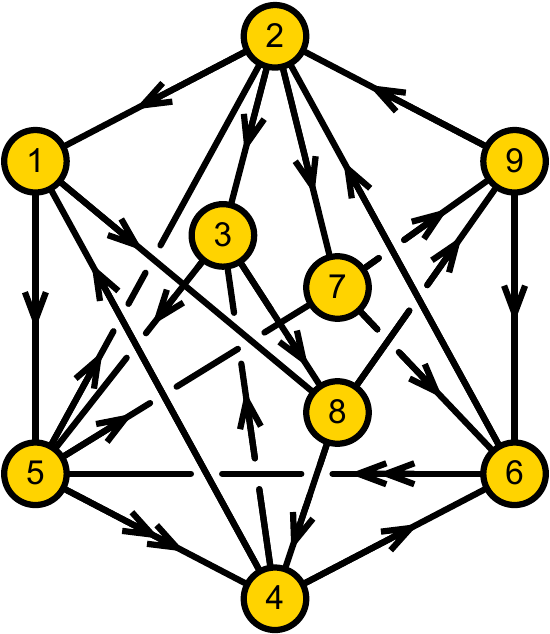} }& 
{\scriptsize $\begin{aligned}
W_{26} &= X_{15} X_{54}^{1} X_{41} + X_{18} X_{89} X_{96} X_{62} X_{21} + X_{23} X_{35} X_{52} \\
&\quad + X_{27} X_{79} X_{92} + X_{38} X_{84} X_{43} + X_{46} X_{65}^{1} X_{54}^{2} \\
&\quad + X_{57} X_{76} X_{65}^{2} - X_{15} X_{52} X_{21} - X_{18} X_{84} X_{41} \\
&\quad - X_{23} X_{38} X_{89} X_{92} - X_{27} X_{76} X_{62} - X_{35} X_{54}^{2} X_{43} \\
&\quad - X_{46} X_{65}^{2} X_{54}^{1} - X_{57} X_{79} X_{96} X_{65}^{1}
\end{aligned}$} & \resizebox{\linewidth}{!}{\adjustbox{valign=m}{
\includegraphics[width=\linewidth]{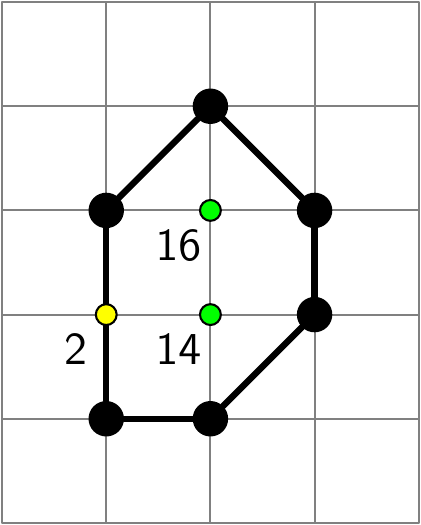}
}}\\ \hline
27 &\adjustbox{valign=m}{ \includegraphics[width=0.9\linewidth]{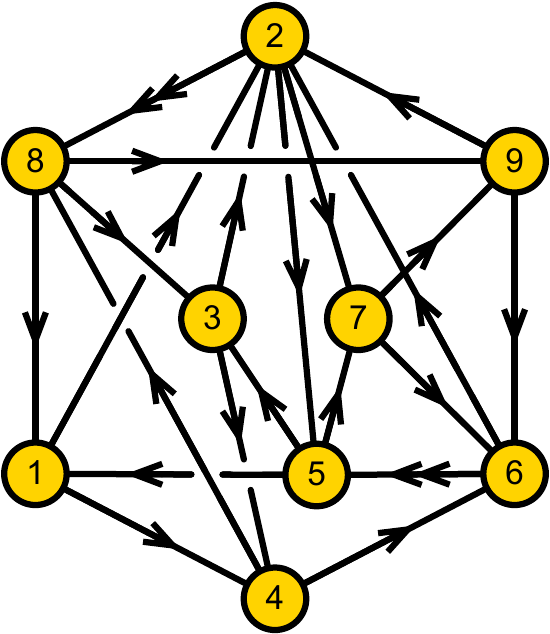}} &
{\scriptsize $\begin{aligned}
W_{27} &= X_{12} X_{25} X_{51} + X_{14} X_{48} X_{81} + X_{27} X_{79} X_{92} \\
&\quad + X_{28}^{2} X_{89} X_{96} X_{62} + X_{28}^{1} X_{83} X_{32} + X_{34} X_{46} X_{65}^{1} X_{53} \\
&\quad + X_{57} X_{76} X_{65}^{2} - X_{12} X_{28}^{2} X_{81} - X_{14} X_{46} X_{65}^{2} X_{51} \\
&\quad - X_{25} X_{53} X_{32} - X_{27} X_{76} X_{62} - X_{28}^{1} X_{89} X_{92} \\
&\quad - X_{34} X_{48} X_{83} - X_{57} X_{79} X_{96} X_{65}^{1}
\end{aligned}$} & \resizebox{\linewidth}{!}{\adjustbox{valign=m}{
\includegraphics[width=\linewidth]{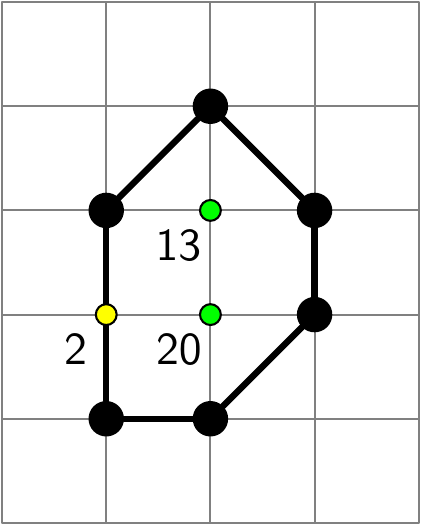}
}}\\ \hline
28 & \adjustbox{valign=m}{\includegraphics[width=0.9\linewidth]{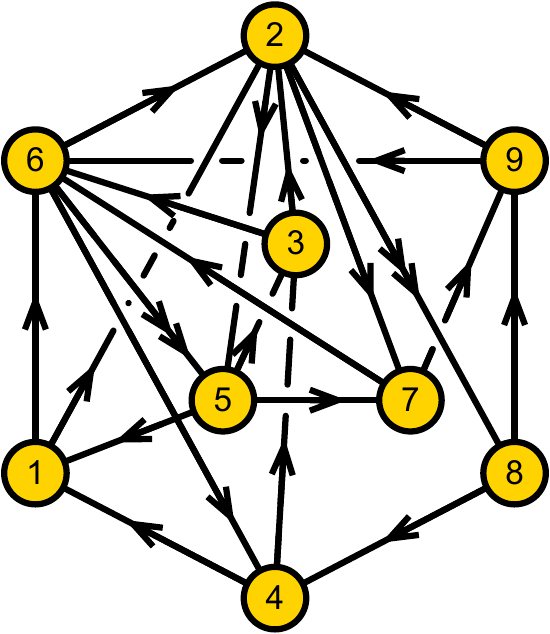} }& 
{\scriptsize $\begin{aligned}
W_{28} &= X_{12} X_{25} X_{51} + X_{16} X_{64} X_{41} + X_{27} X_{79} X_{92} \\
&\quad + X_{28}^{2} X_{89} X_{96} X_{62} + X_{28}^{1} X_{84} X_{43} X_{32} + X_{36} X_{65}^{1} X_{53} \\
&\quad + X_{57} X_{76} X_{65}^{2} - X_{12} X_{28}^{2} X_{84} X_{41} - X_{16} X_{65}^{2} X_{51} \\
&\quad - X_{25} X_{53} X_{32} - X_{27} X_{76} X_{62} - X_{28}^{1} X_{89} X_{92} \\
&\quad - X_{36} X_{64} X_{43} - X_{57} X_{79} X_{96} X_{65}^{1}
\end{aligned}$} & \resizebox{\linewidth}{!}{\adjustbox{valign=m}{
\includegraphics[width=\linewidth]{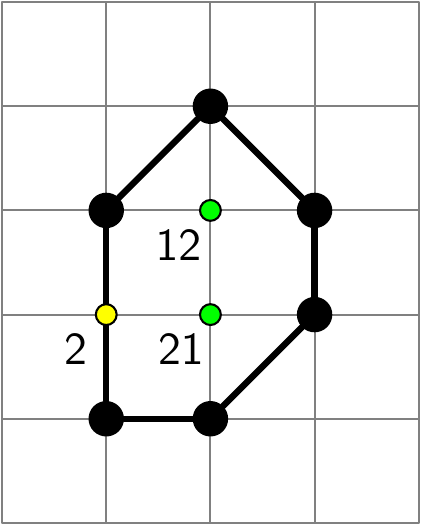}
}}\\ \hline
29 &\adjustbox{valign=m}{ \includegraphics[width=0.9\linewidth]{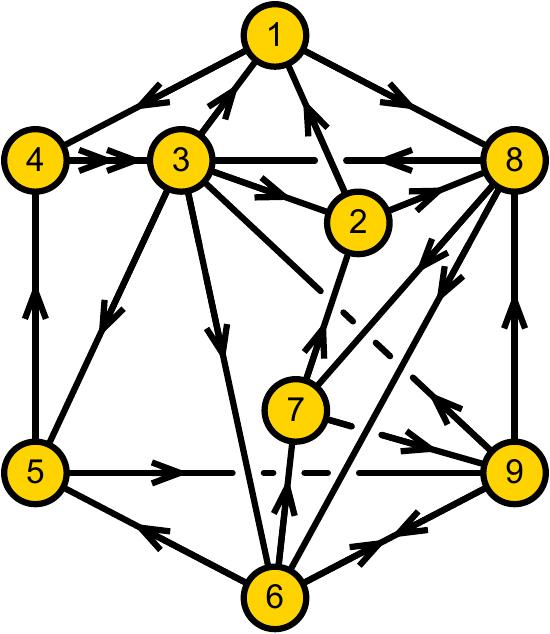} }&
{\scriptsize $\begin{aligned}
W_{29} &= X_{14} X_{43}^{1} X_{31} + X_{18} X_{86} X_{67} X_{72} X_{21} + X_{28} X_{83} X_{32} \\
&\quad + X_{35} X_{54} X_{43}^{2} + X_{36} X_{69} X_{93} + X_{59} X_{96} X_{65} \\
&\quad + X_{79} X_{98} X_{87} - X_{14} X_{43}^{2} X_{32} X_{21} - X_{18} X_{83} X_{31} \\
&\quad - X_{28} X_{87} X_{72} - X_{35} X_{59} X_{93} - X_{36} X_{65} X_{54} X_{43}^{1} \\
&\quad - X_{67} X_{79} X_{96} - X_{69} X_{98} X_{86} 
\end{aligned}$} & \resizebox{\linewidth}{!}{\adjustbox{valign=m}{
\includegraphics[width=\linewidth]{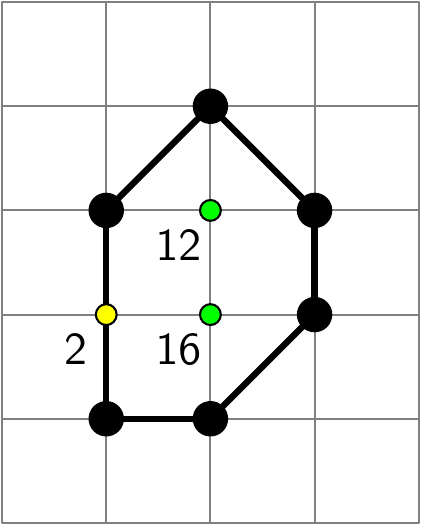}
}}\\ \hline
30 &\adjustbox{valign=m}{ \includegraphics[width=0.9\linewidth]{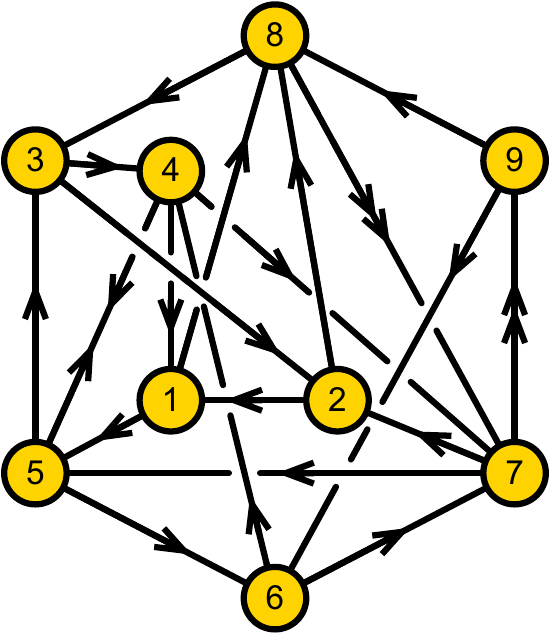} }&
{\scriptsize $\begin{aligned}
W_{30} &= X_{15} X_{54} X_{41} + X_{18} X_{87}^{2} X_{72} X_{21} + X_{28} X_{83} X_{32} \\
&\quad + X_{34} X_{45} X_{53} + X_{47} X_{79}^{2} X_{96} X_{64} + X_{56} X_{67} X_{75} \\
&\quad + X_{79}^{1} X_{98} X_{87}^{1} - X_{15} X_{53} X_{32} X_{21} - X_{18} X_{83} X_{34} X_{41} \\
&\quad - X_{28} X_{87}^{1} X_{72} - X_{45} X_{56} X_{64} - X_{47} X_{75} X_{54} \\
&\quad - X_{67} X_{79}^{1} X_{96} - X_{79}^{2} X_{98} X_{87}^{2} 
\end{aligned}$} & \resizebox{\linewidth}{!}{\adjustbox{valign=m}{
\includegraphics[width=\linewidth]{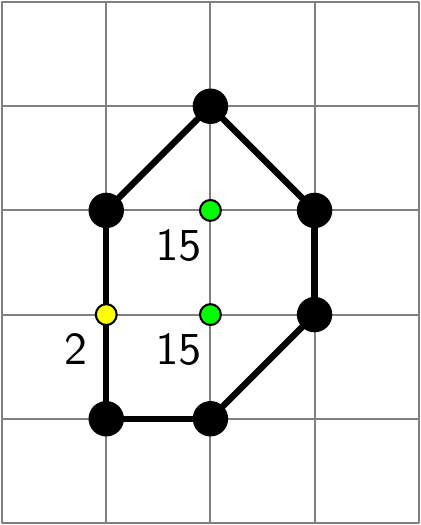}
}}\\ \hline
31 &\adjustbox{valign=m}{ \includegraphics[width=0.9\linewidth]{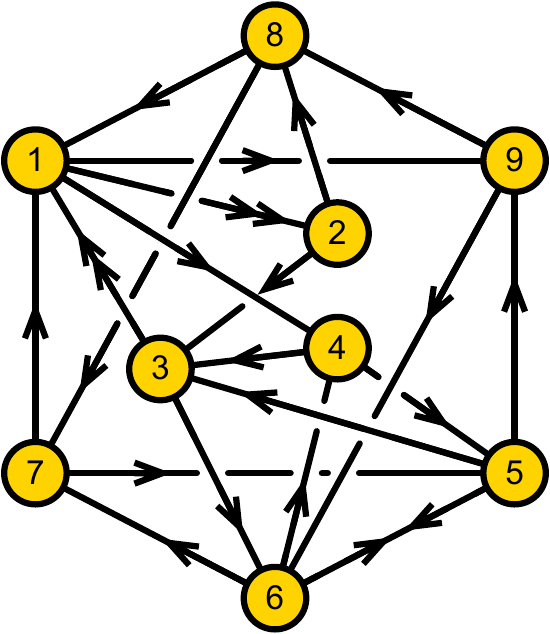} }& 
{\scriptsize $\begin{aligned}
W_{31} &= X_{12}^{2} X_{28} X_{81} + X_{12}^{1} X_{23} X_{31}^{2} + X_{14} X_{43} X_{31}^{1} \\
&\quad + X_{19} X_{96} X_{67} X_{71} + X_{36} X_{65} X_{53} + X_{45} X_{56} X_{64} \\
&\quad + X_{59} X_{98} X_{87} X_{75} - X_{12}^{1} X_{28} X_{87} X_{71} - X_{12}^{2} X_{23} X_{31}^{1} \\
&\quad - X_{14} X_{45} X_{53} X_{31}^{2} - X_{19} X_{98} X_{81} - X_{36} X_{64} X_{43} \\
&\quad - X_{56} X_{67} X_{75} - X_{59} X_{96} X_{65}
\end{aligned}$} & \resizebox{\linewidth}{!}{\adjustbox{valign=m}{
\includegraphics[width=\linewidth]{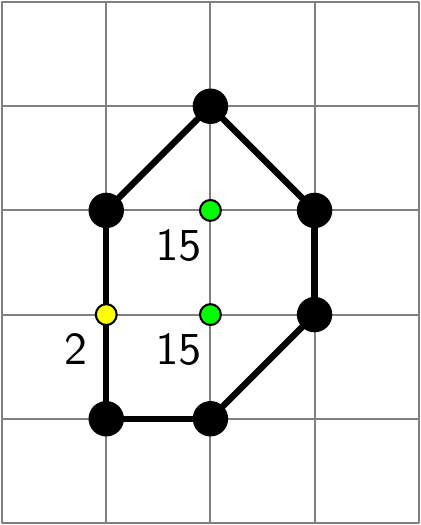}
}}\\ \hline
32 & \adjustbox{valign=m}{\includegraphics[width=0.9\linewidth]{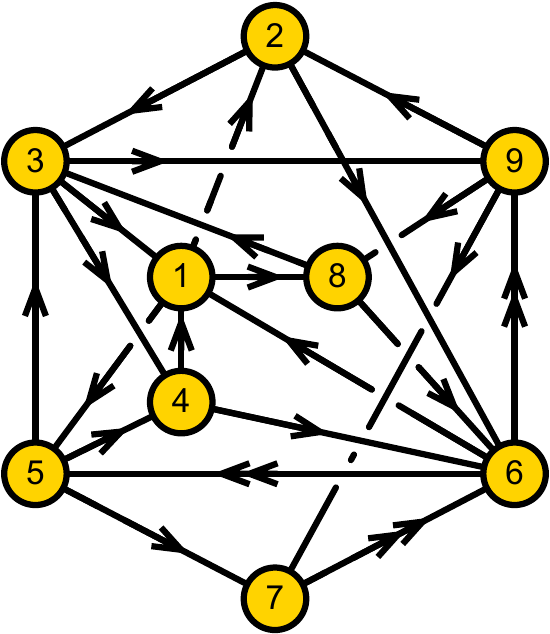} }&
{\scriptsize $\begin{aligned}
W_{32} &= X_{12} X_{23} X_{31} + X_{15} X_{54} X_{41} + X_{18} X_{86} X_{61} \\
&\quad + X_{26} X_{69}^{2} X_{92} + X_{34} X_{46} X_{65}^{1} X_{53} + X_{39} X_{98} X_{83} \\
&\quad + X_{57} X_{76}^{1} X_{65}^{2} + X_{69}^{1} X_{97} X_{76}^{2} - X_{12} X_{26} X_{61} \\
&\quad - X_{15} X_{53} X_{31} - X_{18} X_{83} X_{34} X_{41} - X_{23} X_{39} X_{92} \\
&\quad - X_{46} X_{65}^{2} X_{54} - X_{57} X_{76}^{2} X_{65}^{1} - X_{69}^{1} X_{98} X_{86} \\
&\quad - X_{69}^{2} X_{97} X_{76}^{1}
\end{aligned}$} & \resizebox{\linewidth}{!}{\adjustbox{valign=m}{
\includegraphics[width=\linewidth]{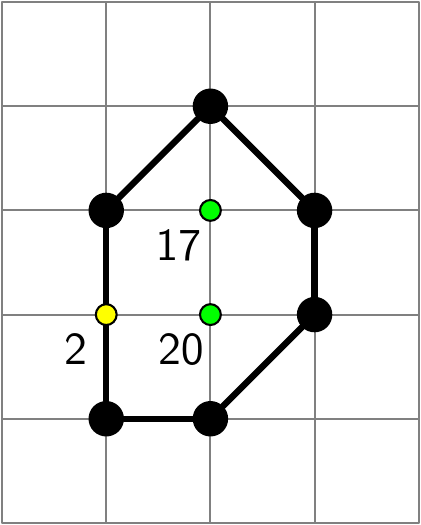}
}}\\ \hline

33 &\adjustbox{valign=m}{ \includegraphics[width=0.9\linewidth]{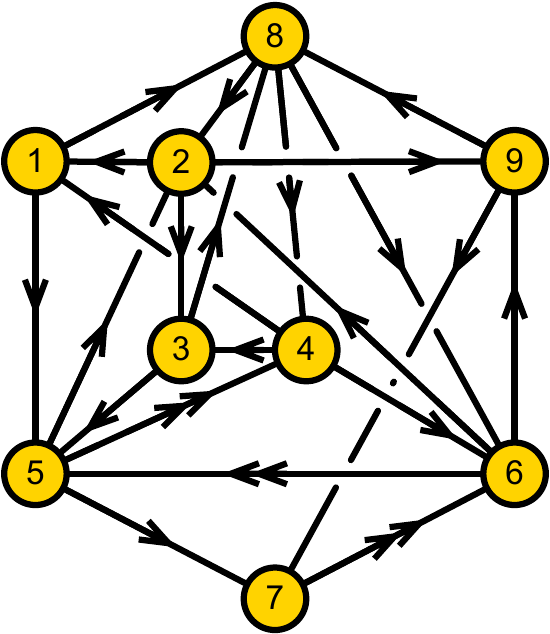} }&
{\scriptsize $\begin{aligned}
W_{33} &= X_{15} X_{54}^{1} X_{41} + X_{18} X_{86} X_{62} X_{21} + X_{23} X_{35} X_{52} \\
&\quad + X_{29} X_{98} X_{82} + X_{38} X_{84} X_{43} + X_{46} X_{65}^{1} X_{54}^{2} \\
&\quad + X_{57} X_{76}^{1} X_{65}^{2} + X_{69} X_{97} X_{76}^{2} - X_{15} X_{52} X_{21} \\
&\quad - X_{18} X_{84} X_{41} - X_{23} X_{38} X_{82} - X_{29} X_{97} X_{76}^{1} X_{62} \\
&\quad - X_{35} X_{54}^{2} X_{43} - X_{46} X_{65}^{2} X_{54}^{1} - X_{57} X_{76}^{2} X_{65}^{1} \\
&\quad - X_{69} X_{98} X_{86}
\end{aligned}$} & \resizebox{\linewidth}{!}{\adjustbox{valign=m}{
\includegraphics[width=\linewidth]{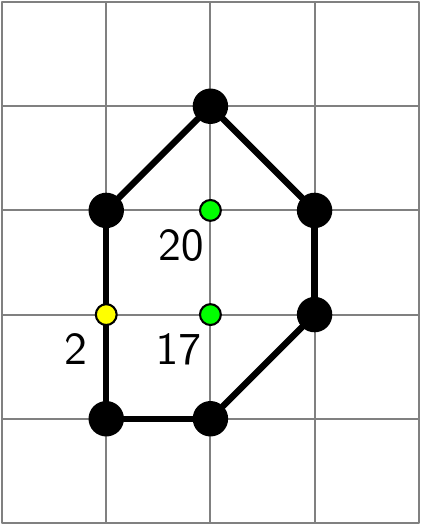}
}}\\ \hline
34 &\adjustbox{valign=m}{ \includegraphics[width=0.9\linewidth]{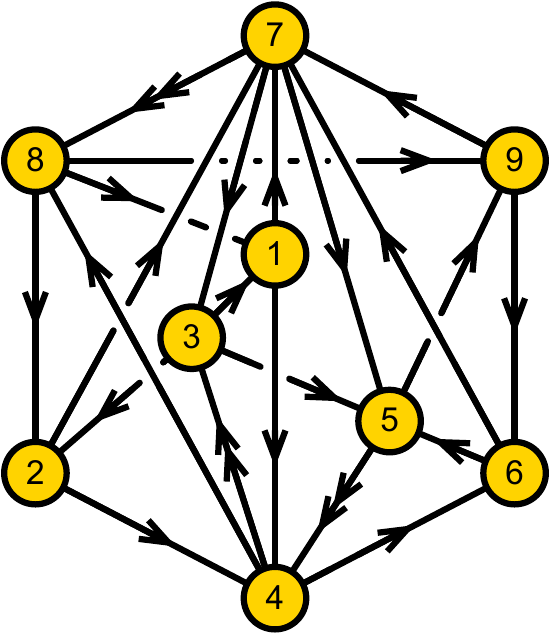} }&
{\scriptsize $\begin{aligned}
W_{34} &= X_{14} X_{48} X_{81} + X_{17} X_{73} X_{31} + X_{24} X_{43}^{1} X_{32} \\
&\quad + X_{27} X_{78}^{1} X_{82} + X_{35} X_{54}^{1} X_{43}^{2} + X_{46} X_{65} X_{54}^{2} \\
&\quad + X_{59} X_{97} X_{75} + X_{67} X_{78}^{2} X_{89} X_{96} - X_{14} X_{43}^{2} X_{31} \\
&\quad - X_{17} X_{78}^{2} X_{81} - X_{24} X_{48} X_{82} - X_{27} X_{73} X_{32} \\
&\quad - X_{35} X_{54}^{2} X_{43}^{1} - X_{46} X_{67} X_{75} X_{54}^{1} - X_{59} X_{96} X_{65} \\
&\quad - X_{78}^{1} X_{89} X_{97}
\end{aligned}$} & \resizebox{\linewidth}{!}{\adjustbox{valign=m}{
\includegraphics[width=\linewidth]{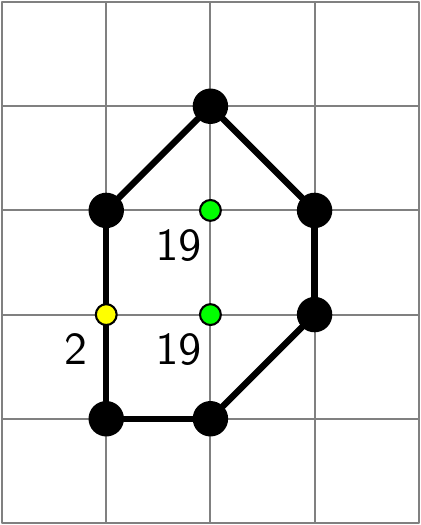}
}}\\ \hline
35 &\adjustbox{valign=m}{ \includegraphics[width=0.9\linewidth]{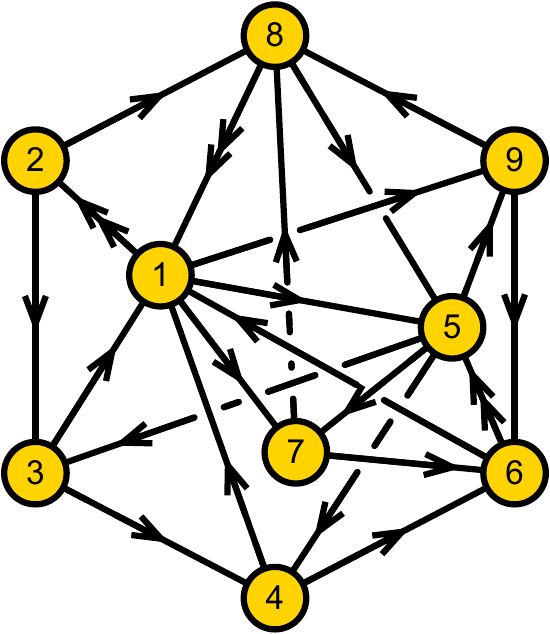} }&
{\scriptsize $\begin{aligned}
W_{35} &= X_{12}^{2} X_{28} X_{81}^{1} + X_{12}^{1} X_{23} X_{31} + X_{15} X_{54} X_{41} \\
&\quad + X_{17} X_{78} X_{81}^{2} + X_{19} X_{96} X_{61} + X_{34} X_{46} X_{65}^{1} X_{53} \\
&\quad + X_{57} X_{76} X_{65}^{2} + X_{59} X_{98} X_{85} - X_{12}^{1} X_{28} X_{81}^{2} \\
&\quad - X_{12}^{2} X_{23} X_{34} X_{41} - X_{15} X_{53} X_{31} - X_{17} X_{76} X_{61} \\
&\quad - X_{19} X_{98} X_{81}^{1} - X_{46} X_{65}^{2} X_{54} - X_{57} X_{78} X_{85} \\
&\quad - X_{59} X_{96} X_{65}^{1}
\end{aligned}$} & \resizebox{\linewidth}{!}{\adjustbox{valign=m}{
\includegraphics[width=\linewidth]{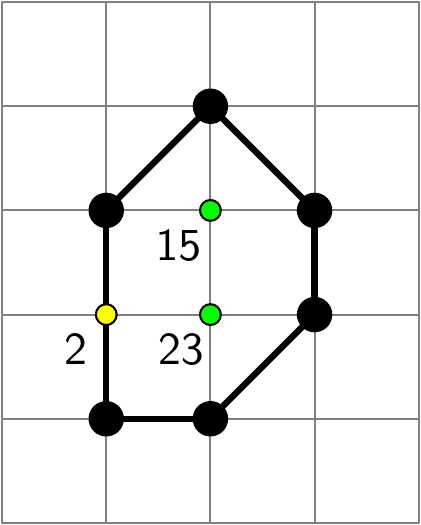}
}}\\ \hline
\end{longtable}
%---------------------------------------------------- 

%======================================================================
\bibliographystyle{JHEP}
\bibliography{mybib}
%======================================================================

\end{document}